\documentclass{article}
\usepackage{ijcai24}

\usepackage{times}
\usepackage{soul}
\usepackage{url}
\usepackage[hidelinks]{hyperref}
\usepackage[utf8]{inputenc}
\usepackage[small]{caption}
\usepackage{graphicx}
\usepackage{amsmath}
\usepackage{amsthm}
\usepackage{amssymb}
\usepackage{booktabs}
\usepackage{algorithm}
\usepackage{cleveref}
\usepackage{color}
\usepackage{nicefrac}
\usepackage{siunitx}
\usepackage{longtable}

\usepackage{bibunits}

\newcommand{\citet}[1]{\citeauthor{#1}~\shortcite{#1}}
\newcommand{\citep}[1]{\cite{#1}}
\newcommand{\citeapp}[1]{\cite{#1}}

\newcommand{\normphi}{\mathrm{norm}\hbox{-}\phi}

\usepackage{todonotes}
\newcommand{\ssnote}[1]{}
\newcommand{\nb}[1]{}

\newcommand{\tocheckbox}[1]{#1}

\usepackage{algorithmic}

\usepackage[utf8]{inputenc}

\usepackage{subcaption}

\usepackage{booktabs}

\newcommand{\omitbecauseappendix}[1]{}

\newtheorem{remark}{Remark}[section]

\usepackage[T1]{fontenc}

\title{Guide to Numerical Experiments on Elections in Computational Social Choice}

\author{
Niclas Boehmer$^1$ \and
Piotr Faliszewski$^2$ \and
Łukasz Janeczko$^2$ \and
Andrzej Kaczmarczyk$^2$ \and \\
Grzegorz Lisowski$^2$ \and
Grzegorz Pierczyński$^2$ \and
Simon Rey$^3$ \and
Dariusz Stolicki$^4$ \and \\
Stanisław Szufa$^{2,5}$ \and
Tomasz W\k{a}s$^5$
\affiliations
$^1$Harvard University,
$^2$AGH University,
$^3$ILLC, University of Amsterdam\\
$^4$Jagiellonian University,
$^5$CNRS, LAMSADE, Université Paris Dauphine-PSL
\emails
nboehmer@g.harvard.edu, 
\{faliszew, andrzej.kaczmarczyk, ljaneczk, glisowski, szufa\}@agh.edu.pl, s.j.rey@uva.nl, dariusz.stolicki@uj.edu.pl, tomasz.was@dauphine.psl.eu
}

\begin{document}
\begin{bibunit}

\maketitle

\begin{abstract}
  We analyze how numerical experiments regarding elections were
  conducted within the computational social choice literature
  (focusing on papers published in the IJCAI, AAAI, and AAMAS
  conferences). We analyze the sizes of the studied elections and the
  methods used for generating preference data, thereby making previously
  hidden standards and practices explicit.  In particular, we survey a
  number of statistical cultures for generating elections and their
  commonly used parameters.
\end{abstract}

\section{Introduction}

Computational social choice is an interdisciplinary area that draws on
artificial intelligence, computer science theory, economics,
operations research, logic, social sciences, and many other
fields~\cite{bra-con-end-lan-pro:b:comsoc-handbook}. Its main goal is the
algorithmic analysis of collective decision making processes, but
over time noncomputational approaches, such as the axiomatic method or
game-theoretic considerations, have also become popular and are
pursued equally vigorously.
Up to a few years ago, results in computational social choice  were
largely theoretical and only recently numerical experiments---not to
mention actual empirical studies---have received more notable
attention.  In this survey,
our goal 
is to encourage further experimental studies on elections and voting,
a prominent subarea of computational social
choice, by presenting a \emph{Guide}. 
Our Guide has two main components:
\begin{enumerate}
\item On the one hand, the Guide surveys how experiments
  were performed so far, what election sizes were considered, how data
  was obtained, and what parameters were considered. Such information
  is helpful when planning one's own experiments, e.g., to stay in
  sync with the literature.  In this sense, the paper is akin to a
  \emph{tourist guide}, which shows the richness of the landscape that one
  would see, e.g., upon visiting a city.
\item On the other hand, we aim to point to good practices and make
  recommendations as to how experiments should be run. While each
  experiment is different and requires specific considerations, there
  are also general rules of thumb that one might want to follow (such
  as using at least several data sources, which in the past has often
  been neglected). In this sense, our guide takes a role of a
  \emph{``how to''} document, giving some specific advice.
\end{enumerate}
To achieve these goals, we have gone over all papers published in the
AAAI, IJCAI, and AAMAS conference series between \tocheckbox{2010} and \tocheckbox{2023} and
collected those that discuss elections and voting (or some very
similar structures; see \Cref{sec:data} for details on the collection
process).

For each of the collected papers, we have analyzed how the authors
obtained preference data for their experiments, which statistical
cultures (i.e., models of generating synthetic data) they used and with
which parameters, and what election sizes they considered. A large
part of the survey is discussing the conclusions from this
analysis. This includes providing general statistics (such as the
number of papers that include experiments in various years, or the
number of data sources used by papers) and an overview of popular
statistical cultures.  We contrast these observations with the
\emph{map of elections}, as introduced by
\citet{szu-fal-sko-sli-tal:c:map-of-elections} and
\citet{boe-bre-fal-nie-szu:c:compass}, which shows relations between
various statistical cultures and real-life data sets, as well as with
the \emph{microscope} of \citet{fal-kac-sor-szu-was:c:div-agr-pol-map},
which visualizes specific elections (and, effectively, specific
synthetic models). We use these tools to give some advice as to which
statistical models are possibly more appealing than others.

We complement our work by providing a Python package with
implementations of the most popular models for sampling approval and
ordinal elections \url{https://github.com/COMSOC-Community/prefsampling} and
a website with access to our database of papers
\url{https://guide.cbip.matinf.uj.edu.pl/}.
Due to limited space, we mostly focus on  ordinal elections.

\section{Collecting Data}\label{sec:data}

We have collected all papers that were published in the AAAI,
IJCAI, and AAMAS conference series between \tocheckbox{2010} and \tocheckbox{2023} (in case of
IJCAI we have also included the papers starting from \tocheckbox{2009}). 
For the Guide, we selected papers that contained experiments on elections
(or very related structures).

By an \emph{election}, we mean a pair $E = (C,V)$, where
$C = \{c_1, \ldots, c_m\}$ is a set of candidates and
$V = (v_1, \ldots, v_n)$ is a sequence of voters that express
preferences over these candidates. 
In an \emph{ordinal election} each voter $v_i$ has a preference order,
i.e., a strict ranking $\succ_{v_i}$ of the candidates, from the one
that $v_i$ likes most to the one that he or she likes least. In an
\emph{approval election}, each voter $v_i$ has a set
$A(v_i) \subseteq C$ of candidates that he or she approves.
Occasionally, authors consider variants of elections where, for
example, the preference orders are either weak or partial, or are
expressed over some combinatorial domain (e.g., see the literature on
CP-nets~\citep{lan-xia:b:combinatorial-domains}). We include papers
that study such elections as well.

We restrict our attention to papers that include elections with at
least three candidates. Indeed, two-candidate
elections are very different from those with at least three.\footnote{Naturally, we include papers that
  consider two candidates as a special case, in addition to larger
  candidate sets.}
As a consequence, we do not include numerous papers that study, e.g.,
a setting where two parties compete (as, e.g., the work of
{\citet{bor-lev-sha-str:c:big-city-great-outdoors}}) or which are
motivated by presidential elections with two candidates (as, e.g.,
the paper of {\citet{wil-vor:c:president-spread-misinformation}}), or
which focus on liquid democracy and voting over two options (as
examples, see the works of {\citet{col-del-gilc:ld-apriori-power} and
  \citet{blo-gro-lac:c:ld-rational-delegations}}).

Occasionally we ran into gray areas and bent (or not) our rules on an
individual basis.\footnote{For example, we did not include the work of
  \citet{pet-pie-sha-sko:c:market-based} in the Guide as in the
  conference versions the authors mention conclusions from
  experiments, but do not describe their details.}
We believe that most readers would agree with most of our
choices. We list and cite all the \tocheckbox{163} papers that we included in the
Guide, together with meta-data about their experiments, in the full
version of the paper.
\omitbecauseappendix{We include the full list of papers from the Guide
  in~\Cref{sec:papers}.}

\paragraph{Collecting Papers.}
We have downloaded the papers from the respective conferences in
September \tocheckbox{2023}, using the links from the DBLP
website.\footnote{Source:
  {https://dblp.org/xml/release/dblp-2023-09-01.xml.gz}}
This way we included all tracks of the conferences, including, e.g.,
demo or doctoral consortium papers, etc. We skipped \tocheckbox{$34$~papers}, whose links were
missing or were corrupted and which could not be downloaded manually from any
official source.
Then, we performed an automated screening to select a long
list of papers that might contain experimental studies of
elections. Specifically, for each paper
we checked whether it included keywords
related to elections and experiments (the keywords were
\texttt{election}, \texttt{vote}, and \texttt{ballot} for elections,
and \texttt{experiment}, \texttt{simulation}, and \texttt{empirical}
for the experiments; to pass the screening, a paper had to include
words from both groups, on at least two pages).  We looked at each
paper that passed the keyword-based screening and checked if it indeed
regarded elections and included experiments.  While our sets of
keywords were selected to limit the number of papers that we had to
analyze manually, they were also meant to not be very restrictive. For
example, \tocheckbox{IJCAI-2023} included \tocheckbox{846} papers of which \tocheckbox{41} passed the initial
screening, but only~\tocheckbox{7} passed manual checking and made it to the Guide.
\omitbecauseappendix{(we discuss some reasons for
 papers not passing the screening in \Cref{sec:false-positives}).}

\paragraph{Recording Experiments.}
Finally, we have
analyzed the experiments that the collected papers included. For each
experiment, we recorded the type of elections used (ordinal or
approval), how the votes were obtained (e.g., if they were generated
from some statistical culture or were based on real-life data), the
sizes of the considered elections (expressed as numbers of candidates
and voters), and the number of samples used to obtain each ``data
point'' (the notion of a data point is paper specific; in most cases
it meant the number of elections generated for each datapoint on some
plot). For each of these parameters we recorded additional notes, if
we felt that some further comments would be helpful.

\begin{remark}
  Authors often consider elections where some parameter---such as the
  number of voters---changes with a particular step (e.g.,  $20$ and $100$ voters, with a step of $5$).  In
  such cases, we recorded the range of election sizes considered, but
  omitted the step parameter. Indeed, we felt that availability of
  such data would not affect our analysis too strongly, but would
  hinder data collection.
\end{remark}

\begin{figure*}
    \centering
    \begin{subfigure}{0.3\textwidth}
        \centering
        \includegraphics[width=5cm]{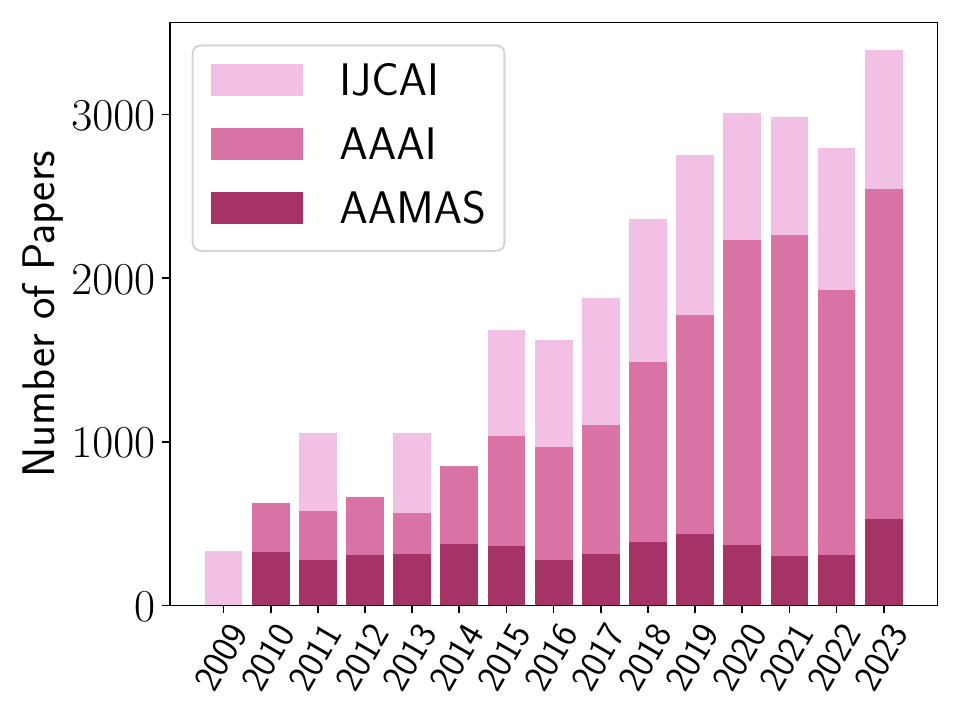}
        \caption{No.\@ of papers from the considered
          conference series downloaded for the Guide.}
        \label{fig:confs}
      \end{subfigure}\quad          
    \begin{subfigure}{0.3\textwidth}
        \centering
        \includegraphics[width=5cm]{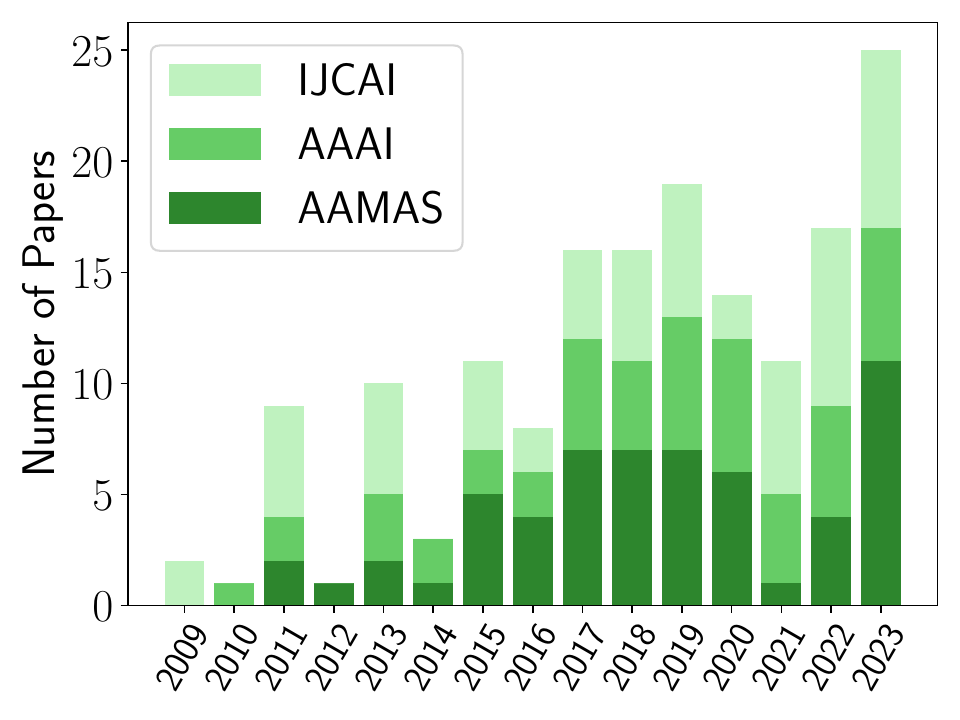}
        \caption{No.\@ of papers in the Guide from AAAI,
          IJCAI, and AAMAS conference series.}
        \label{fig:confyear}
      \end{subfigure}\quad      
    \begin{subfigure}{0.3\textwidth}
        \centering
        \includegraphics[width=5cm]{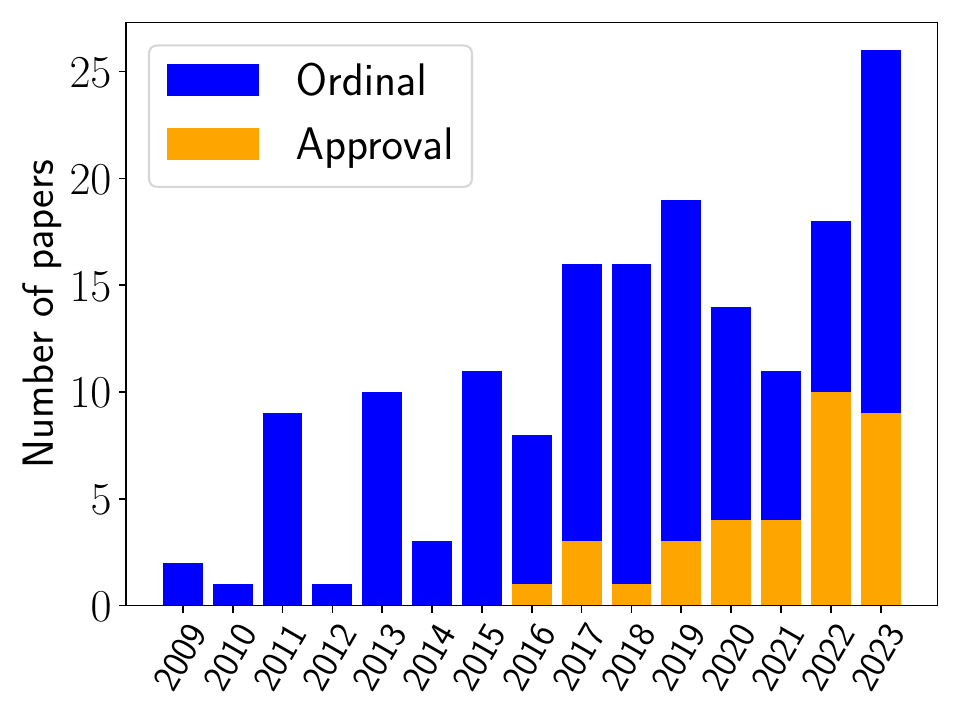}
        \caption{No.\@ of papers in the Guide that consider either
          ordinal or approval elections.}
        \label{fig:appord}
    \end{subfigure}\quad      
    \begin{subfigure}{0.3\textwidth}
        \centering
    \end{subfigure}
    
    \caption{Statistics regarding the numbers of papers in the Guide.}
    \label{fig:overall}
\end{figure*}

We should stress that our notion of what counts as \emph{one}
experiment is quite distinct.  For example, if some hypothetical paper
described two ``experiments,'' where in the former it considered the
running time of some algorithm and in the latter it analyzed whether
some property is satisfied, but it used the same (or, identically
generated) data for both, then we would have recorded this as a single
experiment. Similarly, if a paper included a single ``experiment,''
such as, e.g., testing manipulability of some voting rule, but within
this ``experiment'' it first focused on a particular statistical
culture and a range of election sizes, and then it moved to a
different culture and a different range of sizes, then we would record
this as two experiments.

\section{Bird's Eye View of The Guide}

\begin{figure}[t]
    \centering

    \begin{subfigure}{0.23\textwidth}
        \centering
        \includegraphics[width=4.2cm]{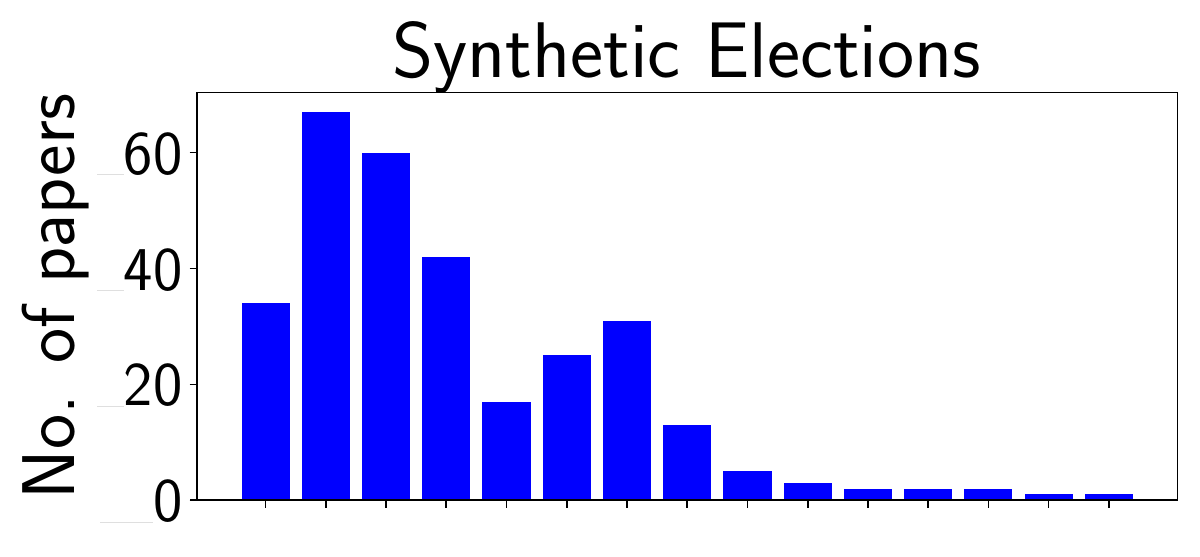}
      \end{subfigure}\quad      
    \begin{subfigure}{0.23\textwidth}
        \centering
        \includegraphics[width=4.2cm]{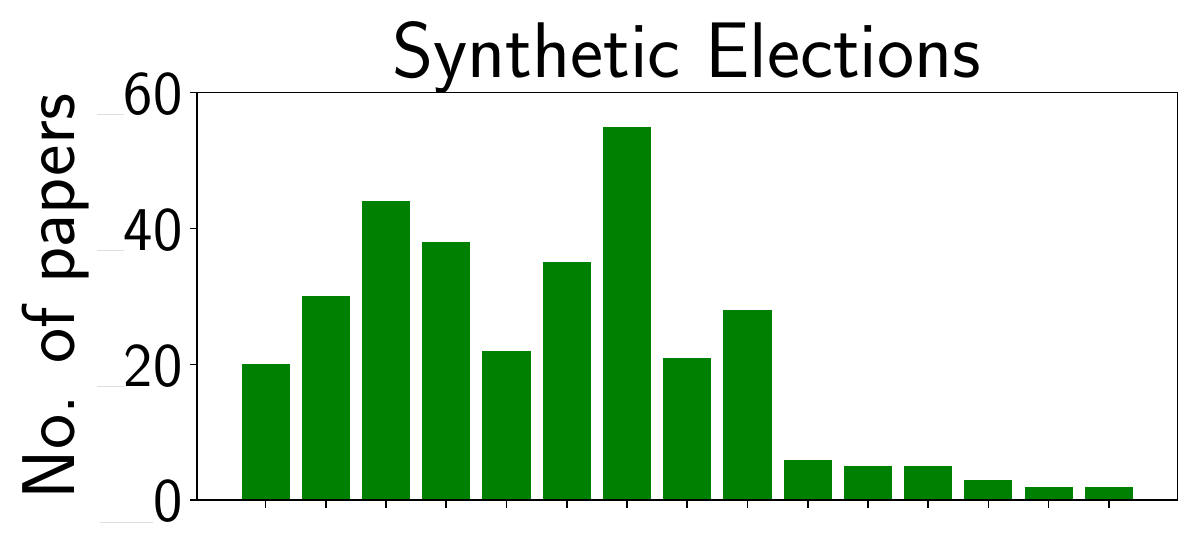}
    \end{subfigure}\quad

    \begin{subfigure}{0.23\textwidth}
        \centering
        \includegraphics[width=4.2cm]{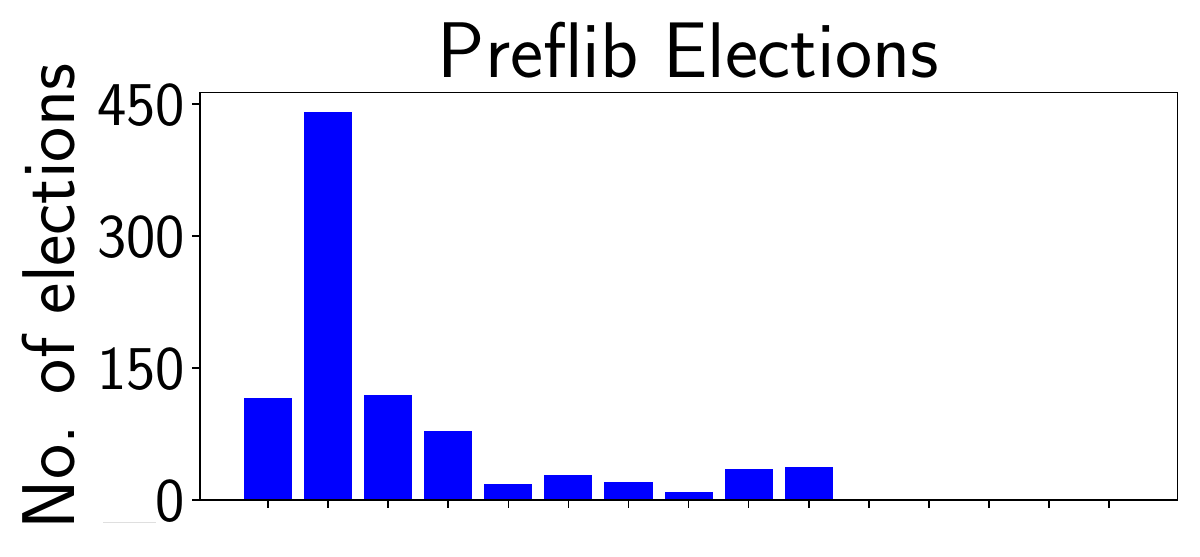}
      \end{subfigure}\quad      
    \begin{subfigure}{0.23\textwidth}
        \centering
        \includegraphics[width=4.2cm]{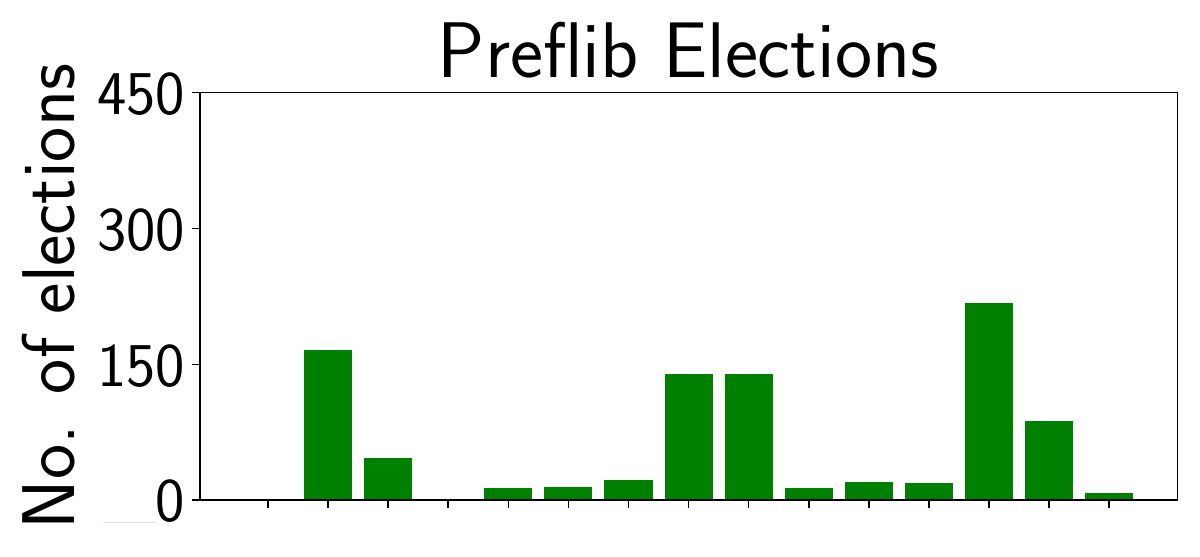}
    \end{subfigure}\quad      

    \begin{subfigure}{0.23\textwidth}
        \centering
        \includegraphics[width=4.2cm]{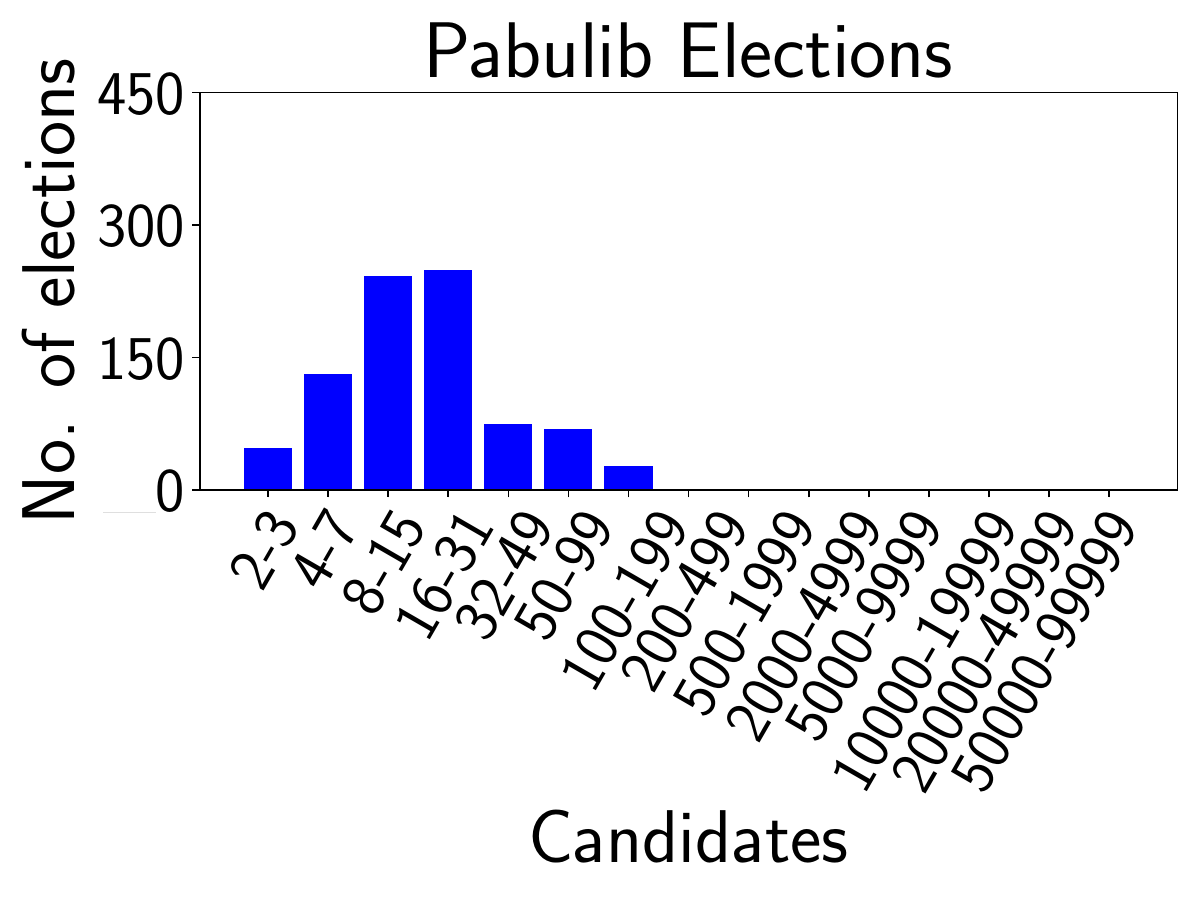}
      \end{subfigure}\quad      
    \begin{subfigure}{0.23\textwidth}
        \centering
        \includegraphics[width=4.2cm]{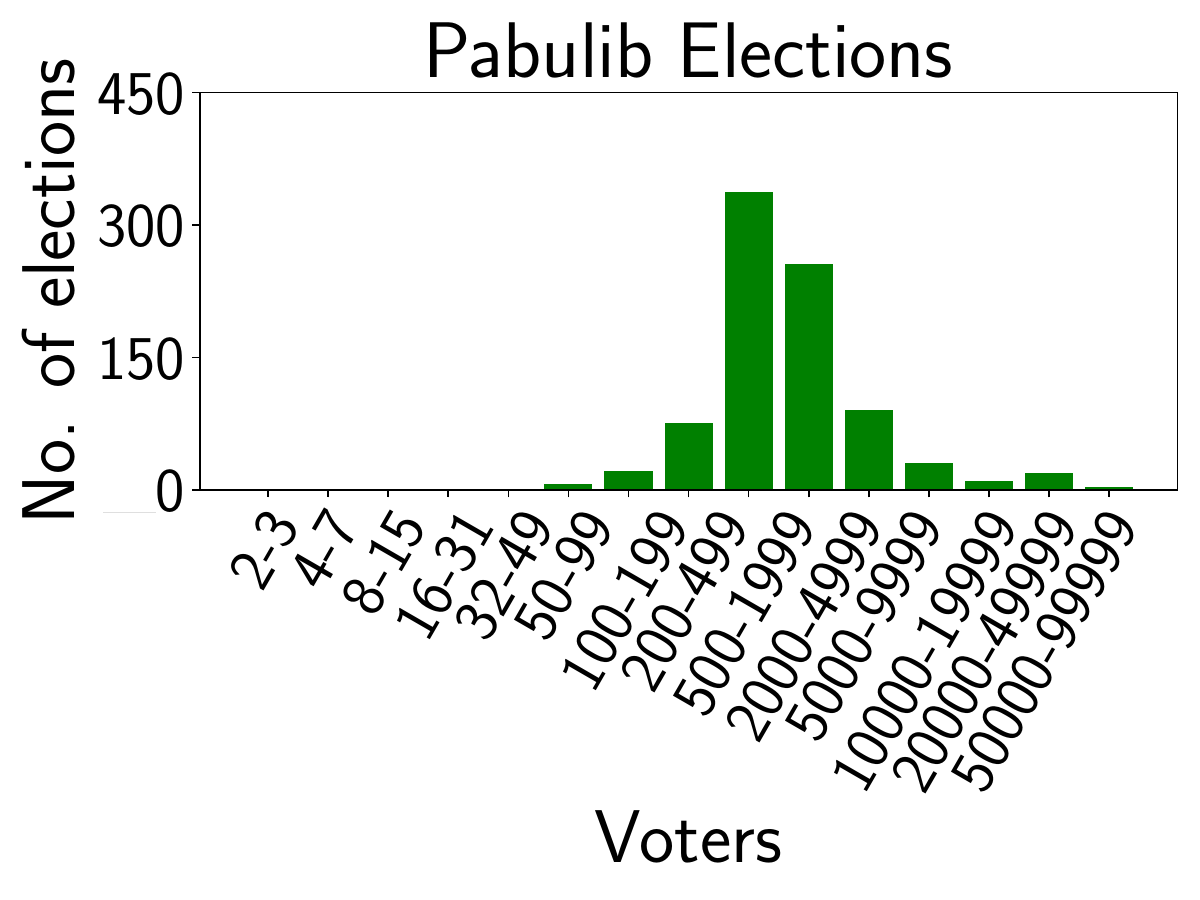}
    \end{subfigure}
    \caption{Histograms of the numbers of candidates and voters of
      synthetic elections used in the papers from the Guide (top),
      and in Preflib (middle) and Pabulib (bottom).}
    \label{fig:size-histogram}
\end{figure}

\begin{figure*}[t]
  \centering
    \noindent
    \begin{subfigure}{0.33\textwidth}
        \centering
        \includegraphics[width=6.53cm]{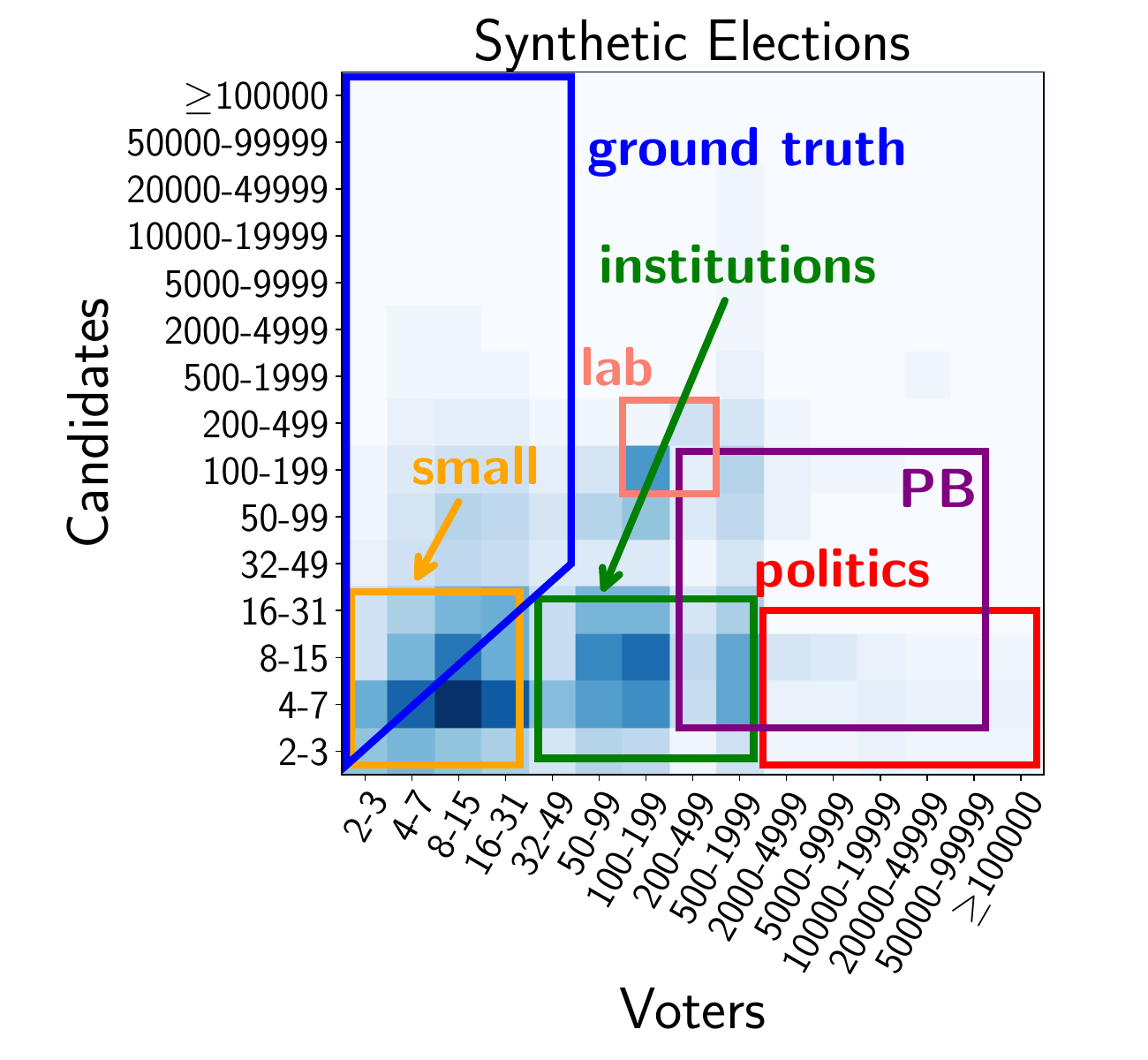}
        \label{fig:cv-heatmap}
    \end{subfigure}
    \begin{subfigure}{0.33\textwidth}
        \centering
        \includegraphics[width=6.53cm]{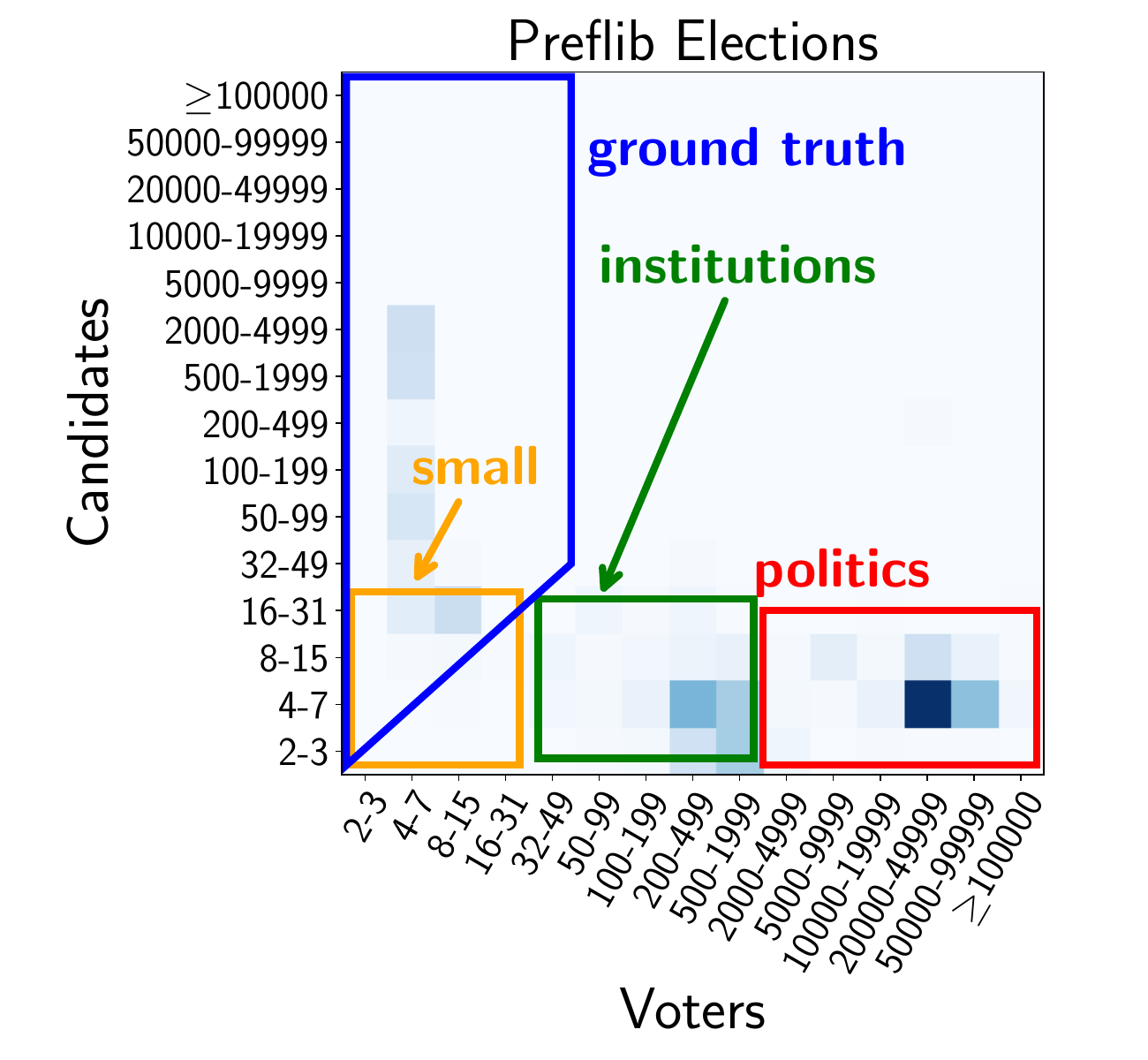}
        \label{fig:preflib-cv-heatmap}
    \end{subfigure}
    \noindent
    \begin{subfigure}{0.33\textwidth}
      \centering
        \includegraphics[width=6.53cm]{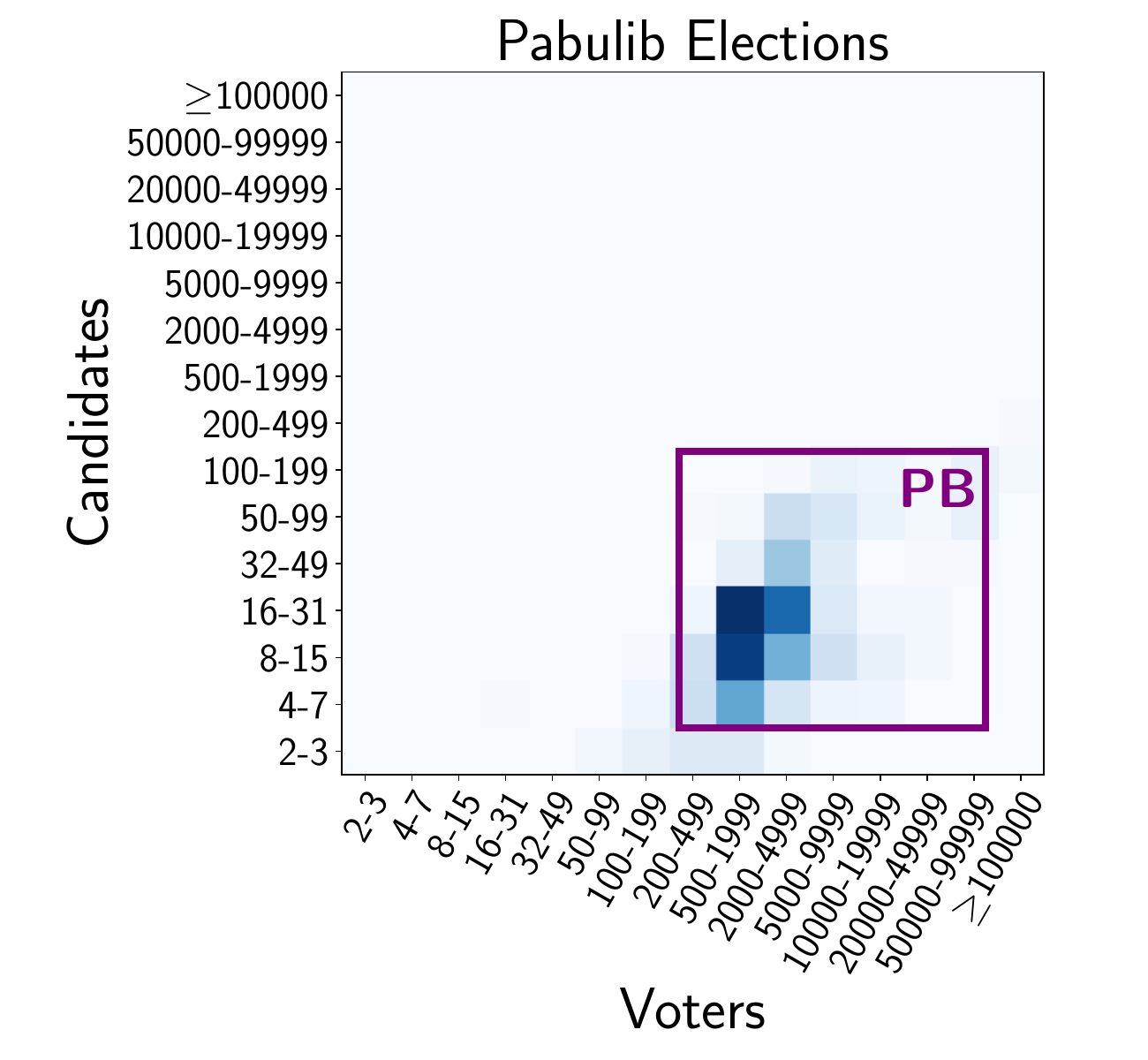}
        \label{fig:pabulib-cv-heatmap}
    \end{subfigure}
    
    \vspace{-0.75cm}
    
    \caption{Heatmaps of the sizes of synthetic elections used in the
      papers from the Guide (left), real-life elections from Preflib
      (middle), and real-life elections from Pabulib (right). Preflib
      plot omits the elections provided by
      \protect\citet{boe-sch:c:real-world-ranking-data} (including
      them would create an overwhelming spike in the area for $8$-$31$ voters and $100$-$499$ candidates).  Darker cells mean
      more papers with elections of a given size.}
    \label{fig:size-heatmap}
\end{figure*}

In this section we present some statistics regarding the papers in the
Guide and the elections that they consider. 

\subsection{Number of Papers}
At the time of writing this survey, the Guide included \tocheckbox{163} papers (we
intend to continue our work and keep collecting papers from future
years and, hopefully, further sources). In \Cref{fig:confs} we plot
the number of papers that we downloaded for each of the considered
conferences, and in \Cref{fig:confyear} we show how many papers in
each of the conferences included numerical experiments on elections. Generally, the trend is that the number
of experimental works is increasing, especially if one compares years
\tocheckbox{2010--2016} and \tocheckbox{2017--2023}, but it is unclear how strong this trend
is. In particular, there was a significant decrease in \tocheckbox{2021} and a
significant increase in \tocheckbox{2023}. It remains to see if 2023 was continuing
the trend, or if it were catching up with the papers ``missing'' in
2021 (it is tempting to speculate that the decrease in 2021 was due to
the COVID-19 pandemics but, as \Cref{fig:confs} shows, the overall
number of papers in the conferences has not decreased as dramatically).

In \Cref{fig:appord} we plot the number of papers in the Guide that
consider either experiments on ordinal or approval elections. While, so far, ordinal
elections have received far greater attention (altogether
\tocheckbox{130} papers consider them, whereas only \tocheckbox{35}
papers include experiments on approval; with some papers including
both types of elections), it is evident that in recent years approval
elections have become popular. One of the reasons for this
partial shift of interest is that approval elections are very natural in the
context of multiwinner
elections~\citep{fal-sko-sli-tal:b:multiwinner-voting,lac-sko:b:multiwinner-approval}
and in participatory
budgeting~\citep{rey-mal:t:participatory-budgeting-survey},
two topics that received a lot of attention in recent years.

\begin{table}
  \centering
  \begin{tabular}{c|c@{\hspace{4pt}}c@{\hspace{-12pt}}c@{\hspace{0pt}}c@{\hspace{4pt}}c@{\hspace{4pt}}c}
    \toprule
    regime                 & \multicolumn{3}{c}{candidates ($m$)} & \multicolumn{3}{c}{voters ($n$)} \\
    \midrule
    small elections        & $2$&$-$&$30$     & $2$&$-$&$30$ \\
    political elections    & $2$&$-$&$20$   & &$\geq$&$2000$ \\
    voting in institutions & $2$&$-$&$30$     & $30$&$-$&$2000$ \\
    participatory budgeting& $4$&$-$&$200$  & $200$&$-$&$100000$ \\
    ground truth           & $m$&$\geq$&$n$ & &$\leq$&$50$ \\
    multiwinner lab        & $100$&$-$&$500$ & $100$&$-$&$500$ \\
    \bottomrule
  \end{tabular}
  \caption{Rough classification of the ranges of numbers of candidates
    and voters in various types of elections in the papers from the
    Guide.}
  \label{tab:sizes}
\end{table}

\subsection{Sizes of Elections in Experiment}

Next, we analyze the sizes of elections studied in the papers from the
Guide. In \Cref{fig:size-histogram} we plot histograms showing how
many papers consider particular numbers of candidates and voters, and
in \Cref{fig:size-heatmap} we show heatmaps illustrating the
popularity of different combinations of these parameters. We also
include analogous data for elections from the
Preflib~\citep{mat-wal:c:preflib} and
Pabulib~\citep{fal-fli-pet-pie-sko-sto-szu-tal:c:participatory-budgeting-tools}
databases of real-life elections (the former mostly contains ordinal
elections, 
whereas the latter mostly includes approval ones, only regarding
participatory budgeting; Pabulib plots omit ``Artificial Mechanical
Turk'' datasets).

\begin{remark}
  In Figures~\ref{fig:size-histogram} and~\ref{fig:size-heatmap}, for
  each paper we record each election size that occurs in its
  experiments only once, even if it appears in several experiments (if
  we recorded each election size once per experiment, the overall
  shape of the figures would not change much).  Further, if an
  experiment considers elections of different sizes (for example,
  analyzing how its result changes as we vary the numbers of
  candidates or voters), then we record an election with a given size
  for each bucket in the histogram/heatmap to which it fits.
\end{remark}

We identify \tocheckbox{six} main regimes in which many of the papers
operate, listed in \Cref{tab:sizes}. The classification is due to us,
but it is inspired by what we have seen in the papers, and it takes
into account the data from Preflib and Pabulib. Hence, the boundaries of the regimes are somewhat
arbitrary and fluid, and papers sometimes mention other motivations for the election
sizes they consider (or often omit such motivation altogether). Further, the classification is naturally not perfectly accurate and rather focuses on capturing general trends and pragmatics.
For example, it is possible
that there is some (fairly atypical) real-life political election with
$30$ candidates and $500$ voters, even though we classify such
elections as having between $2$ and $20$ candidates, and at least
$2000$ voters.
As many papers that consider elections from a given regime do not
mention this explicitly as their motivation or goal, it is 
reassuring that, nonetheless, the community 
focused on elections that match natural, realistic settings (with the
possible exception of the \emph{multiwinner lab} one, which is not
particularly realistic, but has other redeeming features). Below we
discuss the regimes in 
detail.

\paragraph{Small Elections.}
This regime includes the smallest elections and captures, e.g., groups of
friends voting on where to have lunch or small
committees within companies, e.g., deciding who to hire (given a
shortlist).
However, generally, papers using this type of data do not explicitly
state their motivation. Experiments over small elections are sometimes
conducted to provide illustrations for theoretical results, rather
than to get new insights. Notably, small elections are often chosen
due to technical challenges, for instance when the studied problems
are computationally difficult. They also often arise
in studies done on human subjects.

\paragraph{Politics.} Our next type of elections regards various forms
of \emph{political elections}, which contain a limited number of
candidates ($m\leq 20$) and a comparably high number of voters
($n\geq 2000$). Papers that use elections of these sizes
and point to specific motivations indeed typically mention some form
of political elections, such as parliamentary, city board, referendum,
or presidential (nominee) ones.  Accordingly, political elections from
Preflib (such as the Irish dataset) are particularly popular in such
papers.  The only other application scenario that is occasionally
mentioned is crowdsourcing, e.g., in the form of large-scale surveys
(such as the Sushi survey on Preflib) or peer grading.

\paragraph{{Voting in Institutions.}} 

Our next regime involves fairly small groups of up to 30 candidates
and slightly larger numbers of voters (up to 2000), which can be seen
as the sizes of a typical election in an institution such as, e.g.,
professional associations.\footnote{Elections to the IFAAMAS Board of
  Trustees, with over 300 eligible voters, are a possible real-life
  example, and ERS data from Preflib is another. On the other hand,
  presidential elections of the American Psychological Associate (APA) that are available on Preflib have 
  around 5 candidates and 17'000 voters and are thus perhaps
  closer to the political setting.}  However, papers using these
election sizes often do not focus on particular applications and
simply find this setting appealing.
Indeed, elections from this regime are sometimes used due to the
hardness of computational problems studied, 
as they often allow for sufficiently realistic, but manageable
experiments. Papers using such elections  focused on a wide range of
topics, involving matching, party elections, iterative voting, or
randomized voting rules.  It is also worth mentioning that many papers
in this category  included other (smaller or larger) election
sizes.

\paragraph{PB Elections.} Instances in this group are mostly real-life
participatory budgeting elections from Pabulib. They typically contain
hundreds \tocheckbox{(at most~$220$)} of candidates and more than \tocheckbox{$200$}, but up to
\tocheckbox{tens of thousands}, of voters. There is no canonical way of using the
resources from Pabulib. Authors usually consider either (i) all
elections that are available at the time they access Pabulib; (ii)
elections that satisfy certain size criteria (e.g.,\ have at least
$10$~candidates); or (iii) elections that are of high enough quality
(i.e.,\ large-sized elections with a high average number of approvals
per voter), such as PB elections from Warsaw from the years 2020--2023.

\paragraph{Multiwinner Lab.} This type of election contains mid-sized
instances that are characteristic to experimental analyzes of
\emph{multiwinner} voting rules (with very few exceptions).  Papers
often argue that the considered numbers of candidates and voters, both
between~$100$ and~$500$, balance the trade-off between running times
of algorithms and the structural complexity of the
preferences. Briefly put, these elections are big enough to be
interesting in the context of studied properties, but small enough to
be analyzed by the respective computational techniques. Elections with
equal numbers of voters and candidates, specifically $m = n = 100$ and
$m = n = 200$, are particularly prevalent. Sometimes, the
number~$m$~of candidates is determined by the desired committee
size~$k$ with the goal to obtain a certain (e.g.,\ integral) value
of~$\nicefrac{m}{k}$. Naturally, these specific elections are
typically generated using synthetic models.

\paragraph{Search for Ground Truth.} This class of elections is
slightly more vague. It contains elections where there are different
``credible'' sources of information ($n\leq 50$) ranking a variety of
candidates ($m>n$) and typically the goal is to aggregate these
sources to recover an objective quality ranking of the
candidates. These elections appear in many papers with a range of
mentioned application scenarios including aggregating the opinions of
experts (e.g., judges or funding panel members), aggregating rankings
of items according to different criteria (e.g., price, outward
appearance,...), aggregating rankings of athletes in different types
of competitions (e.g., Olympic climbing), aggregating the outputs of
different computer systems (e.g., machine translation systems or
search engines), or deciding which items to select for a small group.
Elections of these sizes are typically generated from the impartial
culture model (even more frequently than in the other regimes),
whereas the Mallows model, which would be a natural choice for such
scenarios, and real-world data are rarely used (see
\Cref{sec:cultures-ordinal} for a discussion of statistical cultures).
Real-world datasets from Preflib that fall into this category include
different sports competitions (such as Formula 1 and speed skating),
criteria-based rankings (e.g., of cities, countries and universities),
and rankings output by different search engines according to the same
query.

\begin{figure}
        \centering
    \begin{subfigure}{0.43\textwidth}
        \centering
        \includegraphics[width=5.7cm]{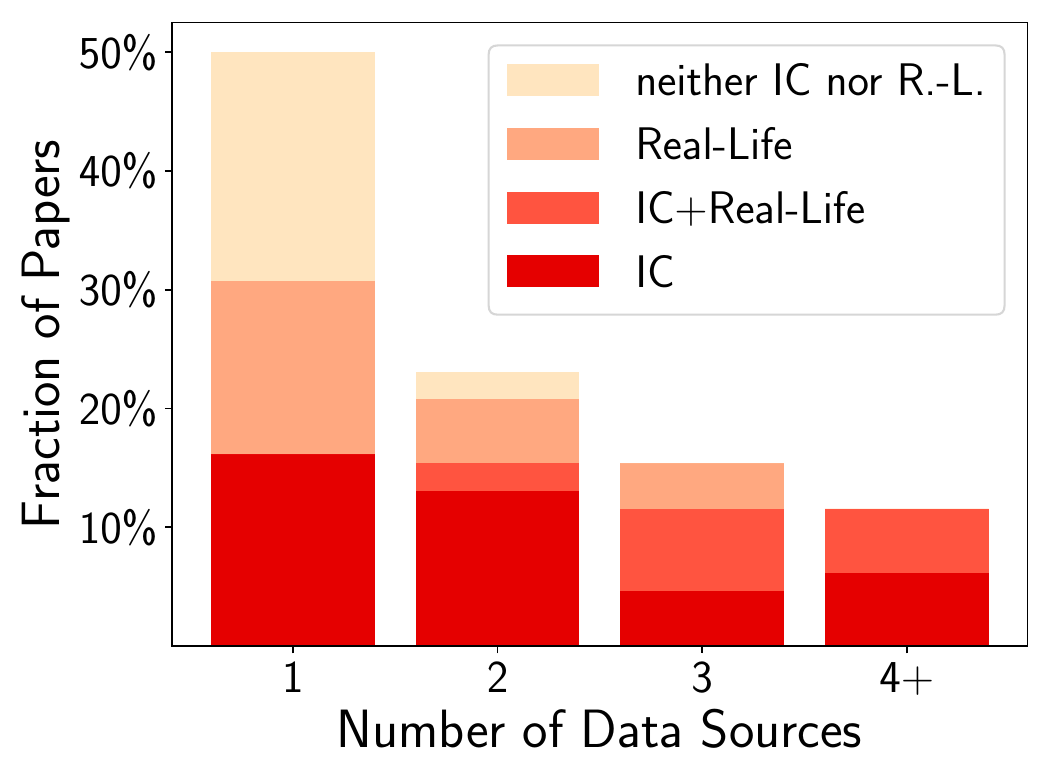}
        \label{fig:cand-hist}
      \end{subfigure}\quad      

      \caption{Numbers of data sources used in the papers that
        consider ordinal elections. ``Neither IC nor R.-L.'' means
        papers that used neither impartial culture (IC) nor real-life
        data, ``Real-Life'' means using real-life data but not IC,
        ``IC + Real-Life'' means using both IC and real-life data, and
        ``IC'' means using ``IC'' but not real-life data.}
      \label{fig:data-sources}

\end{figure}

\begin{figure}
    \centering

    \begin{subfigure}{0.43\textwidth}
        \centering
        \includegraphics[width=6.2cm]{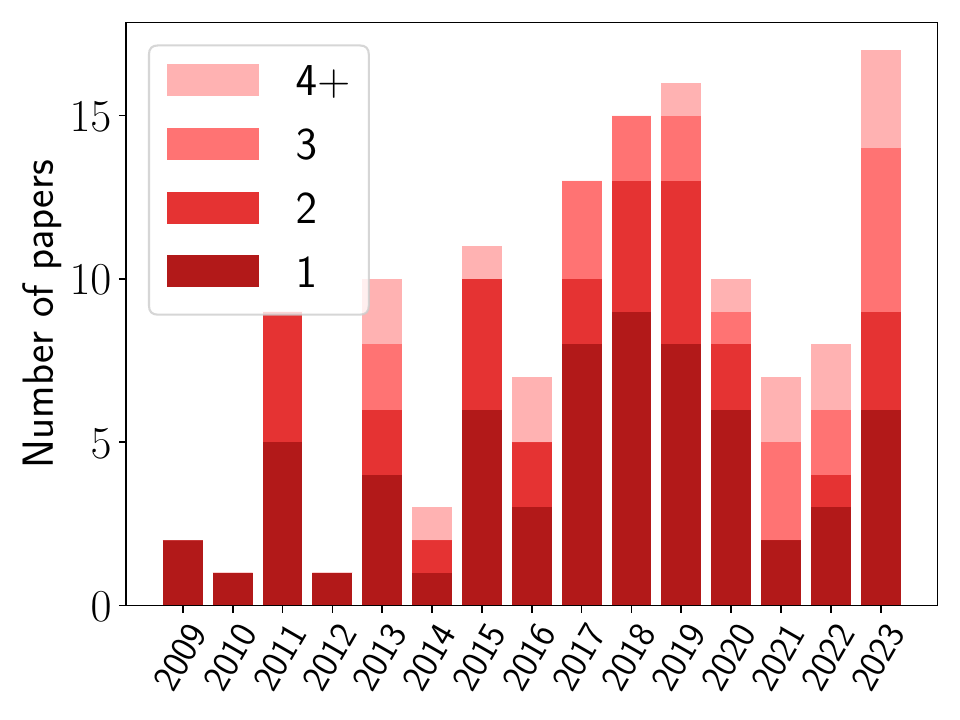}
        \caption{Ordinal Elections}
      \end{subfigure}\quad      
    \begin{subfigure}{0.43\textwidth}
        \centering
        \includegraphics[width=6.2cm]{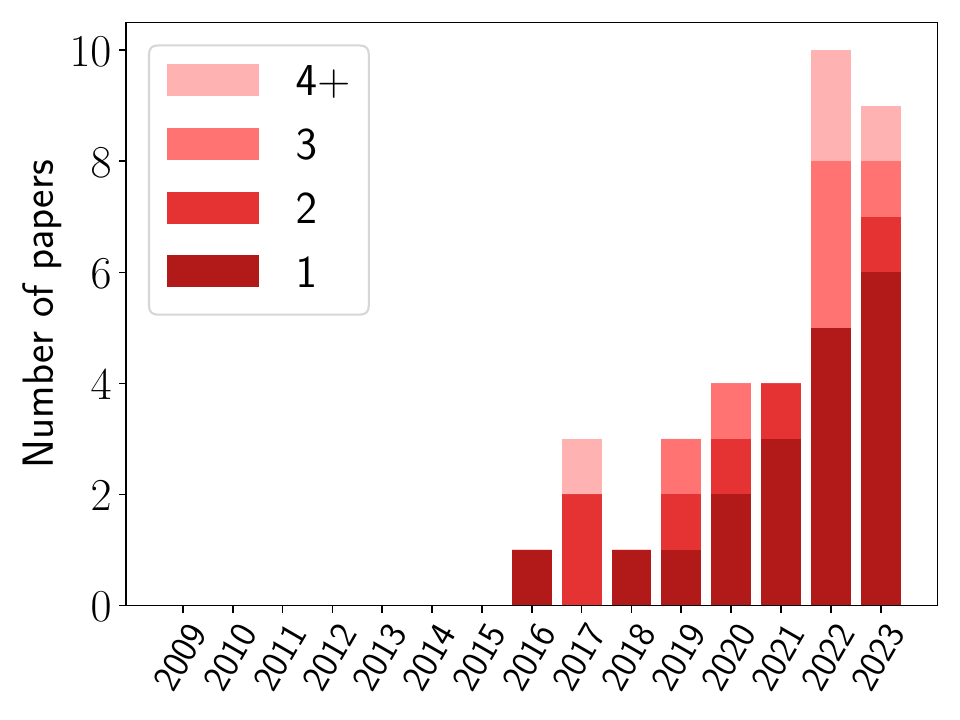}
        \caption{Approval Elections}
    \end{subfigure}\quad

    \caption{Numbers of data sources used in the papers from the Guide
      that consider either ordinal (top) or approval (bottom)
      elections in particular years.}
    \label{fig:add:cultures}
\end{figure}
    
\subsection{Statistics of Data Sources}
Overall, in \tocheckbox{$130$} papers we identified \tocheckbox{$211$}
experiments that were using ordinal elections.  Most of them
\tocheckbox{($62.3\%$)} used only synthetic data. It is a bit
worrisome that \tocheckbox{$16.2\%$} of the papers relied solely on
the highly unrealistic impartial culture model (where we choose votes
uniformly at random).  About \tocheckbox{$13.8\%$} of the papers used
only real-life elections (mostly from Preflib), with the Sushi dataset
being the most popular.  We include aggregated statistics about the
number of data sources for ordinal elections in
\Cref{fig:data-sources}, and in \Cref{fig:add:cultures} (top) we show
the numbers of papers that use a given number of data sources depening
on a year.\footnote{We treat different statistical cultures as
  different data sources, but we view ``real-life data'' as a single
  one, irrespective if a paper is using just a single dataset from
  Preflib (such as the Sushi one) or multiple different ones.}  We see
that in the \tocheckbox{last few years} more and more papers use more
than just a single source of data, which certainly is a positive
trend.

Regarding approval elections, in \tocheckbox{35} papers we recorded
\tocheckbox{46} experiments.  In \Cref{fig:add:cultures} (bottom) we
see how many papers use a given number of data sources. As opposed to
the ordinal case, we see that majority of papers use only a single
source of data. However, in this case it is not as worrisome as most
typically this means using real-life data from Pabulib (or real-life
data from Preflib, adapted to the approval setting). Altogethr,
\tocheckbox{$57.1\%$} of the papers that study the approval setting
used real-life data (see \Cref{fig:approval:data-sources}).

\begin{figure*}
    \centering
    \includegraphics[width=\linewidth]{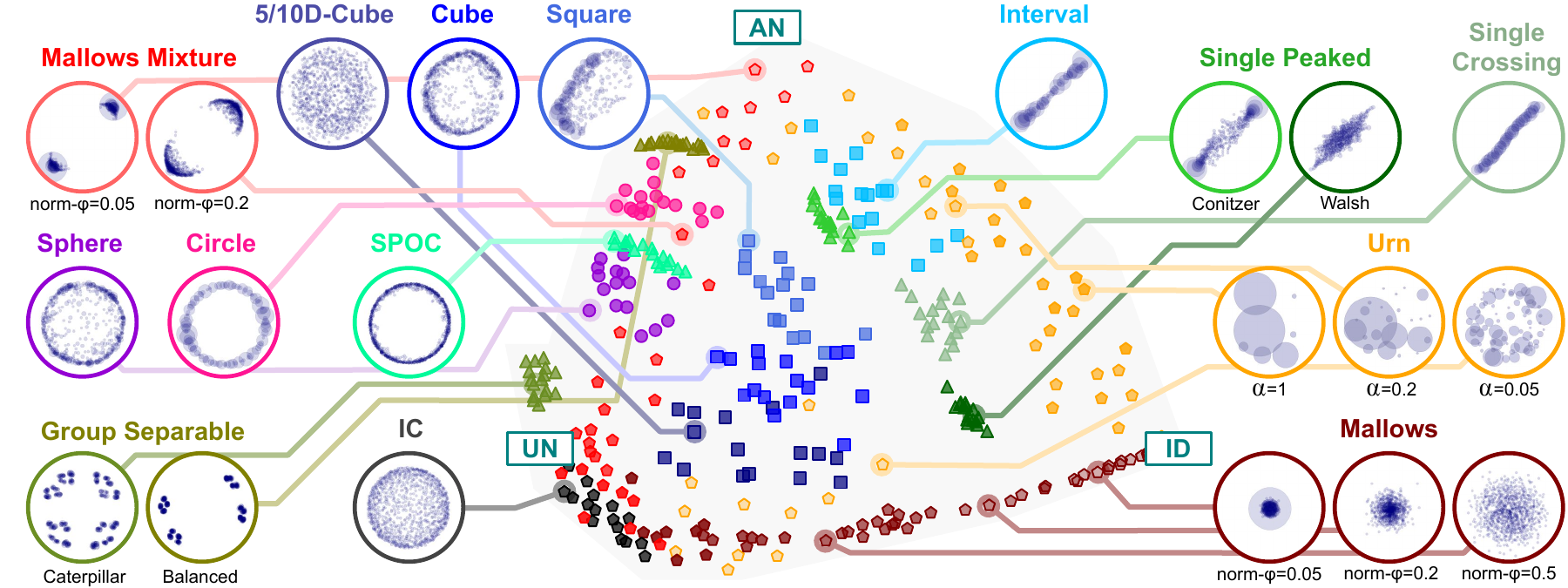}
    \caption{Map of elections and the microscope. Elections in the map have 8 candidates and 96 voters (for computational reasons) and the ones in the microscope have 10 candidates and 1000 voters (for better visualization). Hence, the connections between the elections on the map and their microscopes are meant to show a general behavior, not the exact compositions of the given election.}
    \label{fig:map+microscope}
\end{figure*}

\section{Statistical Cultures for Ordinal Elections}\label{sec:cultures-ordinal}

In this section we take a closer look at the most popular statistical
cultures, i.e., models of generating synthetic preference data, for
ordinal elections (over \tocheckbox{73.8\%} of the papers that study
ordinal elections use at least one of the cultures that we describe,
and this fraction grows to over \tocheckbox{90.7\%} if we include
real-life data). Below we provide their definitions and discuss their
use in the papers from the Guide, including common parameter
settings. Further, in \Cref{fig:map+microscope} we illustrate
elections that these models generate, as well as the relations between
the models, on a map of elections.  The swap distance between two
preference orders $u$ and $v$, denoted $\kappa(u,v)$, is the number of
pairs of candidates $a$ and $b$, such that $u$ and $v$ disagree on
their ranking (i.e., one of them ranks $a$ above $b$, and the other
ranks $b$ above $a$).

Maps of elections are a way to visualize an election dataset and have been
introduced by \citet{szu-fal-sko-sli-tal:c:map-of-elections} and
\citet{boe-bre-fal-nie-szu:c:compass}. Specifically, for each two
elections in the dataset we measure their similarity (using the
isomorphic swap distance~\citep{fal-sko-sli-szu:c:similar-two-elections}) and visualize them as points on a plane, so that the
Euclidean distances between the points resemble these similarities (we
use the MDS embedding~\citep{kru:j:mds}). Crucially, the maps use
distances that are invariant to renaming the candidates and voters
and, hence, illustrate structural similarities between the
elections. Further, our maps include three special elections as
reference points: Identity (ID), where all votes are the same, antagonism
(AN), which has two equal-sized groups of voters with opposite
preference orders, and an approximation of a uniformity (UN) election,
where each possible vote appears once.

Following \citet{fal-kac-sor-szu-was:c:div-agr-pol-map}, we also
include ``microscope'' maps of specific types of elections. To form
such a ``microscope'', we take a single election, measure the swap distance
between each pair of its votes, and then draw a picture where each
disc represents a single vote (with the radii of the discs
representing the number of identical votes) and the Euclidean
distances between the discs resemble the swap distances
between the votes. Such microscope maps allow one to understand
internal structures of the considered elections. Indeed, we recommend
that one looks at them whenever generating data from a new source.

\begin{figure}[t]
    \centering

    \includegraphics[width=8.4cm]{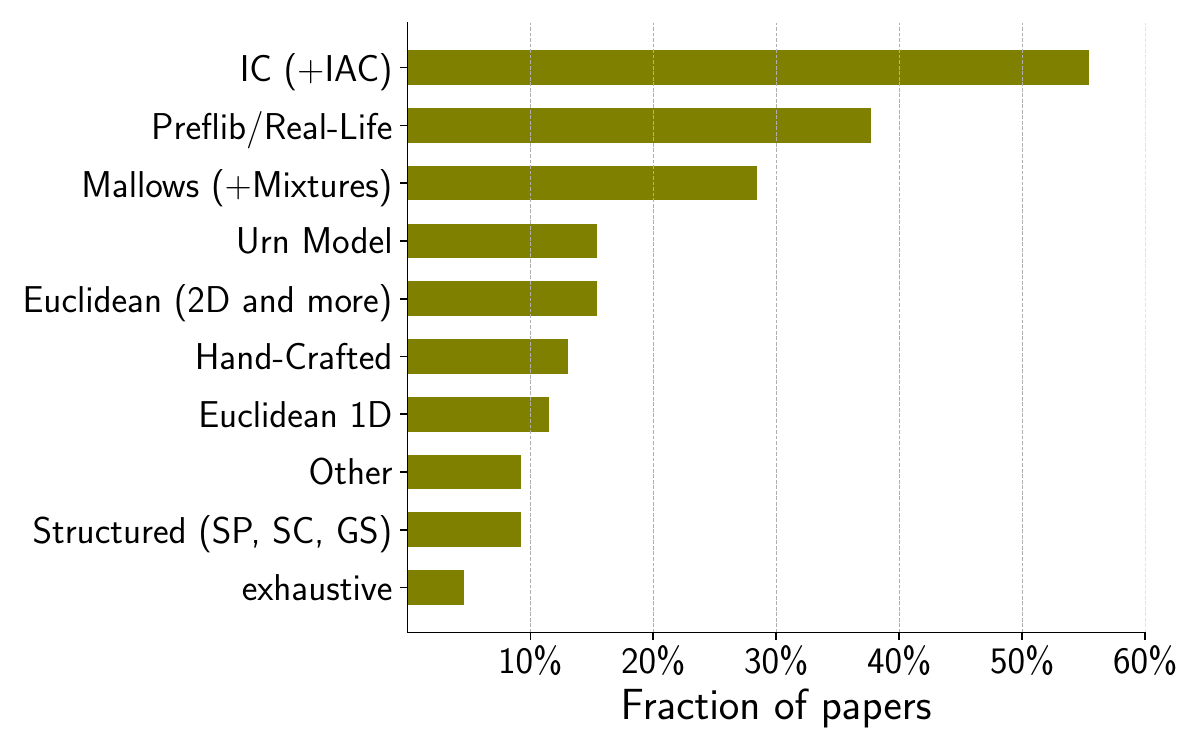}

    \caption{Fractions of papers that use given data sources for
      ordinal  elections. ``Hand-Crafted'' refers to models designed specifically
    for a given paper. ``Exhaustive'' means generating all elections of a given size.}
    \label{fig:ordinal:data-sources}
\end{figure}

\begin{figure}
    \centering
    \includegraphics[width=8.4cm]{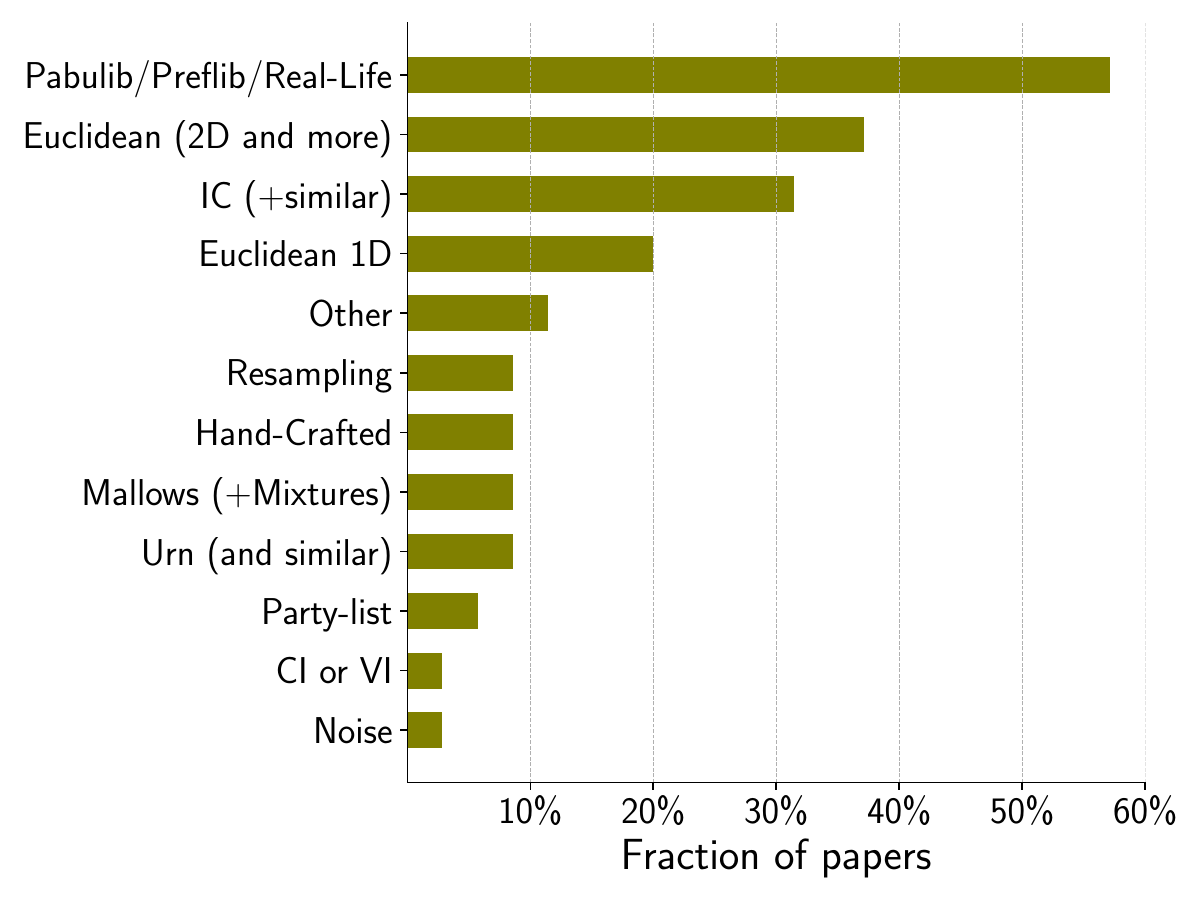}

    \caption{Fractions of papers that use given data sources for
      approval elections. ``Hand-Crafted'' refers to models
      specifically designed for a given paper. Mallows and urn models
      refer to generating ordinal elections according to these models
      and considering some top candidates from each vote as the
      approved ones.}
    \label{fig:approval:data-sources}
\end{figure}

In \Cref{fig:ordinal:data-sources} we give statistics as to how
popular are particular statistical cultures for ordinal elections
(note that the exact fractions given in this plot can be slightly
different than the ones given in the following paragraph headings; for
example, in \Cref{fig:ordinal:data-sources} we distinguish between
1D-Euclidean models and Euclidean models for higher dimensions, but we
do not do so in the text below).

\paragraph{Impartial Culture (Used in \tocheckbox{54.6\%} of the Papers).}

Under the impartial culture (IC) model we generate votes one-by-one,
choosing each preference order uniformly at random. Consequently,
there is no apparent structure among the votes, as seen in
\Cref{fig:map+microscope}. While by now the model is part of the
folklore, its first use dates back to the work of
\citet{gui:j:impartial-culture}, who studied the probability of the
Condorcet paradox. It is commonly agreed that impartial culture does
not generate realistic elections but, nonetheless, it is used in over
\tocheckbox{$54\%$} of the papers. Indeed, the model is extremely
simple and does not require setting any parameters. This means that
every experiment that uses IC, uses the very same
distribution. Consequently, it has become the baseline that many
researchers evaluate their results against.\footnote{This view is
  spelled out, e.g., by \citet{rei-end:c:iterated-poll-condorcet}.} We
largely agree with this use of IC as a common yardstick, but we very
strongly encourage the use of further models in experiments, to get a
broader view of the studied phenomena.

Impartial anonymous culture (IAC), introduced by
\citet{kug-nag:j:impartial-anonymous-culture} and
\citet{fis-geh:j:impartial-anonymous-culture}, is a variant of IC
where the voters are indistinguishable, i.e., it only matters how many
particular votes were cast, but not by whom. Impartial anonymous and
neutral culture (IANC) further abstracts away from candidate
names~\citep{ege-gir:j:isomorphism-ianc}.  Unless there are very few
candidates, IAC and IANC are generally very similar to IC.

\paragraph{Mallows Model (Used in \tocheckbox{28.5\%} of the Papers).}

The Mallows model \cite{mal:j:mallows} is the second most popular way
to generate synthetic elections in the Guide. This is quite positive
as recent work indicated that it provides a good coverage of the space
of real-life elections
\cite{DBLP:phd/dnb/Bohmer23,DBLP:conf/nips/BoehmerBEFS22}.  In
\Cref{fig:map+microscope}, Mallows elections form a line between ID
and UN.  The basic idea is that there is an underlying ``ground
truth'' ordering $v^*$ of the candidates and that the probability of
sampling a vote from the model decreases with the vote's
distance from $v^*$. The expected distance can be controlled by a
dispersion parameter $\phi\in [0,1]$.  Formally, the probability of
sampling a vote $v$ is proportional to $\phi^{\kappa(v,v^*)}$.
(Occasionally authors express the probability of sampling a vote $v$ as
proportional to $e^{-\phi\cdot \kappa(v,v^*)}$, as done, e.g., in the
work of \citet{dou-coh:c:incomplete-preferences-markov}. This is
correct, but yields a different range of $\phi$ values.)

Authors often consider multiple values of the dispersion parameter at
equal distances from each other (e.g.,
$\phi \in \{0.1, 0.2, \ldots \}$), but single values (e.g., $\phi=0.8$
or $\phi=0.5$) appear as well. Generally, there is a trend toward
using larger values.  Another strategy is to not consider specific,
fixed 
values and, instead, generate elections by first sampling a value of
the dispersion parameter uniformly from some pre-specified range and
then drawing votes from the resulting distribution (see e.g., the works of~\citet{bac-lev-lew-zic:c:misrepresentation-in-districts,boe-cai-fal-fan-jan-kac-was:c:position-matrices,fal-kac-sor-szu-was:c:div-agr-pol-map}).
This procedure creates a diverse dataset without the need for separate
evaluations.  Mixtures of Mallows models combining multiple models
with different central orders and dispersion parameters with some
weight function on top have also been used, but less frequently (an
example of such a mixture, with the voters equally split between two
Mallows models with equal noise and opposite central orders, is
visible in \Cref{fig:map+microscope}).

Recently,
\citeauthor{boe-bre-fal-nie-szu:c:compass}~\shortcite{boe-bre-fal-nie-szu:c:compass,DBLP:conf/icml/BoehmerFK23,DBLP:phd/dnb/Bohmer23}
argued that there are certain issues when using the Mallows model.  In
particular, they showed that equally-spaced values of the dispersion
parameter do not provide a uniform coverage of the space between ID
and UN elections: For larger numbers of candidates, parameter values
below, say, $0.8$ will result in elections where votes are fairly
similar to each other (this, indeed, justifies the use of high $\phi$
values in previous works).  Moreover, they argued that fixing a
dispersion parameter and changing the number of candidates
fundamentally changes the nature of the sampled elections, thus
rendering results for different numbers of candidates incomparable.
They provided a new parameter, $\normphi$, that ensures that
uniformly-selected parameter values provide uniform coverage of the
space between ID and UN (indeed, to generate Mallows elections for
\Cref{fig:map+microscope}, we were choosing $\normphi \in [0,1]$
uniformly at random): Given a value of $\normphi \in [0,1]$, one
computes classic $\phi$ so that the expected swap distance between the
central vote and one generated using the Mallows model is
$\normphi = \nicefrac{1}{4}\cdot m(m-1)$ (where $m$ is the number of
candidates).  We point to their paper(s) for further explanations,
intuitions, and ways of computing $\phi$ given $\normphi$.

\paragraph{Pólya-Eggenberger Urn Model (Used in \tocheckbox{15.3\%} of the Papers).}

The P\'olya-Eggenberger urn model
\citep{eggenberger1923statistik,berg1985paradox} uses a nonnegative
parameter of contagion $\alpha\in \mathbb{R}$, which corresponds to the
level of correlation between the votes. Votes are generated
iteratively as follows: We imagine
an urn which initially contains one copy of each possible order; to
generate a vote, we draw one from the urn, include its copy in the
election, and return it to the urn, together with $\alpha \cdot m!$
copies, where $m$ is the number of candidates.\footnote{This
  normalized variant is due to \citet{mcc-sli:j:similarity-rules}; in
  the unnormalized variant the parameter gives the absolute number of
  the additional copies put back into the urn.} For $\alpha=0$ we get
IC, 
and for $\alpha=\nicefrac{1}{m!}$ we get
IAC~\citep{ege-gir:j:isomorphism-ianc}.

Among the considered papers, \tocheckbox{20} conducted experiments on the urn
model. 
Typical values of $\alpha$ were $\nicefrac{10}{m!}$, $0.05$, $0.1$,
$0.2$, $0.5$, and $1$. In a few papers, particularly regarding the map
of elections,
$\alpha$ was derived from the Gamma distribution with shape parameter
$k = 0.8$ and scale parameter $\theta = 1$ (and this is how we
generated the urn elections for \Cref{fig:map+microscope}).

\paragraph{Euclidean Elections (Used in \tocheckbox{20\%} of the Papers).}
Under a Euclidean model, we assume that the candidates and voters are
represented as points in some $d$-dimensional Euclidean space.
Typically, these points are sampled uniformly at random from a
$d$-dimensional cube (usually $[0,1]^d$, for $d=1$ this is the
Interval model, for $d=2$ the Square model, and for $d = 3$ the Cube
model). Occasionally other distributions are considered (such as
various forms of Gaussian distributions and uniform distribution over
a $d$-dimensional sphere; for $d=2$ this is the Circle model and for
$d=3$ the Sphere model).  Each voter's ranking is constructed so
that he or she ranks candidates whose points are closer
to his or hers higher than those whose points are further~away.

Among the considered papers, \tocheckbox{25} conducted experiments on Euclidean
preferences. The most popular choice was the 2D setting (\tocheckbox{18 papers}),
followed by the 1D one (\tocheckbox{12 papers}). Some papers additionally
investigated higher dimensions, reaching up to the 20D model (e.g.,
\citet{boe-cai-fal-fan-jan-kac-was:c:position-matrices},
\citet{boe-bre-fal-nie-szu:c:compass} and
\citet{szu-fal-sko-sli-tal:c:map-of-elections}).

\paragraph{Single-Peaked Elections (Used in \tocheckbox{9.2\%} of the Papers).}
Single-peakedness is one of the most prominent structured domains. An
election is single-peaked~\citep{bla:b:polsci:committees-elections} if
there is an ordering of the candidates---the societal axis---such that
for each voter, sweeping through the axis from left to right, the
position of the corresponding candidates in the voter's ranking first
increases and then decreases. Single-peaked elections are usually
motivated by the fact that they cover applications in which there is an
objective order of candidates; a typical example being the political
left-to-right spectrum.

In practice, authors use two main methods to generate such
elections. Both of them first select an axis 
uniformly at random. The model proposed by~\citet{wal:t:generate-sp}
uses a uniform distribution over the votes that are single-peaked for
the selected axis. In the model proposed
by~\citet{con:j:eliciting-singlepeaked}, to generate a vote we first
pick uniformly at random its top choice. Then, to fill the next
position in the ranking, we flip a symmetric coin and either select
the first unused candidate to the right or to the left of the
top-choice one. We repeat the procedure until all positions are filled
(or the remaining positions are uniquely determined).

While the Walsh approach seems more appealing as a single-peaked
variant of impartial culture, the Conitzer approach is interesting
because it gives elections very similar to the 1D-Euclidean ones
(where the candidate and voter points are sampled uniformly at random
from an interval). Consequently, multiple papers with experiments on
both Walsh and Conitzer models
show that they 
tend to give qualitatively different elections. Thus, when studying
single-peaked elections, we recommend using both approaches.

Single-peakedness on a circle (SPOC) is a variant of single-peakedness
where the axis is cyclic~\citep{pet-lac:j:spoc}. Sampling SPOC
elections using the Conitzer's approach leads to a uniform
distribution of such votes.

\paragraph{Single-Crossing Elections (Used in \tocheckbox{4.6\%} of the Papers).}
An election is single-crossing if we can order all the votes in a
way that for every pair of candidates all the voters either prefer one
of them to the other, or the relative preference between them changes
exactly once when going from the first to the last vote in the
ordering~\citep{mir:j:single-crossing,rob:j:tax}. It is unknown how to
sample such votes uniformly at random in polynomial time (and, indeed,
doing so might be challenging).
\citet{szu-fal-sko-sli-tal:c:map-of-elections} give a sampling
heuristic which seems reasonable, but makes no guarantees about its
distribution (we use it in \Cref{fig:map+microscope}).

\paragraph{Group-Separable Elections (Used in \tocheckbox{3\%} of the Papers).}

A group-separable
election~\cite{ina:j:group-separable,ina:j:simple-majority} can be
characterized by a rooted, ordered tree whose leafs are candidates
(Inada's definition was different, we follow an approach of
\citet{kar:j:group-separable}).
Then, each vote in such an election must be obtainable by, first,
reversing the order of children of arbitrary internal nodes of the
tree (possibly none), and then reading the candidates from leaves from
left to right.
In the considered experiments, only group-separable elections with balanced or caterpillar trees were considered and the votes were drawn uniformly at random.
Such elections do not resemble real-life data, but are different from
elections given by any other culture (which is visible by their
distinct position in the map), thus they can capture unusual
phenomena, which might be hard to spot otherwise.

\paragraph{Which Models to Use?}
There is no clear answer as to which statistical cultures are the \emph{best} in some objective sense. However, there are three natural
approaches to choosing which models to use in a paper: First, one
might want to cover as much of the space of elections as possible
(this might mean including elections from structured domains, in
addition to more common models).
Second, one might know the nature of the real-life data that appears
in a given phenomenon and might want to choose model(s) that generate
similar elections.
Finally, one might want to stick to realistic data, but without
focusing on its specific type. In this case, results on the map of
elections~\citep{boe-bre-fal-nie-szu:c:compass,DBLP:conf/nips/BoehmerBEFS22,fal-kac-sor-szu-was:c:div-agr-pol-map}
suggest choosing cultures that land in a triangle between ID, UN, and
Euclidean elections (for dimension $2$ or higher). This might mean,
e.g., using the Mallows model, urn models with fairly low contagion
parameters, and Euclidean models (such as, e.g., the 5D-Cube).

\section{Statistical Cultures for Approval Elections}\label{app:approval-cultures}

We describe the most common statistical cultures used in approval
elections. In the text below we make some remarks as to how frequently
a particular model or a variant of a particular model is used, but the
reader should take them with a grain of salt: Altogether, only
\tocheckbox{35} papers in the Guide consider approval elections so the
difference between a ``frequently used model'' and a ``rarely used
one'' may be rather small in terms of absolute numbers.  For maps of
approval elections we point to the work of
\citet{szu-fal-jan-lac-sli-sor-tal:c:how-to-sample-approval-elections}.

In \Cref{fig:approval:data-sources} we illustrate fractions of papers
that use particular statistical cultures (or real-life data). As in the
previous section, the cultures are grouped a bit differently in this
plot than in the following text.

\paragraph{Impartial Culture and Variants (Used in \tocheckbox{31\%} of the Papers).}
In the context of approval elections, impartial culture typically
refers to the model where each voter approves each candidate
independently, with some given probability $p$, and is sometimes
referred to as $p$-impartial culture,
$p$-IC~\citep{bre-fal-kac-nie:c:experimental-view-on-ejr,szu-fal-jan-lac-sli-sor-tal:c:how-to-sample-approval-elections}.
While fairly small values of $p$, such as $0.1$ or $0.15$, are more
common, some authors choose $p = 0.5$. Indeed, this latter approach
follows the spirit of impartial culture from ordinal elections more
closely, since each approval vote appears with equal probability (and,
occasionally, this is how this variant of the model is described in
the papers). That said, our view is that smaller values of $p$ are
more realistic, especially for higher numbers of candidates (this is justified by the results of
\citet{szu-fal-jan-lac-sli-sor-tal:c:how-to-sample-approval-elections},
who observe how many projects are approved in participatory budgeting
elections from
Pabulib~\citep{fal-fli-pet-pie-sko-sto-szu-tal:c:participatory-budgeting-tools}).

Some authors use variants of impartial culture with slightly different
definitions. For example,
\citet{mic-pet-sko:c:price-of-fairness-budget-division} use a model
where each voter $v$ has an individual approval probability $p_v$, and
approves each candidate independently, with this
probability.\footnote{To be strict, it is slightly unclear if in this
  paper each voter indeed has an individual approval probability. The
  reason is that the description of the model contains a typo in a
  critical place, differentiating global and individual approval
  probabilities. However, the text around strongly suggests that the
  intention is to use individual approval probabilities.}
\citet{lac-sko:c:quantitative-multiwinner} use a model that is closely
related to impartial culture (but, admittedly, they do not use this
name), where to draw a vote we first choose the number of candidates
that will be approved (in this case, a value between $2$ and $5$
selected uniformly at random), and then we select a subset of
candidates of this size that are approved (uniformly at random, among the sets of a
given size).

\paragraph{Euclidean Models (Used in \tocheckbox{46\%} of the Papers).}

The general idea behind the Euclidean model for approval elections is
analogous to that for ordinal elections: First, each voter and each
candidate is associated with a point sampled from a $d$-dimensional
Euclidean metric space (either a hypercube or a hypersphere) from the
uniform or normal distribution (other distributions can be used as
well, of course). Then, each voter prefers candidates who are closer
to him or her to those that are further away. In the context of
approval preferences, this means that if a voter approves some
candidate $c$, then this voter also approves every candidate that
is closer to him or her than $c$.

However, we also need to specify how many candidates each voter
approves. There is no single well-established answer and the analyzed
papers often differ in the technical details here. However, there are two general approaches:
\begin{enumerate}
\item The first choice (occurring in \tocheckbox{4 papers}) is to fix
  the ballot length of each voter to some $x$ and let the voter approve the $x$
  closest candidates. The value of $x$ can be the same for all the
  voters or chosen randomly (typically, uniformly at random from
  $\{1, \ldots, m\}$ as in the works of
  \citet{god-bat-sko-fal:c:abc-rules-euclidean-2d} and
  \citet{mic-pet-sko:c:price-of-fairness-budget-division}).
\item The second, more popular choice (occurring in \tocheckbox{12
    papers}) is to fix the maximum possible distance (radius) $r$ that voters have to their approved candidates. Again, this value may
  be either the same for all the voters, or chosen randomly for each
  voter from some interval. One can observe a common tendency in the literature to set
  $r$ much lower for 1D-Euclidean elections than for 2D-Euclidean ones
  (e.g., \citet{fal-lac-sor-szu:c:comparison-multiwinner-approval},
  \citet{bre-fal-kac-nie:c:experimental-view-on-ejr} and
  \citet{elk-fal-iga-man-sch-suk:c:price-justified-representation} set
  it to approximately 0.05 for 1D model and 0.2 for 2D model). One
  guiding principle that one could use here is to choose the
  radius so that, on average, the voters approve a given number of
  candidates. Another possibility that has appeared in the works of
  \citet{lac:c:perpetual-voting-fairness} and
  \citet{lac-mal-nar:freeriding-multiissue} is to set $r$ as a
  multiple of the distance to the closest candidate.
\end{enumerate}
A less popular way to generate votes is to apply the aforementioned
approaches to candidates and not the voters; for example, each
candidate is approved by the $x$ closest voters to him or her, as
in the work of
\citet{tal-fal:c:framework-approval-based-budgeting-methods}.

Among the analyzed papers, \tocheckbox{16} conducted experiments on
Euclidean elections. Similarly as in the case of ordinal elections, 2D and
1D models were the most popular ones (used in \tocheckbox{12 and 7
  papers}, respectively).

\paragraph{Resampling Model (Used in \tocheckbox{Three} Papers).}
In the $(p,\phi)$-resampling model, we start by creating a central
ballot~$u$, by approving $\lfloor p \cdot m\rfloor$ candidates
uniformly at random. Then, to generate a vote $v$ we first assume that
$v = u$ and then for every candidate $c \in C$, with probability
$1-\phi$, we keep $c$'s approval as it was in the central ballot, and
with probability $\phi$ we resample it (i.e., approve it with
probability~$p$).  So far, the resampling model has been used only in three
papers from the Guide (and a few others published in venues not currently included in the guide). However, this is not very surprising since it was introduced only very recently
by~\citet{szu-fal-jan-lac-sli-sor-tal:c:how-to-sample-approval-elections}. The
$p\in[0,1]$ parameter denotes the average number of approvals in a
vote, whereas the $\phi\in[0,1]$ parameter denotes the level of
dispersion. While the resampling model is similar to the noise model
based on the Hamming distance (see below; in particular, the two
models have analogous parameters), the resampling model maintains the
same average number of approvals in a vote, irrespective of the
dispersion parameter (under the noise model, changing the dispersion
parameter also changes the average number of approvals). The exact
choice of $p$ and $\phi$ values depends heavily on what is the
motivation for our experiment. For example, if one wants to generate
elections similar to real-life ones (e.g., those from
participatory budgeting scenarios stored in Pabulib),
\citet{szu-fal-jan-lac-sli-sor-tal:c:how-to-sample-approval-elections}
recommend $p$ to be relatively small, i.e., from the $[0, 0.25]$
interval, and for $\phi$ to take values from the $[0.5, 1]$ interval.

\paragraph{Noise Models (Used in \tocheckbox{One} Paper).}

In the $(p,\phi,d)$-noise model, $p$ and $\phi$ are two real numbers
from $[0,1]$ and $d$ is a distance function between two approval votes
that is polynomial-time computable (further, we require that $d(u,v)$
depends only on $|A(u)|$, $|A(v)|$, and $|A(u) \cap A(v)|$ for each
two approval votes $u$ and~$v$).  In this model, we first generate a
central vote $u$ using $p$ (analogously to the resampling model). For
each voter, we generate its vote~$v$ with probability proportional to
$\phi^{d(u,v)}$.  This noise model is somewhat analogous to the
Mallows model for ordinal elections.
Jaccard or Hamming are examples of distance functions satisfying the
above-described conditions for $d$, albeit the model works also for
other functions.  While in the ordinal setting the Mallows model is
very appealing, in the approval setting the noise models have a
certain drawback: As $\phi$ gets close to $1$, such noise models
become more and more similar to $0.5$-IC. Put differently, the average
number of approvals in a generated vote changes together with $\phi$,
which, e.g., is not the case for the resampling model.  More details
can be found in
\citet{szu-fal-jan-lac-sli-sor-tal:c:how-to-sample-approval-elections}.

\paragraph{CI and VI Models (Used in \tocheckbox{One} Paper).}
An approval election has the candidate interval property (CI) if there
is an ordering of the candidates, called the \emph{societal axis},
such that each voter approves candidates that form an interval with
respect to this axis. Similarly, an election has the voter interval
property (VI) if there is an ordering of the voters such that each
candidate is only approved by voters that form an interval in this
ordering. CI is an approval analog of single-peakedness from the
world of ordinal elections, and VI is an approval analog of
single-crossingness; for more details on these structured domains, we
point to the work of \citet{elk-lac:c:approval-sp-sc} where they were
introduced.

We recorded \tocheckbox{a single work} using CI and VI
elections~\citep{bri-isr-mich-pet:c:representation-committee-voting}. The
authors generated them by applying a model for generating
1D~elections: \citet{elk-lac:c:approval-sp-sc} show that the
domain of CI elections is exactly the one of 1D elections, in which
each voter has their own approval radius. We recover VI elections if we assign radii to
candidates, instead of voters, and let each candidate be approved
by the voters in the candidate's radius~\citep{god-bat-sko-fal:c:abc-rules-euclidean-2d}.

\paragraph{Party-List Models (Used in \tocheckbox{Two} Papers).}
In a party-list election, each two voters either approve the same
candidates or their approval sets are disjoint. The candidates that
are jointly approved by a group of voters are said to form a
\emph{party}. One can verify that party-list elections are both CI and~VI.

The \tocheckbox{two papers} from the Guide that consider party-list
elections generate them in two different ways:
\citet{fai-vil-mei-gal:c:welfare-vs-representation-participatory-budgeting}
consider elections with $200$ voters that are split uniformly at
random into groups between $5$ and $20$ voters, and each group
approves between $10$ and $30$ candidates (this number is also selected
uniformly at random). Consequently, all parties have comparable sizes.
\citet{fal-lac-sor-szu:c:comparison-multiwinner-approval} use a
variant of the urn model: To generate an election with $m$ candidates and
$g$ parties, they first partition the candidates into parties of size
$\lfloor \nicefrac{m}{g} \rfloor$ each, and for each of the parties
they put a single vote approving exactly the party members into an urn. Then, they
sample the votes from the urn: For each sampled vote, they include its
copy in the election and return the vote to the urn together with $\alpha g$
copies ($\alpha$ is the contagion parameter, whose meaning
is analogous as in the urn model for ordinal elections). Depending on
 the $\alpha$ parameter, the parties can be of very different sizes
(however \citet{fal-lac-sor-szu:c:comparison-multiwinner-approval} do
not report the $\alpha$ value that they used).

Both above approaches are reasonable under different circumstances.
Indeed,
\citet{fai-vil-mei-gal:c:welfare-vs-representation-participatory-budgeting}
focused on participatory budgeting (and, so, also had a model of
generating the costs for the created ``parties''), whereas
\citet{fal-lac-sor-szu:c:comparison-multiwinner-approval} considered
vanilla multiwinner elections.

\paragraph{Truncated Ordinal Elections.}
Finally, to generate approval elections some authors first generate
ordinal ones and, then, let
voters approve some of their top candidates.
Indeed, the first method of ballot generation is the
Euclidean model, where each voter $v$ approves some $x$ closest
candidates, is an incarnation of this approach. Apart from the
Euclidean setting, such truncation-based methods were also used for the
urn
model~\citep{szu-fal-jan-lac-sli-sor-tal:c:how-to-sample-approval-elections,sko-lac-bri-pet-elk:c:proportional-rankings,bri-isr-mich-pet:c:representation-committee-voting}
and (mixtures of) the Mallows
model~\citep{mic-pet-sko:c:price-of-fairness-budget-division,bri-isr-mich-pet:c:representation-committee-voting,meh-sre-lar:c:deliberation-approval-elections}.

There is no univeral approach to choosing how many of the top
candidates should each voter approve. Two common approaches include
(a)~fixing the number for all the voters (as, e.g., done
by~\citet{szu-fal-jan-lac-sli-sor-tal:c:how-to-sample-approval-elections})
and (b)~choosing this number separately for each voter, either
uniformly at random from some range (as done, e.g., by
\citet{mic-pet-sko:c:price-of-fairness-budget-division} and
\citet{bri-isr-mich-pet:c:representation-committee-voting}) or by
following a (discretized) normal distribution (as done, e.g., by
\citet{meh-sre-lar:c:deliberation-approval-elections}).  We are not
aware of a principled study that would indicate which of the
approaches is more realistic for particular settings, but the last
one---using the normal distribution---appears most appealing.  To make a principled decision on this, one
could evaluate how many projects are approved in participatory
budgeting scenarios by looking at the data from
Pabulib~\citep{fal-fli-pet-pie-sko-sto-szu-tal:c:participatory-budgeting-tools},
but one would need to be careful about the drawn conclusions: In
participatory budgeting, the number of approvals per vote is sometimes
a consequence of procedural regulations, and not purely the choice of
the voters (which, of course, one might want to take into account when
studying participatory budgeting, but less so when using this data for
other purposes).

\paragraph{Which Models to Use?}
For approval elections, real-life data is the most common data source, appearing in a majority of papers with approval elections from the Guide. Pabulib~\citep{fal-fli-pet-pie-sko-sto-szu-tal:c:participatory-budgeting-tools} provides a rich collection of real-life elections from participatory budgeting exercises. 
Thus, especially for works on participatory budgeting, Pabulib is the most attractive and relevant data source. 
However, if one's work is not tailored to participatory budgeting, it might be a good idea to also consider other data sources, such as the real-world data from \citet{polkadot} or suitable elections from Preflib. 
For synthetic data, one may also consult the maps of
\citet{szu-fal-jan-lac-sli-sor-tal:c:how-to-sample-approval-elections}
to check for which parameter choices Euclidean models as well as the
resampling model generate elections that are, in some sense, similar
to those from Pabulib. Indeed, we feel that the resampling model could
be quite an appealing model of generating synthetic approval
elections, but so far there is little evidence to back this view (as
the model was only introduced recently). It would also be interesting
to analyze how elections generated according to various statistical
cultures compare to approval elections from other scenarios than
participatory budgeting.

\section{Conclusions}\label{sec:app-conclusions}

Looking back, we see that impartial culture and real-life data are
popular both in the ordinal and approval settings. While the ordinal
world uses real-life data less frequently and fairly often considers
structured domains, in the approval world the situation is the
opposite. We hope that our analysis will help researchers to see
current trends and approaches, and will allow them to design more
conclusive experiments. For the ordinal setting, we suggest the use of
real-life data, Euclidean models (especially with higher dimensions),
normalized Mallows model, and urn elections (with small contagion
parameter). Impartial culture is a yardstick to measure against
previous papers, and structured domains can give otherwise
difficult-to-spot insights.  For approval elections, Pabulib is a
natural and appealing source of real-life data (for participatory
budgeting). As far as synthetic data goes, Euclidean models and the
resampling models (and, possibly, its mixtures) seem appealing.

\paragraph{Acknowledgments}
This work was funded in part by the French government under management of Agence Nationale de la Recherche as part of the "Investissements d'avenir" program, reference ANR-19-P3IA-0001 (PRAIRIE 3IA Institute).
This project
has received funding from the European Research Council (ERC) under
the European Union’s Horizon 2020 research and innovation programme
(grant agreement No 101002854).
\begin{center}
  \includegraphics[width=2.55cm]{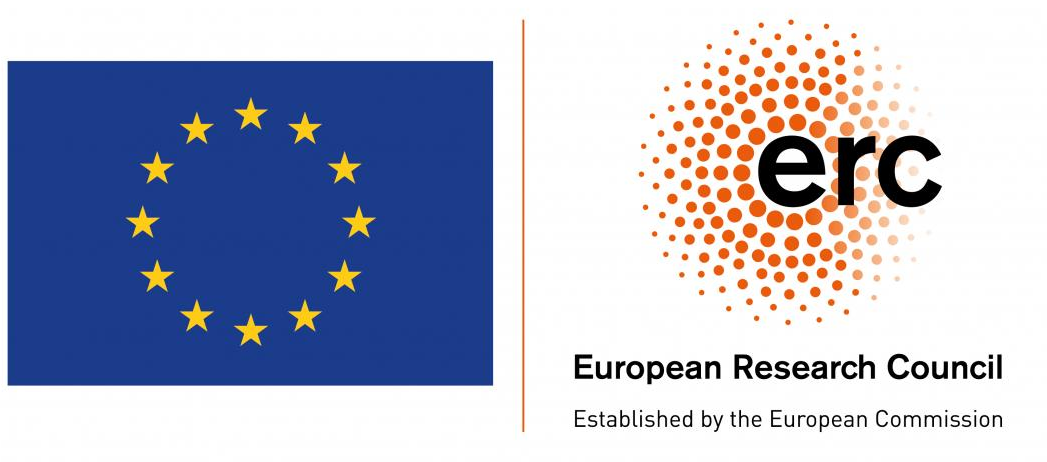}
\end{center}

\end{bibunit}

\appendix

\begin{bibunit}

\clearpage
\clearpage

\begin{center}
     \textbf{\LARGE{Appendix}}
\end{center}
\section{Paper Screening Process}
All the papers downloaded from our conferences had to pass initial
screening.  For a paper to pass this screening it had to include words
related to both elections and experiments. For both categories it had
to include some keyword on at least two pages. We have used the
following keywords:\footnote{Text recovered from PDF files is often
  faulty in the sense that two words may end up glued together, or may
  have some unexpected symbols added before or after. This is why we
  consider prefixes and suffixed and our keywords have somewhat
  specific form.}
\begin{enumerate}
\item[] Elections: \texttt{electi}, \texttt{lection}, \texttt{vot},
	\texttt{ballo}, \texttt{allot}.
\item[] Experiments: \texttt{experime}, \texttt{periment},
  \texttt{empiri}, \texttt{piric}, \texttt{pirical},
  \texttt{simulatio}, \texttt{mulation}, \texttt{mulations}
\end{enumerate}
However, if a matching word also contained as a substring one of the following
forbidden words, then it was disregarded (to decrease the number of false
positives):
\begin{enumerate}
\item[] \texttt{accumul}, \texttt{balloon}, \texttt{vot20}, 
  \texttt{vottir}, \texttt{preselection}, \texttt{flection}, 
  \texttt{formulation}, \texttt{collection}, \texttt{selection}, 
  \texttt{pivot}, \texttt{devot}, \texttt{prelection},
  \texttt{allotment}, \texttt{allotted}.
\end{enumerate}
The somewhat strange form of our keywords is due to the fact that
extracting text from PDF files is sometimes inaccurate and words can
be broken into parts at unexpected places. Thus we selected keywords
that avoided many such problems.

A fairly large number of papers passed our basic screening criterion
without actually considering numerical experiments on elections. This
was intended: We wanted our filter to be fairly nonrestrictive, so
that we would have as few \emph{false negatives} as possible. Below we
list typical reasons why many papers were \emph{false positives}:

\begin{enumerate}
\item Authors make a passing remark to voting.
\item Word ``election'' is recognized as part of ``selection.''
  (Due to inaccurate text extraction, this happens even though we put
  ``selection'' on our list of forbidden words.)
\item Studying text data that includes political discussions.
\item An election-related word appears commonly in another subarea,
  such as, e.g., ``VOTER'' in some community detection papers,
  ``voted-perceptron'' in connection to learning, or the
  ``VOT'' dataset studied in some papers.
\item Using majority voting as a tool for classifiaction or to aggregate data.
\item A form of voting is used by the authors to gather some sort of
  data, or to aggregate data from a questionaire, but in a way that is
  not relevant to our work.
\end{enumerate}
The above reasons typically apply to papers that clearly are out of
scope for our work. Below we mention several reasons to not include
computational social choice papers in the Guide:
\begin{enumerate}
\item Papers looking at two candidates only.
\item Papers that do not actually include experiments, but simply
  discuss their possibility and/or desirability.
\item Papers discussing issues that are close to voting, but
  nonetheless the model used is too far from the kind of elections
  that we consider (examples included aggregating graphs or dependence
  structures in multiissue domains).
\end{enumerate}

\section{Analyzed Papers}\label{sec:papers}

In the table below we list all the papers that made it to the Guide,
together with some notes about the experiments that they include.  For
each paper we include:
\begin{enumerate}
\item The title, authors, year of publication, and the conference
  where it appeared, together with a reference to the bibliography.
\item A list of experiments that we recorded for it (each experiments
  starts either with letter ``O'' for ordinal and ``A'' for approval,
  followed by experiment number and colon).
\item For each experiment we list the statistical cultures/real-life
  data sources used, numbers of samples per data point, followed by
  the considered election sizes (see explanation below for the format
  used). For some experiments we include additional
  notes, e.g., related to the parameters used in various statistical
  cultures, or comments regarding the paper/experiment.
\end{enumerate}

To record the sizes of the considered elections, we write \texttt{CxV},
where \texttt{C} is a string describing the number of candidates and
\texttt{V} is the string describing the number of voters. Each such
string can either be of the form \texttt{\{a,b,c, ...\}}, in which
case it is simply an enumeration of possible values, or of the form
\texttt{[a,b]}, in which case it represents an interval of values
$\{a, a+1, \ldots, b\}$. For example, string
\texttt{\{5,10\}x[20,100]} refers to a set of elections that either
have $5$ or $10$ candidates and between $20$ and $100$ voters. Authors
often consider elections where some parameter---such as the number of
voters---changes with a particular step (e.g., one could consider
between $20$ and $100$ voters, with a step of $5$). We have decided to
omit such details (on the one hand, this simplified the process of
recording data and, on the other hand, we felt that availability of
such data would not affect our analysis too strongly and interested
readers would consult specific papers when needed).

\begin{remark}
  In the table below we present the main contents of the Guide (i.e.,
  our database). For all the papers we tried to find as much relevant
  information as we could, mostly relying on the paper itsefl (but
  occasionally we referred to the full version, if it were available).
  For some of the papers we recorded some details that we found
  interesting, but we did not follow any specific rule in this
  regard. Hence, for some papers we are (most likely) missing such
  comments.  We stress that many comments/details for the papers are
  written in a very concise way. We expect to extend (some of) them as
  the Guide project progresses.

\end{remark}

\begin{remark}
  For some papers we omit certain details, such as the number of
  samples per data points or election sizes. This happens, e.g., if
  such data is not relevant to a given paper (e.g., a paper evaluating
  some property of every real-life election from some set would have
  to list ``one'' as the number of samples, which would feel silly) or
  if it too difficult/cumbersome to obtain this data (e.g., recording
  precisely the sizes of elections from a number of considered
  real-life datasets).

  We occasionally write that some details are unclear in a given
  paper.  This means that we tried to identify the respective bit of
  information and we failed. We will update the Guide as we learn such
  information (provided it is indeed included in the given paper and
  we missed it).
\end{remark}

\begin{remark}
  Whenever we could easily find a journal version of a paper, we
  included a reference to it. In some cases we knew of journal
  versions with different titles than their conference predecessors,
  but we generally did not seek them explicitly. For papers without
  journal versions, we attempted to locate their arXiv versions (but
  we could have failed whenever the authors changed the title).
\end{remark}

\onecolumn
\begin{center}
     \textbf{\LARGE{The Guide}}
\end{center}

\setlength{\tabcolsep}{4pt}
\begin{longtable}{p{0.5cm}|p{3cm}|p{3cm}|p{10cm}}
No. & Title & Authors & Experiments \\
\toprule
\endfirsthead

No. & Title & Authors & Experiments\\
\toprule

\endhead
\endlastfoot
  
\multicolumn{4}{c}{\bf2009}\\ \midrule
1 & \begin{minipage}[t]{3cm}\begin{flushleft}\fontsize{9}{9.5}\selectfont Where Are the Really Hard Manipulation Problems? The Phase Transition
in Manipulating the Veto Rule \vspace{2mm} \newline IJCAI-2009  \vspace{2mm}\newline  \citeapp{wal:c:manipulating-veto} \end{flushleft}\end{minipage} & \begin{minipage}[t]{3cm}\begin{flushleft}\fontsize{9}{9.5}\selectfont T.~Walsh\end{flushleft}\end{minipage} & \fontsize{9}{9.5}\selectfont \begin{minipage}[t]{10cm} \begin{flushleft} O1: Impartial Culture \newline 10000 samples/datapoint, election sizes: \{3\}x[36, 246] \vspace{2mm} \end{flushleft} \end{minipage} \begin{minipage}[t]{10cm} \begin{flushleft}  Sizes note: 3 candidates, {36, 64, 100, 144, 196} “normal” voters and independently [0;50] additional voters (”manipulators”) \newline \vspace{2mm} \end{flushleft} \end{minipage} \begin{minipage}[t]{10cm} \begin{flushleft} \rule{10cm}{0.3pt} \\ O2: Impartial Culture \newline 10000 samples/datapoint, election sizes: Unclear election sizes \vspace{2mm} \end{flushleft} \end{minipage} \begin{minipage}[t]{10cm} \begin{flushleft}  Sizes note: For sure there are here 3 candidates and $\{6, 12, 18, 24\}$ manipulators. The manipulators carry weight drawn uniformly from $(0,k]$ where $\log(k)/m$ is a real number from $[0,2]$. The weight of the remaining voters is also drawn uniformly from $(0,k]$ and we know that the weight of the manipulators is twice greater than the weight of remaining voters. From that, we cannot deduce the exact number of voters. \newline \vspace{2mm} \end{flushleft} \end{minipage} \begin{minipage}[t]{10cm} \begin{flushleft}\rule{10cm}{0.3pt} \\ Comment: There is an arXiv version: https://arxiv.org/abs/0905.3720 \end{flushleft} \end{minipage}  \\ 
\midrule
2 & \begin{minipage}[t]{3cm}\begin{flushleft}\fontsize{9}{9.5}\selectfont Unsupervised Rank Aggregation with Domain-Specific Expertise \vspace{2mm} \newline IJCAI-2009  \vspace{2mm}\newline  \citeapp{kle-rot-sma-tit:c:rank-aggregation-expertise} \end{flushleft}\end{minipage} & \begin{minipage}[t]{3cm}\begin{flushleft}\fontsize{9}{9.5}\selectfont A.~Klementiev, D.~Roth, K.~Small, I.~Titov\end{flushleft}\end{minipage} & \fontsize{9}{9.5}\selectfont \begin{minipage}[t]{10cm} \begin{flushleft} O1: Mallows Mixture \newline 100 samples/datapoint, election sizes: \{30\}x\{10\} \vspace{2mm} \end{flushleft} \end{minipage} \begin{minipage}[t]{10cm} \begin{flushleft}  Mallows: Probability of each mixture component is $\alpha \in (0.4,0.2,0.2,0.1,0.1)$ \vspace{2mm} \end{flushleft} \end{minipage} \begin{minipage}[t]{10cm} \begin{flushleft} \rule{10cm}{0.3pt} \\ O2: Mallows Mixture \newline 100 samples/datapoint, election sizes: \{30\}x\{10\} \vspace{2mm} \end{flushleft} \end{minipage} \begin{minipage}[t]{10cm} \begin{flushleft}  Samples note: Top 30 candidates selected from the pool of 100. \newline \vspace{2mm} \end{flushleft} \end{minipage} \begin{minipage}[t]{10cm} \begin{flushleft}  Mallows: as above  \vspace{2mm} \end{flushleft} \end{minipage}  \\ 
\midrule
\multicolumn{4}{c}{\bf2010}\\ \midrule
3 & \begin{minipage}[t]{3cm}\begin{flushleft}\fontsize{9}{9.5}\selectfont Stackelberg Voting Games: Computational Aspects and Paradoxes \vspace{2mm} \newline AAAI-2010  \vspace{2mm}\newline  \citeapp{xia-con:c:stackelberg-voting-games} \end{flushleft}\end{minipage} & \begin{minipage}[t]{3cm}\begin{flushleft}\fontsize{9}{9.5}\selectfont L.~Xia, V.~Conitzer\end{flushleft}\end{minipage} & \fontsize{9}{9.5}\selectfont \begin{minipage}[t]{10cm} \begin{flushleft} O1: Impartial Culture \newline 25000 samples/datapoint, election sizes: [3, 7]x[2, 20] \vspace{2mm} \end{flushleft} \end{minipage}  \\ 
\midrule
\multicolumn{4}{c}{\bf2011}\\ \midrule
4 & \begin{minipage}[t]{3cm}\begin{flushleft}\fontsize{9}{9.5}\selectfont Solving Election Manipulation Using Integer Partitioning Problems \vspace{2mm} \newline AAMAS-2011  \vspace{2mm}\newline  \citeapp{lin:c:manipulation} \end{flushleft}\end{minipage} & \begin{minipage}[t]{3cm}\begin{flushleft}\fontsize{9}{9.5}\selectfont A.~Lin\end{flushleft}\end{minipage} & \fontsize{9}{9.5}\selectfont \begin{minipage}[t]{10cm} \begin{flushleft} O1: Impartial Culture \newline election sizes: \{4\}x[4, 324], \{5\}x[4, 49], [3, 8]x\{4\}, [4, 12]x\{4\} \vspace{2mm} \end{flushleft} \end{minipage} \begin{minipage}[t]{10cm} \begin{flushleft}  Samples note: Unclear what is the number of samples, but plots show $95\%$ confidence intervals \newline \vspace{2mm} \end{flushleft} \end{minipage}  \\ 
\midrule
5 & \begin{minipage}[t]{3cm}\begin{flushleft}\fontsize{9}{9.5}\selectfont Majority-Rule-Based Preference Aggregation on Multi-Attribute Domains with CP-Nets \vspace{2mm} \newline AAMAS-2011  \vspace{2mm}\newline  \citeapp{li-vo-kow:c:cp-nets} \end{flushleft}\end{minipage} & \begin{minipage}[t]{3cm}\begin{flushleft}\fontsize{9}{9.5}\selectfont M.~Li, Q.~Vo, R.~Kowalczyk\end{flushleft}\end{minipage} & \fontsize{9}{9.5}\selectfont \begin{minipage}[t]{10cm} \begin{flushleft} O1: Impartial Culture \newline 5000 samples/datapoint, election sizes: \{4, 8, 16, 32, 64, 128, 256, 512, 1024\}x\{5, 15\} \vspace{2mm} \end{flushleft} \end{minipage} \begin{minipage}[t]{10cm} \begin{flushleft}  Sizes note: CP-nets with 2 to 10 binary issues \newline \vspace{2mm} \end{flushleft} \end{minipage} \begin{minipage}[t]{10cm} \begin{flushleft}  Note (general): No details on generating the CP-nets: “we generate 5,000 random examples of the agents’ CP-nets.” \vspace{2mm} \end{flushleft} \end{minipage} \begin{minipage}[t]{10cm} \begin{flushleft} \rule{10cm}{0.3pt} \\ O2: Impartial Culture \newline 100 samples/datapoint, election sizes: \{2048\}x\{5, 15\} \vspace{2mm} \end{flushleft} \end{minipage} \begin{minipage}[t]{10cm} \begin{flushleft}  Sizes note: CP-nets with 11 binary issues \newline \vspace{2mm} \end{flushleft} \end{minipage} \begin{minipage}[t]{10cm} \begin{flushleft}  Note (general): No details how the CP-nets are generated \vspace{2mm} \end{flushleft} \end{minipage}  \\ 
\midrule
6 & \begin{minipage}[t]{3cm}\begin{flushleft}\fontsize{9}{9.5}\selectfont Manipulation of Nanson's and Baldwin's Rules \vspace{2mm} \newline AAAI-2011  \vspace{2mm}\newline  \citeapp{nar-wal-xia:c:manipulation-baldwin-nanson} \end{flushleft}\end{minipage} & \begin{minipage}[t]{3cm}\begin{flushleft}\fontsize{9}{9.5}\selectfont N.~Narodytska, T.~Walsh, L.~Xia\end{flushleft}\end{minipage} & \fontsize{9}{9.5}\selectfont \begin{minipage}[t]{10cm} \begin{flushleft} O1: Impartial Culture, Urn Model \newline 3000 samples/datapoint, election sizes: \{5\}x\{5\} \vspace{2mm} \end{flushleft} \end{minipage} \begin{minipage}[t]{10cm} \begin{flushleft}  Samples note: “We threw out the $20\%$ or so of problems generated in which the chosen candidate has already won before the manipulators vote” \newline Sizes note: Election size included non-manipulators only (for manipulators the authors do not generate preference orders) \newline \vspace{2mm} \end{flushleft} \end{minipage} \begin{minipage}[t]{10cm} \begin{flushleft}  Urn: alpha = 1 (contagion parameter, in the paper refrerred to as a = m!) \vspace{2mm} \end{flushleft} \end{minipage} \begin{minipage}[t]{10cm} \begin{flushleft} \rule{10cm}{0.3pt} \\ O2: Impartial Culture, Urn Model \newline 1000 samples/datapoint, election sizes: \{4\}x\{4\}, \{8\}x\{8\}, \{16\}x\{16\}, \{32\}x\{32\}, \{64\}x\{64\}, \{128\}x\{128\} \vspace{2mm} \end{flushleft} \end{minipage} \begin{minipage}[t]{10cm} \begin{flushleft}  Sizes note: Election size included non-manipulators only (for manipulators the authors do not generate preference orders). Equal number of candidates and voters on purpose. \newline \vspace{2mm} \end{flushleft} \end{minipage} \begin{minipage}[t]{10cm} \begin{flushleft}  Urn: as above  \vspace{2mm} \end{flushleft} \end{minipage} \begin{minipage}[t]{10cm} \begin{flushleft}\rule{10cm}{0.3pt} \\ Comment: There is a journal version: Jessica Davies, George Katsirelos, Nina Narodytska, Toby Walsh, Lirong Xia:
Complexity of and algorithms for the manipulation of Borda, Nanson's and Baldwin's voting rules. Artif. Intell. 217: 20-42 (2014) \end{flushleft} \end{minipage}  \\ 
\midrule
7 & \begin{minipage}[t]{3cm}\begin{flushleft}\fontsize{9}{9.5}\selectfont Complexity of and Algorithms for Borda Manipulation \vspace{2mm} \newline AAAI-2011  \vspace{2mm}\newline  \citeapp{dav-kat-nar-wal:c:manipulation-borda} \end{flushleft}\end{minipage} & \begin{minipage}[t]{3cm}\begin{flushleft}\fontsize{9}{9.5}\selectfont J.~Davies, G.~Katsirelos, N.~Narodytska, T.~Walsh\end{flushleft}\end{minipage} & \fontsize{9}{9.5}\selectfont \begin{minipage}[t]{10cm} \begin{flushleft} O1: Impartial Culture, Urn Model \newline 1000 samples/datapoint, election sizes: \{4, 8, 16, 32, 64, 128\}x\{4, 8, 16, 32, 64, 128\} \vspace{2mm} \end{flushleft} \end{minipage} \begin{minipage}[t]{10cm} \begin{flushleft}  Sizes note: Unclear election sizes (the authors mention 1000 elections per problem size, but then they report, e.g., 32679 unique IC elections; this suggests that they try each combination of the number of candidates and voters, and not equal numbers of candidates and voters, as done in the related paper “Manipulation of Nanson's and Baldwin's Rules”) \newline \vspace{2mm} \end{flushleft} \end{minipage} \begin{minipage}[t]{10cm} \begin{flushleft}  Urn: alpha = 1 (b = m! in the language of the paper) \vspace{2mm} \end{flushleft} \end{minipage} \begin{minipage}[t]{10cm} \begin{flushleft}\rule{10cm}{0.3pt} \\ Comment: There is a journal version: Jessica Davies, George Katsirelos, Nina Narodytska, Toby Walsh, Lirong Xia:
Complexity of and algorithms for the manipulation of Borda, Nanson's and Baldwin's voting rules. Artif. Intell. 217: 20-42 (2014) \end{flushleft} \end{minipage}  \\ 
\midrule
8 & \begin{minipage}[t]{3cm}\begin{flushleft}\fontsize{9}{9.5}\selectfont Budgeted Social Choice: From Consensus to Personalized Decision Making \vspace{2mm} \newline IJCAI-2011  \vspace{2mm}\newline  \citeapp{lu-bou:c:budgeted-social-choice} \end{flushleft}\end{minipage} & \begin{minipage}[t]{3cm}\begin{flushleft}\fontsize{9}{9.5}\selectfont T.~Lu, C.~Boutilier\end{flushleft}\end{minipage} & \fontsize{9}{9.5}\selectfont \begin{minipage}[t]{10cm} \begin{flushleft} O1: PrefLib \newline  \vspace{2mm} \end{flushleft} \end{minipage} \begin{minipage}[t]{10cm} \begin{flushleft}  Sizes note: Elections sizes are as in Preflib datasets used (see also Sec. 5 in the paper for details) \newline \vspace{2mm} \end{flushleft} \end{minipage} \begin{minipage}[t]{10cm} \begin{flushleft}  Note (general): Uses APA dataset, Sushi dataset. The tests particular elections, but also evaluates how the results depend on changing item costs. \vspace{2mm} \end{flushleft} \end{minipage}  \\ 
\midrule
9 & \begin{minipage}[t]{3cm}\begin{flushleft}\fontsize{9}{9.5}\selectfont Robust Approximation and Incremental Elicitation in Voting Protocols \vspace{2mm} \newline IJCAI-2011  \vspace{2mm}\newline  \citeapp{lu-bou:c:max-regret} \end{flushleft}\end{minipage} & \begin{minipage}[t]{3cm}\begin{flushleft}\fontsize{9}{9.5}\selectfont T.~Lu, C.~Boutilier\end{flushleft}\end{minipage} & \fontsize{9}{9.5}\selectfont \begin{minipage}[t]{10cm} \begin{flushleft} O1: PrefLib \newline  \vspace{2mm} \end{flushleft} \end{minipage} \begin{minipage}[t]{10cm} \begin{flushleft}  Note (general): Uses Sushi dataset and Irish dataset \vspace{2mm} \end{flushleft} \end{minipage} \begin{minipage}[t]{10cm} \begin{flushleft} \rule{10cm}{0.3pt} \\ O2: Mallows \newline 10 samples/datapoint, election sizes: \{20\}x\{100\} \vspace{2mm} \end{flushleft} \end{minipage} \begin{minipage}[t]{10cm} \begin{flushleft}  Mallows: $\phi \in \{0.1, 0.5, 0.7, 0.9, 1\}$ \vspace{2mm} \end{flushleft} \end{minipage}  \\ 
\midrule
10 & \begin{minipage}[t]{3cm}\begin{flushleft}\fontsize{9}{9.5}\selectfont Choosing Collectively Optimal Sets of Alternatives Based on the Condorcet Criterion \vspace{2mm} \newline IJCAI-2011  \vspace{2mm}\newline  \citeapp{elk-lan-saf:c:condorcet-winning-sets} \end{flushleft}\end{minipage} & \begin{minipage}[t]{3cm}\begin{flushleft}\fontsize{9}{9.5}\selectfont E.~Elkind, J.~Lang, A.~Saffidine\end{flushleft}\end{minipage} & \fontsize{9}{9.5}\selectfont \begin{minipage}[t]{10cm} \begin{flushleft} O1: Impartial Culture \newline election sizes: \{15, 50, 100\}x\{10, 11, 20, 21, 100, 101\},  \{30\}x[1, 100],  \{20\}x\{5, 8, 20, 500, 2000\} \vspace{2mm} \end{flushleft} \end{minipage} \begin{minipage}[t]{10cm} \begin{flushleft}\rule{10cm}{0.3pt} \\ Comment: There is a journal version: 
Edith Elkind, Jérôme Lang, Abdallah Saffidine:
Condorcet winning sets. Soc. Choice Welf. 44(3): 493-517 (2015) \end{flushleft} \end{minipage}  \\ 
\midrule
11 & \begin{minipage}[t]{3cm}\begin{flushleft}\fontsize{9}{9.5}\selectfont On the Fixed-Parameter Tractability of Composition-Consistent Tournament Solutions \vspace{2mm} \newline IJCAI-2011  \vspace{2mm}\newline  \citeapp{bra-bri-see:c:tournaments-composition} \end{flushleft}\end{minipage} & \begin{minipage}[t]{3cm}\begin{flushleft}\fontsize{9}{9.5}\selectfont F.~Brandt, M.~Brill, H.~Seedig\end{flushleft}\end{minipage} & \fontsize{9}{9.5}\selectfont \begin{minipage}[t]{10cm} \begin{flushleft} O1: Euclidean 2D, Mallows \newline 30 samples/datapoint, election sizes: \{10, 50, 100, 150, 200\}x[5, 2000] \vspace{2mm} \end{flushleft} \end{minipage} \begin{minipage}[t]{10cm} \begin{flushleft}  Mallows: Condorcet noise model with $p = 0.55$ \vspace{2mm} \end{flushleft} \end{minipage} \begin{minipage}[t]{10cm} \begin{flushleft}  Euclidean: Euclidean data uses uniform distribution on $[0,1]^2$ \vspace{2mm} \end{flushleft} \end{minipage} \begin{minipage}[t]{10cm} \begin{flushleft} \rule{10cm}{0.3pt} \\ O2: Mallows \newline 30 samples/datapoint, election sizes: \{30\}x[1, 1000] \vspace{2mm} \end{flushleft} \end{minipage} \begin{minipage}[t]{10cm} \begin{flushleft}  Mallows: as above  \vspace{2mm} \end{flushleft} \end{minipage} \begin{minipage}[t]{10cm} \begin{flushleft}\rule{10cm}{0.3pt} \\ Comment: The authors consider tournaments, but use preference profiles to generate them \end{flushleft} \end{minipage}  \\ 
\midrule
12 & \begin{minipage}[t]{3cm}\begin{flushleft}\fontsize{9}{9.5}\selectfont Hypercubewise Preference Aggregation in Multi-Issue Domains \vspace{2mm} \newline IJCAI-2011  \vspace{2mm}\newline  \citeapp{con-lan-xia:c:cp-nets} \end{flushleft}\end{minipage} & \begin{minipage}[t]{3cm}\begin{flushleft}\fontsize{9}{9.5}\selectfont V.~Conitzer, J.~Lang, L.~Xia\end{flushleft}\end{minipage} & \fontsize{9}{9.5}\selectfont \begin{minipage}[t]{10cm} \begin{flushleft} O1: Impartial Culture \newline 20000 samples/datapoint, election sizes: \{4, 8, 16, 32, 64, 128, 256, 512, 1024, 2048, 4096, 8192, 16384, 32768\}x\{?\} \vspace{2mm} \end{flushleft} \end{minipage} \begin{minipage}[t]{10cm} \begin{flushleft}  Sizes note: The authors do not seem to report the number of voters. Or is it the number of samples? \newline \vspace{2mm} \end{flushleft} \end{minipage} \begin{minipage}[t]{10cm} \begin{flushleft}  Note (general): CP-nets are drawn uniformly at random \vspace{2mm} \end{flushleft} \end{minipage} \begin{minipage}[t]{10cm} \begin{flushleft} \rule{10cm}{0.3pt} \\ O2: Impartial Culture \newline 10000 samples/datapoint, election sizes: \{4, 8, 16, 32, 64, 128, 256, 512, 1024, 2048, 4096, 8192, 16384, 32768\}x\{501, 601, 701, 801, 901, 1001\} \vspace{2mm} \end{flushleft} \end{minipage} \begin{minipage}[t]{10cm} \begin{flushleft}\rule{10cm}{0.3pt} \\ Comment: Considers CP-nets \end{flushleft} \end{minipage}  \\ 
\midrule
\multicolumn{4}{c}{\bf2012}\\ \midrule
13 & \begin{minipage}[t]{3cm}\begin{flushleft}\fontsize{9}{9.5}\selectfont Voter Response to Iterated Poll Information \vspace{2mm} \newline AAMAS-2012  \vspace{2mm}\newline  \citeapp{rei-end:c:iterated-poll-condorcet} \end{flushleft}\end{minipage} & \begin{minipage}[t]{3cm}\begin{flushleft}\fontsize{9}{9.5}\selectfont A.~Reijngoud, U.~Endriss\end{flushleft}\end{minipage} & \fontsize{9}{9.5}\selectfont \begin{minipage}[t]{10cm} \begin{flushleft} O1: Impartial Culture \newline 10000 samples/datapoint, election sizes: \{5\}x[10, 100], [3, 15]x\{50\} \vspace{2mm} \end{flushleft} \end{minipage} \begin{minipage}[t]{10cm} \begin{flushleft}  Note (general): IC with Condorcet winner (?) “Nevertheless, the IC assumption is still the de facto standard used in social choice theory; results based on it provide an important base line and allow for direct comparison with a large number of findings documented in the literature.” \vspace{2mm} \end{flushleft} \end{minipage}  \\ 
\midrule
\multicolumn{4}{c}{\bf2013}\\ \midrule
14 & \begin{minipage}[t]{3cm}\begin{flushleft}\fontsize{9}{9.5}\selectfont Empirical analysis of plurality election equilibria \vspace{2mm} \newline AAMAS-2013  \vspace{2mm}\newline  \citeapp{tho-lev-ley-res:c:plurality-election-equlibria} \end{flushleft}\end{minipage} & \begin{minipage}[t]{3cm}\begin{flushleft}\fontsize{9}{9.5}\selectfont D.~Thompson, O.~Lev, K.~Leyton{-}Brown, J.~Rosenschein\end{flushleft}\end{minipage} & \fontsize{9}{9.5}\selectfont \begin{minipage}[t]{10cm} \begin{flushleft} O1: Impartial Culture \newline 1000 samples/datapoint, election sizes: \{5\}x\{10\} \vspace{2mm} \end{flushleft} \end{minipage} \begin{minipage}[t]{10cm} \begin{flushleft}  Note (general): The authors do not write explicitly that they use IC, instead they write the preferences “are chosen randomly” \vspace{2mm} \end{flushleft} \end{minipage} \begin{minipage}[t]{10cm} \begin{flushleft} \rule{10cm}{0.3pt} \\ O2: Impartial Culture \newline 1000 samples/datapoint, election sizes: \{3\}x[3, 96], \{3, 4, 5\}x[3, 20] \vspace{2mm} \end{flushleft} \end{minipage} \begin{minipage}[t]{10cm} \begin{flushleft}  Samples note: Not explicitly stated; the flow of the text suggests that it remains the same as the previous experiment in this paper. \newline \vspace{2mm} \end{flushleft} \end{minipage} \begin{minipage}[t]{10cm} \begin{flushleft}  Note (general): as above \vspace{2mm} \end{flushleft} \end{minipage} \begin{minipage}[t]{10cm} \begin{flushleft} \rule{10cm}{0.3pt} \\ O3: Impartial Culture \newline 1000 samples/datapoint, election sizes: \{5\}x\{10\} \vspace{2mm} \end{flushleft} \end{minipage} \begin{minipage}[t]{10cm} \begin{flushleft}  Samples note: as above \newline \vspace{2mm} \end{flushleft} \end{minipage} \begin{minipage}[t]{10cm} \begin{flushleft}  Note (general): The authors do not write explicitly that they use IC, instead they write the preferences “are chosen randomly”. This experiment differs from the seemingly identical one in this paper by the studied games—this experiments consider imperfect information games \vspace{2mm} \end{flushleft} \end{minipage} \begin{minipage}[t]{10cm} \begin{flushleft}\rule{10cm}{0.3pt} \\ Comment: Game-theoretical approach to manipulation. The paper also contains some approval experiments, but with very limited details (this includes k-approval, veto, but also pure approval) \end{flushleft} \end{minipage}  \\ 
\midrule
15 & \begin{minipage}[t]{3cm}\begin{flushleft}\fontsize{9}{9.5}\selectfont Achieving Fully Proportional Representation Is Easy in Practice \vspace{2mm} \newline AAMAS-2013  \vspace{2mm}\newline  \citeapp{sko-fal-sli:c:cc-monroe-experiments} \end{flushleft}\end{minipage} & \begin{minipage}[t]{3cm}\begin{flushleft}\fontsize{9}{9.5}\selectfont P.~Skowron, P.~Faliszewski, A.~Slinko\end{flushleft}\end{minipage} & \fontsize{9}{9.5}\selectfont \begin{minipage}[t]{10cm} \begin{flushleft} O1: Impartial Culture, Mallows Mixture, PrefLib, Urn Model \newline 500 samples/datapoint, election sizes: \{10\}x\{100\}, \{6\}x\{9\} \vspace{2mm} \end{flushleft} \end{minipage} \begin{minipage}[t]{10cm} \begin{flushleft}  Mallows: Mixture of 5 Mallows models, whose dispersion parameters and central orders are selected uniformly at random. Mixing probabilities are selected uniformly at random as well. \vspace{2mm} \end{flushleft} \end{minipage} \begin{minipage}[t]{10cm} \begin{flushleft}  Urn: {$\alpha = 0.05$} \vspace{2mm} \end{flushleft} \end{minipage} \begin{minipage}[t]{10cm} \begin{flushleft}  Note (parameters): PrefLib datasets: Sushi, Netflix, AGH courses \vspace{2mm} \end{flushleft} \end{minipage} \begin{minipage}[t]{10cm} \begin{flushleft} \rule{10cm}{0.3pt} \\ O2: Impartial Culture, Mallows Mixture, PrefLib, Urn Model \newline 50 samples/datapoint, election sizes: \{10, 100\}x[1000, 10000] \vspace{2mm} \end{flushleft} \end{minipage} \begin{minipage}[t]{10cm} \begin{flushleft}  Mallows: as above  \vspace{2mm} \end{flushleft} \end{minipage} \begin{minipage}[t]{10cm} \begin{flushleft}  Urn: as above  \vspace{2mm} \end{flushleft} \end{minipage} \begin{minipage}[t]{10cm} \begin{flushleft}  Note (parameters): PrefLib datasets: Netflix, Sushi (partially) \vspace{2mm} \end{flushleft} \end{minipage} \begin{minipage}[t]{10cm} \begin{flushleft}  Note (general): Step of 1000 in the number of voters \vspace{2mm} \end{flushleft} \end{minipage} \begin{minipage}[t]{10cm} \begin{flushleft} \rule{10cm}{0.3pt} \\ O3: PrefLib, Urn Model \newline 50 samples/datapoint, election sizes: [30, 300]x\{1000\} \vspace{2mm} \end{flushleft} \end{minipage} \begin{minipage}[t]{10cm} \begin{flushleft}  Urn: as above  \vspace{2mm} \end{flushleft} \end{minipage} \begin{minipage}[t]{10cm} \begin{flushleft}  Note (parameters): PrefLib datasets: Netflix \vspace{2mm} \end{flushleft} \end{minipage} \begin{minipage}[t]{10cm} \begin{flushleft}  Note (general): Step of 30 in the number of candidates \vspace{2mm} \end{flushleft} \end{minipage} \begin{minipage}[t]{10cm} \begin{flushleft} \rule{10cm}{0.3pt} \\ O4: PrefLib, Urn Model \newline 50 samples/datapoint, election sizes: \{10, 30\}x[50, 400] \vspace{2mm} \end{flushleft} \end{minipage} \begin{minipage}[t]{10cm} \begin{flushleft}  Urn: as above  \vspace{2mm} \end{flushleft} \end{minipage} \begin{minipage}[t]{10cm} \begin{flushleft}  Note (parameters): PrefLib datasets: Netflix, Sushi (partially) \vspace{2mm} \end{flushleft} \end{minipage} \begin{minipage}[t]{10cm} \begin{flushleft}  Note (general): This experiment had changing committee size \vspace{2mm} \end{flushleft} \end{minipage} \begin{minipage}[t]{10cm} \begin{flushleft} \rule{10cm}{0.3pt} \\ O5: PrefLib \newline election sizes: \{100\}x\{1000\} \vspace{2mm} \end{flushleft} \end{minipage} \begin{minipage}[t]{10cm} \begin{flushleft}  Samples note: Unclear how many samples per data point were used \newline \vspace{2mm} \end{flushleft} \end{minipage} \begin{minipage}[t]{10cm} \begin{flushleft}  Note (parameters): PrefLib datasets: Netflix \vspace{2mm} \end{flushleft} \end{minipage} \begin{minipage}[t]{10cm} \begin{flushleft}\rule{10cm}{0.3pt} \\ Comment: There is a journal version: Piotr Skowron, Piotr Faliszewski, Arkadii M. Slinko: Achieving fully proportional representation: Approximability results. Artif. Intell. 222: 67-103 (2015) \end{flushleft} \end{minipage}  \\ 
\midrule
16 & \begin{minipage}[t]{3cm}\begin{flushleft}\fontsize{9}{9.5}\selectfont Better Human Computation Through Principled Voting \vspace{2mm} \newline AAAI-2013  \vspace{2mm}\newline  \citeapp{mao-pro-che:c:truth-elicitation} \end{flushleft}\end{minipage} & \begin{minipage}[t]{3cm}\begin{flushleft}\fontsize{9}{9.5}\selectfont A.~Mao, A.~Procaccia, Y.~Chen\end{flushleft}\end{minipage} & \fontsize{9}{9.5}\selectfont \begin{minipage}[t]{10cm} \begin{flushleft} O1: Mallows, Thurstone-Mosteller \newline 100000 samples/datapoint, election sizes: \{4\}x\{10\} \vspace{2mm} \end{flushleft} \end{minipage} \begin{minipage}[t]{10cm} \begin{flushleft}  Mallows: $\phi \in \{0.55, 0.57, 0.59, 0.61, 0.63, 0.65, 0.67\}$ \vspace{2mm} \end{flushleft} \end{minipage} \begin{minipage}[t]{10cm} \begin{flushleft}  Note (general): Thurstone-Mosteller with uniformly spaced strength of candidates with differences 0.1, 0.15, 0.2, 0.25, 0.3, 0.35, and 0.4 - the variance of error is not stated, which probably means it is 1 \vspace{2mm} \end{flushleft} \end{minipage} \begin{minipage}[t]{10cm} \begin{flushleft} \rule{10cm}{0.3pt} \\ O2: PrefLib \newline 40 samples/datapoint, election sizes: \{4\}x\{20\} \vspace{2mm} \end{flushleft} \end{minipage} \begin{minipage}[t]{10cm} \begin{flushleft}  Sizes note: 20 voters is approximate \newline \vspace{2mm} \end{flushleft} \end{minipage} \begin{minipage}[t]{10cm} \begin{flushleft}  Note (general): PrefLib datasets: MTurk Dots and MTurk Puzzle (this paper introduced these datasets) \vspace{2mm} \end{flushleft} \end{minipage}  \\ 
\midrule
17 & \begin{minipage}[t]{3cm}\begin{flushleft}\fontsize{9}{9.5}\selectfont Social Rankings in Human-Computer Committees \vspace{2mm} \newline AAAI-2013  \vspace{2mm}\newline  \citeapp{bit-gal-kra-dok-aza:c:social-rankings-hcc} \end{flushleft}\end{minipage} & \begin{minipage}[t]{3cm}\begin{flushleft}\fontsize{9}{9.5}\selectfont M.~Bitan, Y.~Gal, S.~Kraus, E.~Dokow, A.~Azaria\end{flushleft}\end{minipage} & \fontsize{9}{9.5}\selectfont \begin{minipage}[t]{10cm} \begin{flushleft} O1: Hand-Crafted \newline 80 samples/datapoint, election sizes: \{4\}x\{3\} \vspace{2mm} \end{flushleft} \end{minipage} \begin{minipage}[t]{10cm} \begin{flushleft}  Samples note: The initial voters’ preferences are always the same in this experiment. The goal is to analyze how they evolve in time within a specific game played by both human and computer players (in different proportions). For each such combination, 80 rounds are played. \newline \vspace{2mm} \end{flushleft} \end{minipage}  \\ 
\midrule
18 & \begin{minipage}[t]{3cm}\begin{flushleft}\fontsize{9}{9.5}\selectfont Ties Matter: Complexity of Manipulation when Tie-Breaking with a Random
Vote \vspace{2mm} \newline AAAI-2013  \vspace{2mm}\newline  \citeapp{azi-gas-mat-nar-wal:c:ties-manipulation-random} \end{flushleft}\end{minipage} & \begin{minipage}[t]{3cm}\begin{flushleft}\fontsize{9}{9.5}\selectfont H.~Aziz, S.~Gaspers, N.~Mattei, N.~Narodytska, T.~Walsh\end{flushleft}\end{minipage} & \fontsize{9}{9.5}\selectfont \begin{minipage}[t]{10cm} \begin{flushleft} O1: Impartial Anonymous Culture, Impartial Culture, Single-Peaked (Walsh/Uniform), Urn Model \newline 1000 samples/datapoint, election sizes: \{3, 4, 8, 16, 32, 64, 128\}x\{1, 2, 3, 4, 5, 8, 9,  16, 17, 32, 33, 64, 65, 128, 129\} \vspace{2mm} \end{flushleft} \end{minipage} \begin{minipage}[t]{10cm} \begin{flushleft}  Urn: replacement was set to $m! - 1$ ($b=m!-1$?) \vspace{2mm} \end{flushleft} \end{minipage}  \\ 
\midrule
19 & \begin{minipage}[t]{3cm}\begin{flushleft}\fontsize{9}{9.5}\selectfont Multi-Dimensional Single-Peaked Consistency and Its Approximations \vspace{2mm} \newline IJCAI-2013  \vspace{2mm}\newline  \citeapp{sui-fra-bou:c:multi-dimensional} \end{flushleft}\end{minipage} & \begin{minipage}[t]{3cm}\begin{flushleft}\fontsize{9}{9.5}\selectfont X.~Sui, A.~Francois{-}Nienaber, C.~Boutilier\end{flushleft}\end{minipage} & \fontsize{9}{9.5}\selectfont \begin{minipage}[t]{10cm} \begin{flushleft} O1: PrefLib \newline election sizes: \{9\}x\{3800\} \vspace{2mm} \end{flushleft} \end{minipage} \begin{minipage}[t]{10cm} \begin{flushleft}  Sizes note: Dublin West (only complete rankings) \newline \vspace{2mm} \end{flushleft} \end{minipage} \begin{minipage}[t]{10cm} \begin{flushleft} \rule{10cm}{0.3pt} \\ O2: PrefLib \newline election sizes: \{9\}x\{29989\} \vspace{2mm} \end{flushleft} \end{minipage} \begin{minipage}[t]{10cm} \begin{flushleft}  Sizes note: Dublin West (all rankings) \newline \vspace{2mm} \end{flushleft} \end{minipage} \begin{minipage}[t]{10cm} \begin{flushleft} \rule{10cm}{0.3pt} \\ O3: PrefLib \newline election sizes: \{12\}x\{3662\} \vspace{2mm} \end{flushleft} \end{minipage} \begin{minipage}[t]{10cm} \begin{flushleft}  Sizes note: Dublin North (only complete rankings) \newline \vspace{2mm} \end{flushleft} \end{minipage} \begin{minipage}[t]{10cm} \begin{flushleft} \rule{10cm}{0.3pt} \\ O4: PrefLib \newline election sizes: \{12\}x\{43942\} \vspace{2mm} \end{flushleft} \end{minipage} \begin{minipage}[t]{10cm} \begin{flushleft}  Sizes note: Dublin North (all rankings) \newline \vspace{2mm} \end{flushleft} \end{minipage}  \\ 
\midrule
20 & \begin{minipage}[t]{3cm}\begin{flushleft}\fontsize{9}{9.5}\selectfont Efficient Vote Elicitation under Candidate Uncertainty \vspace{2mm} \newline IJCAI-2013  \vspace{2mm}\newline  \citeapp{ore-fil-bou:c:efficient} \end{flushleft}\end{minipage} & \begin{minipage}[t]{3cm}\begin{flushleft}\fontsize{9}{9.5}\selectfont J.~Oren, Y.~Filmus, C.~Boutilier\end{flushleft}\end{minipage} & \fontsize{9}{9.5}\selectfont \begin{minipage}[t]{10cm} \begin{flushleft} O1: Mallows \newline 10000 samples/datapoint, election sizes: \{10\}x[100, 5000] \vspace{2mm} \end{flushleft} \end{minipage} \begin{minipage}[t]{10cm} \begin{flushleft}  Mallows: $\phi \in \{0.6, 0.7, 0.8, 0.9\}$ \vspace{2mm} \end{flushleft} \end{minipage}  \\ 
\midrule
21 & \begin{minipage}[t]{3cm}\begin{flushleft}\fontsize{9}{9.5}\selectfont Multi-Winner Social Choice with Incomplete Preferences \vspace{2mm} \newline IJCAI-2013  \vspace{2mm}\newline  \citeapp{lu-bou:c:multiwinner-incomplete} \end{flushleft}\end{minipage} & \begin{minipage}[t]{3cm}\begin{flushleft}\fontsize{9}{9.5}\selectfont T.~Lu, C.~Boutilier\end{flushleft}\end{minipage} & \fontsize{9}{9.5}\selectfont \begin{minipage}[t]{10cm} \begin{flushleft} O1: Mallows, PrefLib \newline 20 samples/datapoint, election sizes: \{9, 10\}x\{100\} \vspace{2mm} \end{flushleft} \end{minipage} \begin{minipage}[t]{10cm} \begin{flushleft}  Sizes note: Committee size k=3 and k=4. Sushi and Mallows uses 10 candidates, Irish uses 9 candidates \newline \vspace{2mm} \end{flushleft} \end{minipage} \begin{minipage}[t]{10cm} \begin{flushleft}  Mallows: $\phi \in \{0.3, 0.5, 0.7, 0.9, 1\}$ \vspace{2mm} \end{flushleft} \end{minipage} \begin{minipage}[t]{10cm} \begin{flushleft}  Note (general): Real-life datasets: Sushi, Irish (Dublin West) \vspace{2mm} \end{flushleft} \end{minipage} \begin{minipage}[t]{10cm} \begin{flushleft} \rule{10cm}{0.3pt} \\ O2: Mallows \newline election sizes: \{10, 20, 30, 50\}x\{50\} \vspace{2mm} \end{flushleft} \end{minipage} \begin{minipage}[t]{10cm} \begin{flushleft}  Samples note: No information about the number of samples per data point \newline \vspace{2mm} \end{flushleft} \end{minipage} \begin{minipage}[t]{10cm} \begin{flushleft}  Mallows: $\phi = 0.7$ \vspace{2mm} \end{flushleft} \end{minipage}  \\ 
\midrule
22 & \begin{minipage}[t]{3cm}\begin{flushleft}\fontsize{9}{9.5}\selectfont How to Change a Group's Collective Decision? \vspace{2mm} \newline IJCAI-2013  \vspace{2mm}\newline  \citeapp{haz-lin-kra:c:change-collective-decision} \end{flushleft}\end{minipage} & \begin{minipage}[t]{3cm}\begin{flushleft}\fontsize{9}{9.5}\selectfont N.~Hazon, R.~Lin, S.~Kraus\end{flushleft}\end{minipage} & \fontsize{9}{9.5}\selectfont \begin{minipage}[t]{10cm} \begin{flushleft} O1: Impartial Culture, Urn Model \newline 1000 samples/datapoint, election sizes: \{8\}x\{32\} \vspace{2mm} \end{flushleft} \end{minipage} \begin{minipage}[t]{10cm} \begin{flushleft}  Urn: $\alpha =1$ (expressed as $b=m!$) \vspace{2mm} \end{flushleft} \end{minipage} \begin{minipage}[t]{10cm} \begin{flushleft} \rule{10cm}{0.3pt} \\ O2: Impartial Culture, Urn Model \newline 1000 samples/datapoint, election sizes: \{8\}x\{16, 32, 64, 128\} \vspace{2mm} \end{flushleft} \end{minipage} \begin{minipage}[t]{10cm} \begin{flushleft}  Urn: as above  \vspace{2mm} \end{flushleft} \end{minipage} \begin{minipage}[t]{10cm} \begin{flushleft} \rule{10cm}{0.3pt} \\ O3: Impartial Culture, Urn Model \newline election sizes: \{8, 16, 32, 64\}x\{8, 16, 32, 64\} \vspace{2mm} \end{flushleft} \end{minipage} \begin{minipage}[t]{10cm} \begin{flushleft}  Samples note: In case of the urn model 900 samples were used, whereas 600 samples were included for the impartial culture case. \newline Sizes note: Some of the profiles were excluded from the results of both of the cultures. \newline \vspace{2mm} \end{flushleft} \end{minipage} \begin{minipage}[t]{10cm} \begin{flushleft}  Urn: as above  \vspace{2mm} \end{flushleft} \end{minipage} \begin{minipage}[t]{10cm} \begin{flushleft}\rule{10cm}{0.3pt} \\ Comment: DOI not easily obtainable. Link to the paper: https://www.ijcai.org/Abstract/13/039 \end{flushleft} \end{minipage}  \\ 
\midrule
23 & \begin{minipage}[t]{3cm}\begin{flushleft}\fontsize{9}{9.5}\selectfont Elicitation and Approximately Stable Matching with Partial Preferences \vspace{2mm} \newline IJCAI-2013  \vspace{2mm}\newline  \citeapp{dru-bou:c:elicitation-matching-partial} \end{flushleft}\end{minipage} & \begin{minipage}[t]{3cm}\begin{flushleft}\fontsize{9}{9.5}\selectfont J.~Drummond, C.~Boutilier\end{flushleft}\end{minipage} & \fontsize{9}{9.5}\selectfont \begin{minipage}[t]{10cm} \begin{flushleft} O1: Impartial Culture, Mallows \newline 30 samples/datapoint, election sizes: \{20\}x\{20\} \vspace{2mm} \end{flushleft} \end{minipage} \begin{minipage}[t]{10cm} \begin{flushleft}  Mallows: $\phi = 0.2$ \vspace{2mm} \end{flushleft} \end{minipage} \begin{minipage}[t]{10cm} \begin{flushleft} \rule{10cm}{0.3pt} \\ O2: Mallows \newline 20 samples/datapoint, election sizes: \{250\}x\{250\} \vspace{2mm} \end{flushleft} \end{minipage} \begin{minipage}[t]{10cm} \begin{flushleft}  Mallows: $\phi \in \{0.2, 0.4, 0.6, 0.8, 1 \}$ \vspace{2mm} \end{flushleft} \end{minipage} \begin{minipage}[t]{10cm} \begin{flushleft} \rule{10cm}{0.3pt} \\ O3: Real-Life (beyond PrefLib) \newline 20 samples/datapoint, election sizes: \{250\}x\{250\} \vspace{2mm} \end{flushleft} \end{minipage} \begin{minipage}[t]{10cm} \begin{flushleft}  Samples note: MovieLens dataset. \newline \vspace{2mm} \end{flushleft} \end{minipage} \begin{minipage}[t]{10cm} \begin{flushleft}  Mallows: as above  \vspace{2mm} \end{flushleft} \end{minipage}  \\ 
\midrule
\multicolumn{4}{c}{\bf2014}\\ \midrule
24 & \begin{minipage}[t]{3cm}\begin{flushleft}\fontsize{9}{9.5}\selectfont Empathetic social choice on social networks \vspace{2mm} \newline AAMAS-2014  \vspace{2mm}\newline  \citeapp{sal-bou:c:empathetic-soc-on-networks} \end{flushleft}\end{minipage} & \begin{minipage}[t]{3cm}\begin{flushleft}\fontsize{9}{9.5}\selectfont A.~Salehi{-}Abari, C.~Boutilier\end{flushleft}\end{minipage} & \fontsize{9}{9.5}\selectfont \begin{minipage}[t]{10cm} \begin{flushleft} O1: PrefLib \newline 2500 samples/datapoint, election sizes: \{9\}x\{1000\} \vspace{2mm} \end{flushleft} \end{minipage} \begin{minipage}[t]{10cm} \begin{flushleft}  Samples note: 50 preference profiles for each of 50 random networks \newline \vspace{2mm} \end{flushleft} \end{minipage} \begin{minipage}[t]{10cm} \begin{flushleft}  Note (general): Irish dataset, Dublin West, sampled from the subset of true preferences (only complete votes). Data present in PrefLib, but taken independently from the source. \vspace{2mm} \end{flushleft} \end{minipage} \begin{minipage}[t]{10cm} \begin{flushleft} \rule{10cm}{0.3pt} \\ O2: Impartial Culture \newline 2500 samples/datapoint, election sizes: \{5, 10, 20, 50, 100, 200\}x\{1000\} \vspace{2mm} \end{flushleft} \end{minipage} \begin{minipage}[t]{10cm} \begin{flushleft}  Samples note: as above \newline \vspace{2mm} \end{flushleft} \end{minipage} \begin{minipage}[t]{10cm} \begin{flushleft}\rule{10cm}{0.3pt} \\ Comment: Cardinal utilities are elicited based on intrinsic rankings and social networks. Here, we consider only intrinsic rankings. \end{flushleft} \end{minipage}  \\ 
\midrule
25 & \begin{minipage}[t]{3cm}\begin{flushleft}\fontsize{9}{9.5}\selectfont Voting with Rank Dependent Scoring Rules \vspace{2mm} \newline AAAI-2014  \vspace{2mm}\newline  \citeapp{gol-lan-mat-per:c:rank-dependent-scoring} \end{flushleft}\end{minipage} & \begin{minipage}[t]{3cm}\begin{flushleft}\fontsize{9}{9.5}\selectfont J.~Goldsmith, J.~Lang, N.~Mattei, P.~Perny\end{flushleft}\end{minipage} & \fontsize{9}{9.5}\selectfont \begin{minipage}[t]{10cm} \begin{flushleft} O1: Impartial Anonymous Culture, Impartial Culture, Mallows, Mallows Mixture, Single-Peaked (Walsh/Uniform) \newline 1000 samples/datapoint, election sizes: [3, 8]x\{10\} \vspace{2mm} \end{flushleft} \end{minipage} \begin{minipage}[t]{10cm} \begin{flushleft}  Samples note: Elections were sampled to ensure that the original outcome is different than the manipulator’s preferences (by discarding samples that did not satisfy the condition) \newline Sizes note: The paper mentions experiments for sizes [3,8]x{20,30} as giving similar results. \newline \vspace{2mm} \end{flushleft} \end{minipage} \begin{minipage}[t]{10cm} \begin{flushleft}  Mallows: Mixture model uses 5 components. Unclear how the parameters of the mixtures are selected. \vspace{2mm} \end{flushleft} \end{minipage}  \\ 
\midrule
26 & \begin{minipage}[t]{3cm}\begin{flushleft}\fontsize{9}{9.5}\selectfont Robust Winners and Winner Determination Policies under Candidate Uncertainty \vspace{2mm} \newline AAAI-2014  \vspace{2mm}\newline  \citeapp{bou-lan-ore-pal:c:robust-uncertain} \end{flushleft}\end{minipage} & \begin{minipage}[t]{3cm}\begin{flushleft}\fontsize{9}{9.5}\selectfont C.~Boutilier, J.~Lang, J.~Oren, H.~Palacios\end{flushleft}\end{minipage} & \fontsize{9}{9.5}\selectfont \begin{minipage}[t]{10cm} \begin{flushleft} O1: Mallows \newline 25 samples/datapoint, election sizes: \{10\}x\{100\} \vspace{2mm} \end{flushleft} \end{minipage} \begin{minipage}[t]{10cm} \begin{flushleft}  Mallows: $\phi\in \{0.3, 0.8, 1.0\}$ \vspace{2mm} \end{flushleft} \end{minipage}  \\ 
\midrule
\multicolumn{4}{c}{\bf2015}\\ \midrule
27 & \begin{minipage}[t]{3cm}\begin{flushleft}\fontsize{9}{9.5}\selectfont A Study of Human Behavior in Online Voting \vspace{2mm} \newline AAMAS-2015  \vspace{2mm}\newline  \citeapp{tal-mei-gal:c:human-behavior-online-voting} \end{flushleft}\end{minipage} & \begin{minipage}[t]{3cm}\begin{flushleft}\fontsize{9}{9.5}\selectfont M.~Tal, R.~Meir, Y.~Gal\end{flushleft}\end{minipage} & \fontsize{9}{9.5}\selectfont \begin{minipage}[t]{10cm} \begin{flushleft} O1: Real-Life (beyond PrefLib), not-ordinal \newline 2379 samples/datapoint, election sizes: \{3\}x\{103, 1009, 10007\} \vspace{2mm} \end{flushleft} \end{minipage} \begin{minipage}[t]{10cm} \begin{flushleft}  Samples note: 722, 966, 691 respectively \newline Sizes note: numbers selected to avoid rounding issues \newline \vspace{2mm} \end{flushleft} \end{minipage} \begin{minipage}[t]{10cm} \begin{flushleft}  Note (general): Experiment conducted via Mechanical Turk, where voters were submitting Plurality ballots once, given the poll results designed by the observers. \vspace{2mm} \end{flushleft} \end{minipage} \begin{minipage}[t]{10cm} \begin{flushleft} \rule{10cm}{0.3pt} \\ O2: Real-Life (beyond PrefLib), not-ordinal \newline election sizes: \{3\}x\{3, 5, 7\} \vspace{2mm} \end{flushleft} \end{minipage} \begin{minipage}[t]{10cm} \begin{flushleft}  Samples note: 940 ballots cast (note that this is an iterative voting game, each with unspecified number of rounds but at most 10) \newline Sizes note: numbers selected not to even and small to maximize the number of converged experiments (it is about reaching a stable outcome in iterative voting) \newline \vspace{2mm} \end{flushleft} \end{minipage} \begin{minipage}[t]{10cm} \begin{flushleft}  Note (general): Experiment conducted via Mechanical Turk, where voters were submitting Plurality ballots based on the profile they had assigned. Each voter aim was to have the most preferred candidate win. The voters could change their votes in response to changes of votes from the other agents. \vspace{2mm} \end{flushleft} \end{minipage} \begin{minipage}[t]{10cm} \begin{flushleft}\rule{10cm}{0.3pt} \\ Comment: There is a journal version: Reshef Meir, Kobi Gal, Maor Tal: Strategic voting in the lab: compromise and leader bias behavior. Auton. Agents Multi Agent Syst. 34(1): 31 (2020) \end{flushleft} \end{minipage}  \\ 
\midrule
28 & \begin{minipage}[t]{3cm}\begin{flushleft}\fontsize{9}{9.5}\selectfont Reasoning with PCP-nets in a Multi-Agent Context \vspace{2mm} \newline AAMAS-2015  \vspace{2mm}\newline  \citeapp{cor-gra-gol-mat-ros-ven:c:pcp-nets} \end{flushleft}\end{minipage} & \begin{minipage}[t]{3cm}\begin{flushleft}\fontsize{9}{9.5}\selectfont C.~Cornelio, U.~Grandi, J.~Goldsmith, N.~Mattei, F.~Rossi, K.~Brent Venable\end{flushleft}\end{minipage} & \fontsize{9}{9.5}\selectfont \begin{minipage}[t]{10cm} \begin{flushleft} O1: Impartial Culture \newline 100 samples/datapoint, election sizes: [2, 4, 8, 16, 32, 64, 128, 256, 512, 1024]x\{20\} \vspace{2mm} \end{flushleft} \end{minipage} \begin{minipage}[t]{10cm} \begin{flushleft}  Note (general): Profile size is $2^m \times n$ where $m$ is the number of binary issues and $n$ is the number of votes. IC with (P)CP-nets is challenging and the paper uses a form of its approximation. Each issue depends on at most two others. \vspace{2mm} \end{flushleft} \end{minipage} \begin{minipage}[t]{10cm} \begin{flushleft} \rule{10cm}{0.3pt} \\ O2: Impartial Culture \newline 100 samples/datapoint, election sizes: \{8\}x[1, 30] \vspace{2mm} \end{flushleft} \end{minipage} \begin{minipage}[t]{10cm} \begin{flushleft}  Note (general): Profiles generated as above. Each issue depends on at most one other. \vspace{2mm} \end{flushleft} \end{minipage} \begin{minipage}[t]{10cm} \begin{flushleft} \rule{10cm}{0.3pt} \\ O3: Hand-Crafted \newline 10000 samples/datapoint, election sizes: [0, 1073741824]x\{1\} \vspace{2mm} \end{flushleft} \end{minipage} \begin{minipage}[t]{10cm} \begin{flushleft}  Samples note: 100 PCP-nets tested over 100 outcome pairs each \newline Sizes note: We have up to 30 binary issues, so the number of candidates is $2^30$ \newline \vspace{2mm} \end{flushleft} \end{minipage} \begin{minipage}[t]{10cm} \begin{flushleft}  Note (general): (P)CP nets with polytree structure. Issues can depend on up to m-1, m/2 or m/4 issues (where m is the number of issues) \vspace{2mm} \end{flushleft} \end{minipage} \begin{minipage}[t]{10cm} \begin{flushleft} \rule{10cm}{0.3pt} \\ O4: Hand-Crafted \newline 500 samples/datapoint, election sizes: [2, 4, 8, 16, 32, 64, 128]x\{1\} \vspace{2mm} \end{flushleft} \end{minipage} \begin{minipage}[t]{10cm} \begin{flushleft}  Samples note: 20 PCP-nets, each tested with 25 outcome pairs \newline \vspace{2mm} \end{flushleft} \end{minipage} \begin{minipage}[t]{10cm} \begin{flushleft}  Note (general): as above \vspace{2mm} \end{flushleft} \end{minipage} \begin{minipage}[t]{10cm} \begin{flushleft}\rule{10cm}{0.3pt} \\ Comment: Some parematers in the experiments where not clear to understand from quick reading (e.g. bottom-left paragraph on p. 976 “total mean over 100 case”) \end{flushleft} \end{minipage}  \\ 
\midrule
29 & \begin{minipage}[t]{3cm}\begin{flushleft}\fontsize{9}{9.5}\selectfont On the Susceptibility of the Deferred Acceptance Algorithm \vspace{2mm} \newline AAMAS-2015  \vspace{2mm}\newline  \citeapp{azi-see-wed:c:deferred-acceptence-manipulation} \end{flushleft}\end{minipage} & \begin{minipage}[t]{3cm}\begin{flushleft}\fontsize{9}{9.5}\selectfont H.~Aziz, H.~Seedig, J.~von Wedel\end{flushleft}\end{minipage} & \fontsize{9}{9.5}\selectfont \begin{minipage}[t]{10cm} \begin{flushleft} O1: Impartial Culture, Mallows Mixture \newline 1000 samples/datapoint, election sizes: \{15, 30\}x\{100, 200\}, \{100, 200\}x\{15, 30\} \vspace{2mm} \end{flushleft} \end{minipage} \begin{minipage}[t]{10cm} \begin{flushleft}  Sizes note: Both sides (colleges and students) can be interpreted as both voters and candidates. \newline \vspace{2mm} \end{flushleft} \end{minipage} \begin{minipage}[t]{10cm} \begin{flushleft}  Mallows: Unclear what parameters were used. \vspace{2mm} \end{flushleft} \end{minipage} \begin{minipage}[t]{10cm} \begin{flushleft}  Note (general): PrefLib implementation of models. \vspace{2mm} \end{flushleft} \end{minipage} \begin{minipage}[t]{10cm} \begin{flushleft}\rule{10cm}{0.3pt} \\ Comment: Two-sided matchings. We consider preferences of one side over the other.

There is an arXiv version: https://arxiv.org/abs/1502.06318 \end{flushleft} \end{minipage}  \\ 
\midrule
30 & \begin{minipage}[t]{3cm}\begin{flushleft}\fontsize{9}{9.5}\selectfont Aggregating Partial Rankings with Applications to Peer Grading in Massive Online Open Courses \vspace{2mm} \newline AAMAS-2015  \vspace{2mm}\newline  \citeapp{car-kri-vou:c:peer-grading-rank-aggregation} \end{flushleft}\end{minipage} & \begin{minipage}[t]{3cm}\begin{flushleft}\fontsize{9}{9.5}\selectfont I.~Caragiannis, G.~Krimpas, A.~Voudouris\end{flushleft}\end{minipage} & \fontsize{9}{9.5}\selectfont \begin{minipage}[t]{10cm} \begin{flushleft} O1: Hand-Crafted \newline 50 samples/datapoint, election sizes: [2, 25]x\{1000\} \vspace{2mm} \end{flushleft} \end{minipage} \begin{minipage}[t]{10cm} \begin{flushleft}  Sizes note: Each agent ranks a subset of  2, 3,…,24, or 25 other agents. \newline \vspace{2mm} \end{flushleft} \end{minipage} \begin{minipage}[t]{10cm} \begin{flushleft}  Note (general): Each vote is a partial ranking that is a trimmed version of one ground-truth ranking common for all voters. \vspace{2mm} \end{flushleft} \end{minipage} \begin{minipage}[t]{10cm} \begin{flushleft} \rule{10cm}{0.3pt} \\ O2: Mallows \newline 50 samples/datapoint, election sizes: \{5, 8, 12\}x\{1000\} \vspace{2mm} \end{flushleft} \end{minipage} \begin{minipage}[t]{10cm} \begin{flushleft}  Samples note: not clearly stated, but assumed based on other experiments \newline Sizes note: Each agent ranks a subset of  5, 8, or 12 other agents. \newline \vspace{2mm} \end{flushleft} \end{minipage} \begin{minipage}[t]{10cm} \begin{flushleft}  Mallows: Different value of $\phi$ parameter for different voters. \vspace{2mm} \end{flushleft} \end{minipage} \begin{minipage}[t]{10cm} \begin{flushleft}  Note (general): Each vote is a partial ranking that is based on a ground-truth ranking with some Mallows noise added. \vspace{2mm} \end{flushleft} \end{minipage} \begin{minipage}[t]{10cm} \begin{flushleft}\rule{10cm}{0.3pt} \\ Comment: The paper is about peer grading (not really elections). Agents report preference orders over the subset of other agents.

There is an arXiv version: https://arxiv.org/abs/1411.4619 \end{flushleft} \end{minipage}  \\ 
\midrule
31 & \begin{minipage}[t]{3cm}\begin{flushleft}\fontsize{9}{9.5}\selectfont Adapting the Social Network to Affect Elections \vspace{2mm} \newline AAMAS-2015  \vspace{2mm}\newline  \citeapp{sin-haz-has-kra:c:mainpulation-with-networks} \end{flushleft}\end{minipage} & \begin{minipage}[t]{3cm}\begin{flushleft}\fontsize{9}{9.5}\selectfont S.~Sina, N.~Hazon, A.~Hassidim, S.~Kraus\end{flushleft}\end{minipage} & \fontsize{9}{9.5}\selectfont \begin{minipage}[t]{10cm} \begin{flushleft} O1: Impartial Culture, PrefLib, Single-Peaked (Walsh/Uniform), Urn Model \newline 1000 samples/datapoint, election sizes: [3, 10]x\{25000, 50000, 75000, 100000\} \vspace{2mm} \end{flushleft} \end{minipage} \begin{minipage}[t]{10cm} \begin{flushleft}  Samples note: 100 runs for each of 10 different network configurations \newline Sizes note: A poll on a random subset of 500 voters was also considered in each run of the experiment. \newline \vspace{2mm} \end{flushleft} \end{minipage} \begin{minipage}[t]{10cm} \begin{flushleft}  Urn: “Urn model (with 2-urns and with 3-urns)”. Unclear what that means. \vspace{2mm} \end{flushleft} \end{minipage} \begin{minipage}[t]{10cm} \begin{flushleft}  Note (parameters): PrefLib datasets: Sushi. \vspace{2mm} \end{flushleft} \end{minipage}  \\ 
\midrule
32 & \begin{minipage}[t]{3cm}\begin{flushleft}\fontsize{9}{9.5}\selectfont Voting Rules as Error-Correcting Codes \vspace{2mm} \newline AAAI-2015  \vspace{2mm}\newline  \citeapp{pro-sha-zic:c:voting-rules-error-correcting-codes} \end{flushleft}\end{minipage} & \begin{minipage}[t]{3cm}\begin{flushleft}\fontsize{9}{9.5}\selectfont A.~Procaccia, N.~Shah, Y.~Zick\end{flushleft}\end{minipage} & \fontsize{9}{9.5}\selectfont \begin{minipage}[t]{10cm} \begin{flushleft} O1: PrefLib \newline election sizes: \{4\}x\{20\} \vspace{2mm} \end{flushleft} \end{minipage} \begin{minipage}[t]{10cm} \begin{flushleft}  Sizes note: 20 is an approximate value \newline \vspace{2mm} \end{flushleft} \end{minipage} \begin{minipage}[t]{10cm} \begin{flushleft}  Note (general): PrefLib datasets: Dots and Puzzle \vspace{2mm} \end{flushleft} \end{minipage} \begin{minipage}[t]{10cm} \begin{flushleft}\rule{10cm}{0.3pt} \\ Comment: There is a journal version: Ariel D. Procaccia, Nisarg Shah, Yair Zick:
Voting rules as error-correcting codes. Artif. Intell. 231: 1-16 (2016) \end{flushleft} \end{minipage}  \\ 
\midrule
33 & \begin{minipage}[t]{3cm}\begin{flushleft}\fontsize{9}{9.5}\selectfont Conventional Machine Learning for Social Choice \vspace{2mm} \newline AAAI-2015  \vspace{2mm}\newline  \citeapp{dou-lar-coh:c:machine-learning-social-choice} \end{flushleft}\end{minipage} & \begin{minipage}[t]{3cm}\begin{flushleft}\fontsize{9}{9.5}\selectfont J.~Doucette, K.~Larson, R.~Cohen\end{flushleft}\end{minipage} & \fontsize{9}{9.5}\selectfont \begin{minipage}[t]{10cm} \begin{flushleft} O1: PrefLib \newline 100 samples/datapoint, election sizes: [4, 8]x[50, 400], [8, 14]x\{4000\} \vspace{2mm} \end{flushleft} \end{minipage} \begin{minipage}[t]{10cm} \begin{flushleft}  Samples note: For each dataset ballots with missing information were completed randomly, 100 times per dataset \newline Sizes note: Irish and Debian datasets, values are approximate \newline \vspace{2mm} \end{flushleft} \end{minipage}  \\ 
\midrule
34 & \begin{minipage}[t]{3cm}\begin{flushleft}\fontsize{9}{9.5}\selectfont Lie on the Fly: Iterative Voting Center with Manipulative Voters \vspace{2mm} \newline IJCAI-2015  \vspace{2mm}\newline  \citeapp{naa-obr-rab-kal:c:manipulative-voters} \end{flushleft}\end{minipage} & \begin{minipage}[t]{3cm}\begin{flushleft}\fontsize{9}{9.5}\selectfont L.~Naamani-Dery, S.~Obraztsova, Z.~Rabinovich, M.~Kalech\end{flushleft}\end{minipage} & \fontsize{9}{9.5}\selectfont \begin{minipage}[t]{10cm} \begin{flushleft} O1: PrefLib \newline 20 samples/datapoint, election sizes: \{10\}x\{10, 50, 100, 150, 200, 250\} \vspace{2mm} \end{flushleft} \end{minipage} \begin{minipage}[t]{10cm} \begin{flushleft}  Sizes note: Sampled uniformly from Sushi dataset \newline \vspace{2mm} \end{flushleft} \end{minipage}  \\ 
\midrule
35 & \begin{minipage}[t]{3cm}\begin{flushleft}\fontsize{9}{9.5}\selectfont Parliamentary Voting Procedures: Agenda Control, Manipulation, and Uncertainty \vspace{2mm} \newline IJCAI-2015  \vspace{2mm}\newline  \citeapp{bre-che-nie-wal:c:parliamentary} \end{flushleft}\end{minipage} & \begin{minipage}[t]{3cm}\begin{flushleft}\fontsize{9}{9.5}\selectfont R.~Bredereck, J.~Chen, R.~Niedermeier, T.~Walsh\end{flushleft}\end{minipage} & \fontsize{9}{9.5}\selectfont \begin{minipage}[t]{10cm} \begin{flushleft} O1: PrefLib \newline election sizes: Unclear \vspace{2mm} \end{flushleft} \end{minipage} \begin{minipage}[t]{10cm} \begin{flushleft}  Sizes note: All data from Preflib \newline \vspace{2mm} \end{flushleft} \end{minipage} \begin{minipage}[t]{10cm} \begin{flushleft}\rule{10cm}{0.3pt} \\ Comment: There is a journal version: Robert Bredereck, Jiehua Chen, Rolf Niedermeier, Toby Walsh:
Parliamentary Voting Procedures: Agenda Control, Manipulation, and Uncertainty. J. Artif. Intell. Res. 59: 133-173 (2017) \end{flushleft} \end{minipage}  \\ 
\midrule
36 & \begin{minipage}[t]{3cm}\begin{flushleft}\fontsize{9}{9.5}\selectfont Characterization of Scoring Rules with Distances: Application to the Clustering of Rankings \vspace{2mm} \newline IJCAI-2015  \vspace{2mm}\newline  \citeapp{via:c:scoring-rules-distances} \end{flushleft}\end{minipage} & \begin{minipage}[t]{3cm}\begin{flushleft}\fontsize{9}{9.5}\selectfont P.~Viappiani\end{flushleft}\end{minipage} & \fontsize{9}{9.5}\selectfont \begin{minipage}[t]{10cm} \begin{flushleft} O1: PrefLib \newline election sizes: \{100\}x\{5000\} \vspace{2mm} \end{flushleft} \end{minipage} \begin{minipage}[t]{10cm} \begin{flushleft}  Sizes note: Full Sushi dataset \newline \vspace{2mm} \end{flushleft} \end{minipage}  \\ 
\midrule
37 & \begin{minipage}[t]{3cm}\begin{flushleft}\fontsize{9}{9.5}\selectfont The Power of Local Manipulation Strategies in Assignment Mechanisms \vspace{2mm} \newline IJCAI-2015  \vspace{2mm}\newline  \citeapp{men-wei-phi-seu:c:power-local-manipulation-strategies} \end{flushleft}\end{minipage} & \begin{minipage}[t]{3cm}\begin{flushleft}\fontsize{9}{9.5}\selectfont T.~Mennle, M.~Weiss, B.~Philipp, S.~Seuken\end{flushleft}\end{minipage} & \fontsize{9}{9.5}\selectfont \begin{minipage}[t]{10cm} \begin{flushleft} O1: Real-Life (beyond PrefLib) \newline 339 samples/datapoint, election sizes: \{5\}x\{5\} \vspace{2mm} \end{flushleft} \end{minipage} \begin{minipage}[t]{10cm} \begin{flushleft}  Note (general): Instances were generated by humans recruited within Amazon Mechanical Turk \vspace{2mm} \end{flushleft} \end{minipage}  \\ 
\midrule
\multicolumn{4}{c}{\bf2016}\\ \midrule
38 & \begin{minipage}[t]{3cm}\begin{flushleft}\fontsize{9}{9.5}\selectfont Convergence and Quality of Iterative Voting Under Non-Scoring Rules \vspace{2mm} \newline AAMAS-2016  \vspace{2mm}\newline  \citeapp{koo-lev-ros:c:iterative-voting} \end{flushleft}\end{minipage} & \begin{minipage}[t]{3cm}\begin{flushleft}\fontsize{9}{9.5}\selectfont A.~Koolyk, O.~Lev, J.~Rosenschein\end{flushleft}\end{minipage} & \fontsize{9}{9.5}\selectfont \begin{minipage}[t]{10cm} \begin{flushleft} O1: Impartial Culture \newline 10000 samples/datapoint, election sizes: \{4\} x \{10, 25\} \vspace{2mm} \end{flushleft} \end{minipage} \begin{minipage}[t]{10cm} \begin{flushleft}\rule{10cm}{0.3pt} \\ Comment: Short paper, later published in IJCAI-2017 \end{flushleft} \end{minipage}  \\ 
\midrule
39 & \begin{minipage}[t]{3cm}\begin{flushleft}\fontsize{9}{9.5}\selectfont Achieving Fully Proportional Representation by Clustering Voters \vspace{2mm} \newline AAMAS-2016  \vspace{2mm}\newline  \citeapp{fal-sli-sta-tal:c:multiwinner-clustering} \end{flushleft}\end{minipage} & \begin{minipage}[t]{3cm}\begin{flushleft}\fontsize{9}{9.5}\selectfont P.~Faliszewski, A.~Slinko, K.~Stahl, N.~Talmon\end{flushleft}\end{minipage} & \fontsize{9}{9.5}\selectfont \begin{minipage}[t]{10cm} \begin{flushleft} O1: Urn Model \newline 500 samples/datapoint, election sizes: \{100\}x\{100\} \vspace{2mm} \end{flushleft} \end{minipage} \begin{minipage}[t]{10cm} \begin{flushleft}  Sizes note: committee sizes 3, 5, 27, 47 \newline \vspace{2mm} \end{flushleft} \end{minipage} \begin{minipage}[t]{10cm} \begin{flushleft}  Urn: {0.05, 0.25} \vspace{2mm} \end{flushleft} \end{minipage} \begin{minipage}[t]{10cm} \begin{flushleft}\rule{10cm}{0.3pt} \\ Comment: There is a journal version: Piotr Faliszewski, Arkadii Slinko, Kolja Stahl, Nimrod Talmon:
Achieving fully proportional representation by clustering voters. J. Heuristics 24(5): 725-756 (2018) \end{flushleft} \end{minipage}  \\ 
\midrule
40 & \begin{minipage}[t]{3cm}\begin{flushleft}\fontsize{9}{9.5}\selectfont Minimising the Rank Aggregation Error \vspace{2mm} \newline AAMAS-2016  \vspace{2mm}\newline  \citeapp{wee-ger-ste:c:minimising-rank-aggregation-error} \end{flushleft}\end{minipage} & \begin{minipage}[t]{3cm}\begin{flushleft}\fontsize{9}{9.5}\selectfont M.~M.~de Weerdt, E.~H.~Gerding, S.~Stein\end{flushleft}\end{minipage} & \fontsize{9}{9.5}\selectfont \begin{minipage}[t]{10cm} \begin{flushleft} O1: Mallows, PrefLib \newline election sizes: \{6\}x\{5, 10, 15, 20, 25, 30, 45, 50\}, \{4\}x\{10, 20, 30, 40, 50\} \vspace{2mm} \end{flushleft} \end{minipage} \begin{minipage}[t]{10cm} \begin{flushleft}  Samples note: The number of samples per datapoint is not clearly stated (possibly 1?). \newline Sizes note: 6 candidates for Mallows, 4 candidates for PrefLib (dataset Dots) \newline \vspace{2mm} \end{flushleft} \end{minipage} \begin{minipage}[t]{10cm} \begin{flushleft}  Mallows: $\phi = 0.6666$ \vspace{2mm} \end{flushleft} \end{minipage} \begin{minipage}[t]{10cm} \begin{flushleft}  Note (parameters): The Mallow’s model is modified to have probability q that a vote is missing a candidate. IN the experiments q=0.7 \vspace{2mm} \end{flushleft} \end{minipage} \begin{minipage}[t]{10cm} \begin{flushleft}  Note (general): Selecting votes from the Dots dataset is not described. \vspace{2mm} \end{flushleft} \end{minipage} \begin{minipage}[t]{10cm} \begin{flushleft}\rule{10cm}{0.3pt} \\ Comment: Extended Abstract (we did not find long version) \end{flushleft} \end{minipage}  \\ 
\midrule
41 & \begin{minipage}[t]{3cm}\begin{flushleft}\fontsize{9}{9.5}\selectfont Analyzing the Practical Relevance of Voting Paradoxes via Ehrhart Theory, Computer Simulations, and Empirical Data \vspace{2mm} \newline AAMAS-2016  \vspace{2mm}\newline  \citeapp{bra-gei-str:c:practical-relevance-of-paradoxes} \end{flushleft}\end{minipage} & \begin{minipage}[t]{3cm}\begin{flushleft}\fontsize{9}{9.5}\selectfont F.~Brandt, C.~Geist, M.~Strobel\end{flushleft}\end{minipage} & \fontsize{9}{9.5}\selectfont \begin{minipage}[t]{10cm} \begin{flushleft} O1: Euclidean 2D, Impartial Anonymous Culture, Impartial Culture, Mallows, PrefLib, Urn Model \newline 100000 samples/datapoint, election sizes: \{4\}x[1, 30], \{4\}x[49, 51], \{4\}x[99, 101], \{4\}x[199, 201], \{4\}x[499, 501], \{4\}x[999, 1001] \vspace{2mm} \end{flushleft} \end{minipage} \begin{minipage}[t]{10cm} \begin{flushleft}  Mallows: 0.8 \vspace{2mm} \end{flushleft} \end{minipage} \begin{minipage}[t]{10cm} \begin{flushleft}  Urn: $\alpha = 10/m!$ \vspace{2mm} \end{flushleft} \end{minipage} \begin{minipage}[t]{10cm} \begin{flushleft}  Euclidean: Uniform 2D \vspace{2mm} \end{flushleft} \end{minipage} \begin{minipage}[t]{10cm} \begin{flushleft}  Note (parameters): The Mallows model parameter is selected to be a worst-case scenario based on the theoretical analysis presented earlier in the paper. The urn model assumes we put back exactly $\alpha$ copies of a drawn vote. \vspace{2mm} \end{flushleft} \end{minipage} \begin{minipage}[t]{10cm} \begin{flushleft}  Note (general): The whole preflib at that time gave 314 meaningful elections. Additionally from the Netflix Prize dataset (nowadays also preflib) 11 million suitable elections derived by Mattei et al (An empiri-
cal study of voting rules and manipulation with large
datasets) were used. \vspace{2mm} \end{flushleft} \end{minipage} \begin{minipage}[t]{10cm} \begin{flushleft} \rule{10cm}{0.3pt} \\ O2: Euclidean 2D, Impartial Anonymous Culture, Impartial Culture, Mallows, PrefLib, Urn Model \newline 1000000 samples/datapoint, election sizes: [2, 10] x \{50, 51\} \vspace{2mm} \end{flushleft} \end{minipage} \begin{minipage}[t]{10cm} \begin{flushleft}  Mallows: as above  \vspace{2mm} \end{flushleft} \end{minipage} \begin{minipage}[t]{10cm} \begin{flushleft}  Urn: as above  \vspace{2mm} \end{flushleft} \end{minipage} \begin{minipage}[t]{10cm} \begin{flushleft}  Euclidean: as above  \vspace{2mm} \end{flushleft} \end{minipage} \begin{minipage}[t]{10cm} \begin{flushleft}  Note (parameters): as above \vspace{2mm} \end{flushleft} \end{minipage} \begin{minipage}[t]{10cm} \begin{flushleft} \rule{10cm}{0.3pt} \\ O3:  \newline  \vspace{2mm} \end{flushleft} \end{minipage} \begin{minipage}[t]{10cm} \begin{flushleft}\rule{10cm}{0.3pt} \\ Comment: The authors use $\alpha$ to refer to an absolute and not normalized parameter of contagion for the urn model (we here report the value as if it were normalized)

Has a journal version with different title and behind a paywall: Analyzing the Practical Relevance of the Condorcet Loser Paradox and the Agenda Contraction Paradox (\url{https://link.springer.com/chapter/10.1007/978-3-030-48598-6_5}) \end{flushleft} \end{minipage}  \\ 
\midrule
42 & \begin{minipage}[t]{3cm}\begin{flushleft}\fontsize{9}{9.5}\selectfont Optimal Aggregation of Uncertain Preferences \vspace{2mm} \newline AAAI-2016  \vspace{2mm}\newline  \citeapp{pro-shah:c:optimal-aggregation} \end{flushleft}\end{minipage} & \begin{minipage}[t]{3cm}\begin{flushleft}\fontsize{9}{9.5}\selectfont A.~Procaccia, N.~Shah\end{flushleft}\end{minipage} & \fontsize{9}{9.5}\selectfont \begin{minipage}[t]{10cm} \begin{flushleft} O1: PrefLib \newline 1000 samples/datapoint \vspace{2mm} \end{flushleft} \end{minipage} \begin{minipage}[t]{10cm} \begin{flushleft}  Sizes note: AGH Course Selection, Netflix,
Skate, Sushi and T-Shirt \newline \vspace{2mm} \end{flushleft} \end{minipage}  \\ 
\midrule
43 & \begin{minipage}[t]{3cm}\begin{flushleft}\fontsize{9}{9.5}\selectfont Judgment Aggregation under Issue Dependencies \vspace{2mm} \newline AAAI-2016  \vspace{2mm}\newline  \citeapp{cos-gro-end:c:aggregation-issue-dependencies} \end{flushleft}\end{minipage} & \begin{minipage}[t]{3cm}\begin{flushleft}\fontsize{9}{9.5}\selectfont M.~Costantini, C.~Groenland, U.~Endriss\end{flushleft}\end{minipage} & \fontsize{9}{9.5}\selectfont \begin{minipage}[t]{10cm} \begin{flushleft} A1: PrefLib \newline 1850 samples/datapoint, election sizes: \{6\}x\{68\} \vspace{2mm} \end{flushleft} \end{minipage} \begin{minipage}[t]{10cm} \begin{flushleft}  Samples note: Data is based on the hotel reviews dataset. \newline Sizes note: Average number of ballots was provided. \newline \vspace{2mm} \end{flushleft} \end{minipage} \begin{minipage}[t]{10cm} \begin{flushleft} \rule{10cm}{0.3pt} \\ A2: PrefLib \newline 31 samples/datapoint, election sizes: \{6\}x\{32\} \vspace{2mm} \end{flushleft} \end{minipage} \begin{minipage}[t]{10cm} \begin{flushleft}  Samples note: Highly polarised profiles from the previous dataset. \newline Sizes note: Average number of ballots was provided \newline \vspace{2mm} \end{flushleft} \end{minipage} \begin{minipage}[t]{10cm} \begin{flushleft}\rule{10cm}{0.3pt} \\ Comment: Paper is about judgement aggregation. \end{flushleft} \end{minipage}  \\ 
\midrule
44 & \begin{minipage}[t]{3cm}\begin{flushleft}\fontsize{9}{9.5}\selectfont Misrepresentation in District Voting \vspace{2mm} \newline IJCAI-2016  \vspace{2mm}\newline  \citeapp{bac-lev-lew-zic:c:misrepresentation-in-districts} \end{flushleft}\end{minipage} & \begin{minipage}[t]{3cm}\begin{flushleft}\fontsize{9}{9.5}\selectfont Y.~Bachrach, O.~Lev, Y.~Lewenberg, Y.~Zick\end{flushleft}\end{minipage} & \fontsize{9}{9.5}\selectfont \begin{minipage}[t]{10cm} \begin{flushleft} O1: Impartial Culture, Mallows \newline election sizes: [3-7]x[1500-75000] \vspace{2mm} \end{flushleft} \end{minipage} \begin{minipage}[t]{10cm} \begin{flushleft}  Samples note: Unclear (“Each point is average (…) obtained for many elections sampled using this probabilistic model”) \newline Sizes note: the voters are split into 15 regions with equal number of voters \newline \vspace{2mm} \end{flushleft} \end{minipage} \begin{minipage}[t]{10cm} \begin{flushleft}  Mallows: $\phi$ drawn uniformly at random from (1/2,1) \vspace{2mm} \end{flushleft} \end{minipage}  \\ 
\midrule
45 & \begin{minipage}[t]{3cm}\begin{flushleft}\fontsize{9}{9.5}\selectfont Subset Selection via Implicit Utilitarian Voting \vspace{2mm} \newline IJCAI-2016  \vspace{2mm}\newline  \citeapp{car-nat-pro-sha:c:subset-implicit-voting} \end{flushleft}\end{minipage} & \begin{minipage}[t]{3cm}\begin{flushleft}\fontsize{9}{9.5}\selectfont I.~Caragiannis, S.~Nath, A.~Procaccia, N.~Shah\end{flushleft}\end{minipage} & \fontsize{9}{9.5}\selectfont \begin{minipage}[t]{10cm} \begin{flushleft} O1: Impartial Culture, PrefLib, Real-Life (beyond PrefLib) \newline 10000 samples/datapoint, election sizes: \{10\}x\{8\} \vspace{2mm} \end{flushleft} \end{minipage} \begin{minipage}[t]{10cm} \begin{flushleft}  Sizes note: Beyond PrefLib: Jester dataset \newline \vspace{2mm} \end{flushleft} \end{minipage} \begin{minipage}[t]{10cm} \begin{flushleft} \rule{10cm}{0.3pt} \\ O2:  \newline 10000 samples/datapoint, election sizes: [10, 50]x\{15\} \vspace{2mm} \end{flushleft} \end{minipage} \begin{minipage}[t]{10cm} \begin{flushleft}  Sizes note: Statistical culture is not clear \newline \vspace{2mm} \end{flushleft} \end{minipage} \begin{minipage}[t]{10cm} \begin{flushleft}\rule{10cm}{0.3pt} \\ Comment: There is a journal version: Ioannis Caragiannis, Swaprava Nath, Ariel D. Procaccia, Nisarg Shah:
Subset Selection Via Implicit Utilitarian Voting. J. Artif. Intell. Res. 58: 123-152 (2017) \end{flushleft} \end{minipage}  \\ 
\midrule
\multicolumn{4}{c}{\bf2017}\\ \midrule
46 & \begin{minipage}[t]{3cm}\begin{flushleft}\fontsize{9}{9.5}\selectfont New Approximation for Borda Coalitional Manipulation \vspace{2mm} \newline AAMAS-2017  \vspace{2mm}\newline  \citeapp{kel-has-haz:c:manipulation-approx-borda} \end{flushleft}\end{minipage} & \begin{minipage}[t]{3cm}\begin{flushleft}\fontsize{9}{9.5}\selectfont O.~Keller, A.~Hassidim, N.~Hazon\end{flushleft}\end{minipage} & \fontsize{9}{9.5}\selectfont \begin{minipage}[t]{10cm} \begin{flushleft} O1: Impartial Culture \newline 8 samples/datapoint, election sizes: \{5, 15, 25, 35, 45, 55\}x\{6, 12, 24\} \vspace{2mm} \end{flushleft} \end{minipage} \begin{minipage}[t]{10cm} \begin{flushleft}  Sizes note: number of voters includes both manipulators and non-manipulators \newline \vspace{2mm} \end{flushleft} \end{minipage} \begin{minipage}[t]{10cm} \begin{flushleft}  Note (general): m+1 candidates, k manipulators, n = 2k (non-manipulators) \vspace{2mm} \end{flushleft} \end{minipage} \begin{minipage}[t]{10cm} \begin{flushleft}\rule{10cm}{0.3pt} \\ Comment: There is a journal version: Orgad Keller, Avinatan Hassidim, Noam Hazon: New Approximation for Borda Coalitional Manipulation. AAMAS 2017: 606-614 \end{flushleft} \end{minipage}  \\ 
\midrule
47 & \begin{minipage}[t]{3cm}\begin{flushleft}\fontsize{9}{9.5}\selectfont Majority Graphs of Assignment Problems and Properties of Popular Random Assignments \vspace{2mm} \newline AAMAS-2017  \vspace{2mm}\newline  \citeapp{bra-hof-sud:c:majority-graphs} \end{flushleft}\end{minipage} & \begin{minipage}[t]{3cm}\begin{flushleft}\fontsize{9}{9.5}\selectfont F.~Brandt, J.~Hofbauer, M.~Suderland\end{flushleft}\end{minipage} & \fontsize{9}{9.5}\selectfont \begin{minipage}[t]{10cm} \begin{flushleft} O1: Euclidean 2D, Impartial Culture \newline 10000 samples/datapoint, election sizes: [1, 7]x[1, 7] \vspace{2mm} \end{flushleft} \end{minipage} \begin{minipage}[t]{10cm} \begin{flushleft}  Euclidean: Unclear (but most probably uniform) 2D ($[0,1]^2$) \vspace{2mm} \end{flushleft} \end{minipage} \begin{minipage}[t]{10cm} \begin{flushleft} \rule{10cm}{0.3pt} \\ O2: exhaustive \newline election sizes: [1, 4]x[1, 4] \vspace{2mm} \end{flushleft} \end{minipage} \begin{minipage}[t]{10cm} \begin{flushleft}  Samples note: Brute-force on all profiles \newline \vspace{2mm} \end{flushleft} \end{minipage} \begin{minipage}[t]{10cm} \begin{flushleft}\rule{10cm}{0.3pt} \\ Comment: The paper is about application of majority graphs from voting theory to matching. \end{flushleft} \end{minipage}  \\ 
\midrule
48 & \begin{minipage}[t]{3cm}\begin{flushleft}\fontsize{9}{9.5}\selectfont Divide and Conquer: Using Geographic Manipulation to Win District-Based
Elections \vspace{2mm} \newline AAMAS-2017  \vspace{2mm}\newline  \citeapp{lew-lev-ros:c:district-manipulation} \end{flushleft}\end{minipage} & \begin{minipage}[t]{3cm}\begin{flushleft}\fontsize{9}{9.5}\selectfont Y.~Lewenberg, O.~Lev, J.~Rosenschein\end{flushleft}\end{minipage} & \fontsize{9}{9.5}\selectfont \begin{minipage}[t]{10cm} \begin{flushleft} O1: Mallows Mixture \newline 1000 samples/datapoint, election sizes: \{6\}x\{1000000\} \vspace{2mm} \end{flushleft} \end{minipage} \begin{minipage}[t]{10cm} \begin{flushleft}  Samples note: According to the paper: “Each setting was run several thousand times” \newline \vspace{2mm} \end{flushleft} \end{minipage} \begin{minipage}[t]{10cm} \begin{flushleft}  Mallows: Complicated mixture simulating division into urban and rural areas, with components whose dispersion parameter depends on the location in a 2D space (this is geographical space, not ideology-one, like in Euclidean models). Central preferences of the mixtures were (apparently) hand-designed. \vspace{2mm} \end{flushleft} \end{minipage} \begin{minipage}[t]{10cm} \begin{flushleft} \rule{10cm}{0.3pt} \\ O2: Real-Life (beyond PrefLib) \newline election sizes: \{10\}x\{?\}, \{11\}x\{?\} \vspace{2mm} \end{flushleft} \end{minipage} \begin{minipage}[t]{10cm} \begin{flushleft}  Sizes note: Information about the number of voters does not seem to be reported in the paper (should be in the referenced web pages, but those were down at the time of making this entry) \newline \vspace{2mm} \end{flushleft} \end{minipage} \begin{minipage}[t]{10cm} \begin{flushleft}  Note (general): 2015 UK and Israeli elections. Data sets are reported to be at http://votes20.gov.il/ and http://www.electoralcommission.org.uk/our-work/ but neither of the pages was available while recording this entry. Generally, this is not exactly ordinal data (the experiment was recorded in the database for the sake of completeness) \vspace{2mm} \end{flushleft} \end{minipage}  \\ 
\midrule
49 & \begin{minipage}[t]{3cm}\begin{flushleft}\fontsize{9}{9.5}\selectfont Bisimulations for Verifying Strategic Abilities with an Application to ThreeBallot \vspace{2mm} \newline AAMAS-2017  \vspace{2mm}\newline  \citeapp{bel-con-dim-jam-jon:c:bisimulation-threeballot} \end{flushleft}\end{minipage} & \begin{minipage}[t]{3cm}\begin{flushleft}\fontsize{9}{9.5}\selectfont F.~Belardinelli, R.~Condurache, C.~Dima, W.~Jamroga, A.~Jones\end{flushleft}\end{minipage} & \fontsize{9}{9.5}\selectfont \begin{minipage}[t]{10cm} \begin{flushleft}\rule{10cm}{0.3pt} \\ Comment: There is a journal version: Francesco Belardinelli, Rodica Condurache, Catalin Dima, Wojciech Jamroga, Michal Knapik:
Bisimulations for verifying strategic abilities with an application to the ThreeBallot voting protocol. Inf. Comput. 276: 104552 (2021) \end{flushleft} \end{minipage}  \\ 
\midrule
50 & \begin{minipage}[t]{3cm}\begin{flushleft}\fontsize{9}{9.5}\selectfont Proxy Voting for Better Outcomes \vspace{2mm} \newline AAMAS-2017  \vspace{2mm}\newline  \citeapp{coh-man-mei-mei-ord:c:proxy-better-outcomes} \end{flushleft}\end{minipage} & \begin{minipage}[t]{3cm}\begin{flushleft}\fontsize{9}{9.5}\selectfont G.~Cohensius, S.~Mannor, R.~Meir, E.~Meirom, A.~Orda\end{flushleft}\end{minipage} & \fontsize{9}{9.5}\selectfont \begin{minipage}[t]{10cm} \begin{flushleft} O1: PrefLib \newline 1000 samples/datapoint, election sizes: \{16\}x\{?\} \vspace{2mm} \end{flushleft} \end{minipage} \begin{minipage}[t]{10cm} \begin{flushleft} \rule{10cm}{0.3pt} \\ A1: PrefLib \newline 1000 samples/datapoint, election sizes: \{16\}x\{?\} \vspace{2mm} \end{flushleft} \end{minipage} \begin{minipage}[t]{10cm} \begin{flushleft} \rule{10cm}{0.3pt} \\ A2: specific own model \newline 1000 samples/datapoint, election sizes: \{?\}x[0, 100] \vspace{2mm} \end{flushleft} \end{minipage} \begin{minipage}[t]{10cm} \begin{flushleft}\rule{10cm}{0.3pt} \\ Comment: There is an arXiv version: https://arxiv.org/abs/1611.08308 \end{flushleft} \end{minipage}  \\ 
\midrule
51 & \begin{minipage}[t]{3cm}\begin{flushleft}\fontsize{9}{9.5}\selectfont Manipulation of Hamming-based Approval Voting for Multiple Referenda
and Committee Elections \vspace{2mm} \newline AAMAS-2017  \vspace{2mm}\newline  \citeapp{bar-lan-yok:c:approval-hamming-manipulation} \end{flushleft}\end{minipage} & \begin{minipage}[t]{3cm}\begin{flushleft}\fontsize{9}{9.5}\selectfont N.~Barrot, J.~Lang, M.~Yokoo\end{flushleft}\end{minipage} & \fontsize{9}{9.5}\selectfont \begin{minipage}[t]{10cm} \begin{flushleft} A1: Impartial Culture, Impartial Culture (Variant) \newline 10000 samples/datapoint, election sizes: \{3, 4, 5\}x\{25\} \vspace{2mm} \end{flushleft} \end{minipage} \begin{minipage}[t]{10cm} \begin{flushleft}  Note (general): Under IC each voter approves each candidate with probability 0.5. The authors also use biased model, where they choose p1 and p2 as two randomly selected probabilities. 40\% of the voters approve each candidate with probability p1, 40\% approve each candidate with probability p2, and 20\% approve each candidate with probability 0.5. \vspace{2mm} \end{flushleft} \end{minipage} \begin{minipage}[t]{10cm} \begin{flushleft} \rule{10cm}{0.3pt} \\ A2: Impartial Culture \newline election sizes: \{5\}x[5, 70] \vspace{2mm} \end{flushleft} \end{minipage} \begin{minipage}[t]{10cm} \begin{flushleft}  Samples note: Unclear number of samples per datapoint (likely 10000 as in the previous experiment) \newline Sizes note: Step of 5 for the number of voters \newline \vspace{2mm} \end{flushleft} \end{minipage}  \\ 
\midrule
52 & \begin{minipage}[t]{3cm}\begin{flushleft}\fontsize{9}{9.5}\selectfont A Restricted Markov Tree Model for Inference and Generation in Social
Choice with Incomplete Preferences \vspace{2mm} \newline AAMAS-2017  \vspace{2mm}\newline  \citeapp{dou-coh:c:incomplete-preferences-markov} \end{flushleft}\end{minipage} & \begin{minipage}[t]{3cm}\begin{flushleft}\fontsize{9}{9.5}\selectfont J.~Doucette, R.~Cohen\end{flushleft}\end{minipage} & \fontsize{9}{9.5}\selectfont \begin{minipage}[t]{10cm} \begin{flushleft} O1: Mallows \newline election sizes: \{5\}x[8,  65536] \vspace{2mm} \end{flushleft} \end{minipage} \begin{minipage}[t]{10cm} \begin{flushleft}  Samples note: “Samples per data point” does not seem to be meaningful in this experiment (one could also view it as the number of voters) \newline Sizes note: The authors mention that for 10 and 20 candidates results are similar. Number of voters (draws from the Mallows model) deduced from Figure 3 in the paper. \newline \vspace{2mm} \end{flushleft} \end{minipage} \begin{minipage}[t]{10cm} \begin{flushleft}  Mallows: $\phi \in \{0.5, 0.8\}$ \vspace{2mm} \end{flushleft} \end{minipage} \begin{minipage}[t]{10cm} \begin{flushleft}  Note (general): The authors mention that they obtained similar results for RUMs (random utility models). \vspace{2mm} \end{flushleft} \end{minipage} \begin{minipage}[t]{10cm} \begin{flushleft} \rule{10cm}{0.3pt} \\ O2: PrefLib \newline election sizes: [4, 14]x[100, 4000] \vspace{2mm} \end{flushleft} \end{minipage} \begin{minipage}[t]{10cm} \begin{flushleft}  Sizes note: The paper only covers a subset of the election sizes indicated above. \newline \vspace{2mm} \end{flushleft} \end{minipage} \begin{minipage}[t]{10cm} \begin{flushleft}  Note (general): Preflib data: Debian Project, Irish (Dublin North, Dublin West, Meath) \vspace{2mm} \end{flushleft} \end{minipage} \begin{minipage}[t]{10cm} \begin{flushleft} \rule{10cm}{0.3pt} \\ O3: Hand-Crafted \newline 100 samples/datapoint, election sizes: [15]x[100, 6400] \vspace{2mm} \end{flushleft} \end{minipage} \begin{minipage}[t]{10cm} \begin{flushleft}  Note (general): Hand-crafted model: Elaborate scheme to simulate a swarm of robots in a mining scenario. \vspace{2mm} \end{flushleft} \end{minipage}  \\ 
\midrule
53 & \begin{minipage}[t]{3cm}\begin{flushleft}\fontsize{9}{9.5}\selectfont What Do Multiwinner Voting Rules Do? An Experiment over the Two-Dimensional Euclidean Domain \vspace{2mm} \newline AAAI-2017  \vspace{2mm}\newline  \citeapp{elk-fal-las-sko-sli-tal:c:multiwinner-voting} \end{flushleft}\end{minipage} & \begin{minipage}[t]{3cm}\begin{flushleft}\fontsize{9}{9.5}\selectfont E.~Elkind, P.~Faliszewski, J.~Laslier, P.~Skowron, A.~Slinko, N.~Talmon\end{flushleft}\end{minipage} & \fontsize{9}{9.5}\selectfont \begin{minipage}[t]{10cm} \begin{flushleft} O1: Euclidean 2D \newline 10000 samples/datapoint, election sizes: \{200\}x\{200\} \vspace{2mm} \end{flushleft} \end{minipage} \begin{minipage}[t]{10cm} \begin{flushleft}  Euclidean: Gaussian 2D ($(0,0)$ $sd=1$); Uniform 2D ($[-3, 3]^2$);
Uniform 2D Sphere ($(0, 0)$ $r=3$); Specific(Gaussian 2D) \vspace{2mm} \end{flushleft} \end{minipage} \begin{minipage}[t]{10cm} \begin{flushleft}\rule{10cm}{0.3pt} \\ Comment: There is an arXiv version: http://arxiv.org/abs/1901.09217 \end{flushleft} \end{minipage}  \\ 
\midrule
54 & \begin{minipage}[t]{3cm}\begin{flushleft}\fontsize{9}{9.5}\selectfont The Positronic Economist: A Computational System for Analyzing Economic Mechanisms \vspace{2mm} \newline AAAI-2017  \vspace{2mm}\newline  \citeapp{tho-new-ley:c:positronic-economist} \end{flushleft}\end{minipage} & \begin{minipage}[t]{3cm}\begin{flushleft}\fontsize{9}{9.5}\selectfont D.~Thompson, N.~Newman, K.~Leyton-Brown\end{flushleft}\end{minipage} & \fontsize{9}{9.5}\selectfont \begin{minipage}[t]{10cm} \begin{flushleft} O1: Impartial Culture \newline 10 samples/datapoint, election sizes: \{5\}x[2, 10] \vspace{2mm} \end{flushleft} \end{minipage}  \\ 
\midrule
55 & \begin{minipage}[t]{3cm}\begin{flushleft}\fontsize{9}{9.5}\selectfont Optimizing Positional Scoring Rules for Rank Aggregation \vspace{2mm} \newline AAAI-2017  \vspace{2mm}\newline  \citeapp{car-cha-kri-vou:c:optimizing-scoring-rules} \end{flushleft}\end{minipage} & \begin{minipage}[t]{3cm}\begin{flushleft}\fontsize{9}{9.5}\selectfont I.~Caragiannis, X.~Chatzigeorgiou, G.~Krimpas, A.~Voudouris\end{flushleft}\end{minipage} & \fontsize{9}{9.5}\selectfont \begin{minipage}[t]{10cm} \begin{flushleft} O1: Bradley-Terry, Plackett-Luce, Real-Life (beyond PrefLib) \newline 500 samples/datapoint, election sizes: \{6\}x\{392\} \vspace{2mm} \end{flushleft} \end{minipage} \begin{minipage}[t]{10cm} \begin{flushleft}  Samples note: For synthetic data \newline Sizes note: A bit tricky: there are 48 countries and 36 cities, yet they are used to generate data where the number of alternatives is 6. \newline \vspace{2mm} \end{flushleft} \end{minipage} \begin{minipage}[t]{10cm} \begin{flushleft}\rule{10cm}{0.3pt} \\ Comment: There is a journal version: Ioannis Caragiannis, Xenophon Chatzigeorgiou, George A. Krimpas, Alexandros A. Voudouris:
Optimizing positional scoring rules for rank aggregation. Artif. Intell. 267: 58-77 (2019) \end{flushleft} \end{minipage}  \\ 
\midrule
56 & \begin{minipage}[t]{3cm}\begin{flushleft}\fontsize{9}{9.5}\selectfont Preference Elicitation For Participatory Budgeting \vspace{2mm} \newline AAAI-2017  \vspace{2mm}\newline  \citeapp{ben-nat-pro-sha:c:pb-preference-elicitation} \end{flushleft}\end{minipage} & \begin{minipage}[t]{3cm}\begin{flushleft}\fontsize{9}{9.5}\selectfont G.~Benade, S.~Nath, A.~Procaccia, N.~Shah\end{flushleft}\end{minipage} & \fontsize{9}{9.5}\selectfont \begin{minipage}[t]{10cm} \begin{flushleft} O1: Real-Life (beyond PrefLib) \newline 80 samples/datapoint, election sizes: \{10\}x\{5, 7, 10\} \vspace{2mm} \end{flushleft} \end{minipage} \begin{minipage}[t]{10cm} \begin{flushleft}  Samples note: Somewhat nonstandard process of generating profiles with partial reuse of data. Depending on the setting, we have 50, 80, or 100 profiles, with some correlation (see Sec. 4, “Datasets” in the paper) \newline Sizes note: Original data is much larger (2600 and 4430 votes) but sampling is used to get final profiles \newline \vspace{2mm} \end{flushleft} \end{minipage} \begin{minipage}[t]{10cm} \begin{flushleft}  Note (general): Participatory budgeting data from Boston 2015 and 2016 (4 approval and knapsack voting). The paper is as much about ordinal as approval preferences. \vspace{2mm} \end{flushleft} \end{minipage} \begin{minipage}[t]{10cm} \begin{flushleft}\rule{10cm}{0.3pt} \\ Comment: There is a journal version: Gerdus Benadè, Swaprava Nath, Ariel D. Procaccia, Nisarg Shah:
Preference Elicitation for Participatory Budgeting. Manag. Sci. 67(5): 2813-2827 (2021) \end{flushleft} \end{minipage}  \\ 
\midrule
57 & \begin{minipage}[t]{3cm}\begin{flushleft}\fontsize{9}{9.5}\selectfont Winner Determination in Huge Elections with MapReduce \vspace{2mm} \newline AAAI-2017  \vspace{2mm}\newline  \citeapp{csa-lac-pic-sal:c:determination-mapreduce} \end{flushleft}\end{minipage} & \begin{minipage}[t]{3cm}\begin{flushleft}\fontsize{9}{9.5}\selectfont T.~Csar, M.~Lackner, R.~Pichler, E.~Sallinger\end{flushleft}\end{minipage} & \fontsize{9}{9.5}\selectfont \begin{minipage}[t]{10cm} \begin{flushleft} O1: Hand-Crafted \newline 5 samples/datapoint, election sizes: \{1000, 3000, 5000, 7000\} \vspace{2mm} \end{flushleft} \end{minipage} \begin{minipage}[t]{10cm} \begin{flushleft}  Sizes note: They work directly on dominance graphs, so there are no voters explicitly generated. \newline \vspace{2mm} \end{flushleft} \end{minipage} \begin{minipage}[t]{10cm} \begin{flushleft}  Note (general): They generate dominance graphs not full elections. \vspace{2mm} \end{flushleft} \end{minipage} \begin{minipage}[t]{10cm} \begin{flushleft} \rule{10cm}{0.3pt} \\ O2: Hand-Crafted \newline 1 samples/datapoint, election sizes: \{1000, 3000, 5000, 7000\} \vspace{2mm} \end{flushleft} \end{minipage} \begin{minipage}[t]{10cm} \begin{flushleft}  Sizes note: as above \newline  \vspace{2mm} \end{flushleft} \end{minipage} \begin{minipage}[t]{10cm} \begin{flushleft}  Note (general): as above \vspace{2mm} \end{flushleft} \end{minipage}  \\ 
\midrule
58 & \begin{minipage}[t]{3cm}\begin{flushleft}\fontsize{9}{9.5}\selectfont Multiwinner Rules on Paths From $k$-Borda to Chamberlin-Courant \vspace{2mm} \newline IJCAI-2017  \vspace{2mm}\newline  \citeapp{fal-sko-sli-tal:c:multiwinner-cc-kborda-paths} \end{flushleft}\end{minipage} & \begin{minipage}[t]{3cm}\begin{flushleft}\fontsize{9}{9.5}\selectfont P.~Faliszewski, P.~Skowron, A.~Slinko, N.~Talmon\end{flushleft}\end{minipage} & \fontsize{9}{9.5}\selectfont \begin{minipage}[t]{10cm} \begin{flushleft} O1: Euclidean 2D \newline 5000 samples/datapoint, election sizes: \{200\}x\{200\} \vspace{2mm} \end{flushleft} \end{minipage} \begin{minipage}[t]{10cm} \begin{flushleft}  Samples note: This is the number of elections generated to obtain a single histogram in the visualization studied in the paper. \newline Sizes note: Committee size k=20 \newline \vspace{2mm} \end{flushleft} \end{minipage} \begin{minipage}[t]{10cm} \begin{flushleft}  Euclidean: Uniform 2D ($[-3,3]^2$) \vspace{2mm} \end{flushleft} \end{minipage}  \\ 
\midrule
59 & \begin{minipage}[t]{3cm}\begin{flushleft}\fontsize{9}{9.5}\selectfont Convergence and Quality of Iterative Voting Under Non-Scoring Rules \vspace{2mm} \newline IJCAI-2017  \vspace{2mm}\newline  \citeapp{koo-str-lev-ros:c:iterative-voting-convergence} \end{flushleft}\end{minipage} & \begin{minipage}[t]{3cm}\begin{flushleft}\fontsize{9}{9.5}\selectfont A.~Koolyk, T.~Strangway, O.~Lev, J.~Rosenschein\end{flushleft}\end{minipage} & \fontsize{9}{9.5}\selectfont \begin{minipage}[t]{10cm} \begin{flushleft} O1: Impartial Culture, Single-Peaked (Conitzer/Random Peak) \newline 100000 samples/datapoint, election sizes: \{4\}x\{10\} \vspace{2mm} \end{flushleft} \end{minipage} \begin{minipage}[t]{10cm} \begin{flushleft}  Samples note: 1000 games (elections), each tested 100 times \newline \vspace{2mm} \end{flushleft} \end{minipage} \begin{minipage}[t]{10cm} \begin{flushleft}  Note (general): The authors do not specify which model of generating single-peaked elections they chose (recorded Conitzer/random peak as an educated guess) \vspace{2mm} \end{flushleft} \end{minipage}  \\ 
\midrule
60 & \begin{minipage}[t]{3cm}\begin{flushleft}\fontsize{9}{9.5}\selectfont Proportional Rankings \vspace{2mm} \newline IJCAI-2017  \vspace{2mm}\newline  \citeapp{sko-lac-bri-pet-elk:c:proportional-rankings} \end{flushleft}\end{minipage} & \begin{minipage}[t]{3cm}\begin{flushleft}\fontsize{9}{9.5}\selectfont P.~Skowron, M.~Lackner, M.~Brill, D.~Peters, E.~Elkind\end{flushleft}\end{minipage} & \fontsize{9}{9.5}\selectfont \begin{minipage}[t]{10cm} \begin{flushleft} A1: Euclidean 2D, Hand-Crafted, Impartial Culture, Impartial Culture (Variant), PrefLib, Urn Model \newline 315500 samples/datapoint, election sizes: [4, 15]x[20, 600] \vspace{2mm} \end{flushleft} \end{minipage} \begin{minipage}[t]{10cm} \begin{flushleft}  Samples note: Each plot is obtained from 315500 profiles from various sources \newline Sizes note: Preflib data: 246 profiles with at most 25 candidates and 2000 voters \newline \vspace{2mm} \end{flushleft} \end{minipage} \begin{minipage}[t]{10cm} \begin{flushleft}  Euclidean: [0,1]x[0,1] square with three points placed uniformly at random, describing three districs. Each district is a center of a Gaussian distribution (with std. dev. 0.2 in both dimensions). “For each district, we then sample a number of points representing voters and alternatives according to this distribution.” Each voter approves all candidates within distance 0.4 \vspace{2mm} \end{flushleft} \end{minipage} \begin{minipage}[t]{10cm} \begin{flushleft}  Note (general): The conference paper contains very little details about the experiment. The data was taken from the full version (the conference paper mentions using the Mallows model, but this does not appear in the full version) \vspace{2mm} \end{flushleft} \end{minipage} \begin{minipage}[t]{10cm} \begin{flushleft}\rule{10cm}{0.3pt} \\ Comment: There is an arXiv version: https://arxiv.org/abs/1612.01434 \end{flushleft} \end{minipage}  \\ 
\midrule
61 & \begin{minipage}[t]{3cm}\begin{flushleft}\fontsize{9}{9.5}\selectfont Learning a Ground Truth Ranking Using Noisy Approval Votes \vspace{2mm} \newline IJCAI-2017  \vspace{2mm}\newline  \citeapp{car-mic:c:learning-ground-truth} \end{flushleft}\end{minipage} & \begin{minipage}[t]{3cm}\begin{flushleft}\fontsize{9}{9.5}\selectfont I.~Caragiannis, E.~Micha\end{flushleft}\end{minipage} & \fontsize{9}{9.5}\selectfont \begin{minipage}[t]{10cm} \begin{flushleft} O1: Mallows \newline 1000 samples/datapoint, election sizes: \{50\}x[100, 3000] \vspace{2mm} \end{flushleft} \end{minipage} \begin{minipage}[t]{10cm} \begin{flushleft}  Sizes note: The experiment considered three values of $\phi$ (for the Mallows model) with three different ranges of voters, but altogether they give the recorded range \newline \vspace{2mm} \end{flushleft} \end{minipage} \begin{minipage}[t]{10cm} \begin{flushleft}  Mallows: $\phi \in \{0.052, 0.333, 0.666\}$ (note that the authors use the parameter $p$, which translates to $\phi = (1-p)/p$ (see the Preliminaries section) \vspace{2mm} \end{flushleft} \end{minipage} \begin{minipage}[t]{10cm} \begin{flushleft}  Note (general): While the paper considers approval voting, data generation is more ordinal in spirit and, hence, is recorded as such. \vspace{2mm} \end{flushleft} \end{minipage}  \\ 
\midrule
\multicolumn{4}{c}{\bf2018}\\ \midrule
62 & \begin{minipage}[t]{3cm}\begin{flushleft}\fontsize{9}{9.5}\selectfont Controlling Elections through Social Influence \vspace{2mm} \newline AAMAS-2018  \vspace{2mm}\newline  \citeapp{wil-vor:c:control-social-influence} \end{flushleft}\end{minipage} & \begin{minipage}[t]{3cm}\begin{flushleft}\fontsize{9}{9.5}\selectfont B.~Wilder, Y.~Vorobeychik\end{flushleft}\end{minipage} & \fontsize{9}{9.5}\selectfont \begin{minipage}[t]{10cm} \begin{flushleft} O1: Impartial Culture \newline 30 samples/datapoint, election sizes: \{2, 5, 10\}x\{1224, 1461, 1889, 2888\} \vspace{2mm} \end{flushleft} \end{minipage} \begin{minipage}[t]{10cm} \begin{flushleft}  Samples note: The numbers of voters (nodes) come from real-life networks. \newline \vspace{2mm} \end{flushleft} \end{minipage} \begin{minipage}[t]{10cm} \begin{flushleft}\rule{10cm}{0.3pt} \\ Comment: There is an arXiv version: https://arxiv.org/abs/1711.08615 \end{flushleft} \end{minipage}  \\ 
\midrule
63 & \begin{minipage}[t]{3cm}\begin{flushleft}\fontsize{9}{9.5}\selectfont Boundedly Rational Voters in Large(r) Networks \vspace{2mm} \newline AAMAS-2018  \vspace{2mm}\newline  \citeapp{tsa-sal-lar:c:large-net} \end{flushleft}\end{minipage} & \begin{minipage}[t]{3cm}\begin{flushleft}\fontsize{9}{9.5}\selectfont A.~Tsang, A.~Salehi-Abari, K.~Larson\end{flushleft}\end{minipage} & \fontsize{9}{9.5}\selectfont \begin{minipage}[t]{10cm} \begin{flushleft} O1: Euclidean 1D \newline 1000 samples/datapoint, election sizes: \{4\}x\{500\}, \{5\}x\{100, 500\}, \{6\}x\{100\} \vspace{2mm} \end{flushleft} \end{minipage} \begin{minipage}[t]{10cm} \begin{flushleft}  Euclidean: Uniform 1D {[0,100]} \vspace{2mm} \end{flushleft} \end{minipage}  \\ 
\midrule
64 & \begin{minipage}[t]{3cm}\begin{flushleft}\fontsize{9}{9.5}\selectfont Gerrymandering Over Graphs \vspace{2mm} \newline AAMAS-2018  \vspace{2mm}\newline  \citeapp{coh-lew-ros:c:gerrymandering-graphs} \end{flushleft}\end{minipage} & \begin{minipage}[t]{3cm}\begin{flushleft}\fontsize{9}{9.5}\selectfont A.~Cohen-Zemach, Y.~Lewenberg, J.~Rosenschein\end{flushleft}\end{minipage} & \fontsize{9}{9.5}\selectfont \begin{minipage}[t]{10cm} \begin{flushleft} O1: Euclidean 1D \newline 2000 samples/datapoint, election sizes: \{5\}x\{100\} \vspace{2mm} \end{flushleft} \end{minipage} \begin{minipage}[t]{10cm} \begin{flushleft}  Samples note: 2000 per each of 10 models of generating network connections over voters, so in total 20000 for fixed election parameters \newline \vspace{2mm} \end{flushleft} \end{minipage} \begin{minipage}[t]{10cm} \begin{flushleft}  Euclidean: Unclear (but most probably uniform) 1D {[0,1]} \vspace{2mm} \end{flushleft} \end{minipage} \begin{minipage}[t]{10cm} \begin{flushleft}  Note (general): Voters might be here connected to one another and there are a few models to generate these connections; I skipped that part and focused only on the way how preferences are generated. \vspace{2mm} \end{flushleft} \end{minipage}  \\ 
\midrule
65 & \begin{minipage}[t]{3cm}\begin{flushleft}\fontsize{9}{9.5}\selectfont Between Proportionality and Diversity: Balancing District Sizes under the Chamberlin-Courant Rule \vspace{2mm} \newline AAMAS-2018  \vspace{2mm}\newline  \citeapp{fal-tal:c:balancing-district-sizes-cc} \end{flushleft}\end{minipage} & \begin{minipage}[t]{3cm}\begin{flushleft}\fontsize{9}{9.5}\selectfont P.~Faliszewski, N.~Talmon\end{flushleft}\end{minipage} & \fontsize{9}{9.5}\selectfont \begin{minipage}[t]{10cm} \begin{flushleft} O1: Euclidean 2D \newline 5000 samples/datapoint, election sizes: \{100\}x\{100\} \vspace{2mm} \end{flushleft} \end{minipage} \begin{minipage}[t]{10cm} \begin{flushleft}  Euclidean: Specific(Uniform 2D Sphere, Gaussian 2D); Specific(Gaussian 2D); Uniform 2D Sphere ($(0,0)$ $r=3$); Uniform 2D ($[-3, 3]^2$) \vspace{2mm} \end{flushleft} \end{minipage} \begin{minipage}[t]{10cm} \begin{flushleft}  Note (parameters): 7 different models combining uniform and gaussian distributions on various squares/discs in various ways \vspace{2mm} \end{flushleft} \end{minipage} \begin{minipage}[t]{10cm} \begin{flushleft}  Note (general): Visualisations \vspace{2mm} \end{flushleft} \end{minipage} \begin{minipage}[t]{10cm} \begin{flushleft} \rule{10cm}{0.3pt} \\ O2: Impartial Culture, Urn Model \newline 500 samples/datapoint, election sizes: \{100\}x\{100\} \vspace{2mm} \end{flushleft} \end{minipage} \begin{minipage}[t]{10cm} \begin{flushleft}  Urn: $\alpha \in \{0, 0.05, 0.10, 0.25, 0.5, 1\}$ \vspace{2mm} \end{flushleft} \end{minipage} \begin{minipage}[t]{10cm} \begin{flushleft} \rule{10cm}{0.3pt} \\ O3: Euclidean 2D \newline 500 samples/datapoint, election sizes: \{100\}x\{100\} \vspace{2mm} \end{flushleft} \end{minipage} \begin{minipage}[t]{10cm} \begin{flushleft}  Euclidean: Specific(Uniform 2D Sphere, Gaussian 2D); Specific(Gaussian 2D); Uniform 2D Sphere ($(0,0)$ $r=3$); Uniform 2D ($[-3, 3]^2$) \vspace{2mm} \end{flushleft} \end{minipage} \begin{minipage}[t]{10cm} \begin{flushleft}  Note (parameters): 7 different models combining uniform and gaussian distributions on various squares/discs in various ways \vspace{2mm} \end{flushleft} \end{minipage}  \\ 
\midrule
66 & \begin{minipage}[t]{3cm}\begin{flushleft}\fontsize{9}{9.5}\selectfont On the Distance Between CP-nets \vspace{2mm} \newline AAMAS-2018  \vspace{2mm}\newline  \citeapp{lor-mat-ros-ven:c:distance-cp-nets} \end{flushleft}\end{minipage} & \begin{minipage}[t]{3cm}\begin{flushleft}\fontsize{9}{9.5}\selectfont A.~Loreggia, N.~Mattei, F.~Rossi, K.~Venable\end{flushleft}\end{minipage} & \fontsize{9}{9.5}\selectfont \begin{minipage}[t]{10cm} \begin{flushleft} O1: Impartial Culture \newline 1 samples/datapoint, election sizes: [4, 8, 16, 32, 64, 128, 256, 512]x\{1000\} \vspace{2mm} \end{flushleft} \end{minipage} \begin{minipage}[t]{10cm} \begin{flushleft}  Samples note: Experiment is not entirely clear to analyze. \newline Sizes note: Experiment on CP-nets and corresponding preorders. \newline \vspace{2mm} \end{flushleft} \end{minipage}  \\ 
\midrule
67 & \begin{minipage}[t]{3cm}\begin{flushleft}\fontsize{9}{9.5}\selectfont Optimization-Based Voting Rule Design: The Closer to Utopia the Better \vspace{2mm} \newline AAMAS-2018  \vspace{2mm}\newline  \citeapp{fal-szu-tal:c:closer-to-utopia-better} \end{flushleft}\end{minipage} & \begin{minipage}[t]{3cm}\begin{flushleft}\fontsize{9}{9.5}\selectfont P.~Faliszewski, S.~Szufa, N.~Talmon\end{flushleft}\end{minipage} & \fontsize{9}{9.5}\selectfont \begin{minipage}[t]{10cm} \begin{flushleft} O1: Euclidean 1D, Euclidean 2D \newline 1000 samples/datapoint, election sizes: \{100\}x\{100\} \vspace{2mm} \end{flushleft} \end{minipage} \begin{minipage}[t]{10cm} \begin{flushleft}  Samples note: For 2D elections: 2000. \newline Sizes note: Committee size k=10 \newline \vspace{2mm} \end{flushleft} \end{minipage} \begin{minipage}[t]{10cm} \begin{flushleft}  Euclidean: Uniform 1D {[0,1]}; Uniform 2D Sphere {(0.5, 0.5) r=0.5} \vspace{2mm} \end{flushleft} \end{minipage} \begin{minipage}[t]{10cm} \begin{flushleft}\rule{10cm}{0.3pt} \\ Comment: Has a full version: Faliszewski, P., Szufa, S., Talmon, N. (2022). Optimization-Based Voting Rule Design: The Closer to Utopia the Better. In: Szapiro, T., Kacprzyk, J. (eds) Collective Decisions: Theory, Algorithms And Decision Support Systems. Studies in Systems, Decision and Control, vol 392. Springer, Cham.  \end{flushleft} \end{minipage}  \\ 
\midrule
68 & \begin{minipage}[t]{3cm}\begin{flushleft}\fontsize{9}{9.5}\selectfont Collective Schedules: Scheduling Meets Computational Social Choice \vspace{2mm} \newline AAMAS-2018  \vspace{2mm}\newline  \citeapp{pas-rza-sko:c:scheduling-social-choice} \end{flushleft}\end{minipage} & \begin{minipage}[t]{3cm}\begin{flushleft}\fontsize{9}{9.5}\selectfont F.~Pascual, K.~Rzadca, P.~Skowron\end{flushleft}\end{minipage} & \fontsize{9}{9.5}\selectfont \begin{minipage}[t]{10cm} \begin{flushleft} O1: Impartial Culture, Mallows, PrefLib \newline 100 samples/datapoint, election sizes: \{9\}x\{146\}, \{7\}x\{153\}, \{10\}x\{5000\}, \{10\}x\{500\} \vspace{2mm} \end{flushleft} \end{minipage} \begin{minipage}[t]{10cm} \begin{flushleft}  Samples note: Preflib: AGH and Sushi; \newline \vspace{2mm} \end{flushleft} \end{minipage} \begin{minipage}[t]{10cm} \begin{flushleft}\rule{10cm}{0.3pt} \\ Comment: Has an arXiv version: https://arxiv.org/abs/1803.07484 \end{flushleft} \end{minipage}  \\ 
\midrule
69 & \begin{minipage}[t]{3cm}\begin{flushleft}\fontsize{9}{9.5}\selectfont Rank Maximal Equal Contribution: A Probabilistic Social Choice Function \vspace{2mm} \newline AAAI-2018  \vspace{2mm}\newline  \citeapp{azi-luo-riz:c:rmax-equal-contribution} \end{flushleft}\end{minipage} & \begin{minipage}[t]{3cm}\begin{flushleft}\fontsize{9}{9.5}\selectfont H.~Aziz, P.~Luo, C.~Rizkallah\end{flushleft}\end{minipage} & \fontsize{9}{9.5}\selectfont \begin{minipage}[t]{10cm} \begin{flushleft} O1: exhaustive \newline election sizes: \{4\}x\{4\} \vspace{2mm} \end{flushleft} \end{minipage} \begin{minipage}[t]{10cm} \begin{flushleft}  Samples note: All possible elections \newline \vspace{2mm} \end{flushleft} \end{minipage} \begin{minipage}[t]{10cm} \begin{flushleft} \rule{10cm}{0.3pt} \\ O2: Impartial Culture \newline 1000 samples/datapoint, election sizes: [4, 8]x[4, 8] \vspace{2mm} \end{flushleft} \end{minipage} \begin{minipage}[t]{10cm} \begin{flushleft} \rule{10cm}{0.3pt} \\ O3: Impartial Culture \newline 10000 samples/datapoint, election sizes: [4, 8]x[4, 8] \vspace{2mm} \end{flushleft} \end{minipage} \begin{minipage}[t]{10cm} \begin{flushleft}\rule{10cm}{0.3pt} \\ Comment: Has an arXiv version: https://arxiv.org/abs/1705.00544 \end{flushleft} \end{minipage}  \\ 
\midrule
70 & \begin{minipage}[t]{3cm}\begin{flushleft}\fontsize{9}{9.5}\selectfont The Conference Paper Assignment Problem: Using Order Weighted Averages to Assign Indivisible Goods \vspace{2mm} \newline AAAI-2018  \vspace{2mm}\newline  \citeapp{lia-mat-nob-wal:c:conf-paper-assign-owa} \end{flushleft}\end{minipage} & \begin{minipage}[t]{3cm}\begin{flushleft}\fontsize{9}{9.5}\selectfont J.~Lian, N.~Mattei, R.~Noble, T.~Walsh\end{flushleft}\end{minipage} & \fontsize{9}{9.5}\selectfont \begin{minipage}[t]{10cm} \begin{flushleft} O1: PrefLib \newline election sizes: \{175\}x\{146\} \vspace{2mm} \end{flushleft} \end{minipage} \begin{minipage}[t]{10cm} \begin{flushleft}  Sizes note: MD-00002-00000003 dataset \newline \vspace{2mm} \end{flushleft} \end{minipage} \begin{minipage}[t]{10cm} \begin{flushleft}\rule{10cm}{0.3pt} \\ Comment: Has an arXiv version: https://arxiv.org/abs/1705.06840 \end{flushleft} \end{minipage}  \\ 
\midrule
71 & \begin{minipage}[t]{3cm}\begin{flushleft}\fontsize{9}{9.5}\selectfont Effective Heuristics for Committee Scoring Rules \vspace{2mm} \newline AAAI-2018  \vspace{2mm}\newline  \citeapp{fal-lac-pet-tal:c:eff-heur-committee-scoring} \end{flushleft}\end{minipage} & \begin{minipage}[t]{3cm}\begin{flushleft}\fontsize{9}{9.5}\selectfont P.~Faliszewski, M.~Lackner, D.~Peters, N.~Talmon\end{flushleft}\end{minipage} & \fontsize{9}{9.5}\selectfont \begin{minipage}[t]{10cm} \begin{flushleft} O1: Euclidean 2D, Impartial Culture \newline 5000 samples/datapoint, election sizes: \{100\}x\{100\} \vspace{2mm} \end{flushleft} \end{minipage} \begin{minipage}[t]{10cm} \begin{flushleft}  Euclidean: Uniform 2D ($[-3,3]^2$) \vspace{2mm} \end{flushleft} \end{minipage} \begin{minipage}[t]{10cm} \begin{flushleft}  Note (general): Experiments for t-Borda \vspace{2mm} \end{flushleft} \end{minipage} \begin{minipage}[t]{10cm} \begin{flushleft} \rule{10cm}{0.3pt} \\ O2: Euclidean 2D, Impartial Culture \newline 1000 samples/datapoint, election sizes: \{100\}x\{100\} \vspace{2mm} \end{flushleft} \end{minipage} \begin{minipage}[t]{10cm} \begin{flushleft}  Euclidean: as above  \vspace{2mm} \end{flushleft} \end{minipage} \begin{minipage}[t]{10cm} \begin{flushleft}  Note (general): Experiments for CC \vspace{2mm} \end{flushleft} \end{minipage}  \\ 
\midrule
72 & \begin{minipage}[t]{3cm}\begin{flushleft}\fontsize{9}{9.5}\selectfont A Voting-Based System for Ethical Decision Making \vspace{2mm} \newline AAAI-2018  \vspace{2mm}\newline  \citeapp{noo-gai-awa-dso-rah-rav-pro:2018:c:system-for-ethical-decisions} \end{flushleft}\end{minipage} & \begin{minipage}[t]{3cm}\begin{flushleft}\fontsize{9}{9.5}\selectfont R.~Noothigattu, S.~Gaikwad, E.~Awad, S.~Dsouza, I.~Rahwan, P.~Ravikumar, A.~Procaccia\end{flushleft}\end{minipage} & \fontsize{9}{9.5}\selectfont \begin{minipage}[t]{10cm} \begin{flushleft} O1: Hand-Crafted \newline 50 samples/datapoint, election sizes: \{5\}x\{20\}, \{5\}x[1, 175] \vspace{2mm} \end{flushleft} \end{minipage} \begin{minipage}[t]{10cm} \begin{flushleft}  Note (general): First cardinal utilities are sampled from Gaussian distribution (mean is chosen uniformly at random), then rankings are obtained from utilities \vspace{2mm} \end{flushleft} \end{minipage} \begin{minipage}[t]{10cm} \begin{flushleft} \rule{10cm}{0.3pt} \\ O2: Real-Life (beyond PrefLib) \newline election sizes: [2, 10]x\{1303778\} \vspace{2mm} \end{flushleft} \end{minipage} \begin{minipage}[t]{10cm} \begin{flushleft}  Note (general): Data from Moral Machine \vspace{2mm} \end{flushleft} \end{minipage} \begin{minipage}[t]{10cm} \begin{flushleft}\rule{10cm}{0.3pt} \\ Comment: Has an arXiv version: https://arxiv.org/abs/1709.06692 \end{flushleft} \end{minipage}  \\ 
\midrule
73 & \begin{minipage}[t]{3cm}\begin{flushleft}\fontsize{9}{9.5}\selectfont Approval-Based Multi-Winner Rules and Strategic Voting \vspace{2mm} \newline IJCAI-2018  \vspace{2mm}\newline  \citeapp{lac-sko:c:approval-multi-winner-strategic} \end{flushleft}\end{minipage} & \begin{minipage}[t]{3cm}\begin{flushleft}\fontsize{9}{9.5}\selectfont M.~Lackner, P.~Skowron\end{flushleft}\end{minipage} & \fontsize{9}{9.5}\selectfont \begin{minipage}[t]{10cm} \begin{flushleft} A1: Impartial Culture \newline 1000 samples/datapoint, election sizes: \{8\}x\{24\}, \{6\}x\{24\} \vspace{2mm} \end{flushleft} \end{minipage} \begin{minipage}[t]{10cm} \begin{flushleft}  Samples note: boe-cai-fal-fan-jan-kac-was:c:position-matrices \newline Sizes note: committee sizes 4, and 3 respectively \newline \vspace{2mm} \end{flushleft} \end{minipage}  \\ 
\midrule
74 & \begin{minipage}[t]{3cm}\begin{flushleft}\fontsize{9}{9.5}\selectfont An Analytical and Experimental Comparison of Maximal Lottery Schemes \vspace{2mm} \newline IJCAI-2018  \vspace{2mm}\newline  \citeapp{bra-bra-str:c:maximal-lottery} \end{flushleft}\end{minipage} & \begin{minipage}[t]{3cm}\begin{flushleft}\fontsize{9}{9.5}\selectfont F.~Brandl, F.~Brandt, C.~Stricker\end{flushleft}\end{minipage} & \fontsize{9}{9.5}\selectfont \begin{minipage}[t]{10cm} \begin{flushleft} O1: Impartial Anonymous Culture \newline 100000 samples/datapoint, election sizes: \{3, 5, 11, 21\}x\{3, 5, 11, 21, 51, 151, 501\} \vspace{2mm} \end{flushleft} \end{minipage} \begin{minipage}[t]{10cm} \begin{flushleft} \rule{10cm}{0.3pt} \\ O2: Impartial Anonymous Culture \newline 2000 samples/datapoint, election sizes: \{3, 4, 5, 6\}x[3, 29] \vspace{2mm} \end{flushleft} \end{minipage} \begin{minipage}[t]{10cm} \begin{flushleft}\rule{10cm}{0.3pt} \\ Comment: Has a journal version: Florian Brandl, Felix Brandt, Christian Stricker:
An analytical and experimental comparison of maximal lottery schemes. Soc. Choice Welf. 58(1): 5-38 (2022) \end{flushleft} \end{minipage}  \\ 
\midrule
75 & \begin{minipage}[t]{3cm}\begin{flushleft}\fontsize{9}{9.5}\selectfont Computing the Schulze Method for Large-Scale Preference Data Sets \vspace{2mm} \newline IJCAI-2018  \vspace{2mm}\newline  \citeapp{csa-lac-pic:c:schulze-method} \end{flushleft}\end{minipage} & \begin{minipage}[t]{3cm}\begin{flushleft}\fontsize{9}{9.5}\selectfont T.~Csar, M.~Lackner, R.~Pichler\end{flushleft}\end{minipage} & \fontsize{9}{9.5}\selectfont \begin{minipage}[t]{10cm} \begin{flushleft} O1: PrefLib \newline 1 samples/datapoint, election sizes: \{9698, 12250\}x\{7481\}, \{14187, 18407\}x\{15553\} \vspace{2mm} \end{flushleft} \end{minipage} \begin{minipage}[t]{10cm} \begin{flushleft}  Samples note: Each profile is considered once \newline Sizes note: top-truncated preferences; every voter ranks 150 or 200 top candidates \newline \vspace{2mm} \end{flushleft} \end{minipage} \begin{minipage}[t]{10cm} \begin{flushleft}  Note (general): Spotify 2017 data. Although not implicit, it seems it is the same data as in PrefLib. Obtained separately, but probably from the same source (Kaggle: https://www.kaggle.com/datasets/edumucelli/spotifys-worldwide-daily-song-ranking) \vspace{2mm} \end{flushleft} \end{minipage} \begin{minipage}[t]{10cm} \begin{flushleft}\rule{10cm}{0.3pt} \\ Comment: the code used in the experiments is part of the greater project https://github.com/theresacsar/CloudVoting \end{flushleft} \end{minipage}  \\ 
\midrule
76 & \begin{minipage}[t]{3cm}\begin{flushleft}\fontsize{9}{9.5}\selectfont Egalitarian Committee Scoring Rules \vspace{2mm} \newline IJCAI-2018  \vspace{2mm}\newline  \citeapp{azi-fal-gro-sli-tal:c:egalitarian} \end{flushleft}\end{minipage} & \begin{minipage}[t]{3cm}\begin{flushleft}\fontsize{9}{9.5}\selectfont H.~Aziz, P.~Faliszewski, B.~Grofman, A.~Slinko, N.~Talmon\end{flushleft}\end{minipage} & \fontsize{9}{9.5}\selectfont \begin{minipage}[t]{10cm} \begin{flushleft} O1: Euclidean 2D \newline 10000 samples/datapoint, election sizes: \{100\}x\{100\} \vspace{2mm} \end{flushleft} \end{minipage} \begin{minipage}[t]{10cm} \begin{flushleft}  Euclidean: Uniform 2D Sphere {(0, 0) r=3}; Specific(Uniform 2D, Gaussian 2D); Specific(Uniform 2D) \vspace{2mm} \end{flushleft} \end{minipage} \begin{minipage}[t]{10cm} \begin{flushleft}  Note (parameters): Three variants considered: 1) Both candidates and voters drawn uniformly from the same disc; 2) Candidates drawn uniformly from a square, voters from a mixture of two gaussian distributions; 3) Candidates drawn uniformly from a square and voters uniformly from a square that partially intersects with the candidate square \vspace{2mm} \end{flushleft} \end{minipage}  \\ 
\midrule
77 & \begin{minipage}[t]{3cm}\begin{flushleft}\fontsize{9}{9.5}\selectfont Multiwinner Voting with Fairness Constraints \vspace{2mm} \newline IJCAI-2018  \vspace{2mm}\newline  \citeapp{cel-hua-vis:c:multiwinner-fairness} \end{flushleft}\end{minipage} & \begin{minipage}[t]{3cm}\begin{flushleft}\fontsize{9}{9.5}\selectfont L.~Celis, L.~Huang, N.~Vishnoi\end{flushleft}\end{minipage} & \fontsize{9}{9.5}\selectfont \begin{minipage}[t]{10cm} \begin{flushleft} O1: Euclidean 2D \newline 1000 samples/datapoint, election sizes: \{120\}x\{400\} \vspace{2mm} \end{flushleft} \end{minipage} \begin{minipage}[t]{10cm} \begin{flushleft}  Euclidean: Specific(Uniform 2D) \vspace{2mm} \end{flushleft} \end{minipage} \begin{minipage}[t]{10cm} \begin{flushleft}  Note (parameters): Voters are drawn uniformly from a square, candidates are drawn uniformly from four quadrants of the big square in varying proportions (1/3, 1/4, 1/6, 1/4) \vspace{2mm} \end{flushleft} \end{minipage}  \\ 
\midrule
\multicolumn{4}{c}{\bf2019}\\ \midrule
78 & \begin{minipage}[t]{3cm}\begin{flushleft}\fontsize{9}{9.5}\selectfont Fall if it Lifts your Teammate: {A} Novel Type of Candidate Manipulation \vspace{2mm} \newline AAMAS-2019  \vspace{2mm}\newline  \citeapp{kru-sch:c:team-manipulation} \end{flushleft}\end{minipage} & \begin{minipage}[t]{3cm}\begin{flushleft}\fontsize{9}{9.5}\selectfont J.~Kruger, S.~Schneckenburger\end{flushleft}\end{minipage} & \fontsize{9}{9.5}\selectfont \begin{minipage}[t]{10cm} \begin{flushleft} O1: Impartial Culture, Plackett-Luce \newline 100000 samples/datapoint, election sizes: \{6, 20\}x\{3\} \vspace{2mm} \end{flushleft} \end{minipage} \begin{minipage}[t]{10cm} \begin{flushleft}  Sizes note: The votes correspond to Olympic sports climbing competition \newline \vspace{2mm} \end{flushleft} \end{minipage} \begin{minipage}[t]{10cm} \begin{flushleft}  Note (parameters): Plackett-Luce with initial odds $2^1:2^2:2^3:2^4:2^5:2^6$, and Plackett-Luce as above, but for one vote the initial odds are reversed \vspace{2mm} \end{flushleft} \end{minipage} \begin{minipage}[t]{10cm} \begin{flushleft}  Note (general): Three ways of generating profiles: IC, Plackett-Luce (two variants). \vspace{2mm} \end{flushleft} \end{minipage}  \\ 
\midrule
79 & \begin{minipage}[t]{3cm}\begin{flushleft}\fontsize{9}{9.5}\selectfont Approximation Algorithms for BalancedCC Multiwinner Rules \vspace{2mm} \newline AAMAS-2019  \vspace{2mm}\newline  \citeapp{bri-fal-som-tal:c:approx-multiwinner-cc} \end{flushleft}\end{minipage} & \begin{minipage}[t]{3cm}\begin{flushleft}\fontsize{9}{9.5}\selectfont M.~Brill, P.~Faliszewski, F.~Sommer, N.~Talmon\end{flushleft}\end{minipage} & \fontsize{9}{9.5}\selectfont \begin{minipage}[t]{10cm} \begin{flushleft} O1: Impartial Culture, Urn Model \newline 150 samples/datapoint, election sizes: \{100\}x\{100\} \vspace{2mm} \end{flushleft} \end{minipage} \begin{minipage}[t]{10cm} \begin{flushleft}  Urn: $\alpha \in \{0,0.1, 0.25, 0.5\}$ \vspace{2mm} \end{flushleft} \end{minipage}  \\ 
\midrule
80 & \begin{minipage}[t]{3cm}\begin{flushleft}\fontsize{9}{9.5}\selectfont Single Transferable Vote: Incomplete Knowledge and Communication Issues \vspace{2mm} \newline AAMAS-2019  \vspace{2mm}\newline  \citeapp{aya-ben-lan-pet:c:stv-truncations-and-communication} \end{flushleft}\end{minipage} & \begin{minipage}[t]{3cm}\begin{flushleft}\fontsize{9}{9.5}\selectfont M.~Ayadi, N.~Ben Amor, J.~Lang, D.~Peters\end{flushleft}\end{minipage} & \fontsize{9}{9.5}\selectfont \begin{minipage}[t]{10cm} \begin{flushleft} O1: Impartial Culture, Mallows \newline 1000 samples/datapoint, election sizes: \{7\}x\{100\} \vspace{2mm} \end{flushleft} \end{minipage} \begin{minipage}[t]{10cm} \begin{flushleft}  Sizes note: Truncation of votes to $1,2,\ldots,5$, or $6$ top candidates also considered in each run. \newline \vspace{2mm} \end{flushleft} \end{minipage} \begin{minipage}[t]{10cm} \begin{flushleft}  Mallows: $\phi \in \{0.7, 0.8, 0.9, 1\}$ \vspace{2mm} \end{flushleft} \end{minipage} \begin{minipage}[t]{10cm} \begin{flushleft} \rule{10cm}{0.3pt} \\ O2: PrefLib \newline election sizes: \{10\}x\{5000\}, \{12\}x\{3662\}, \{10\}x\{43\}, \{9\}x\{548\}, \{7\}x\{327\} \vspace{2mm} \end{flushleft} \end{minipage} \begin{minipage}[t]{10cm} \begin{flushleft}  Samples note: Unclear \newline Sizes note: the number of voters increasing, exact values unclear, only the maximal reported \newline \vspace{2mm} \end{flushleft} \end{minipage} \begin{minipage}[t]{10cm} \begin{flushleft}  Note (parameters): PrefLib datasets: Sushi, Dublin, ERS, Glasgow City Council, and Debian. \vspace{2mm} \end{flushleft} \end{minipage} \begin{minipage}[t]{10cm} \begin{flushleft} \rule{10cm}{0.3pt} \\ O3: PrefLib \newline 1000 samples/datapoint, election sizes: \{12\}x[10, 1500] \vspace{2mm} \end{flushleft} \end{minipage} \begin{minipage}[t]{10cm} \begin{flushleft}  Sizes note: step 10 for voters \newline \vspace{2mm} \end{flushleft} \end{minipage} \begin{minipage}[t]{10cm} \begin{flushleft}  Note (parameters): PrefLib datasets: Dublin \vspace{2mm} \end{flushleft} \end{minipage} \begin{minipage}[t]{10cm} \begin{flushleft} \rule{10cm}{0.3pt} \\ O4: Impartial Culture, Mallows \newline 1000 samples/datapoint, election sizes: \{5\}x\{30, 500\} \vspace{2mm} \end{flushleft} \end{minipage} \begin{minipage}[t]{10cm} \begin{flushleft}  Sizes note: Truncation of votes to $1,2,3$, or $4$ top candidates also considered in each run. \newline \vspace{2mm} \end{flushleft} \end{minipage} \begin{minipage}[t]{10cm} \begin{flushleft}  Mallows: $\phi \in \{0.7, 0.8, 0.9, 1\}$ \vspace{2mm} \end{flushleft} \end{minipage}  \\ 
\midrule
81 & \begin{minipage}[t]{3cm}\begin{flushleft}\fontsize{9}{9.5}\selectfont Exploring the No-Show Paradox for Condorcet Extensions Using Ehrhart Theory and Computer Simulations \vspace{2mm} \newline AAMAS-2019  \vspace{2mm}\newline  \citeapp{bra-hof-str:c:no-show-paradox} \end{flushleft}\end{minipage} & \begin{minipage}[t]{3cm}\begin{flushleft}\fontsize{9}{9.5}\selectfont F.~Brandt, J.~Hofbauer, M.~Strobel\end{flushleft}\end{minipage} & \fontsize{9}{9.5}\selectfont \begin{minipage}[t]{10cm} \begin{flushleft} O1: Impartial Anonymous Culture \newline 1000000 samples/datapoint, election sizes: [1, 30]x[1, 50] \vspace{2mm} \end{flushleft} \end{minipage} \begin{minipage}[t]{10cm} \begin{flushleft} \rule{10cm}{0.3pt} \\ O2: Euclidean 2D, Impartial Anonymous Culture, Impartial Culture, Mallows, Urn Model \newline 1000000 samples/datapoint, election sizes: \{4\}x[1, 1000], \{30\}x[1, 200] \vspace{2mm} \end{flushleft} \end{minipage} \begin{minipage}[t]{10cm} \begin{flushleft}  Samples note: only 50000 samples for m = 30 and n >= 99 \newline \vspace{2mm} \end{flushleft} \end{minipage} \begin{minipage}[t]{10cm} \begin{flushleft}  Mallows: $\phi = {0.8}$ \vspace{2mm} \end{flushleft} \end{minipage} \begin{minipage}[t]{10cm} \begin{flushleft}  Urn: $\alpha = 10/m!$ (we add 10 copies of the chosen ranking to the urn after each draw) \vspace{2mm} \end{flushleft} \end{minipage} \begin{minipage}[t]{10cm} \begin{flushleft}  Euclidean: Uniform 2D ($[0,1]^2$) \vspace{2mm} \end{flushleft} \end{minipage} \begin{minipage}[t]{10cm} \begin{flushleft} \rule{10cm}{0.3pt} \\ O3: PrefLib \newline 315 samples/datapoint \vspace{2mm} \end{flushleft} \end{minipage} \begin{minipage}[t]{10cm} \begin{flushleft}  Samples note: Each of 315 datasets is considered once \newline Sizes note: no details \newline \vspace{2mm} \end{flushleft} \end{minipage} \begin{minipage}[t]{10cm} \begin{flushleft}  Note (general): The experiment is conducted on all 315 strict profiles in PrefLib taken as they are \vspace{2mm} \end{flushleft} \end{minipage}  \\ 
\midrule
82 & \begin{minipage}[t]{3cm}\begin{flushleft}\fontsize{9}{9.5}\selectfont Testing Preferential Domains Using Sampling \vspace{2mm} \newline AAMAS-2019  \vspace{2mm}\newline  \citeapp{dey-nat-sha:c:domain-test-sampling} \end{flushleft}\end{minipage} & \begin{minipage}[t]{3cm}\begin{flushleft}\fontsize{9}{9.5}\selectfont P.~Dey, S.~Nath, G.~Shakya\end{flushleft}\end{minipage} & \fontsize{9}{9.5}\selectfont \begin{minipage}[t]{10cm} \begin{flushleft} O1: Impartial Culture \newline 100 samples/datapoint, election sizes: \{3, 5\}x\{10000\} \vspace{2mm} \end{flushleft} \end{minipage} \begin{minipage}[t]{10cm} \begin{flushleft}  Samples note: For each of the 100 generated profiles, one makes 100 sampling tests \newline \vspace{2mm} \end{flushleft} \end{minipage} \begin{minipage}[t]{10cm} \begin{flushleft}  Note (general): Experiment from Section 4.1 \vspace{2mm} \end{flushleft} \end{minipage} \begin{minipage}[t]{10cm} \begin{flushleft} \rule{10cm}{0.3pt} \\ O2: Impartial Culture \newline election sizes: \{9\}x\{10000\} \vspace{2mm} \end{flushleft} \end{minipage} \begin{minipage}[t]{10cm} \begin{flushleft}  Samples note: “The alternatives of size $\ell$ are sampled 100 times”. Unclear how many profiles are generated per data point \newline \vspace{2mm} \end{flushleft} \end{minipage} \begin{minipage}[t]{10cm} \begin{flushleft}  Note (general): Experiment from Section 4.2 \vspace{2mm} \end{flushleft} \end{minipage} \begin{minipage}[t]{10cm} \begin{flushleft}\rule{10cm}{0.3pt} \\ Comment: Has an arXiv version: https://arxiv.org/abs/1902.08930 \end{flushleft} \end{minipage}  \\ 
\midrule
83 & \begin{minipage}[t]{3cm}\begin{flushleft}\fontsize{9}{9.5}\selectfont Avoiding Social Disappointment in Elections \vspace{2mm} \newline AAMAS-2019  \vspace{2mm}\newline  \citeapp{jav-jam-ram:c:avoiding-social-disappointment} \end{flushleft}\end{minipage} & \begin{minipage}[t]{3cm}\begin{flushleft}\fontsize{9}{9.5}\selectfont M.~A.~Javidian, P.~Jamshidi, R.~Ramezanian\end{flushleft}\end{minipage} & \fontsize{9}{9.5}\selectfont \begin{minipage}[t]{10cm} \begin{flushleft} O1: Impartial Culture \newline 1000 samples/datapoint, election sizes: \{3, 4, 5, 6\}x\{6, 7, 8, 9, 10\} \vspace{2mm} \end{flushleft} \end{minipage} \begin{minipage}[t]{10cm} \begin{flushleft}  Note (general): The voter numbers and the number of samples taken form arXiv \vspace{2mm} \end{flushleft} \end{minipage} \begin{minipage}[t]{10cm} \begin{flushleft} \rule{10cm}{0.3pt} \\ O2: Impartial Culture \newline 30 samples/datapoint, election sizes: \{3, 4, 5, 6, 7, 8, 9, 10\}x\{10, 100, 1000\} \vspace{2mm} \end{flushleft} \end{minipage} \begin{minipage}[t]{10cm} \begin{flushleft}  Note (general): as above \vspace{2mm} \end{flushleft} \end{minipage} \begin{minipage}[t]{10cm} \begin{flushleft}\rule{10cm}{0.3pt} \\ Comment: Extended Abstract. Has a long version on arXiv under a different title: https://arxiv.org/abs/1801.05911 \end{flushleft} \end{minipage}  \\ 
\midrule
84 & \begin{minipage}[t]{3cm}\begin{flushleft}\fontsize{9}{9.5}\selectfont Manipulating Elections by Selecting Issues \vspace{2mm} \newline AAMAS-2019  \vspace{2mm}\newline  \citeapp{lu-zha-rab-obr-vor:c:manipulation-issue-selection} \end{flushleft}\end{minipage} & \begin{minipage}[t]{3cm}\begin{flushleft}\fontsize{9}{9.5}\selectfont J.~Lu, D.~Zhang, Z.~Rabinovich, S.~Obraztsova, Y.~Vorobeychik\end{flushleft}\end{minipage} & \fontsize{9}{9.5}\selectfont \begin{minipage}[t]{10cm} \begin{flushleft} O1: Hand-Crafted \newline 100 samples/datapoint, election sizes: [3, 100]x\{100\}, \{3\}x[10-1000] \vspace{2mm} \end{flushleft} \end{minipage} \begin{minipage}[t]{10cm} \begin{flushleft}  Sizes note: {3}x{100} were also considered for different parameter values of the culture \newline \vspace{2mm} \end{flushleft} \end{minipage} \begin{minipage}[t]{10cm} \begin{flushleft}  Note (general): Two cultures were considered. Both resemble multidimensional Euclidean (interpreted as closeness of stance of some issues between voters and candidates), but the details are missing to confirm that it is this (and most likely not) \vspace{2mm} \end{flushleft} \end{minipage} \begin{minipage}[t]{10cm} \begin{flushleft}\rule{10cm}{0.3pt} \\ Comment: Has an arxiv version: https://arxiv.org/abs/1911.12472 \end{flushleft} \end{minipage}  \\ 
\midrule
85 & \begin{minipage}[t]{3cm}\begin{flushleft}\fontsize{9}{9.5}\selectfont How Similar Are Two Elections? \vspace{2mm} \newline AAAI-2019  \vspace{2mm}\newline  \citeapp{fal-sko-sli-szu:c:similar-two-elections} \end{flushleft}\end{minipage} & \begin{minipage}[t]{3cm}\begin{flushleft}\fontsize{9}{9.5}\selectfont P.~Faliszewski, P.~Skowron, A.~Slinko, S.~Szufa, N.~Talmon\end{flushleft}\end{minipage} & \fontsize{9}{9.5}\selectfont \begin{minipage}[t]{10cm} \begin{flushleft} O1: Euclidean 1D, Euclidean 2D, Impartial Culture \newline 100 samples/datapoint, election sizes: \{6\}x[6, 16] \vspace{2mm} \end{flushleft} \end{minipage} \begin{minipage}[t]{10cm} \begin{flushleft}  Euclidean: Uniform 1D {[0,1]}; Uniform 2D Sphere {(0, 0), r=1} \vspace{2mm} \end{flushleft} \end{minipage}  \\ 
\midrule
86 & \begin{minipage}[t]{3cm}\begin{flushleft}\fontsize{9}{9.5}\selectfont "Reverse Gerrymandering": Manipulation in Multi-Group Decision Making \vspace{2mm} \newline AAAI-2019  \vspace{2mm}\newline  \citeapp{lev-lew:c:reverse-gerrymandering} \end{flushleft}\end{minipage} & \begin{minipage}[t]{3cm}\begin{flushleft}\fontsize{9}{9.5}\selectfont O.~Lev, Y.~Lewenberg\end{flushleft}\end{minipage} & \fontsize{9}{9.5}\selectfont \begin{minipage}[t]{10cm} \begin{flushleft} O1: Euclidean 1D, Impartial Culture \newline 1000 samples/datapoint, election sizes: \{5, 8\}x\{53\} \vspace{2mm} \end{flushleft} \end{minipage}  \\ 
\midrule
87 & \begin{minipage}[t]{3cm}\begin{flushleft}\fontsize{9}{9.5}\selectfont Poll-Confident Voters in Iterative Voting \vspace{2mm} \newline AAAI-2019  \vspace{2mm}\newline  \citeapp{wil:c:poll-confident-voters} \end{flushleft}\end{minipage} & \begin{minipage}[t]{3cm}\begin{flushleft}\fontsize{9}{9.5}\selectfont A.~Wilczynski\end{flushleft}\end{minipage} & \fontsize{9}{9.5}\selectfont \begin{minipage}[t]{10cm} \begin{flushleft} O1: Impartial Culture \newline 10000 samples/datapoint, election sizes: \{10\}x\{100\} \vspace{2mm} \end{flushleft} \end{minipage}  \\ 
\midrule
88 & \begin{minipage}[t]{3cm}\begin{flushleft}\fontsize{9}{9.5}\selectfont Low-Distortion Social Welfare Functions \vspace{2mm} \newline AAAI-2019  \vspace{2mm}\newline  \citeapp{ben-pro-qia:c:low-distortion-swf} \end{flushleft}\end{minipage} & \begin{minipage}[t]{3cm}\begin{flushleft}\fontsize{9}{9.5}\selectfont G.~Benad\`{e}, A.~Procaccia, M.~Qiao\end{flushleft}\end{minipage} & \fontsize{9}{9.5}\selectfont \begin{minipage}[t]{10cm} \begin{flushleft} O1: Hand-Crafted \newline 100 samples/datapoint, election sizes: [3, 10]x\{10\} \vspace{2mm} \end{flushleft} \end{minipage} \begin{minipage}[t]{10cm} \begin{flushleft}  Note (general): Experiments on distortion. Model: cardinal utilities sampled from normal distribution, then rankings obtained from utilities \vspace{2mm} \end{flushleft} \end{minipage} \begin{minipage}[t]{10cm} \begin{flushleft} \rule{10cm}{0.3pt} \\ O2: Hand-Crafted \newline 200 samples/datapoint, election sizes: [3, 30]x\{10\} \vspace{2mm} \end{flushleft} \end{minipage} \begin{minipage}[t]{10cm} \begin{flushleft}  Note (general): Experiments on social welfare. Model: cardinal utilities sampled from normal distribution, then rankings obtained from utilities \vspace{2mm} \end{flushleft} \end{minipage}  \\ 
\midrule
89 & \begin{minipage}[t]{3cm}\begin{flushleft}\fontsize{9}{9.5}\selectfont Practical Algorithms for Multi-Stage Voting Rules with Parallel Universes Tiebreaking \vspace{2mm} \newline AAAI-2019  \vspace{2mm}\newline  \citeapp{wan-sik-she-zha-jia-xia:c:multi-stage-rules-parallel-universes-tb} \end{flushleft}\end{minipage} & \begin{minipage}[t]{3cm}\begin{flushleft}\fontsize{9}{9.5}\selectfont J.~Wang, S.~Sikdar, T.~Shepherd, Z.~Zhao, C.~Jiang, L.~Xia\end{flushleft}\end{minipage} & \fontsize{9}{9.5}\selectfont \begin{minipage}[t]{10cm} \begin{flushleft} O1: Impartial Culture \newline 10000 samples/datapoint, election sizes: \{30\}x\{30\} \vspace{2mm} \end{flushleft} \end{minipage} \begin{minipage}[t]{10cm} \begin{flushleft}  Note (general): Experiments for STV \vspace{2mm} \end{flushleft} \end{minipage} \begin{minipage}[t]{10cm} \begin{flushleft} \rule{10cm}{0.3pt} \\ O2: Impartial Culture \newline 14875 samples/datapoint, election sizes: \{10\}x\{10\} \vspace{2mm} \end{flushleft} \end{minipage} \begin{minipage}[t]{10cm} \begin{flushleft}  Note (general): Experiments for Ranked Pairs \vspace{2mm} \end{flushleft} \end{minipage} \begin{minipage}[t]{10cm} \begin{flushleft} \rule{10cm}{0.3pt} \\ O3: PrefLib \newline  \vspace{2mm} \end{flushleft} \end{minipage} \begin{minipage}[t]{10cm} \begin{flushleft}  Sizes note: No precise information, only approximate: “They
represent several real world settings including political elections, movies and sports competitions. For political elections, the number of candidates is often not more than 30.
For example, 76.1
has no more than 10 candidates, and 98.5
than 30 candidates.” \newline \vspace{2mm} \end{flushleft} \end{minipage} \begin{minipage}[t]{10cm} \begin{flushleft}  Note (general): 315 profiles from Strict Order-Complete Lists (SOC), and
275 profiles from Strict Order-Incomplete Lists (SOI) \vspace{2mm} \end{flushleft} \end{minipage} \begin{minipage}[t]{10cm} \begin{flushleft}\rule{10cm}{0.3pt} \\ Comment: Has an arXiv version: https://arxiv.org/abs/1901.09791 \end{flushleft} \end{minipage}  \\ 
\midrule
90 & \begin{minipage}[t]{3cm}\begin{flushleft}\fontsize{9}{9.5}\selectfont A Framework for Approval-Based Budgeting Methods \vspace{2mm} \newline AAAI-2019  \vspace{2mm}\newline  \citeapp{tal-fal:c:framework-approval-based-budgeting-methods} \end{flushleft}\end{minipage} & \begin{minipage}[t]{3cm}\begin{flushleft}\fontsize{9}{9.5}\selectfont N.~Talmon, P.~Faliszewski\end{flushleft}\end{minipage} & \fontsize{9}{9.5}\selectfont \begin{minipage}[t]{10cm} \begin{flushleft} A1: Euclidean 2D \newline 100 samples/datapoint, election sizes: \{50, 100\}x\{50\} \vspace{2mm} \end{flushleft} \end{minipage} \begin{minipage}[t]{10cm} \begin{flushleft}  Euclidean: [0,1]x[0,1] plane

1. voters, positioned uniformly on a disc of radius 0.3, centered at position (0.5, 0.5); we have 50 such voters for
Experiment 1 and 100 for Experiment 2;
2. 50 cheap items, positioned uniformly on a disc of radius
0.2, centered at (0.3, 0.5), and 50 expensive items, positioned uniformly on a disc of radius 0.2, centered at
(0.7, 0.5); \vspace{2mm} \end{flushleft} \end{minipage} \begin{minipage}[t]{10cm} \begin{flushleft}  Note (general): Experiment 1 and 2 (depicted in Table 2 and 3, resp.) \vspace{2mm} \end{flushleft} \end{minipage} \begin{minipage}[t]{10cm} \begin{flushleft} \rule{10cm}{0.3pt} \\ A2: Euclidean 2D \newline 100 samples/datapoint, election sizes: \{20\}x\{35\} \vspace{2mm} \end{flushleft} \end{minipage} \begin{minipage}[t]{10cm} \begin{flushleft}  Euclidean: [0,1]x[0,1] plane

1. 20 voters, positioned uniformly on the whole $1 \times 1$ square;
2. 5 items, termed global items, which are also positioned uniformly on the square; and another 30 items, termed local
items, also positioned uniformly on the square; \vspace{2mm} \end{flushleft} \end{minipage} \begin{minipage}[t]{10cm} \begin{flushleft}  Note (general): Experiment 3 (depicted in Table 4) \vspace{2mm} \end{flushleft} \end{minipage} \begin{minipage}[t]{10cm} \begin{flushleft}\rule{10cm}{0.3pt} \\ Comment: Has an arXiv version: https://arxiv.org/abs/1809.04382 \end{flushleft} \end{minipage}  \\ 
\midrule
91 & \begin{minipage}[t]{3cm}\begin{flushleft}\fontsize{9}{9.5}\selectfont How Hard Is the Manipulative Design of Scoring Systems? \vspace{2mm} \newline IJCAI-2019  \vspace{2mm}\newline  \citeapp{bau-hog:c:manipulative-design-scoring-rules} \end{flushleft}\end{minipage} & \begin{minipage}[t]{3cm}\begin{flushleft}\fontsize{9}{9.5}\selectfont D.~Baumeister, T.~Hogrebe\end{flushleft}\end{minipage} & \fontsize{9}{9.5}\selectfont \begin{minipage}[t]{10cm} \begin{flushleft} O1: PrefLib \newline election sizes: [22, 62]x[19, 22] \vspace{2mm} \end{flushleft} \end{minipage} \begin{minipage}[t]{10cm} \begin{flushleft}  Samples note: The experiments are not randomized so a single test per instance is natural. \newline Sizes note: Not all sizes exist in the data. \newline \vspace{2mm} \end{flushleft} \end{minipage} \begin{minipage}[t]{10cm} \begin{flushleft}  Note (general): Formula 1 data from PrefLib, 48 instances \vspace{2mm} \end{flushleft} \end{minipage}  \\ 
\midrule
92 & \begin{minipage}[t]{3cm}\begin{flushleft}\fontsize{9}{9.5}\selectfont A Parameterized Perspective on Protecting Elections \vspace{2mm} \newline IJCAI-2019  \vspace{2mm}\newline  \citeapp{dey-mis-nat-sha:c:parameterized-protecting} \end{flushleft}\end{minipage} & \begin{minipage}[t]{3cm}\begin{flushleft}\fontsize{9}{9.5}\selectfont P.~Dey, N.~Misra, S.~Nath, G.~Shakya\end{flushleft}\end{minipage} & \fontsize{9}{9.5}\selectfont \begin{minipage}[t]{10cm} \begin{flushleft} O1: Hand-Crafted, Impartial Culture \newline 1000 samples/datapoint, election sizes: $\{5\}x\{12000\}$ \vspace{2mm} \end{flushleft} \end{minipage} \begin{minipage}[t]{10cm} \begin{flushleft}  Note (general): The hand-crafted model is as follows: (1) choose two candidates, a and b. (2) 40\% of the voters have a on top, following by preference order chosen uniformly at random, another 40\% votes are generated similarly, but with b on top. (3) Final 20\% votes are generated from impartial culture. \vspace{2mm} \end{flushleft} \end{minipage} \begin{minipage}[t]{10cm} \begin{flushleft}\rule{10cm}{0.3pt} \\ Comment: Has a journal version: Palash Dey, Neeldhara Misra, Swaprava Nath, Garima Shakya:
A parameterized perspective on protecting elections. Theor. Comput. Sci. 874: 15-31 (2021) \end{flushleft} \end{minipage}  \\ 
\midrule
93 & \begin{minipage}[t]{3cm}\begin{flushleft}\fontsize{9}{9.5}\selectfont Multigoal Committee Selection \vspace{2mm} \newline IJCAI-2019  \vspace{2mm}\newline  \citeapp{koc-kol-elk-fal-tal:c:multigoal-committees} \end{flushleft}\end{minipage} & \begin{minipage}[t]{3cm}\begin{flushleft}\fontsize{9}{9.5}\selectfont M.~Kocot, A.~Kolonko, E.~Elkind, P.~Faliszewski, N.~Talmon\end{flushleft}\end{minipage} & \fontsize{9}{9.5}\selectfont \begin{minipage}[t]{10cm} \begin{flushleft} O1: Euclidean 2D \newline 100 samples/datapoint, election sizes: \{100\}x\{100\} \vspace{2mm} \end{flushleft} \end{minipage} \begin{minipage}[t]{10cm} \begin{flushleft}  Sizes note: Committee size k = 10 \newline \vspace{2mm} \end{flushleft} \end{minipage} \begin{minipage}[t]{10cm} \begin{flushleft}  Euclidean: Uniform 2D ($[-3,3]^2$) \vspace{2mm} \end{flushleft} \end{minipage} \begin{minipage}[t]{10cm} \begin{flushleft}  Note (general): Experiment from Figure 1 (comparison of two rules) \vspace{2mm} \end{flushleft} \end{minipage} \begin{minipage}[t]{10cm} \begin{flushleft} \rule{10cm}{0.3pt} \\ O2: Euclidean 2D \newline 5000 samples/datapoint, election sizes: \{100\}x\{100\} \vspace{2mm} \end{flushleft} \end{minipage} \begin{minipage}[t]{10cm} \begin{flushleft}  Sizes note: as above \newline  \vspace{2mm} \end{flushleft} \end{minipage} \begin{minipage}[t]{10cm} \begin{flushleft}  Euclidean: as above  \vspace{2mm} \end{flushleft} \end{minipage} \begin{minipage}[t]{10cm} \begin{flushleft}  Note (general): Experiment from Figure 2 (drawing histograms). \vspace{2mm} \end{flushleft} \end{minipage}  \\ 
\midrule
94 & \begin{minipage}[t]{3cm}\begin{flushleft}\fontsize{9}{9.5}\selectfont Preferences Single-Peaked on a Tree: Sampling and Tree Recognition \vspace{2mm} \newline IJCAI-2019  \vspace{2mm}\newline  \citeapp{sli-elk:c:single-peaked-on-a-tree} \end{flushleft}\end{minipage} & \begin{minipage}[t]{3cm}\begin{flushleft}\fontsize{9}{9.5}\selectfont J.~Sliwinski, E.~Elkind\end{flushleft}\end{minipage} & \fontsize{9}{9.5}\selectfont \begin{minipage}[t]{10cm} \begin{flushleft} O1: Hand-Crafted \newline 2000 samples/datapoint, election sizes: [5, 127]x\{?\} \vspace{2mm} \end{flushleft} \end{minipage} \begin{minipage}[t]{10cm} \begin{flushleft}  Samples note: for some large candidate sizes, less data points were considered \newline Sizes note: The experiment aims to establish the minimal number of voters needed to be able to identify underlying societal-tree (analogue of societal axis in standard single-peaked elections) \newline \vspace{2mm} \end{flushleft} \end{minipage} \begin{minipage}[t]{10cm} \begin{flushleft}  Note (general): Single-peaked on a tree elections (sampled uniformly), with underlying trees being: stars, balanced binary trees, caterpillars, and paths \vspace{2mm} \end{flushleft} \end{minipage}  \\ 
\midrule
95 & \begin{minipage}[t]{3cm}\begin{flushleft}\fontsize{9}{9.5}\selectfont A Quantitative Analysis of Multi-Winner Rules \vspace{2mm} \newline IJCAI-2019  \vspace{2mm}\newline  \citeapp{lac-sko:c:quantitative-multiwinner} \end{flushleft}\end{minipage} & \begin{minipage}[t]{3cm}\begin{flushleft}\fontsize{9}{9.5}\selectfont M.~Lackner, P.~Skowron\end{flushleft}\end{minipage} & \fontsize{9}{9.5}\selectfont \begin{minipage}[t]{10cm} \begin{flushleft} A1: PrefLib \newline 243 samples/datapoint, election sizes: ? \vspace{2mm} \end{flushleft} \end{minipage} \begin{minipage}[t]{10cm} \begin{flushleft}  Sizes note: size dependent on the chosen preflib instance \newline \vspace{2mm} \end{flushleft} \end{minipage} \begin{minipage}[t]{10cm} \begin{flushleft}  Note (general): “for $i\in \{1,\dots, k - 1\}$, we generated an approval profile assuming that voters approve all candidates that are ranked in the top $i$ positions” \vspace{2mm} \end{flushleft} \end{minipage} \begin{minipage}[t]{10cm} \begin{flushleft} \rule{10cm}{0.3pt} \\ A2: Impartial Culture (Variant) \newline 500 samples/datapoint, election sizes: \{20\}x\{50\} \vspace{2mm} \end{flushleft} \end{minipage} \begin{minipage}[t]{10cm} \begin{flushleft}  Note (general): IC where voters’ approval sets are of size 2–5 (chosen uniformly at random) \vspace{2mm} \end{flushleft} \end{minipage} \begin{minipage}[t]{10cm} \begin{flushleft}\rule{10cm}{0.3pt} \\ Comment: Has a journal version: Martin Lackner, Piotr Skowron:
Utilitarian welfare and representation guarantees of approval-based multiwinner rules. Artif. Intell. 288: 103366 (2020) \end{flushleft} \end{minipage}  \\ 
\midrule
96 & \begin{minipage}[t]{3cm}\begin{flushleft}\fontsize{9}{9.5}\selectfont An Experimental View on Committees Providing Justified Representation \vspace{2mm} \newline IJCAI-2019  \vspace{2mm}\newline  \citeapp{bre-fal-kac-nie:c:experimental-view-on-ejr} \end{flushleft}\end{minipage} & \begin{minipage}[t]{3cm}\begin{flushleft}\fontsize{9}{9.5}\selectfont R.~Bredereck, P.~Faliszewski, A.~Kaczmarczyk, R.~Niedermeier\end{flushleft}\end{minipage} & \fontsize{9}{9.5}\selectfont \begin{minipage}[t]{10cm} \begin{flushleft} A1: Euclidean 1D, Euclidean 2D, Impartial Culture \newline 500 samples/datapoint, election sizes: \{100\}x\{100\}, \{100, 200\}, \{200, 100\} \vspace{2mm} \end{flushleft} \end{minipage} \begin{minipage}[t]{10cm} \begin{flushleft}  Sizes note: committee size k = 10 for {100}x{100}, and 20 otherwise \newline \vspace{2mm} \end{flushleft} \end{minipage} \begin{minipage}[t]{10cm} \begin{flushleft}  Euclidean: 1D: Uniform Interval (r-UI) [0,1]
2D: Uniform Square (r-US) [0,1]x[0,1]

For r-UI, r between 0.01 and 0.25 with step 0.01
For r-US, r between 0.02 and 0.5 with step 0.02 \vspace{2mm} \end{flushleft} \end{minipage} \begin{minipage}[t]{10cm} \begin{flushleft}  Note (parameters): For IC, p between 0.015 and 0.375 with step 0.15) \vspace{2mm} \end{flushleft} \end{minipage} \begin{minipage}[t]{10cm} \begin{flushleft}  Note (general): Section 3, Subsection “Experimental Evaluation” \vspace{2mm} \end{flushleft} \end{minipage} \begin{minipage}[t]{10cm} \begin{flushleft} \rule{10cm}{0.3pt} \\ A2: Euclidean 1D, Euclidean 2D, Impartial Culture \newline 1000 samples/datapoint, election sizes: \{100, 100\} \vspace{2mm} \end{flushleft} \end{minipage} \begin{minipage}[t]{10cm} \begin{flushleft}  Sizes note: committee size in {2,…,15} \newline \vspace{2mm} \end{flushleft} \end{minipage} \begin{minipage}[t]{10cm} \begin{flushleft}  Euclidean: 1D: Uniform Interval (r-UI) [0,1]
2D: Uniform Square (r-US) [0,1]x[0,1]

For r-UI, r = 0.06
For r-US, r = 0.2 \vspace{2mm} \end{flushleft} \end{minipage} \begin{minipage}[t]{10cm} \begin{flushleft}  Note (parameters): For p-IC, p = 0.1 \vspace{2mm} \end{flushleft} \end{minipage} \begin{minipage}[t]{10cm} \begin{flushleft}  Note (general): Section 4, Subsection “Experimental Evaluation” \vspace{2mm} \end{flushleft} \end{minipage}  \\ 
\midrule
\multicolumn{4}{c}{\bf2020}\\ \midrule
97 & \begin{minipage}[t]{3cm}\begin{flushleft}\fontsize{9}{9.5}\selectfont Automated Justification of Collective Decisions via Constraint Solving \vspace{2mm} \newline AAMAS-2020  \vspace{2mm}\newline  \citeapp{boi-end:c:automated-justification} \end{flushleft}\end{minipage} & \begin{minipage}[t]{3cm}\begin{flushleft}\fontsize{9}{9.5}\selectfont A.~Boixel, U.~Endriss\end{flushleft}\end{minipage} & \fontsize{9}{9.5}\selectfont \begin{minipage}[t]{10cm} \begin{flushleft} O1: exhaustive \newline election sizes: \{3\}x\{3\} \vspace{2mm} \end{flushleft} \end{minipage} \begin{minipage}[t]{10cm} \begin{flushleft}  Samples note: Checking all profiles \newline \vspace{2mm} \end{flushleft} \end{minipage}  \\ 
\midrule
98 & \begin{minipage}[t]{3cm}\begin{flushleft}\fontsize{9}{9.5}\selectfont Manipulation of Opinion Polls to Influence Iterative Elections \vspace{2mm} \newline AAMAS-2020  \vspace{2mm}\newline  \citeapp{bau-sel-wil:c:manipulation} \end{flushleft}\end{minipage} & \begin{minipage}[t]{3cm}\begin{flushleft}\fontsize{9}{9.5}\selectfont D.~Baumeister, A.~Selker, A.~Wilczynski\end{flushleft}\end{minipage} & \fontsize{9}{9.5}\selectfont \begin{minipage}[t]{10cm} \begin{flushleft} O1: Impartial Culture \newline 2000 samples/datapoint, election sizes: \{5\}x\{50\} \vspace{2mm} \end{flushleft} \end{minipage} \begin{minipage}[t]{10cm} \begin{flushleft} \rule{10cm}{0.3pt} \\ O2: Impartial Culture \newline 2000 samples/datapoint, election sizes: \{5\}x[10,  50] \vspace{2mm} \end{flushleft} \end{minipage}  \\ 
\midrule
99 & \begin{minipage}[t]{3cm}\begin{flushleft}\fontsize{9}{9.5}\selectfont Game Theoretic Analysis for Two-Sided Matching with Resource Allocation \vspace{2mm} \newline AAMAS-2020  \vspace{2mm}\newline  \citeapp{yah-yok:c:game-matching-allocation} \end{flushleft}\end{minipage} & \begin{minipage}[t]{3cm}\begin{flushleft}\fontsize{9}{9.5}\selectfont K.~Yahiro, M.~Yokoo\end{flushleft}\end{minipage} & \fontsize{9}{9.5}\selectfont \begin{minipage}[t]{10cm} \begin{flushleft} O1: Impartial Culture, Mallows \newline election sizes: \{10\}x\{200\},  \{200\}x\{10\} \vspace{2mm} \end{flushleft} \end{minipage} \begin{minipage}[t]{10cm} \begin{flushleft}  Sizes note: It is a matching instance, with 200 students having preferences over 10 projects and vice versa  \newline \vspace{2mm} \end{flushleft} \end{minipage} \begin{minipage}[t]{10cm} \begin{flushleft}  Mallows: $\phi \in [0.1,0.7]$ \vspace{2mm} \end{flushleft} \end{minipage}  \\ 
\midrule
100 & \begin{minipage}[t]{3cm}\begin{flushleft}\fontsize{9}{9.5}\selectfont Objective Social Choice: Using Auxiliary Information to Improve Voting Outcomes \vspace{2mm} \newline AAMAS-2020  \vspace{2mm}\newline  \citeapp{pit-zha:c:objective-social-choice} \end{flushleft}\end{minipage} & \begin{minipage}[t]{3cm}\begin{flushleft}\fontsize{9}{9.5}\selectfont S.~Pitis, M.~R.~Zhang\end{flushleft}\end{minipage} & \fontsize{9}{9.5}\selectfont \begin{minipage}[t]{10cm} \begin{flushleft} O1: Hand-Crafted \newline 100000 samples/datapoint, election sizes: \{10\}x\{3, 10, 30, 100, 300\} \vspace{2mm} \end{flushleft} \end{minipage} \begin{minipage}[t]{10cm} \begin{flushleft}  Note (general): Each voter i “samples” a candidate j from a one-armed bandit, then computes the average for each candidate, and then returns the ranking of the candidates based on the computed averages, decreasingly. \vspace{2mm} \end{flushleft} \end{minipage} \begin{minipage}[t]{10cm} \begin{flushleft}\rule{10cm}{0.3pt} \\ Comment: Studies collective estimation: where the goal of voters is to recover the ground truth. 

Has an arXiv version: https://arxiv.org/abs/2001.10092 \end{flushleft} \end{minipage}  \\ 
\midrule
101 & \begin{minipage}[t]{3cm}\begin{flushleft}\fontsize{9}{9.5}\selectfont Heuristic Strategies in Uncertain Approval Voting Environments \vspace{2mm} \newline AAMAS-2020  \vspace{2mm}\newline  \citeapp{sch-har-mat-ven:c:strategies-uncertain-approval} \end{flushleft}\end{minipage} & \begin{minipage}[t]{3cm}\begin{flushleft}\fontsize{9}{9.5}\selectfont J.~Scheuerman, J.~Harman, N.~Mattei, K.~Venable\end{flushleft}\end{minipage} & \fontsize{9}{9.5}\selectfont \begin{minipage}[t]{10cm} \begin{flushleft} A1: Real-Life (beyond PrefLib) \newline election sizes: 5x? \vspace{2mm} \end{flushleft} \end{minipage} \begin{minipage}[t]{10cm} \begin{flushleft}  Sizes note: a few voters, but exact number is unclear to us \newline \vspace{2mm} \end{flushleft} \end{minipage} \begin{minipage}[t]{10cm} \begin{flushleft}  Note (general): same as sch-har-mat-ven:c:voter-model; Their six real-world elections are in a very non-standard format. They present 104 study participants with partial election results (i.e., how many votes candidates received and how many votes are still missing) as well as how much utility the voter would derive from getting the candidate elected. As such they only track single-votes and do not analyze full elections. \vspace{2mm} \end{flushleft} \end{minipage} \begin{minipage}[t]{10cm} \begin{flushleft} \rule{10cm}{0.3pt} \\ A2:  \newline  \vspace{2mm} \end{flushleft} \end{minipage} \begin{minipage}[t]{10cm} \begin{flushleft}\rule{10cm}{0.3pt} \\ Comment: Has an arXiv version: https://arxiv.org/abs/1912.00011 \end{flushleft} \end{minipage}  \\ 
\midrule
102 & \begin{minipage}[t]{3cm}\begin{flushleft}\fontsize{9}{9.5}\selectfont Drawing a Map of Elections in the Space of Statistical Cultures \vspace{2mm} \newline AAMAS-2020  \vspace{2mm}\newline  \citeapp{szu-fal-sko-sli-tal:c:map-of-elections} \end{flushleft}\end{minipage} & \begin{minipage}[t]{3cm}\begin{flushleft}\fontsize{9}{9.5}\selectfont S.~Szufa, P.~Faliszewski, P.~Skowron, A.~Slinko, N.~Talmon\end{flushleft}\end{minipage} & \fontsize{9}{9.5}\selectfont \begin{minipage}[t]{10cm} \begin{flushleft} O1: Euclidean 1D, Euclidean 2D, Euclidean 3D-or-more, Impartial Culture, Mallows, Single-Crossing, Single-Peaked (Conitzer/Random Peak), Single-Peaked (Walsh/Uniform), Urn Model \newline election sizes: \{100\}x\{100\} \vspace{2mm} \end{flushleft} \end{minipage} \begin{minipage}[t]{10cm} \begin{flushleft}  Samples note: See Table 1 in the paper for the composition of the dataset \newline \vspace{2mm} \end{flushleft} \end{minipage} \begin{minipage}[t]{10cm} \begin{flushleft}  Mallows: $\phi \in \{ 0.999, 0.99, 0.95, 0.75, 0.5, 0.25, 0.1, 0.05, 0.01, 0.001\}$ \vspace{2mm} \end{flushleft} \end{minipage} \begin{minipage}[t]{10cm} \begin{flushleft}  Urn: $\alpha \in \{0.01, 0.02, 0.05, 0.1, 0.2, 0.5\}$ \vspace{2mm} \end{flushleft} \end{minipage} \begin{minipage}[t]{10cm} \begin{flushleft}  Euclidean: Uniform 1D {[0,1]}; Uniform 2D ($[0,1]^2$); Uniform 3D ($[0,1]^3$); Uniform 5D ($[0,1]^5$); Uniform 10D ($[0,1]^10$); Uniform 20D ($[0,1]^20$); Uniform 2D Sphere ($(0, 0)$ $r=1$); Uniform 3D Sphere ($(0, 0)$ $r=1$); Uniform 5D Sphere ($(0, 0)$ $r=1$) \vspace{2mm} \end{flushleft} \end{minipage} \begin{minipage}[t]{10cm} \begin{flushleft} \rule{10cm}{0.3pt} \\ O2: Euclidean 1D, Euclidean 2D, Impartial Culture, Mallows, Single-Crossing, Single-Peaked (Conitzer/Random Peak), Urn Model \newline 10 samples/datapoint, election sizes: \{8\}x\{8\} \vspace{2mm} \end{flushleft} \end{minipage} \begin{minipage}[t]{10cm} \begin{flushleft}  Samples note: 10 elections from each distribution \newline \vspace{2mm} \end{flushleft} \end{minipage} \begin{minipage}[t]{10cm} \begin{flushleft}  Mallows: $\phi \in \{ 0.01, 0.05, 0.1\}$ \vspace{2mm} \end{flushleft} \end{minipage} \begin{minipage}[t]{10cm} \begin{flushleft}  Urn: $\alpha \in \{0.1, 0.2\}$ \vspace{2mm} \end{flushleft} \end{minipage} \begin{minipage}[t]{10cm} \begin{flushleft}  Euclidean: Uniform 1D ($[0,1]$); Uniform 2D ($[0,1]^2$); Uniform 1D Sphere; Uniform 2D Sphere \vspace{2mm} \end{flushleft} \end{minipage} \begin{minipage}[t]{10cm} \begin{flushleft} \rule{10cm}{0.3pt} \\ O3: Mallows \newline 100 samples/datapoint, election sizes: \{100\}x\{100\} \vspace{2mm} \end{flushleft} \end{minipage} \begin{minipage}[t]{10cm} \begin{flushleft}  Samples note: Two sets of 100 elections (?) with $\phi$ parameter sampled from two different distributions \newline Sizes note: The size is not given explicitly in the paper, but follows from context. \newline \vspace{2mm} \end{flushleft} \end{minipage} \begin{minipage}[t]{10cm} \begin{flushleft}  Mallows: See paragraph “Mallows” in Section 4.1 of the paper (a somewhat involved variant of probing the $\phi$ parameter, later replaced by the normalized Mallows model in the paper “Putting a Compass on the Map of Elections”) \vspace{2mm} \end{flushleft} \end{minipage}  \\ 
\midrule
103 & \begin{minipage}[t]{3cm}\begin{flushleft}\fontsize{9}{9.5}\selectfont Did That Lost Ballot Box Cost Me a Seat? Computing Manipulations of STV Elections \vspace{2mm} \newline AAAI-2020  \vspace{2mm}\newline  \citeapp{blo-con-stu-tea:c:manipulations-stv} \end{flushleft}\end{minipage} & \begin{minipage}[t]{3cm}\begin{flushleft}\fontsize{9}{9.5}\selectfont M.~Blom, A.~Conway, P.~Stuckey, V.~Teague\end{flushleft}\end{minipage} & \fontsize{9}{9.5}\selectfont \begin{minipage}[t]{10cm} \begin{flushleft} O1: Real-Life (beyond PrefLib) \newline election sizes: \{17\}x\{270231\}, \{42\}x\{1094823\}, \{67\}x\{1446623\}, \{82\}x\{3739443\}, \{105\}x\{4695326\}, \{83\}x\{2901464\}, \{44\}x\{351988\}, \{22\}x\{254767\}, \{64\}x\{1061165\}, \{79\}x\{1366182\}, \{116\}x\{3963992\}, \{151\}x\{4492197\}, \{122\}x\{2723166\}, \{58\}x\{339159\} \vspace{2mm} \end{flushleft} \end{minipage} \begin{minipage}[t]{10cm} \begin{flushleft}  Sizes note: Australian Senate Elections from 2016 and 2019 \newline \vspace{2mm} \end{flushleft} \end{minipage}  \\ 
\midrule
104 & \begin{minipage}[t]{3cm}\begin{flushleft}\fontsize{9}{9.5}\selectfont Bounded Incentives in Manipulating the Probabilistic Serial Rule \vspace{2mm} \newline AAAI-2020  \vspace{2mm}\newline  \citeapp{wan-wei-zha:c:bounded-incentives-psr} \end{flushleft}\end{minipage} & \begin{minipage}[t]{3cm}\begin{flushleft}\fontsize{9}{9.5}\selectfont Z.~Wang, Z.~Wei, J.~Zhang\end{flushleft}\end{minipage} & \fontsize{9}{9.5}\selectfont \begin{minipage}[t]{10cm} \begin{flushleft} O1: Impartial Culture \newline 10000 samples/datapoint, election sizes: [8, 20]x[8, 20] \vspace{2mm} \end{flushleft} \end{minipage} \begin{minipage}[t]{10cm} \begin{flushleft}  Sizes note: n=m \newline \vspace{2mm} \end{flushleft} \end{minipage} \begin{minipage}[t]{10cm} \begin{flushleft}\rule{10cm}{0.3pt} \\ Comment: Has a  journal version: Haoqiang Huang, Zihe Wang, Zhide Wei, Jie Zhang:
Bounded incentives in manipulating the probabilistic serial rule. J. Comput. Syst. Sci. 140: 103491 (2024) \end{flushleft} \end{minipage}  \\ 
\midrule
105 & \begin{minipage}[t]{3cm}\begin{flushleft}\fontsize{9}{9.5}\selectfont Price of Fairness in Budget Division and Probabilistic Social Choice \vspace{2mm} \newline AAAI-2020  \vspace{2mm}\newline  \citeapp{mic-pet-sko:c:price-of-fairness-budget-division} \end{flushleft}\end{minipage} & \begin{minipage}[t]{3cm}\begin{flushleft}\fontsize{9}{9.5}\selectfont M.~Michorzewski, D.~Peters, P.~Skowron\end{flushleft}\end{minipage} & \fontsize{9}{9.5}\selectfont \begin{minipage}[t]{10cm} \begin{flushleft} A1: Euclidean 2D, Impartial Culture, Mallows Mixture \newline 500 samples/datapoint, election sizes: 
		
		$\{2, 3, 4, 5, 10, 30, 50, 100\}\times \{10, 20, 30, 40, 50, 60, 70, 80,90, 100, 150, 200,$ $ 250, 300, 350, 400, 450, 500, 550, 600, 650, 700, 750, 800, 850, 900, 950, 1000\}$ \vspace{2mm} \end{flushleft} \end{minipage} \begin{minipage}[t]{10cm} \begin{flushleft}  Euclidean: Voters and candidates uniformly at random from a square; then each voters approves k between 1 and m (inclusive) candidates, where k is chosen uniformly at random. \vspace{2mm} \end{flushleft} \end{minipage} \begin{minipage}[t]{10cm} \begin{flushleft}  Note (parameters): For IC, each voters selects k, k selected uniformly at random, and then approves each project with probability k/m. If so obtained vote is empty, then the voter approves a single candidate selected uniformly at random. The Mixed Mallows is over three rankings. For each vote one reference ranking is selected uniformly at random.The, the number of approvals k is drawn uniformly at random from 1 to m, and k first entries in the ranking are approved. Dispersion parameters of Mallows is not given. \vspace{2mm} \end{flushleft} \end{minipage} \begin{minipage}[t]{10cm} \begin{flushleft}  Note (general): Experiments for Euclidean variants where the number of approvals was the same for all voters, or voters approved candidates from a uniformly at random chosen were conducted but not presented as they gave qualitatively the same results. \vspace{2mm} \end{flushleft} \end{minipage}  \\ 
\midrule
106 & \begin{minipage}[t]{3cm}\begin{flushleft}\fontsize{9}{9.5}\selectfont Can We Predict the Election Outcome from Sampled Votes? \vspace{2mm} \newline AAAI-2020  \vspace{2mm}\newline  \citeapp{mic-sha:c:predict-election-outcome-from-votes} \end{flushleft}\end{minipage} & \begin{minipage}[t]{3cm}\begin{flushleft}\fontsize{9}{9.5}\selectfont E.~Micha, N.~Shah\end{flushleft}\end{minipage} & \fontsize{9}{9.5}\selectfont \begin{minipage}[t]{10cm} \begin{flushleft} O1: Impartial Culture, Mallows \newline 1000000 samples/datapoint, election sizes: \{5\}x\{1000\} \vspace{2mm} \end{flushleft} \end{minipage} \begin{minipage}[t]{10cm} \begin{flushleft}  Mallows: $\phi = 1/3$ \vspace{2mm} \end{flushleft} \end{minipage} \begin{minipage}[t]{10cm} \begin{flushleft}  Note (general): Section 5. “Experiments” \vspace{2mm} \end{flushleft} \end{minipage}  \\ 
\midrule
107 & \begin{minipage}[t]{3cm}\begin{flushleft}\fontsize{9}{9.5}\selectfont Comparing Election Methods Where Each Voter Ranks Only Few Candidates \vspace{2mm} \newline AAAI-2020  \vspace{2mm}\newline  \citeapp{ben-sko:c:comparing-elections-methods-only-a-few-candidates} \end{flushleft}\end{minipage} & \begin{minipage}[t]{3cm}\begin{flushleft}\fontsize{9}{9.5}\selectfont M.~Bentert, P.~Skowron\end{flushleft}\end{minipage} & \fontsize{9}{9.5}\selectfont \begin{minipage}[t]{10cm} \begin{flushleft} O1: Euclidean 1D, Impartial Culture, Mallows \newline 500 samples/datapoint, election sizes: \{50\}x[10, 1000] \vspace{2mm} \end{flushleft} \end{minipage} \begin{minipage}[t]{10cm} \begin{flushleft}  Mallows: not specified \vspace{2mm} \end{flushleft} \end{minipage} \begin{minipage}[t]{10cm} \begin{flushleft}  Euclidean: Unclear \vspace{2mm} \end{flushleft} \end{minipage} \begin{minipage}[t]{10cm} \begin{flushleft}  Note (general): Section 5. “Experimental Evaluation” \vspace{2mm} \end{flushleft} \end{minipage} \begin{minipage}[t]{10cm} \begin{flushleft}\rule{10cm}{0.3pt} \\ Comment: Has an arXiv version: https://arxiv.org/abs/1901.10848 \end{flushleft} \end{minipage}  \\ 
\midrule
108 & \begin{minipage}[t]{3cm}\begin{flushleft}\fontsize{9}{9.5}\selectfont Perpetual Voting: Fairness in Long-Term Decision Making \vspace{2mm} \newline AAAI-2020  \vspace{2mm}\newline  \citeapp{lac:c:perpetual-voting-fairness} \end{flushleft}\end{minipage} & \begin{minipage}[t]{3cm}\begin{flushleft}\fontsize{9}{9.5}\selectfont M.~Lackner\end{flushleft}\end{minipage} & \fontsize{9}{9.5}\selectfont \begin{minipage}[t]{10cm} \begin{flushleft} A1: Euclidean 2D \newline 200000 samples/datapoint, election sizes: \{20\}x\{5\} \vspace{2mm} \end{flushleft} \end{minipage} \begin{minipage}[t]{10cm} \begin{flushleft}  Samples note: 10000 instances x 20 decisions (sequences) \newline \vspace{2mm} \end{flushleft} \end{minipage} \begin{minipage}[t]{10cm} \begin{flushleft}  Euclidean: Voters are split in two groups and are placed on the 2d plane by a bivariate normal distribution. For
the first group (6 voters) both x- and y-coordinates are  independently drawn from N (-0.5, 0.2); for the second group
(14 voters) x- and y-coordinates are from N (0.5, 0.2). That
is, the first, smaller group is centered around (-0.5, -0.5),
the second, larger group around (0.5, 0.5).
Alternatives are distributed uniformly in the rectangle
$[-1, 1] \times [-1, 1]$. Voters approve all alternatives that have
a distance of at most 1.5 times the distance to the closest alternative. This yields approval sets of size 1.8 on average. It
is important to note that alternatives change in every round
and thus even voters that are close to each other do not necessarily have the same approval sets each round. \vspace{2mm} \end{flushleft} \end{minipage} \begin{minipage}[t]{10cm} \begin{flushleft}  Note (general): Section 6. “Experiments”

Experiments on perpetual voting model, 20 sequential decisions with changing preferences. \vspace{2mm} \end{flushleft} \end{minipage}  \\ 
\midrule
109 & \begin{minipage}[t]{3cm}\begin{flushleft}\fontsize{9}{9.5}\selectfont Evaluating Committees for Representative Democracies: the Distortion and Beyond \vspace{2mm} \newline IJCAI-2020  \vspace{2mm}\newline  \citeapp{jaw-sko:c:evaluating-committees-distortion} \end{flushleft}\end{minipage} & \begin{minipage}[t]{3cm}\begin{flushleft}\fontsize{9}{9.5}\selectfont M.~Jaworski, P.~Skowron\end{flushleft}\end{minipage} & \fontsize{9}{9.5}\selectfont \begin{minipage}[t]{10cm} \begin{flushleft} O1: Hand-Crafted \newline 500 samples/datapoint, election sizes: \{100\}x\{500\} \vspace{2mm} \end{flushleft} \end{minipage} \begin{minipage}[t]{10cm} \begin{flushleft}  Sizes note: 100 binary issues, each candidate and voter reports a 0/1 opinion for each issue, voters prefer candidates that agree with them on more issues (ties broken uniformly at random); committee size k = 31 (the paper mentions that for k=11 results are similar) \newline \vspace{2mm} \end{flushleft} \end{minipage} \begin{minipage}[t]{10cm} \begin{flushleft}  Note (general): Four different models of generating the preferences, each designed to take into account the issues that the voters care about. There are also results for weighted issues (Figure 3). \vspace{2mm} \end{flushleft} \end{minipage}  \\ 
\midrule
110 & \begin{minipage}[t]{3cm}\begin{flushleft}\fontsize{9}{9.5}\selectfont Concentration of Distortion: The Value of Extra Voters in Randomized Social Choice \vspace{2mm} \newline IJCAI-2020  \vspace{2mm}\newline  \citeapp{fai-fan-mun:c:extra-voters} \end{flushleft}\end{minipage} & \begin{minipage}[t]{3cm}\begin{flushleft}\fontsize{9}{9.5}\selectfont B.~Fain, W.~Fan, K.~Munagala\end{flushleft}\end{minipage} & \fontsize{9}{9.5}\selectfont \begin{minipage}[t]{10cm} \begin{flushleft} A1: Real-Life PB (beyond Pabulib) \newline election sizes: \{23\}x\{945\} \vspace{2mm} \end{flushleft} \end{minipage} \begin{minipage}[t]{10cm} \begin{flushleft}  Note (general): The authors use a single real-world PB election with knapsack voting from Cambridge 2015 \vspace{2mm} \end{flushleft} \end{minipage} \begin{minipage}[t]{10cm} \begin{flushleft}\rule{10cm}{0.3pt} \\ Comment: Has an arXiv version: https://arxiv.org/abs/2004.13153 \end{flushleft} \end{minipage}  \\ 
\midrule
\multicolumn{4}{c}{\bf2021}\\ \midrule
111 & \begin{minipage}[t]{3cm}\begin{flushleft}\fontsize{9}{9.5}\selectfont Approval-Based Shortlisting \vspace{2mm} \newline AAMAS-2021  \vspace{2mm}\newline  \citeapp{lac-mal:c:approval-shortlisting} \end{flushleft}\end{minipage} & \begin{minipage}[t]{3cm}\begin{flushleft}\fontsize{9}{9.5}\selectfont M.~Lackner, J.~Maly\end{flushleft}\end{minipage} & \fontsize{9}{9.5}\selectfont \begin{minipage}[t]{10cm} \begin{flushleft} A1: Impartial Culture (Variant) \newline 1000 samples/datapoint, election sizes: \{30\}x\{100\} \vspace{2mm} \end{flushleft} \end{minipage} \begin{minipage}[t]{10cm} \begin{flushleft}  Note (general): Each candidate is generated a quality from [0,1] using a Gaussian distribution. Each voter approves a candidate with probability proportional to the candidate’s quality. The authors also consider a noise model, where the voters’ perception of the quality is distorted. \vspace{2mm} \end{flushleft} \end{minipage} \begin{minipage}[t]{10cm} \begin{flushleft}\rule{10cm}{0.3pt} \\ Comment: Has an arXiv version: https://arxiv.org/abs/2005.07094 \end{flushleft} \end{minipage}  \\ 
\midrule
112 & \begin{minipage}[t]{3cm}\begin{flushleft}\fontsize{9}{9.5}\selectfont An Analysis of Approval-Based Committee Rules for 2D-Euclidean Elections \vspace{2mm} \newline AAAI-2021  \vspace{2mm}\newline  \citeapp{god-bat-sko-fal:c:abc-rules-euclidean-2d} \end{flushleft}\end{minipage} & \begin{minipage}[t]{3cm}\begin{flushleft}\fontsize{9}{9.5}\selectfont M.~T.~Godziszewski, P.~Batko, P.~Skowron, P.~Faliszewski\end{flushleft}\end{minipage} & \fontsize{9}{9.5}\selectfont \begin{minipage}[t]{10cm} \begin{flushleft} A1: Euclidean 2D \newline 2000 samples/datapoint, election sizes: \{100\}x\{100\} \vspace{2mm} \end{flushleft} \end{minipage} \begin{minipage}[t]{10cm} \begin{flushleft}  Samples note: For each point generation model, 2000 election of size {100}x{100} were sampled \newline \vspace{2mm} \end{flushleft} \end{minipage} \begin{minipage}[t]{10cm} \begin{flushleft}  Euclidean: More details in Visualization and Discussion sections.

Point generation models: a) uniform square, b) asymmetric Gaussian Model with two centres, c) overlapping squares.
Voters \& candidates types:
1.1) Fixed voters’ radii; zero candidate radii
1.2) Fixed voters’ ballot length; zero candidate radii
2.1) Randomly selected voters’ radii from [0,3]; zero candidate radii
2.2) Randomly selected voters’ ballot length from {1,…,100}; zero candidate radii
3) (Only for asymmetric model): Voters as above; candidate radii 1 for noncenter-going gaussian side and 1.5 for center-going gaussian side \vspace{2mm} \end{flushleft} \end{minipage} \begin{minipage}[t]{10cm} \begin{flushleft}  Note (parameters): All of the studied models come from Euclidean 2D, but some of them are 2D-CVR (Voter Candidate Range), some 2D-CR (Candidate Range), and some 2D-VR (Voter Range) \vspace{2mm} \end{flushleft} \end{minipage} \begin{minipage}[t]{10cm} \begin{flushleft}  Note (general): More details in Visualization and Discussion sections.

The authors plot histograms for the studied ABC rules (AV, CC, MAV, PAV, Phragmen, MES under name Rule X) and ways of generating elections. The numbers in parentheses over each column provide the
average number of candidates approved by a single voter. \vspace{2mm} \end{flushleft} \end{minipage} \begin{minipage}[t]{10cm} \begin{flushleft}\rule{10cm}{0.3pt} \\ Comment: Doesn’t have arxiv version \end{flushleft} \end{minipage}  \\ 
\midrule
113 & \begin{minipage}[t]{3cm}\begin{flushleft}\fontsize{9}{9.5}\selectfont Preference Elicitation as Average-Case Sorting \vspace{2mm} \newline AAAI-2021  \vspace{2mm}\newline  \citeapp{pet-pro:c:elicitation-average-case} \end{flushleft}\end{minipage} & \begin{minipage}[t]{3cm}\begin{flushleft}\fontsize{9}{9.5}\selectfont D.~Peters, A.~Procaccia\end{flushleft}\end{minipage} & \fontsize{9}{9.5}\selectfont \begin{minipage}[t]{10cm} \begin{flushleft} O1: Mallows \newline election sizes: [1, 10]x\{1\} \vspace{2mm} \end{flushleft} \end{minipage} \begin{minipage}[t]{10cm} \begin{flushleft}  Samples note: Number of samples is not clear. \newline \vspace{2mm} \end{flushleft} \end{minipage} \begin{minipage}[t]{10cm} \begin{flushleft}  Mallows: $\phi \in (0,1)$ \vspace{2mm} \end{flushleft} \end{minipage} \begin{minipage}[t]{10cm} \begin{flushleft}  Note (general): Experiments on values of phi \vspace{2mm} \end{flushleft} \end{minipage} \begin{minipage}[t]{10cm} \begin{flushleft} \rule{10cm}{0.3pt} \\ O2: PrefLib, Real-Life (beyond PrefLib) \newline 10 samples/datapoint, election sizes: \{19\}x\{8169\} \vspace{2mm} \end{flushleft} \end{minipage} \begin{minipage}[t]{10cm} \begin{flushleft}  Sizes note: Provided sizes are for the jokes dataset. Preflib sizes are not clear. \newline \vspace{2mm} \end{flushleft} \end{minipage} \begin{minipage}[t]{10cm} \begin{flushleft}  Mallows: as above  \vspace{2mm} \end{flushleft} \end{minipage} \begin{minipage}[t]{10cm} \begin{flushleft}  Note (general): as above \vspace{2mm} \end{flushleft} \end{minipage}  \\ 
\midrule
114 & \begin{minipage}[t]{3cm}\begin{flushleft}\fontsize{9}{9.5}\selectfont Modeling Voters in Multi-Winner Approval Voting \vspace{2mm} \newline AAAI-2021  \vspace{2mm}\newline  \citeapp{sch-har-mat-ven:c:voter-model} \end{flushleft}\end{minipage} & \begin{minipage}[t]{3cm}\begin{flushleft}\fontsize{9}{9.5}\selectfont J.~Scheuerman, J.~Harman, N.~Mattei, K.~Venable\end{flushleft}\end{minipage} & \fontsize{9}{9.5}\selectfont \begin{minipage}[t]{10cm} \begin{flushleft} A1: Real-Life (beyond PrefLib) \newline election sizes: 5x? \vspace{2mm} \end{flushleft} \end{minipage} \begin{minipage}[t]{10cm} \begin{flushleft}  Sizes note: a few voters, but exact number is unclear to us \newline \vspace{2mm} \end{flushleft} \end{minipage} \begin{minipage}[t]{10cm} \begin{flushleft}  Note (general): Their  two real-world elections are in a very non-standard format. They present 104 study participants with partial election results (i.e., how many votes candidates received and how many votes are still missing) as well as how much utility the voter would derive from getting the candidate elected. As such they only track single-votes and do not analyze full elections. \vspace{2mm} \end{flushleft} \end{minipage} \begin{minipage}[t]{10cm} \begin{flushleft}\rule{10cm}{0.3pt} \\ Comment: Has an arXiv version: https://arxiv.org/abs/2012.02811 \end{flushleft} \end{minipage}  \\ 
\midrule
115 & \begin{minipage}[t]{3cm}\begin{flushleft}\fontsize{9}{9.5}\selectfont Margin of Victory in Tournaments: Structural and Experimental Results \vspace{2mm} \newline AAAI-2021  \vspace{2mm}\newline  \citeapp{bri-sch-suk:c:mov-tournaments} \end{flushleft}\end{minipage} & \begin{minipage}[t]{3cm}\begin{flushleft}\fontsize{9}{9.5}\selectfont M.~Brill, U.~Schmidt{-}Kraepelin, W.~Suksompong\end{flushleft}\end{minipage} & \fontsize{9}{9.5}\selectfont \begin{minipage}[t]{10cm} \begin{flushleft} O1: Impartial Culture, Mallows, Urn Model \newline 100 samples/datapoint, election sizes: \{5, 10, 15, 20, 25, 30\}x\{51\} \vspace{2mm} \end{flushleft} \end{minipage} \begin{minipage}[t]{10cm} \begin{flushleft}  Sizes note: The number of voters is available in the full version. \newline \vspace{2mm} \end{flushleft} \end{minipage} \begin{minipage}[t]{10cm} \begin{flushleft}  Mallows: $\phi = 0.95$ \vspace{2mm} \end{flushleft} \end{minipage} \begin{minipage}[t]{10cm} \begin{flushleft}  Urn: adding 10 copies of a vote after sampling it ($\alpha = \frac{10}{m!}$) \vspace{2mm} \end{flushleft} \end{minipage} \begin{minipage}[t]{10cm} \begin{flushleft}  Note (general): Some details of the experiments are available in the arXiv technical report (https://arxiv.org/pdf/2012.02657.pdf). The conference version does not show results for IC and Mallows (but points to the full version); the results for these models are included in the analysis though (Sec. 4, Observations). \vspace{2mm} \end{flushleft} \end{minipage} \begin{minipage}[t]{10cm} \begin{flushleft}\rule{10cm}{0.3pt} \\ Comment: The paper is about tournaments but uses preference-like data to generate them. 

Has a journal version: Markus Brill, Ulrike Schmidt-Kraepelin, Warut Suksompong: Margin of victory for tournament solutions. Artif. Intell. 302: 103600 (2022) \end{flushleft} \end{minipage}  \\ 
\midrule
116 & \begin{minipage}[t]{3cm}\begin{flushleft}\fontsize{9}{9.5}\selectfont Surprisingly Popular Voting Recovers Rankings, Surprisingly! \vspace{2mm} \newline IJCAI-2021  \vspace{2mm}\newline  \citeapp{hos-man-sha-shi:c:popular-coting-recovering} \end{flushleft}\end{minipage} & \begin{minipage}[t]{3cm}\begin{flushleft}\fontsize{9}{9.5}\selectfont H.~Hosseini, D.~Mandal, N.~Shah, K.~Shi\end{flushleft}\end{minipage} & \fontsize{9}{9.5}\selectfont \begin{minipage}[t]{10cm} \begin{flushleft} O1: Real-Life (beyond PrefLib) \newline 60 samples/datapoint, election sizes: \{4\}x\{60\} \vspace{2mm} \end{flushleft} \end{minipage} \begin{minipage}[t]{10cm} \begin{flushleft}  Note (general): Behavioral experiment with rankings based on MTurkers replies to questions. \vspace{2mm} \end{flushleft} \end{minipage} \begin{minipage}[t]{10cm} \begin{flushleft}\rule{10cm}{0.3pt} \\ Comment: Has an arXiv version: https://arxiv.org/abs/2105.09386 \end{flushleft} \end{minipage}  \\ 
\midrule
117 & \begin{minipage}[t]{3cm}\begin{flushleft}\fontsize{9}{9.5}\selectfont Combining Fairness and Optimality when Selecting and Allocating Projects \vspace{2mm} \newline IJCAI-2021  \vspace{2mm}\newline  \citeapp{bel-mou-wil:c:project-allocations} \end{flushleft}\end{minipage} & \begin{minipage}[t]{3cm}\begin{flushleft}\fontsize{9}{9.5}\selectfont K.~Belahc{\`{e}}ne, V.~Mousseau, A.~Wilczynski\end{flushleft}\end{minipage} & \fontsize{9}{9.5}\selectfont \begin{minipage}[t]{10cm} \begin{flushleft} O1: Impartial Culture, Single-Peaked (Conitzer/Random Peak), Single-Peaked (Walsh/Uniform) \newline 100 samples/datapoint, election sizes: \{15\}x\{5\} \vspace{2mm} \end{flushleft} \end{minipage} \begin{minipage}[t]{10cm} \begin{flushleft}\rule{10cm}{0.3pt} \\ Comment: Election-like structure (project allocations). \end{flushleft} \end{minipage}  \\ 
\midrule
118 & \begin{minipage}[t]{3cm}\begin{flushleft}\fontsize{9}{9.5}\selectfont Social Ranking Manipulability for the CP-Majority, Banzhaf and Lexicographic Excellence Solutions \vspace{2mm} \newline IJCAI-2021  \vspace{2mm}\newline  \citeapp{all-esc-mor-ozt:c:social-ranking-manipulability} \end{flushleft}\end{minipage} & \begin{minipage}[t]{3cm}\begin{flushleft}\fontsize{9}{9.5}\selectfont T.~Allouche, B.~Escoffier, S.~Moretti, M.~{\"O}zt{\"u}rk\end{flushleft}\end{minipage} & \fontsize{9}{9.5}\selectfont \begin{minipage}[t]{10cm} \begin{flushleft} O1: Impartial Culture \newline 1000 samples/datapoint, election sizes: [2, 32]x[2, 5] \vspace{2mm} \end{flushleft} \end{minipage} \begin{minipage}[t]{10cm} \begin{flushleft}  Sizes note: [2,5] individuals, voting is over the power set of individuals \newline \vspace{2mm} \end{flushleft} \end{minipage}  \\ 
\midrule
119 & \begin{minipage}[t]{3cm}\begin{flushleft}\fontsize{9}{9.5}\selectfont Dynamic Proportional Rankings \vspace{2mm} \newline IJCAI-2021  \vspace{2mm}\newline  \citeapp{isr-bri:c:dynamic-ranking} \end{flushleft}\end{minipage} & \begin{minipage}[t]{3cm}\begin{flushleft}\fontsize{9}{9.5}\selectfont J.~Israel, M.~Brill\end{flushleft}\end{minipage} & \fontsize{9}{9.5}\selectfont \begin{minipage}[t]{10cm} \begin{flushleft} A1: Euclidean 2D, specific own model \newline 100 samples/datapoint, election sizes: \{20\}x\{60\} \vspace{2mm} \end{flushleft} \end{minipage} \begin{minipage}[t]{10cm} \begin{flushleft}  Euclidean: there are multiple Gaussians with equidistant center points  and both voters and candidates are sampled points from one of these Gaussians \vspace{2mm} \end{flushleft} \end{minipage} \begin{minipage}[t]{10cm} \begin{flushleft}  Note (general): Both models create elections that are somewhat close to party-list elections. \vspace{2mm} \end{flushleft} \end{minipage} \begin{minipage}[t]{10cm} \begin{flushleft}\rule{10cm}{0.3pt} \\ Comment: Has an arXiv version: https://arxiv.org/abs/2105.08043 \end{flushleft} \end{minipage}  \\ 
\midrule
120 & \begin{minipage}[t]{3cm}\begin{flushleft}\fontsize{9}{9.5}\selectfont Winner Robustness via Swap- and Shift-Bribery: Parameterized Counting Complexity and Experiments \vspace{2mm} \newline IJCAI-2021  \vspace{2mm}\newline  \citeapp{boe-bre-fal-nie:c:winner-robustness-swap-shift-bribery-counting-complexity} \end{flushleft}\end{minipage} & \begin{minipage}[t]{3cm}\begin{flushleft}\fontsize{9}{9.5}\selectfont N.~Boehmer, R.~Bredereck, P.~Faliszewski, R.~Niedermeier\end{flushleft}\end{minipage} & \fontsize{9}{9.5}\selectfont \begin{minipage}[t]{10cm} \begin{flushleft} O1: Euclidean 1D, Euclidean 2D, Euclidean 3D-or-more, Impartial Culture, Mallows, Single-Crossing, Single-Peaked (Conitzer/Random Peak), Single-Peaked (Walsh/Uniform), Urn Model \newline 10000 samples/datapoint, election sizes: \{10\}x\{100\} \vspace{2mm} \end{flushleft} \end{minipage} \begin{minipage}[t]{10cm} \begin{flushleft}  Samples note: The same 800 elections as in “Putting the Compass on the Map of Elections”.

For each election (except those
with tied winners) and each normalized swap distance $r \in
\{0.05, 0.1, \ldots , 1\}$ we sampled 500 elections at this distance
and for each candidate recorded the proportion of elections
where he or she won.
Therefore, we sampled 8000000 = 800 x 20 x 500 in total, 20 x 500 = 10000 per data point. \newline \vspace{2mm} \end{flushleft} \end{minipage} \begin{minipage}[t]{10cm} \begin{flushleft}  Mallows: The same as in “Putting the Compass on the Map of Elections” \vspace{2mm} \end{flushleft} \end{minipage} \begin{minipage}[t]{10cm} \begin{flushleft}  Urn: The same as in “Putting the Compass on the Map of Elections” \vspace{2mm} \end{flushleft} \end{minipage} \begin{minipage}[t]{10cm} \begin{flushleft}  Euclidean: Uniform 1D {[0,1]}; Uniform 2D ($[0,1]^2$); Uniform 3D ($[0,1]^3$); Uniform 5D ($[0,1]^5$); Uniform 10D ($[0,1]^10$); Uniform 20D ($[0,1]^20$); Uniform 2D Sphere ($(0, 0)$ $r=1$); Uniform 3D Sphere ($(0, 0)$ $r=1$); Uniform 5D Sphere ($(0, 0)$ $r=1$) \vspace{2mm} \end{flushleft} \end{minipage} \begin{minipage}[t]{10cm} \begin{flushleft}  Note (parameters): The same as in “Putting the Compass on the Map of Elections” \vspace{2mm} \end{flushleft} \end{minipage} \begin{minipage}[t]{10cm} \begin{flushleft}  Note (general): Section 4. Experiments \vspace{2mm} \end{flushleft} \end{minipage} \begin{minipage}[t]{10cm} \begin{flushleft} \rule{10cm}{0.3pt} \\ O2: Urn Model \newline 800000 samples/datapoint, election sizes: \{10\}x\{100\} \vspace{2mm} \end{flushleft} \end{minipage} \begin{minipage}[t]{10cm} \begin{flushleft}  Samples note: Two particular elections out of 800 elections from “Putting the Compass on the Map of Elections”.

For each of these two elections, each $r \in \{0.0125, 0.025, \ldots , 0.5\}$, the authors used 10’000 samples.

2 * 40 * 10000 = 800000 \newline \vspace{2mm} \end{flushleft} \end{minipage} \begin{minipage}[t]{10cm} \begin{flushleft}  Urn: $\alpha \in \{0.01, 0.02\}$ \vspace{2mm} \end{flushleft} \end{minipage} \begin{minipage}[t]{10cm} \begin{flushleft}  Note (parameters): To exemplify this phenomenon, we visualize PE,c(r) under Borda for two particular elections from the 800-elections dataset in Figure 2 (for these two elections, we estimated PE,c(r) for $r \in \{0.0125, 0.025, . . . , 0.5\}$ using 10000 samples in each case). We want to emphasize that these elections
are not artificial extreme examples but were generated both as part of 180 elections generated by the Urn model with different parameters. In fact, there exist several other elections
in the dataset exhibiting a similar behavior. \vspace{2mm} \end{flushleft} \end{minipage} \begin{minipage}[t]{10cm} \begin{flushleft}  Note (general): as above \vspace{2mm} \end{flushleft} \end{minipage}  \\ 
\midrule
121 & \begin{minipage}[t]{3cm}\begin{flushleft}\fontsize{9}{9.5}\selectfont Putting a Compass on the Map of Elections \vspace{2mm} \newline IJCAI-2021  \vspace{2mm}\newline  \citeapp{boe-bre-fal-nie-szu:c:compass} \end{flushleft}\end{minipage} & \begin{minipage}[t]{3cm}\begin{flushleft}\fontsize{9}{9.5}\selectfont N.~Boehmer, R.~Bredereck, P.~Faliszewski, R.~Niedermeier, S.~Szufa\end{flushleft}\end{minipage} & \fontsize{9}{9.5}\selectfont \begin{minipage}[t]{10cm} \begin{flushleft} O1: PrefLib \newline election sizes: \{10\}x\{100\} \vspace{2mm} \end{flushleft} \end{minipage} \begin{minipage}[t]{10cm} \begin{flushleft}  Sizes note: Glasgow, Aspen, Irish, ERS, TDF, GDI, speed skating, figure skating, Sushi, T-Shirt, cities. \newline \vspace{2mm} \end{flushleft} \end{minipage} \begin{minipage}[t]{10cm} \begin{flushleft} \rule{10cm}{0.3pt} \\ O2: Euclidean 1D, Euclidean 2D, Euclidean 3D-or-more, Group-Separable, Impartial Culture, Mallows, Single-Crossing, Single-Peaked (Conitzer/Random Peak), Single-Peaked (Walsh/Uniform), Urn Model \newline election sizes: \{10\}x\{100\} \vspace{2mm} \end{flushleft} \end{minipage} \begin{minipage}[t]{10cm} \begin{flushleft}  Samples note: 800 points in total \newline \vspace{2mm} \end{flushleft} \end{minipage} \begin{minipage}[t]{10cm} \begin{flushleft}  Euclidean: Uniform 1D {[0,1]}; Uniform 2D ($[0,1]^2$); Uniform 3D ($[0,1]^3$); Uniform 5D ($[0,1]^5$); Uniform 10D ($[0,1]^10$); Uniform 20D ($[0,1]^20$); Uniform 2D Sphere ($(0, 0)$ $r=1$); Uniform 3D Sphere ($(0, 0)$ $r=1$); Uniform 5D Sphere ($(0, 0)$ $r=1$) \vspace{2mm} \end{flushleft} \end{minipage} \begin{minipage}[t]{10cm} \begin{flushleft} \rule{10cm}{0.3pt} \\ O3: Mallows, Urn Model \newline election sizes: \{20\}x\{100\} \vspace{2mm} \end{flushleft} \end{minipage} \begin{minipage}[t]{10cm} \begin{flushleft} \rule{10cm}{0.3pt} \\ O4: Mallows \newline election sizes: \{5, 10, 20, 50, 100\}x\{100\} \vspace{2mm} \end{flushleft} \end{minipage} \begin{minipage}[t]{10cm} \begin{flushleft}\rule{10cm}{0.3pt} \\ Comment: Has an arXiv version: https://arxiv.org/abs/2105.07815 \end{flushleft} \end{minipage}  \\ 
\midrule
\multicolumn{4}{c}{\bf2022}\\ \midrule
122 & \begin{minipage}[t]{3cm}\begin{flushleft}\fontsize{9}{9.5}\selectfont A Graph-Based Algorithm for the Automated Justification of Collective
Decisions \vspace{2mm} \newline AAMAS-2022  \vspace{2mm}\newline  \citeapp{nar-boi-end:c:graph-automated-decisions} \end{flushleft}\end{minipage} & \begin{minipage}[t]{3cm}\begin{flushleft}\fontsize{9}{9.5}\selectfont O.~Nardi, A.~Boixel, U.~Endriss\end{flushleft}\end{minipage} & \fontsize{9}{9.5}\selectfont \begin{minipage}[t]{10cm} \begin{flushleft} O1: exhaustive \newline election sizes: \{3\}x[2, 8], \{4\}x[2, 4] \vspace{2mm} \end{flushleft} \end{minipage} \begin{minipage}[t]{10cm} \begin{flushleft}  Samples note: All possible profiles \newline \vspace{2mm} \end{flushleft} \end{minipage} \begin{minipage}[t]{10cm} \begin{flushleft} \rule{10cm}{0.3pt} \\ O2: Impartial Culture, PrefLib \newline 180 samples/datapoint, election sizes: \{3\}x[10, 12], \{4\}x[5, 8] \vspace{2mm} \end{flushleft} \end{minipage}  \\ 
\midrule
123 & \begin{minipage}[t]{3cm}\begin{flushleft}\fontsize{9}{9.5}\selectfont Justifying Social-Choice Mechanism Outcome for Improving Participant Satisfaction \vspace{2mm} \newline AAMAS-2022  \vspace{2mm}\newline  \citeapp{sur-sar-kra:c:justifying-outcome} \end{flushleft}\end{minipage} & \begin{minipage}[t]{3cm}\begin{flushleft}\fontsize{9}{9.5}\selectfont S.~Suryanarayana, D.~Sarne, S.~Kraus\end{flushleft}\end{minipage} & \fontsize{9}{9.5}\selectfont \begin{minipage}[t]{10cm} \begin{flushleft} O1: Hand-Crafted \newline 465 samples/datapoint, election sizes: \{3\}x\{29\} \vspace{2mm} \end{flushleft} \end{minipage} \begin{minipage}[t]{10cm} \begin{flushleft}  Note (general): Experiment on humans interacting with elections. 4 different profiles were considered, and they were randomly presented to each of 465 participants of the study. \vspace{2mm} \end{flushleft} \end{minipage} \begin{minipage}[t]{10cm} \begin{flushleft}\rule{10cm}{0.3pt} \\ Comment: Has an arXiv version: https://arxiv.org/abs/2205.15863 \end{flushleft} \end{minipage}  \\ 
\midrule
124 & \begin{minipage}[t]{3cm}\begin{flushleft}\fontsize{9}{9.5}\selectfont A Graph-Based Algorithm for the Automated Justification of Collective Decisions \vspace{2mm} \newline AAMAS-2022  \vspace{2mm}\newline  \citeapp{nar-boi-end:c:graph-justification-decisions} \end{flushleft}\end{minipage} & \begin{minipage}[t]{3cm}\begin{flushleft}\fontsize{9}{9.5}\selectfont O.~Nardi, A.~Boixel, U.~Endriss\end{flushleft}\end{minipage} & \fontsize{9}{9.5}\selectfont \begin{minipage}[t]{10cm} \begin{flushleft} O1: Impartial Culture, PrefLib, exhaustive \newline 180 samples/datapoint, election sizes: [3, 4]x[2, 12] \vspace{2mm} \end{flushleft} \end{minipage} \begin{minipage}[t]{10cm} \begin{flushleft}  Samples note: For Preflib sampling, bootstraping was used due to large number of voters in the samples. \newline Sizes note: For profiles with 3 alternatives [2,8] voters were used for the exhaustive case, while [10-12] were used for other methods. For 4 alternatives, [2,4] voters were used in the exhaustive case, while [5,8] were used in other methods.  \newline \vspace{2mm} \end{flushleft} \end{minipage}  \\ 
\midrule
125 & \begin{minipage}[t]{3cm}\begin{flushleft}\fontsize{9}{9.5}\selectfont Welfare vs. Representation in Participatory Budgeting \vspace{2mm} \newline AAMAS-2022  \vspace{2mm}\newline  \citeapp{fai-vil-mei-gal:c:welfare-vs-representation-participatory-budgeting} \end{flushleft}\end{minipage} & \begin{minipage}[t]{3cm}\begin{flushleft}\fontsize{9}{9.5}\selectfont R.~Fairstein, D.~Vilenchik, R.~Meir, K.~Gal\end{flushleft}\end{minipage} & \fontsize{9}{9.5}\selectfont \begin{minipage}[t]{10cm} \begin{flushleft} A1: Euclidean 2D, PabuLib, Party-list \newline 1000 samples/datapoint, election sizes: \{100\}x\{1000\}, \{420\}x\{200\} \vspace{2mm} \end{flushleft} \end{minipage} \begin{minipage}[t]{10cm} \begin{flushleft}  Samples note: Except for pabulib, where 130 instances were considered . \newline Sizes note: Number 420 is estimated (as a mean of the number of candidates chosen uniformly at random) \newline \vspace{2mm} \end{flushleft} \end{minipage} \begin{minipage}[t]{10cm} \begin{flushleft}  Euclidean: [0,1]x[0,1] square with voters and candidates sampled from a multivariate normal distribution with  uncorrelated distributions with means in 0.5 and 0.5 and std 0.2 and 0.2. Each voter approves max(x,1) closest candidates, where x is sampled from a normal distribution with mean 10 and std 3. \vspace{2mm} \end{flushleft} \end{minipage} \begin{minipage}[t]{10cm} \begin{flushleft}  Note (parameters): Party-list: voters split at random into groups of sizes 5 to 20, then each group supports between  x in [10, 30] unique candidates, where x is chosen uniformly at random, \vspace{2mm} \end{flushleft} \end{minipage} \begin{minipage}[t]{10cm} \begin{flushleft}  Note (general): Pabulib: instances from Warsaw, Poland, years 2017-2021, each instance included between 50-10,000 voters (2,982 on average) and between 20-100 projects (36 on average). \vspace{2mm} \end{flushleft} \end{minipage} \begin{minipage}[t]{10cm} \begin{flushleft}\rule{10cm}{0.3pt} \\ Comment: There is an arXiv version: https://arxiv.org/abs/2201.07546 \end{flushleft} \end{minipage}  \\ 
\midrule
126 & \begin{minipage}[t]{3cm}\begin{flushleft}\fontsize{9}{9.5}\selectfont Worst-Case Voting When the Stakes Are High \vspace{2mm} \newline AAAI-2022  \vspace{2mm}\newline  \citeapp{kah-keh:c:worst-case-voting} \end{flushleft}\end{minipage} & \begin{minipage}[t]{3cm}\begin{flushleft}\fontsize{9}{9.5}\selectfont A.~Kahng, G.~Kehne\end{flushleft}\end{minipage} & \fontsize{9}{9.5}\selectfont \begin{minipage}[t]{10cm} \begin{flushleft} O1: PrefLib \newline election sizes: [3, 6]x[532, 1960], \{13\}x[5199, 12744], [4, 9]x[142, 504], \{5\}x[13318, 20239] \vspace{2mm} \end{flushleft} \end{minipage} \begin{minipage}[t]{10cm} \begin{flushleft}  Sizes note:  * Vermont consists of data from public office elections in 2014 (15 different races, with 3 to 6 candidates and 532
to 1960 voters per race)
 * Glasgow consists of data from the 2007 Glasgow City Council elections (21 wards, with 13 candidates and 5199 to 12744 voters per ward)
 * Debian consists of votes for the Debian logo (8 elections, with 4 to
9 alternatives and 142 to 504 voters per election)
 * APA consists of election data from the American Psychological
Association between 1998 and 2009 (12 elections, with 5
alternatives and 13318 and 20239 voters). \newline \vspace{2mm} \end{flushleft} \end{minipage}  \\ 
\midrule
127 & \begin{minipage}[t]{3cm}\begin{flushleft}\fontsize{9}{9.5}\selectfont The Complexity of Subelection Isomorphism Problems \vspace{2mm} \newline AAAI-2022  \vspace{2mm}\newline  \citeapp{fal-sor-szu:c:subelection-complexity} \end{flushleft}\end{minipage} & \begin{minipage}[t]{3cm}\begin{flushleft}\fontsize{9}{9.5}\selectfont P.~Faliszewski, K.~Sornat, S.~Szufa\end{flushleft}\end{minipage} & \fontsize{9}{9.5}\selectfont \begin{minipage}[t]{10cm} \begin{flushleft} O1: Euclidean 1D, Identity, Impartial Culture, Mallows, Single-Peaked (Conitzer/Random Peak), Single-Peaked (Walsh/Uniform), Urn Model \newline 1000 samples/datapoint, election sizes: \{4, 7, 10\}x\{50\} \vspace{2mm} \end{flushleft} \end{minipage} \begin{minipage}[t]{10cm} \begin{flushleft}  Mallows: $\phi \in \{0.33,0.67\}$ \vspace{2mm} \end{flushleft} \end{minipage} \begin{minipage}[t]{10cm} \begin{flushleft}  Urn: $\alpha \in \{0.1,0.5\}$ \vspace{2mm} \end{flushleft} \end{minipage} \begin{minipage}[t]{10cm} \begin{flushleft}  Euclidean: Uniform 1D {[0,1]} \vspace{2mm} \end{flushleft} \end{minipage} \begin{minipage}[t]{10cm} \begin{flushleft}  Note (parameters): Norm-Mallows model was used, and the Mallows parameter value is norm-phi. \vspace{2mm} \end{flushleft} \end{minipage} \begin{minipage}[t]{10cm} \begin{flushleft} \rule{10cm}{0.3pt} \\ O2: Euclidean 1D, Identity, Impartial Culture, Mallows, Single-Peaked (Conitzer/Random Peak), Single-Peaked (Walsh/Uniform) \newline 1000 samples/datapoint, election sizes: \{10\}x\{5, 10, 15, 20, 25, 30, 35, 40, 45, 50\}, [3, 10]x\{50\} \vspace{2mm} \end{flushleft} \end{minipage} \begin{minipage}[t]{10cm} \begin{flushleft}  Mallows: $\phi = {0.5}$ \vspace{2mm} \end{flushleft} \end{minipage} \begin{minipage}[t]{10cm} \begin{flushleft}  Euclidean: as above  \vspace{2mm} \end{flushleft} \end{minipage} \begin{minipage}[t]{10cm} \begin{flushleft}  Note (parameters): as above \vspace{2mm} \end{flushleft} \end{minipage} \begin{minipage}[t]{10cm} \begin{flushleft}\rule{10cm}{0.3pt} \\ Comment: Has an arXiv version: https://arxiv.org/abs/2105.11923 \end{flushleft} \end{minipage}  \\ 
\midrule
128 & \begin{minipage}[t]{3cm}\begin{flushleft}\fontsize{9}{9.5}\selectfont The Price of Justified Representation \vspace{2mm} \newline AAAI-2022  \vspace{2mm}\newline  \citeapp{elk-fal-iga-man-sch-suk:c:price-justified-representation} \end{flushleft}\end{minipage} & \begin{minipage}[t]{3cm}\begin{flushleft}\fontsize{9}{9.5}\selectfont E.~Elkind, P.~Faliszewski, A.~Igarashi, P.~Manurangsi, U.~Schmidt{-}Kraepelin, W.~Suksompong\end{flushleft}\end{minipage} & \fontsize{9}{9.5}\selectfont \begin{minipage}[t]{10cm} \begin{flushleft} A1: Euclidean 1D, Euclidean 2D, Impartial Culture \newline 100 samples/datapoint, election sizes: \{100\}x\{100\} \vspace{2mm} \end{flushleft} \end{minipage} \begin{minipage}[t]{10cm} \begin{flushleft}  Euclidean: 1D: r=0.054, 2D: r=0.195 \vspace{2mm} \end{flushleft} \end{minipage} \begin{minipage}[t]{10cm} \begin{flushleft}  Note (parameters): In all cultures voters approve 10 candidates on average. \vspace{2mm} \end{flushleft} \end{minipage} \begin{minipage}[t]{10cm} \begin{flushleft}\rule{10cm}{0.3pt} \\ Comment: There is an arXiv version: https://arxiv.org/abs/2112.05994 \end{flushleft} \end{minipage}  \\ 
\midrule
129 & \begin{minipage}[t]{3cm}\begin{flushleft}\fontsize{9}{9.5}\selectfont Individual Representation in Approval-Based Committee Voting \vspace{2mm} \newline AAAI-2022  \vspace{2mm}\newline  \citeapp{bri-isr-mich-pet:c:representation-committee-voting} \end{flushleft}\end{minipage} & \begin{minipage}[t]{3cm}\begin{flushleft}\fontsize{9}{9.5}\selectfont M.~Brill, J.~Israel, E.~Micha, J.~Peters\end{flushleft}\end{minipage} & \fontsize{9}{9.5}\selectfont \begin{minipage}[t]{10cm} \begin{flushleft} A1: Euclidean 2D, Impartial Culture, Mallows, Urn Model, candidate interval (CI), voter interval (VI) \newline 500 samples/datapoint, election sizes: \{50\}x\{100\} \vspace{2mm} \end{flushleft} \end{minipage} \begin{minipage}[t]{10cm} \begin{flushleft}\rule{10cm}{0.3pt} \\ Comment: There is an arXiv version: https://arxiv.org/abs/2112.05193 \end{flushleft} \end{minipage}  \\ 
\midrule
130 & \begin{minipage}[t]{3cm}\begin{flushleft}\fontsize{9}{9.5}\selectfont Truth-Tracking via Approval Voting: Size Matters \vspace{2mm} \newline AAAI-2022  \vspace{2mm}\newline  \citeapp{all-lan-yge:c:truth-tracking-av} \end{flushleft}\end{minipage} & \begin{minipage}[t]{3cm}\begin{flushleft}\fontsize{9}{9.5}\selectfont T.~Allouche, J.~Lang, F.~Yger\end{flushleft}\end{minipage} & \fontsize{9}{9.5}\selectfont \begin{minipage}[t]{10cm} \begin{flushleft} A1: Real-Life (beyond PrefLib) \newline 16.25 samples/datapoint, election sizes: \{6\}x\{203, 213\}, \{8\}x\{203\} \vspace{2mm} \end{flushleft} \end{minipage} \begin{minipage}[t]{10cm} \begin{flushleft}  Samples note: Not clear if the number of samples indeed is recorded correctly (it refers to the number of questions in the dataset, where each question can be seen as a separate election) \newline \vspace{2mm} \end{flushleft} \end{minipage} \begin{minipage}[t]{10cm} \begin{flushleft}  Note (general): Data comes from the work of Shah, Zhou and Peres (Approval voting and incentives in crowdsourcing. ICML 2015). \url{https://www.cs.cmu.edu/%E2%88%BCnihars/data/data_approval.zip} \vspace{2mm} \end{flushleft} \end{minipage} \begin{minipage}[t]{10cm} \begin{flushleft}\rule{10cm}{0.3pt} \\ Comment: There is an arXiv version: https://arxiv.org/abs/2112.04387 \end{flushleft} \end{minipage}  \\ 
\midrule
131 & \begin{minipage}[t]{3cm}\begin{flushleft}\fontsize{9}{9.5}\selectfont DiRe Committee : Diversity and Representation Constraints in Multiwinner Elections \vspace{2mm} \newline IJCAI-2022  \vspace{2mm}\newline  \citeapp{rel:c:dire} \end{flushleft}\end{minipage} & \begin{minipage}[t]{3cm}\begin{flushleft}\fontsize{9}{9.5}\selectfont K.~Relia\end{flushleft}\end{minipage} & \fontsize{9}{9.5}\selectfont \begin{minipage}[t]{10cm} \begin{flushleft} O1: Real-Life (beyond PrefLib) \newline election sizes: \{10\}x\{26\} \vspace{2mm} \end{flushleft} \end{minipage} \begin{minipage}[t]{10cm} \begin{flushleft}  Note (general): Eurovision song contest finalist rankings (https://www.kaggle.com/datasets/datagraver/eurovision-song-contest-scores-19752019) \vspace{2mm} \end{flushleft} \end{minipage} \begin{minipage}[t]{10cm} \begin{flushleft} \rule{10cm}{0.3pt} \\ O2: Mallows \newline 10 samples/datapoint, election sizes: \{50\}x\{100\} \vspace{2mm} \end{flushleft} \end{minipage} \begin{minipage}[t]{10cm} \begin{flushleft}  Mallows: $\phi = {0.5}$ \vspace{2mm} \end{flushleft} \end{minipage} \begin{minipage}[t]{10cm} \begin{flushleft} \rule{10cm}{0.3pt} \\ O3: Mallows \newline 5 samples/datapoint, election sizes: \{50\}x\{100\} \vspace{2mm} \end{flushleft} \end{minipage} \begin{minipage}[t]{10cm} \begin{flushleft}  Mallows: $\phi \in [0.1,1]$, step 0.1 \vspace{2mm} \end{flushleft} \end{minipage} \begin{minipage}[t]{10cm} \begin{flushleft} \rule{10cm}{0.3pt} \\ A1: Real-Life (beyond PrefLib) \newline election sizes: \{81\}x\{193\} \vspace{2mm} \end{flushleft} \end{minipage} \begin{minipage}[t]{10cm} \begin{flushleft}  Note (general): United Nations Resolutions dataset [Voeten, 2014] \vspace{2mm} \end{flushleft} \end{minipage} \begin{minipage}[t]{10cm} \begin{flushleft}\rule{10cm}{0.3pt} \\ Comment: Has an arxiv version: https://arxiv.org/abs/2107.07356 \end{flushleft} \end{minipage}  \\ 
\midrule
132 & \begin{minipage}[t]{3cm}\begin{flushleft}\fontsize{9}{9.5}\selectfont Let's Agree to Agree: Targeting Consensus for Incomplete Preferences through Majority Dynamics \vspace{2mm} \newline IJCAI-2022  \vspace{2mm}\newline  \citeapp{bot-rey-ter:c:majority-dynamics} \end{flushleft}\end{minipage} & \begin{minipage}[t]{3cm}\begin{flushleft}\fontsize{9}{9.5}\selectfont S.~Botan, S.~Rey, Z.~Terzopoulou\end{flushleft}\end{minipage} & \fontsize{9}{9.5}\selectfont \begin{minipage}[t]{10cm} \begin{flushleft} O1: Hand-Crafted \newline 5000000 samples/datapoint, election sizes: \{5\}x[1-25] \vspace{2mm} \end{flushleft} \end{minipage} \begin{minipage}[t]{10cm} \begin{flushleft}  Note (general): Partial orders are considered. For each pair of candidates a,b they choose a>b, b>a, or a~b, with equal probability independently. If the resulting relation is nontransitive, they repeat. \vspace{2mm} \end{flushleft} \end{minipage} \begin{minipage}[t]{10cm} \begin{flushleft}\rule{10cm}{0.3pt} \\ Comment: Considers partial orders

There is an arXiv version: https://arxiv.org/abs/2205.00881 \end{flushleft} \end{minipage}  \\ 
\midrule
133 & \begin{minipage}[t]{3cm}\begin{flushleft}\fontsize{9}{9.5}\selectfont Phragm{\'{e}}n Rules for Degressive and Regressive Proportionality \vspace{2mm} \newline IJCAI-2022  \vspace{2mm}\newline  \citeapp{jaw-sko:c:phrag-degr-regr-proportionality} \end{flushleft}\end{minipage} & \begin{minipage}[t]{3cm}\begin{flushleft}\fontsize{9}{9.5}\selectfont M.~Jaworski, P.~Skowron\end{flushleft}\end{minipage} & \fontsize{9}{9.5}\selectfont \begin{minipage}[t]{10cm} \begin{flushleft} A1: Euclidean 1D \newline 1000 samples/datapoint, election sizes: 150x200 \vspace{2mm} \end{flushleft} \end{minipage} \begin{minipage}[t]{10cm} \begin{flushleft}  Samples note: No direct information about the number of samples, I put there the number of simulations for each scenario \newline Sizes note: committee size k=25

p=100, modified Bernoulli distribution $p_{\eta}$, see Subsection 5.3 \newline \vspace{2mm} \end{flushleft} \end{minipage} \begin{minipage}[t]{10cm} \begin{flushleft}  Euclidean: Voters

We draw the individuals independently at random from
beta distributions, scaled into [-1, 1]. We consider: Beta(1/2,1/2), Beta(1/2, 2), Beta(2, 2), Beta(2, 4). The voters’ preferences are constructed from their positions as follows. We fix the approval radius $\xi \in \{0.1, 0.2, 0.3, 0.4, 0.5\}$,
and assume that a voter $v$ approves a candidate $c$ if and only
if $|v - c| \leq \xi$. We set a threshold of 0.5 for the approval radius.

acceptance radius $\tau = 0.2$ and the parameters of the probability function $p\mu$ to: $\tau = 30$, $\delta = 120$. \vspace{2mm} \end{flushleft} \end{minipage} \begin{minipage}[t]{10cm} \begin{flushleft}  Note (parameters): n = 200 voters, m = 150 candidates
and for the committee size k = 25. 

We also checked
several others sets of parameters (e.g. {$\tau = 5$, $\delta  = 20$},
{$\tau = 10$, $\delta = 60$}), but we found that the key observations
and regularities stay the same. We ran 1000 simulations for
each scenario. \vspace{2mm} \end{flushleft} \end{minipage} \begin{minipage}[t]{10cm} \begin{flushleft}  Note (general): Section 5: Degressive and Regressive Proportionality
in the Euclidean Model

Results: table 1, figure 4, figure 5 \vspace{2mm} \end{flushleft} \end{minipage} \begin{minipage}[t]{10cm} \begin{flushleft}\rule{10cm}{0.3pt} \\ Comment: There is an arXiv version: https://arxiv.org/abs/2201.04248 \end{flushleft} \end{minipage}  \\ 
\midrule
134 & \begin{minipage}[t]{3cm}\begin{flushleft}\fontsize{9}{9.5}\selectfont Time-Constrained Participatory Budgeting Under Uncertain Project Costs \vspace{2mm} \newline IJCAI-2022  \vspace{2mm}\newline  \citeapp{bau-boe-lau:c:pb-uncertain-costs} \end{flushleft}\end{minipage} & \begin{minipage}[t]{3cm}\begin{flushleft}\fontsize{9}{9.5}\selectfont D.~Baumeister, L.~Boes, C.~Lau{\ss}mann\end{flushleft}\end{minipage} & \fontsize{9}{9.5}\selectfont \begin{minipage}[t]{10cm} \begin{flushleft} A1: PabuLib \newline election sizes: ? \vspace{2mm} \end{flushleft} \end{minipage} \begin{minipage}[t]{10cm} \begin{flushleft}  Note (general): it only says that “We test both algorithms using real data from [PabuLib]” They modify these instances to fit their uncertain cost setting but don’t specify how exactly they are modified. \vspace{2mm} \end{flushleft} \end{minipage}  \\ 
\midrule
135 & \begin{minipage}[t]{3cm}\begin{flushleft}\fontsize{9}{9.5}\selectfont How to Sample Approval Elections? \vspace{2mm} \newline IJCAI-2022  \vspace{2mm}\newline  \citeapp{szu-fal-jan-lac-sli-sor-tal:c:how-to-sample-approval-elections} \end{flushleft}\end{minipage} & \begin{minipage}[t]{3cm}\begin{flushleft}\fontsize{9}{9.5}\selectfont S.~Szufa, P.~Faliszewski, L.~Janeczko, M.~Lackner, A.~Slinko, K.~Sornat, N.~Talmon\end{flushleft}\end{minipage} & \fontsize{9}{9.5}\selectfont \begin{minipage}[t]{10cm} \begin{flushleft} A1: Disjoint, Euclidean 1D, Euclidean 2D, Identity (ID), Impartial Culture, Noise, Resampling, Truncated Urn \newline election sizes: \{100\}x\{1000\}, \{50\}x\{100\} \vspace{2mm} \end{flushleft} \end{minipage} \begin{minipage}[t]{10cm} \begin{flushleft}  Sizes note: If applicable, committee size is 10 \newline \vspace{2mm} \end{flushleft} \end{minipage} \begin{minipage}[t]{10cm} \begin{flushleft}  Euclidean: for 1D-Uniform,
radius in (0.0025, 0.25)

for 2D-Square, radius
in (0.005, 0.5)) \vspace{2mm} \end{flushleft} \end{minipage} \begin{minipage}[t]{10cm} \begin{flushleft}  Note (parameters): the Disjoint Model (50 for each $g \in \{2, 3, 4, 5, 6\}$ with $\phi \in (0.05, 1/g)$)

Noise Model with Hamming distance (25 for each $p \in \{0.1, 0.2, . . . , 0.9\}$ with $\phi \in (0, 1)$)

Truncated Urn Model (25 for each $p \in \{0.1, 0.2, . . . , 0.9\}$ with $\alpha \in (0, 1)$)

Euclidean Model (100 for 1D-Uniform, with radius in (0.0025, 0.25), and 100 for 2D-Square, with radius in (0.005, 0.5))

$(p, \phi)$-resampling elections, with $p$ and $\phi$ parameters as follows:
1. $p$ is chosen from $\{0, 0.1, 0.2, . . . , 0.9, 1\}$ and $\phi$ is chosen from the interval $(0, 1)$, $1$ or 
2. $\phi$ is chosen from $\{0, 0.25, 0.5, 0.75, 1\}$ and $p$ is chosen from the interval $(0, 1)$. \vspace{2mm} \end{flushleft} \end{minipage} \begin{minipage}[t]{10cm} \begin{flushleft}  Note (general): Sections 4, 5, 6

Experiment involved 1) computing max approval score, 2) computing max cohesiveness level, 3) computing the number of voters belonging to at least one cohesive group, 4) measuring PAV runtime.

Disjoint model is based on resampling model \vspace{2mm} \end{flushleft} \end{minipage} \begin{minipage}[t]{10cm} \begin{flushleft} \rule{10cm}{0.3pt} \\ A2: PabuLib \newline election sizes: \{50\}x\{1000\} \vspace{2mm} \end{flushleft} \end{minipage} \begin{minipage}[t]{10cm} \begin{flushleft}  Samples note: The sixth dataset uses real-life participatory budgeting data and contains 44 elections from Pabulib, where for each (large enough) election we randomly selected a subset of 50 candidates and 1000 voters.

44 elections \newline Sizes note: for each (large enough) election we randomly selected a subset of 50 candidates and 1000 voters \newline \vspace{2mm} \end{flushleft} \end{minipage} \begin{minipage}[t]{10cm} \begin{flushleft}  Note (general): Sections 5, 6

Experiment involved 1) computing max approval score, 2) computing max cohesiveness level, 3) computing the number of voters belonging to at least one cohesive group, 4) measuring PAV runtime. However, the experiment was conducted on truncated real-world data from Pabulib \vspace{2mm} \end{flushleft} \end{minipage} \begin{minipage}[t]{10cm} \begin{flushleft} \rule{10cm}{0.3pt} \\ A3: Disjoint, Euclidean 1D, Euclidean 2D, Identity (ID), Impartial Culture, Noise, Resampling, Truncated Urn \newline 363 samples/datapoint, election sizes: \{10\}x\{50\} \vspace{2mm} \end{flushleft} \end{minipage} \begin{minipage}[t]{10cm} \begin{flushleft}  Note (general): Sections 4, 5, 6, mainly 6.3

Measuring correlation between approvalwise distance and isomorphic Hamming distance \vspace{2mm} \end{flushleft} \end{minipage} \begin{minipage}[t]{10cm} \begin{flushleft}\rule{10cm}{0.3pt} \\ Comment: There is an arXiv version: https://arxiv.org/abs/2207.01140 \end{flushleft} \end{minipage}  \\ 
\midrule
136 & \begin{minipage}[t]{3cm}\begin{flushleft}\fontsize{9}{9.5}\selectfont Understanding Distance Measures Among Elections \vspace{2mm} \newline IJCAI-2022  \vspace{2mm}\newline  \citeapp{boe-fal-nie-szu-was:c:understanding-distance-measures-among-elections} \end{flushleft}\end{minipage} & \begin{minipage}[t]{3cm}\begin{flushleft}\fontsize{9}{9.5}\selectfont N.~Boehmer, P.~Faliszewski, R.~Niedermeier, S.~Szufa, T.~Was\end{flushleft}\end{minipage} & \fontsize{9}{9.5}\selectfont \begin{minipage}[t]{10cm} \begin{flushleft} O1: exhaustive \newline election sizes: \{3, 4\} x \{3, 4, 5\} \vspace{2mm} \end{flushleft} \end{minipage} \begin{minipage}[t]{10cm} \begin{flushleft}  Samples note: All elections of given sizes \newline \vspace{2mm} \end{flushleft} \end{minipage} \begin{minipage}[t]{10cm} \begin{flushleft}  Note (parameters): Section 4

To get a feeling as to how much precision is lost due to
various aggregate representations, in Table 1 we compare the
number of ANECs and the numbers of equivalence classes of
the positionwise, pairwise, and Bordawise metrics, for small
elections; we computed the table using exhaustive search. \vspace{2mm} \end{flushleft} \end{minipage} \begin{minipage}[t]{10cm} \begin{flushleft} \rule{10cm}{0.3pt} \\ O2: Euclidean 1D, Euclidean 2D, Euclidean 3D-or-more, Group-Separable, Impartial Culture, Mallows, Single-Crossing, Single-Peaked (Conitzer/Random Peak), Single-Peaked (Walsh/Uniform), Urn Model \newline election sizes: \{10\}x\{50\} \vspace{2mm} \end{flushleft} \end{minipage} \begin{minipage}[t]{10cm} \begin{flushleft}  Samples note: The number of samples is:
(20 + 60 + 60) + (3 * 20 + 20) + 4 * 20 + (20 + 20) = 340 \newline \vspace{2mm} \end{flushleft} \end{minipage} \begin{minipage}[t]{10cm} \begin{flushleft}  Mallows: classic Mallows model (we used the same sampling protocol as
Boehmer et al. [2021b] - Putting a compass on the map of elections) \vspace{2mm} \end{flushleft} \end{minipage} \begin{minipage}[t]{10cm} \begin{flushleft}  Urn: classic urn
model (we used the same sampling protocol as
Boehmer et al. [2021b] - Putting a compass on the map of elections) \vspace{2mm} \end{flushleft} \end{minipage} \begin{minipage}[t]{10cm} \begin{flushleft}  Euclidean: Uniform 1D {[0,1]}; Uniform 2D ($[0,1]^2$); Uniform 3D ($[0,1]^3$); Uniform 5D ($[0,1]^5$); Uniform 10D ($[0,1]^10$); Uniform 20D ($[0,1]^20$); Uniform 2D Sphere ($(0, 0)$ $r=1$); Uniform 3D Sphere ($(0, 0)$ $r=1$); Uniform 5D Sphere ($(0, 0)$ $r=1$) \vspace{2mm} \end{flushleft} \end{minipage} \begin{minipage}[t]{10cm} \begin{flushleft}  Note (parameters): Section 6

To get a feeling as to how much precision is lost due to
various aggregate representations, in Table 1 we compare the
number of ANECs and the numbers of equivalence classes of
the positionwise, pairwise, and Bordawise metrics, for small
elections; we computed the table using exhaustive search. \vspace{2mm} \end{flushleft} \end{minipage} \begin{minipage}[t]{10cm} \begin{flushleft} \rule{10cm}{0.3pt} \\ O3: Euclidean 1D, Euclidean 2D, Euclidean 3D-or-more, Group-Separable, Impartial Culture, Mallows, Single-Crossing, Single-Peaked (Conitzer/Random Peak), Single-Peaked (Walsh/Uniform), Urn Model \newline election sizes: \{3, 4\} x \{3, 4, 5\} \vspace{2mm} \end{flushleft} \end{minipage} \begin{minipage}[t]{10cm} \begin{flushleft}  Samples note: as above \newline \vspace{2mm} \end{flushleft} \end{minipage} \begin{minipage}[t]{10cm} \begin{flushleft}  Mallows: as above  \vspace{2mm} \end{flushleft} \end{minipage} \begin{minipage}[t]{10cm} \begin{flushleft}  Urn: as above  \vspace{2mm} \end{flushleft} \end{minipage} \begin{minipage}[t]{10cm} \begin{flushleft}  Euclidean: as above  \vspace{2mm} \end{flushleft} \end{minipage} \begin{minipage}[t]{10cm} \begin{flushleft}  Note (parameters): Section 6

To compute PCC between different metrics and then visualize elections. \vspace{2mm} \end{flushleft} \end{minipage} \begin{minipage}[t]{10cm} \begin{flushleft}\rule{10cm}{0.3pt} \\ Comment: There is an arXiv version:
https://arxiv.org/abs/2205.00492 \end{flushleft} \end{minipage}  \\ 
\midrule
137 & \begin{minipage}[t]{3cm}\begin{flushleft}\fontsize{9}{9.5}\selectfont Approval with Runoff \vspace{2mm} \newline IJCAI-2022  \vspace{2mm}\newline  \citeapp{del-lan-las-san:c:approval-with-runoff} \end{flushleft}\end{minipage} & \begin{minipage}[t]{3cm}\begin{flushleft}\fontsize{9}{9.5}\selectfont T.~Delemazure, J.~Lang, J.-F.~Laslier, M.~R.~Sanver\end{flushleft}\end{minipage} & \fontsize{9}{9.5}\selectfont \begin{minipage}[t]{10cm} \begin{flushleft} A1: Euclidean 1D \newline election sizes: \{1000\}x\{20000\} \vspace{2mm} \end{flushleft} \end{minipage} \begin{minipage}[t]{10cm} \begin{flushleft}  Samples note: We didn’t find the number of samples in the paper \newline \vspace{2mm} \end{flushleft} \end{minipage} \begin{minipage}[t]{10cm} \begin{flushleft}  Euclidean: triangular distribution (voters and candidates located on [-1,1], details in Subsection 5.1)

gaussian distribution (with center 0 and standard deviation 1/2) \vspace{2mm} \end{flushleft} \end{minipage} \begin{minipage}[t]{10cm} \begin{flushleft}  Note (general): Section 5 “Statistical Analysis with One-Dimensional
Euclidean Preferences” (with 5.1 and 5.2) \vspace{2mm} \end{flushleft} \end{minipage} \begin{minipage}[t]{10cm} \begin{flushleft} \rule{10cm}{0.3pt} \\ A2: PrefLib, Real-Life (beyond PrefLib) \newline election sizes: \{11\}x\{1000\}, \{17\}x\{60\} \vspace{2mm} \end{flushleft} \end{minipage} \begin{minipage}[t]{10cm} \begin{flushleft}  Samples note: as above \newline Sizes note: {11}x{1000} (French presidential elections, Grenoble 2017), {17}x{60} (poster competition, San Sebastian) \newline \vspace{2mm} \end{flushleft} \end{minipage} \begin{minipage}[t]{10cm} \begin{flushleft}  Note (general): Section 6 “Experiments” \vspace{2mm} \end{flushleft} \end{minipage} \begin{minipage}[t]{10cm} \begin{flushleft}\rule{10cm}{0.3pt} \\ Comment: There is an arXiv version: https://arxiv.org/abs/2203.02343 \end{flushleft} \end{minipage}  \\ 
\midrule
138 & \begin{minipage}[t]{3cm}\begin{flushleft}\fontsize{9}{9.5}\selectfont Maxmin Participatory Budgeting \vspace{2mm} \newline IJCAI-2022  \vspace{2mm}\newline  \citeapp{sre-bha-nar:c:maximin-pb} \end{flushleft}\end{minipage} & \begin{minipage}[t]{3cm}\begin{flushleft}\fontsize{9}{9.5}\selectfont G.~Sreedurga, M.~Bhardwaj, Y.~Narahari\end{flushleft}\end{minipage} & \fontsize{9}{9.5}\selectfont \begin{minipage}[t]{10cm} \begin{flushleft} A1: PabuLib \newline  \vspace{2mm} \end{flushleft} \end{minipage} \begin{minipage}[t]{10cm} \begin{flushleft}  Note (general): Tried all datasets from Pabulib (as available at the time of writing) \vspace{2mm} \end{flushleft} \end{minipage} \begin{minipage}[t]{10cm} \begin{flushleft}\rule{10cm}{0.3pt} \\ Comment: There is an arXiv version: https://arxiv.org/abs/2204.13923 \end{flushleft} \end{minipage}  \\ 
\midrule
\multicolumn{4}{c}{\bf2023}\\ \midrule
139 & \begin{minipage}[t]{3cm}\begin{flushleft}\fontsize{9}{9.5}\selectfont Learning to Explain Voting Rules \vspace{2mm} \newline AAMAS-2023  \vspace{2mm}\newline  \citeapp{kan-han-xia:c:learning-voting-rules} \end{flushleft}\end{minipage} & \begin{minipage}[t]{3cm}\begin{flushleft}\fontsize{9}{9.5}\selectfont I.~Kang, Q.~Han, L.~Xia\end{flushleft}\end{minipage} & \fontsize{9}{9.5}\selectfont \begin{minipage}[t]{10cm} \begin{flushleft} O1: Impartial Culture \newline 10000 samples/datapoint, election sizes: [3, 5]x[10, 100] \vspace{2mm} \end{flushleft} \end{minipage} \begin{minipage}[t]{10cm} \begin{flushleft}  Samples note: Statistical culture is not very precisely defined. \newline Sizes note: Numbers of voters are in increments of 10. \newline \vspace{2mm} \end{flushleft} \end{minipage} \begin{minipage}[t]{10cm} \begin{flushleft}  Note (general): The paper also mentions results for greater values of m and n, but they are not directly defined. \vspace{2mm} \end{flushleft} \end{minipage}  \\ 
\midrule
140 & \begin{minipage}[t]{3cm}\begin{flushleft}\fontsize{9}{9.5}\selectfont Collecting, Classifying, Analyzing, and Using Real-World Ranking Data \vspace{2mm} \newline AAMAS-2023  \vspace{2mm}\newline  \citeapp{boe-sch:c:real-world-ranking-data} \end{flushleft}\end{minipage} & \begin{minipage}[t]{3cm}\begin{flushleft}\fontsize{9}{9.5}\selectfont N.~Boehmer, N.~Schaar\end{flushleft}\end{minipage} & \fontsize{9}{9.5}\selectfont \begin{minipage}[t]{10cm} \begin{flushleft} O1: Real-Life (beyond PrefLib) \newline  \vspace{2mm} \end{flushleft} \end{minipage} \begin{minipage}[t]{10cm} \begin{flushleft}  Sizes note: A lot of real-life datasets. No precise information about their sizes given (only approximations over all instances per dataset) \newline \vspace{2mm} \end{flushleft} \end{minipage}  \\ 
\midrule
141 & \begin{minipage}[t]{3cm}\begin{flushleft}\fontsize{9}{9.5}\selectfont SAT-based Judgment Aggregation \vspace{2mm} \newline AAMAS-2023  \vspace{2mm}\newline  \citeapp{con-nis-jar:c:sat-judgment-aggregation} \end{flushleft}\end{minipage} & \begin{minipage}[t]{3cm}\begin{flushleft}\fontsize{9}{9.5}\selectfont A.~Conati, A.~Niskanen, M.~J{\"{a}}rvisalo\end{flushleft}\end{minipage} & \fontsize{9}{9.5}\selectfont \begin{minipage}[t]{10cm} \begin{flushleft} O1: PrefLib \newline 405 samples/datapoint \vspace{2mm} \end{flushleft} \end{minipage} \begin{minipage}[t]{10cm} \begin{flushleft}  Sizes note: All profiles in the database that do not have a Condorcet winner. \newline \vspace{2mm} \end{flushleft} \end{minipage} \begin{minipage}[t]{10cm} \begin{flushleft} \rule{10cm}{0.3pt} \\ O2: Impartial Culture \newline 1000 samples/datapoint, election sizes: [5, 100]x\{1000\} \vspace{2mm} \end{flushleft} \end{minipage} \begin{minipage}[t]{10cm} \begin{flushleft}  Sizes note: Used to test Kemeny and Slater rules (candidates are given by increments of 5).  \newline \vspace{2mm} \end{flushleft} \end{minipage} \begin{minipage}[t]{10cm} \begin{flushleft} \rule{10cm}{0.3pt} \\ O3: Impartial Culture \newline 20 samples/datapoint, election sizes: \{10, 20, 30, 40, 50, 100, 150\}x\{10, 25, 50, 100\} \vspace{2mm} \end{flushleft} \end{minipage} \begin{minipage}[t]{10cm} \begin{flushleft}  Sizes note: Used to test Young and Dodgeson rules. \newline \vspace{2mm} \end{flushleft} \end{minipage} \begin{minipage}[t]{10cm} \begin{flushleft}\rule{10cm}{0.3pt} \\ Comment: Paper on judgement aggregation. \end{flushleft} \end{minipage}  \\ 
\midrule
142 & \begin{minipage}[t]{3cm}\begin{flushleft}\fontsize{9}{9.5}\selectfont Voting with Limited Energy: A Study of Plurality and Borda \vspace{2mm} \newline AAMAS-2023  \vspace{2mm}\newline  \citeapp{ter:c:limited-energy} \end{flushleft}\end{minipage} & \begin{minipage}[t]{3cm}\begin{flushleft}\fontsize{9}{9.5}\selectfont Z.~Terzopoulou\end{flushleft}\end{minipage} & \fontsize{9}{9.5}\selectfont \begin{minipage}[t]{10cm} \begin{flushleft} O1: Plackett-Luce \newline 3000 samples/datapoint, election sizes: \{5\}x[10-1000] \vspace{2mm} \end{flushleft} \end{minipage} \begin{minipage}[t]{10cm} \begin{flushleft}  Note (general): The weights in Plackett-Luce model are based on certain Gaussian distribution, with varying variance. \vspace{2mm} \end{flushleft} \end{minipage} \begin{minipage}[t]{10cm} \begin{flushleft}\rule{10cm}{0.3pt} \\ Comment: there is a repository: https://github.com/zoi-ter/voting-with-limited-energy, with limited resources \end{flushleft} \end{minipage}  \\ 
\midrule
143 & \begin{minipage}[t]{3cm}\begin{flushleft}\fontsize{9}{9.5}\selectfont Revisiting the Distortion of Distributed Voting \vspace{2mm} \newline AAMAS-2023  \vspace{2mm}\newline  \citeapp{fil-vou:c:distributed-distortion} \end{flushleft}\end{minipage} & \begin{minipage}[t]{3cm}\begin{flushleft}\fontsize{9}{9.5}\selectfont A.~Filos-Ratsikas, A.~Voudouris\end{flushleft}\end{minipage} & \fontsize{9}{9.5}\selectfont \begin{minipage}[t]{10cm} \begin{flushleft} O1: Hand-Crafted, Impartial Culture, Real-Life (beyond PrefLib) \newline 1000 samples/datapoint, election sizes: \{8\}x\{100\} \vspace{2mm} \end{flushleft} \end{minipage} \begin{minipage}[t]{10cm} \begin{flushleft}  Note (general): All profiles are created as cardinal utilities profiles, which are then changed to ordinal to compute distortion. Real-Life data: Jester Joke Dataset. \vspace{2mm} \end{flushleft} \end{minipage} \begin{minipage}[t]{10cm} \begin{flushleft}\rule{10cm}{0.3pt} \\ Comment: There is an arXiv version: https://arxiv.org/abs/2301.03279 \end{flushleft} \end{minipage}  \\ 
\midrule
144 & \begin{minipage}[t]{3cm}\begin{flushleft}\fontsize{9}{9.5}\selectfont Computational Complexity of Verifying the Group No-show Paradox \vspace{2mm} \newline AAMAS-2023  \vspace{2mm}\newline  \citeapp{moh-han-rua-che-ros-xia:c:complexity-group-paradox} \end{flushleft}\end{minipage} & \begin{minipage}[t]{3cm}\begin{flushleft}\fontsize{9}{9.5}\selectfont F.~Mohsin, Q.~Han, S.~Ruan, P.~Chen, F.~Rossi, L.~Xia\end{flushleft}\end{minipage} & \fontsize{9}{9.5}\selectfont \begin{minipage}[t]{10cm} \begin{flushleft} O1: PrefLib \newline 306 samples/datapoint \vspace{2mm} \end{flushleft} \end{minipage} \begin{minipage}[t]{10cm} \begin{flushleft} \rule{10cm}{0.3pt} \\ O2: Impartial Culture, Mallows \newline election sizes: \{?\}x[10, 100] \vspace{2mm} \end{flushleft} \end{minipage} \begin{minipage}[t]{10cm} \begin{flushleft}  Mallows: $\phi \in \{0.5, 0.8, 0.9\}$ \vspace{2mm} \end{flushleft} \end{minipage} \begin{minipage}[t]{10cm} \begin{flushleft}\rule{10cm}{0.3pt} \\ Comment: Short paper. \end{flushleft} \end{minipage}  \\ 
\midrule
145 & \begin{minipage}[t]{3cm}\begin{flushleft}\fontsize{9}{9.5}\selectfont Free-Riding in Multi-Issue Decisions \vspace{2mm} \newline AAMAS-2023  \vspace{2mm}\newline  \citeapp{lac-mal-nar:freeriding-multiissue} \end{flushleft}\end{minipage} & \begin{minipage}[t]{3cm}\begin{flushleft}\fontsize{9}{9.5}\selectfont M.~Lackner, J.~Maly, O.~Nardi\end{flushleft}\end{minipage} & \fontsize{9}{9.5}\selectfont \begin{minipage}[t]{10cm} \begin{flushleft} A1: Euclidean 2D \newline 1000 samples/datapoint, election sizes: \{4\}x\{20\} \vspace{2mm} \end{flushleft} \end{minipage} \begin{minipage}[t]{10cm} \begin{flushleft}  Sizes note: they have 20 issues and for each issue 4 candidates. So either 20 classic parallel elections with 4 candidates or one multi-issue election with 80 candidates \newline \vspace{2mm} \end{flushleft} \end{minipage} \begin{minipage}[t]{10cm} \begin{flushleft}  Euclidean: points sampled uniformly at random; approve closest candidate plus those which are at most 20\% further away \vspace{2mm} \end{flushleft} \end{minipage} \begin{minipage}[t]{10cm} \begin{flushleft}  Note (general): they describe the influence of varying the number of voters, candidates, and issues; but no details even in arxiv version provided just trends \vspace{2mm} \end{flushleft} \end{minipage} \begin{minipage}[t]{10cm} \begin{flushleft}\rule{10cm}{0.3pt} \\ Comment: There is an arXiv version: https://arxiv.org/abs/2310.08194 \end{flushleft} \end{minipage}  \\ 
\midrule
146 & \begin{minipage}[t]{3cm}\begin{flushleft}\fontsize{9}{9.5}\selectfont A Hotelling-Downs Game for Strategic Candidacy with Binary Issues \vspace{2mm} \newline AAMAS-2023  \vspace{2mm}\newline  \citeapp{maa-mou-wil:c:hotelling-candidacy-binary} \end{flushleft}\end{minipage} & \begin{minipage}[t]{3cm}\begin{flushleft}\fontsize{9}{9.5}\selectfont J.~Maass, V.~Mousseau, A.~Wilczynski\end{flushleft}\end{minipage} & \fontsize{9}{9.5}\selectfont \begin{minipage}[t]{10cm} \begin{flushleft} O1: Hand-Crafted \newline 1000 samples/datapoint, election sizes: [2, 4]x\{5000\} \vspace{2mm} \end{flushleft} \end{minipage} \begin{minipage}[t]{10cm} \begin{flushleft}  Samples note: Preferences are based on hypercubes of binary issues, which are selected uniformly at random. \newline \vspace{2mm} \end{flushleft} \end{minipage} \begin{minipage}[t]{10cm} \begin{flushleft} \rule{10cm}{0.3pt} \\ O2: Impartial Culture \newline 1000 samples/datapoint, election sizes: [2, 4]x\{5000\} \vspace{2mm} \end{flushleft} \end{minipage}  \\ 
\midrule
147 & \begin{minipage}[t]{3cm}\begin{flushleft}\fontsize{9}{9.5}\selectfont Fairness in Participatory Budgeting via Equality of Resources \vspace{2mm} \newline AAMAS-2023  \vspace{2mm}\newline  \citeapp{mal-rey-end-lac:c:fairness-pb-equality} \end{flushleft}\end{minipage} & \begin{minipage}[t]{3cm}\begin{flushleft}\fontsize{9}{9.5}\selectfont J.~Maly, S.~Rey, U.~Endriss, M.~Lackner\end{flushleft}\end{minipage} & \fontsize{9}{9.5}\selectfont \begin{minipage}[t]{10cm} \begin{flushleft} A1: PabuLib \newline election sizes: ? \vspace{2mm} \end{flushleft} \end{minipage} \begin{minipage}[t]{10cm} \begin{flushleft}  Samples note: A total of 353 PB instances are covered by their analysis. \newline Sizes note: They used all instances from Pabulib with up to 65 projects, except for trivial instances, where either no project or the set of all projects are affordable.   \newline \vspace{2mm} \end{flushleft} \end{minipage}  \\ 
\midrule
148 & \begin{minipage}[t]{3cm}\begin{flushleft}\fontsize{9}{9.5}\selectfont The Distortion of Approval Voting with Runoff \vspace{2mm} \newline AAMAS-2023  \vspace{2mm}\newline  \citeapp{eva-lat-sha:c:approval-distortion} \end{flushleft}\end{minipage} & \begin{minipage}[t]{3cm}\begin{flushleft}\fontsize{9}{9.5}\selectfont S.~Ebadian, M.~Latifian, N.~Shah\end{flushleft}\end{minipage} & \fontsize{9}{9.5}\selectfont \begin{minipage}[t]{10cm} \begin{flushleft} A1: Impartial Culture (Variant) \newline 1000 samples/datapoint, election sizes: [5, 100]x\{200\} \vspace{2mm} \end{flushleft} \end{minipage} \begin{minipage}[t]{10cm} \begin{flushleft}  Note (general): The authors also generate utilities (in addition to approval votes). Utilities follow Dirichlet distribution. \vspace{2mm} \end{flushleft} \end{minipage} \begin{minipage}[t]{10cm} \begin{flushleft} \rule{10cm}{0.3pt} \\ A2: PrefLib \newline  \vspace{2mm} \end{flushleft} \end{minipage} \begin{minipage}[t]{10cm} \begin{flushleft}  Sizes note: The authors write: “We also run our experiment on 134 different real-world datasets from PrefLib (all datasets of type SOC, i.e., Strict Orders Complete List). We exclude the instances with less than 3 candidates.” \newline \vspace{2mm} \end{flushleft} \end{minipage} \begin{minipage}[t]{10cm} \begin{flushleft}  Note (general): Utilities generated from Dirichlet distribution. \vspace{2mm} \end{flushleft} \end{minipage}  \\ 
\midrule
149 & \begin{minipage}[t]{3cm}\begin{flushleft}\fontsize{9}{9.5}\selectfont A Map of Diverse Synthetic Stable Roommates Instances \vspace{2mm} \newline AAMAS-2023  \vspace{2mm}\newline  \citeapp{boe-hee-szu:c:map-stable-roommates} \end{flushleft}\end{minipage} & \begin{minipage}[t]{3cm}\begin{flushleft}\fontsize{9}{9.5}\selectfont N.~Boehmer, K.~Heeger, S.~Szufa\end{flushleft}\end{minipage} & \fontsize{9}{9.5}\selectfont \begin{minipage}[t]{10cm} \begin{flushleft} O1: Euclidean 1D, Euclidean 2D, Hand-Crafted, Impartial Culture, Mallows \newline 1 samples/datapoint, election sizes: \{200\}x\{200\} \vspace{2mm} \end{flushleft} \end{minipage} \begin{minipage}[t]{10cm} \begin{flushleft}  Samples note: Drawing a map of profiles (see the paper for the composition of the dataset and numbers of instances of given types) \newline \vspace{2mm} \end{flushleft} \end{minipage} \begin{minipage}[t]{10cm} \begin{flushleft}  Mallows: $\phi \in \{0.2, 0.4, 0.6, 0.8 \}$ \vspace{2mm} \end{flushleft} \end{minipage} \begin{minipage}[t]{10cm} \begin{flushleft}  Euclidean: Uniform 1D ($[0,1]$); Uniform 2D ($[0,1]^2$) \vspace{2mm} \end{flushleft} \end{minipage} \begin{minipage}[t]{10cm} \begin{flushleft}  Note (parameters): $p \in \{0.05, 0.15, 0.25 \}$ \vspace{2mm} \end{flushleft} \end{minipage} \begin{minipage}[t]{10cm} \begin{flushleft}  Note (general): Experiment on the stable roommate problem. \vspace{2mm} \end{flushleft} \end{minipage} \begin{minipage}[t]{10cm} \begin{flushleft}\rule{10cm}{0.3pt} \\ Comment: There is an arXiv version: https://arxiv.org/abs/2208.04041 \end{flushleft} \end{minipage}  \\ 
\midrule
150 & \begin{minipage}[t]{3cm}\begin{flushleft}\fontsize{9}{9.5}\selectfont Rank Aggregation Using Scoring Rules \vspace{2mm} \newline AAAI-2023  \vspace{2mm}\newline  \citeapp{boe-bre-pet:c:rank-aggregation-scoring-rules} \end{flushleft}\end{minipage} & \begin{minipage}[t]{3cm}\begin{flushleft}\fontsize{9}{9.5}\selectfont N.~Boehmer, R.~Bredereck, D.~Peters\end{flushleft}\end{minipage} & \fontsize{9}{9.5}\selectfont \begin{minipage}[t]{10cm} \begin{flushleft} O1: Mallows \newline 10000 samples/datapoint, election sizes: \{10\}x\{100\} \vspace{2mm} \end{flushleft} \end{minipage} \begin{minipage}[t]{10cm} \begin{flushleft}  Mallows: $\phi \in \{0,0.1,0.2,0.3,0.4,0.5,0.6,0.7,0.8,0.9,1\}$ \vspace{2mm} \end{flushleft} \end{minipage} \begin{minipage}[t]{10cm} \begin{flushleft}  Note (parameters): It is normalized Mallows \vspace{2mm} \end{flushleft} \end{minipage} \begin{minipage}[t]{10cm} \begin{flushleft}\rule{10cm}{0.3pt} \\ Comment: There is an arXiv version: https://arxiv.org/abs/2209.08856 \end{flushleft} \end{minipage}  \\ 
\midrule
151 & \begin{minipage}[t]{3cm}\begin{flushleft}\fontsize{9}{9.5}\selectfont Ballot Length in Instant Runoff Voting \vspace{2mm} \newline AAAI-2023  \vspace{2mm}\newline  \citeapp{tom-uga-kle:c:ballot-length-irv} \end{flushleft}\end{minipage} & \begin{minipage}[t]{3cm}\begin{flushleft}\fontsize{9}{9.5}\selectfont K.~Tomlinson, J.~Ugander, J.~Kleinberg\end{flushleft}\end{minipage} & \fontsize{9}{9.5}\selectfont \begin{minipage}[t]{10cm} \begin{flushleft} O1: Impartial Culture \newline 10000 samples/datapoint, election sizes: [3, 40]x\{1000\} \vspace{2mm} \end{flushleft} \end{minipage} \begin{minipage}[t]{10cm} \begin{flushleft} \rule{10cm}{0.3pt} \\ O2: Euclidean 1D \newline 10000 samples/datapoint, election sizes: [3, 40]x\{inf\} \vspace{2mm} \end{flushleft} \end{minipage} \begin{minipage}[t]{10cm} \begin{flushleft}  Sizes note: Simulation of an infinite population of voters \newline \vspace{2mm} \end{flushleft} \end{minipage} \begin{minipage}[t]{10cm} \begin{flushleft}  Euclidean: Uniform 1D {[0,1]} \vspace{2mm} \end{flushleft} \end{minipage} \begin{minipage}[t]{10cm} \begin{flushleft} \rule{10cm}{0.3pt} \\ O3: PrefLib \newline election sizes: [3, 29]x\{?\} \vspace{2mm} \end{flushleft} \end{minipage} \begin{minipage}[t]{10cm} \begin{flushleft}  Sizes note: Among others: 
* 12 American Psychological Association (APA) presidential elections,
 * 14 San Francisco local elections
(h = 3),
 * 21 Glasgow local elections (h = k).

The number of voters is described approximately as “from tens to hundreds of thousands” \newline \vspace{2mm} \end{flushleft} \end{minipage} \begin{minipage}[t]{10cm} \begin{flushleft}\rule{10cm}{0.3pt} \\ Comment: There is an arXiv version: https://arxiv.org/abs/2207.08958 \end{flushleft} \end{minipage}  \\ 
\midrule
152 & \begin{minipage}[t]{3cm}\begin{flushleft}\fontsize{9}{9.5}\selectfont Participatory Budgeting Designs for the Real World \vspace{2mm} \newline AAAI-2023  \vspace{2mm}\newline  \citeapp{far-ben-gal:c:mechanical-turk-pb} \end{flushleft}\end{minipage} & \begin{minipage}[t]{3cm}\begin{flushleft}\fontsize{9}{9.5}\selectfont R.~Fairstein, G.~Benad{\`e}, K.~Gal\end{flushleft}\end{minipage} & \fontsize{9}{9.5}\selectfont \begin{minipage}[t]{10cm} \begin{flushleft} O1: Semi-Real-Life \newline election sizes: \{10, 20\}x\{75\} \vspace{2mm} \end{flushleft} \end{minipage} \begin{minipage}[t]{10cm} \begin{flushleft}  Samples note: In total there are 8 instances \newline Sizes note: In each instance there was around 75 voters. \newline \vspace{2mm} \end{flushleft} \end{minipage} \begin{minipage}[t]{10cm} \begin{flushleft}  Note (general): Data comes from the experiment related to Participatory Budgeting conducted on Amazon Mechanical Turk. \vspace{2mm} \end{flushleft} \end{minipage} \begin{minipage}[t]{10cm} \begin{flushleft} \rule{10cm}{0.3pt} \\ A1:  \newline election sizes: \{10, 20\}x\{75\} \vspace{2mm} \end{flushleft} \end{minipage} \begin{minipage}[t]{10cm} \begin{flushleft}  Samples note: In total there are 16 instances \newline Sizes note: In each instance there was around 75 voters. \newline \vspace{2mm} \end{flushleft} \end{minipage} \begin{minipage}[t]{10cm} \begin{flushleft}  Note (general): Data comes from the experiment related to Participatory Budgeting conducted on Amazon Mechanical Turk. \vspace{2mm} \end{flushleft} \end{minipage}  \\ 
\midrule
153 & \begin{minipage}[t]{3cm}\begin{flushleft}\fontsize{9}{9.5}\selectfont Properties of Position Matrices and Their Elections \vspace{2mm} \newline AAAI-2023  \vspace{2mm}\newline  \citeapp{boe-cai-fal-fan-jan-kac-was:c:position-matrices} \end{flushleft}\end{minipage} & \begin{minipage}[t]{3cm}\begin{flushleft}\fontsize{9}{9.5}\selectfont N.~Boehmer, J.{-}Y.~Cai, P.~Faliszewski, A.~Z.~Fan, L.~Janeczko, A.~Kaczmarczyk, T.~W{\k{a}}s\end{flushleft}\end{minipage} & \fontsize{9}{9.5}\selectfont \begin{minipage}[t]{10cm} \begin{flushleft} O1: Euclidean 1D, Euclidean 2D, Euclidean 3D-or-more, Group-Separable, Impartial Culture, Mallows, Single-Crossing, Single-Peaked (Conitzer/Random Peak), Single-Peaked (Walsh/Uniform), Urn Model \newline 4000 samples/datapoint, election sizes: \{8\}x\{80\} \vspace{2mm} \end{flushleft} \end{minipage} \begin{minipage}[t]{10cm} \begin{flushleft}  Samples note: 480 elections, for each E its position matrix P was computed, and then there were sampled 100 pairs of elections between whom its isomorphic swap distance was computed. Map shows 480 points (elections) with colour dependent on the maximum isomorphic swap distance for the sampled elections realising a position matrix of the original election. Approximately 20 elections per culture

4000 = 20 x 100 x 2

Exact quantities of 480 elections (Appendix A.3) \newline Sizes note: 480 elections of size 8x80 was later used for sampling \newline \vspace{2mm} \end{flushleft} \end{minipage} \begin{minipage}[t]{10cm} \begin{flushleft}  Mallows: For the 80 elections generated using the urn model and
the normalized Mallows model, we followed the protocol
of Boehmer et al. [2021b, 2022b]. Hence, for each of the
elections generated with the normalized Mallows Model,
we drew the value of rel-$\phi$ uniformly at random from the
[0, 1] interval. \vspace{2mm} \end{flushleft} \end{minipage} \begin{minipage}[t]{10cm} \begin{flushleft}  Urn: The parameter for the urn model elections
was drawn according to the Gamma distribution with the
shape parameter k = 0.8 and the scale parameter $\theta =
1$. \vspace{2mm} \end{flushleft} \end{minipage} \begin{minipage}[t]{10cm} \begin{flushleft}  Euclidean: Uniform 1D {[0,1]}; Uniform 2D ($[0,1]^2$); Uniform 3D ($[0,1]^3$); Uniform 5D ($[0,1]^5$); Uniform 10D ($[0,1]^10$); Uniform 20D ($[0,1]^20$); Uniform 2D Sphere ($(0, 0)$ $r=1$); Uniform 3D Sphere ($(0, 0)$ $r=1$); Uniform 5D Sphere ($(0, 0)$ $r=1$) \vspace{2mm} \end{flushleft} \end{minipage} \begin{minipage}[t]{10cm} \begin{flushleft}  Note (general): Dataset comes from the paper introducing ordinal map of elections

See Section 3.3 and Appendix A.3, B.1 for more details. \vspace{2mm} \end{flushleft} \end{minipage} \begin{minipage}[t]{10cm} \begin{flushleft} \rule{10cm}{0.3pt} \\ O2: Euclidean 1D, Euclidean 2D, Euclidean 3D-or-more, Group-Separable, Impartial Culture, Mallows, Single-Crossing, Single-Peaked (Conitzer/Random Peak), Single-Peaked (Walsh/Uniform), Urn Model \newline 20 samples/datapoint, election sizes: \{4\}x\{16\} \vspace{2mm} \end{flushleft} \end{minipage} \begin{minipage}[t]{10cm} \begin{flushleft}  Samples note: Exact quantities of 480 elections (Appendix A.3). Approximately 20 elections per culture. \newline Sizes note: 480 elections of size 4x16 \newline \vspace{2mm} \end{flushleft} \end{minipage} \begin{minipage}[t]{10cm} \begin{flushleft}  Mallows: as above  \vspace{2mm} \end{flushleft} \end{minipage} \begin{minipage}[t]{10cm} \begin{flushleft}  Urn: as above  \vspace{2mm} \end{flushleft} \end{minipage} \begin{minipage}[t]{10cm} \begin{flushleft}  Euclidean: as above  \vspace{2mm} \end{flushleft} \end{minipage} \begin{minipage}[t]{10cm} \begin{flushleft}  Note (general): for each election of the dataset
we computed its position matrix and all elections realizing this matrix

See Appendix B.2 for more details \vspace{2mm} \end{flushleft} \end{minipage} \begin{minipage}[t]{10cm} \begin{flushleft} \rule{10cm}{0.3pt} \\ O3: Euclidean 1D, Euclidean 2D, Euclidean 3D-or-more, Group-Separable, Impartial Culture, Mallows, Single-Crossing, Single-Peaked (Conitzer/Random Peak), Single-Peaked (Walsh/Uniform), Urn Model \newline 20 samples/datapoint, election sizes: \{8\}x\{80\} \vspace{2mm} \end{flushleft} \end{minipage} \begin{minipage}[t]{10cm} \begin{flushleft}  Samples note: Exact quantities of 480 elections (Appendix A.3) Approximately 20 elections per culture \newline Sizes note: 480 elections of size 8x80 \newline \vspace{2mm} \end{flushleft} \end{minipage} \begin{minipage}[t]{10cm} \begin{flushleft}  Mallows: as above  \vspace{2mm} \end{flushleft} \end{minipage} \begin{minipage}[t]{10cm} \begin{flushleft}  Urn: as above  \vspace{2mm} \end{flushleft} \end{minipage} \begin{minipage}[t]{10cm} \begin{flushleft}  Euclidean: as above  \vspace{2mm} \end{flushleft} \end{minipage} \begin{minipage}[t]{10cm} \begin{flushleft}  Note (general): See Section 4. (An Experiment) for more details of frequency matrix realizability in restricted domain.

See Section 5. (Experiment 1.) for more details of condorcet condition efficiency.

See Section 5. (Experiment 2.) for more details of possible condorcet winners number. \vspace{2mm} \end{flushleft} \end{minipage} \begin{minipage}[t]{10cm} \begin{flushleft}\rule{10cm}{0.3pt} \\ Comment: There is an arXiv version: https://arxiv.org/abs/2303.02538 \end{flushleft} \end{minipage}  \\ 
\midrule
154 & \begin{minipage}[t]{3cm}\begin{flushleft}\fontsize{9}{9.5}\selectfont Efficient and Accurate Learning of Mixtures of Plackett-Luce Models \vspace{2mm} \newline AAAI-2023  \vspace{2mm}\newline  \citeapp{ngu-zha:c:plackett-luce} \end{flushleft}\end{minipage} & \begin{minipage}[t]{3cm}\begin{flushleft}\fontsize{9}{9.5}\selectfont D.~Nguyen, A.~Zhang\end{flushleft}\end{minipage} & \fontsize{9}{9.5}\selectfont \begin{minipage}[t]{10cm} \begin{flushleft} O1: Plackett-Luce \newline 25 samples/datapoint, election sizes: \{100\}x[100-2000] \vspace{2mm} \end{flushleft} \end{minipage} \begin{minipage}[t]{10cm} \begin{flushleft}  Sizes note: lower end for number of votes unspecified, but from plot it seems to be 100 \newline \vspace{2mm} \end{flushleft} \end{minipage} \begin{minipage}[t]{10cm} \begin{flushleft}  Note (general): In fact a mixture of Plackett-Luce with 2 or 10 groups of voters (the voters are assigned to a group uniformly at random). In each group, the rankings are drawn from Plackett-Luce in which first 5 candidates (the same for all groups) have weights drawn from a log-normal distribution, and the remaining candidates have weight equal to 1. \vspace{2mm} \end{flushleft} \end{minipage} \begin{minipage}[t]{10cm} \begin{flushleft} \rule{10cm}{0.3pt} \\ O2: PrefLib \newline  \vspace{2mm} \end{flushleft} \end{minipage} \begin{minipage}[t]{10cm} \begin{flushleft}  Samples note: Unclear, but probably each profile considered once \newline Sizes note: Unclear. Voters split into 80-20 training-test datasets \newline \vspace{2mm} \end{flushleft} \end{minipage} \begin{minipage}[t]{10cm} \begin{flushleft}  Note (general): Preflib data: APA, Irish (West Dublin, North Dublin, Meath), and Sushi \vspace{2mm} \end{flushleft} \end{minipage} \begin{minipage}[t]{10cm} \begin{flushleft} \rule{10cm}{0.3pt} \\ O3: Real-Life (beyond PrefLib) \newline election sizes: \{25, 50, 100, 200\}x[100-14000] \vspace{2mm} \end{flushleft} \end{minipage} \begin{minipage}[t]{10cm} \begin{flushleft}  Sizes note: lower end for number of votes unspecified, but from plot it seems to be 100 \newline \vspace{2mm} \end{flushleft} \end{minipage} \begin{minipage}[t]{10cm} \begin{flushleft}  Note (general): ML-10M movie ratings datasets [Harper and Konstan, 2015] with some preprocessing and then sampling \vspace{2mm} \end{flushleft} \end{minipage} \begin{minipage}[t]{10cm} \begin{flushleft}\rule{10cm}{0.3pt} \\ Comment: There is an arXiv version: https://arxiv.org/abs/2302.05343 \end{flushleft} \end{minipage}  \\ 
\midrule
155 & \begin{minipage}[t]{3cm}\begin{flushleft}\fontsize{9}{9.5}\selectfont Representation with Incomplete Votes \vspace{2mm} \newline AAAI-2023  \vspace{2mm}\newline  \citeapp{hal-keh-pro-tuc-wut:c:repr-with-incomplete-votes} \end{flushleft}\end{minipage} & \begin{minipage}[t]{3cm}\begin{flushleft}\fontsize{9}{9.5}\selectfont D.~Halpern, G.~Kehne, A.~D.~Procaccia, J.~Tucker{-}Foltz, M.~W{\"{u}}thrich\end{flushleft}\end{minipage} & \fontsize{9}{9.5}\selectfont \begin{minipage}[t]{10cm} \begin{flushleft} A1: Real-Life (beyond PrefLib) \newline 390 samples/datapoint, election sizes: [31, 1719]x[87, 1000] \vspace{2mm} \end{flushleft} \end{minipage} \begin{minipage}[t]{10cm} \begin{flushleft}  Samples note: 13 x 3 (target committee sizes k = 5, 7, 10) x 10 (random seeds) \newline Sizes note: I’ve put the number of queries as number of voters \newline \vspace{2mm} \end{flushleft} \end{minipage} \begin{minipage}[t]{10cm} \begin{flushleft}  Note (parameters): These elections assume incomplete votes, so these are not the standard ones \vspace{2mm} \end{flushleft} \end{minipage} \begin{minipage}[t]{10cm} \begin{flushleft}  Note (general): Section 5. “Experiments” \vspace{2mm} \end{flushleft} \end{minipage} \begin{minipage}[t]{10cm} \begin{flushleft}\rule{10cm}{0.3pt} \\ Comment: There is an arXiv version: https://arxiv.org/abs/2211.15608 \end{flushleft} \end{minipage}  \\ 
\midrule
156 & \begin{minipage}[t]{3cm}\begin{flushleft}\fontsize{9}{9.5}\selectfont Measuring and Controlling Divisiveness in Rank Aggregation \vspace{2mm} \newline IJCAI-2023  \vspace{2mm}\newline  \citeapp{col-gra-hid-mac-nav:c:controlling-rank-aggregation} \end{flushleft}\end{minipage} & \begin{minipage}[t]{3cm}\begin{flushleft}\fontsize{9}{9.5}\selectfont R.~Colley, U.~Grandi, C.~Hidalgo, M.~Macedo, C.~Navarrete\end{flushleft}\end{minipage} & \fontsize{9}{9.5}\selectfont \begin{minipage}[t]{10cm} \begin{flushleft} O1: Impartial Culture, PrefLib, Urn Model \newline 100 samples/datapoint, election sizes: [3, 18]x\{100\} \vspace{2mm} \end{flushleft} \end{minipage} \begin{minipage}[t]{10cm} \begin{flushleft}  Urn: 10\% and 50\% correlation (not completely clear what this means). \vspace{2mm} \end{flushleft} \end{minipage} \begin{minipage}[t]{10cm} \begin{flushleft}  Note (general): Rankings generated using preflib. \vspace{2mm} \end{flushleft} \end{minipage} \begin{minipage}[t]{10cm} \begin{flushleft} \rule{10cm}{0.3pt} \\ O2: Impartial Culture, Urn Model \newline 100 samples/datapoint, election sizes: [3, 18]x\{100\} \vspace{2mm} \end{flushleft} \end{minipage} \begin{minipage}[t]{10cm} \begin{flushleft}  Urn: as above  \vspace{2mm} \end{flushleft} \end{minipage} \begin{minipage}[t]{10cm} \begin{flushleft} \rule{10cm}{0.3pt} \\ O3: Impartial Culture, Urn Model \newline 100 samples/datapoint, election sizes: [2, 11]x\{100\} \vspace{2mm} \end{flushleft} \end{minipage} \begin{minipage}[t]{10cm} \begin{flushleft}  Urn: as above  \vspace{2mm} \end{flushleft} \end{minipage} \begin{minipage}[t]{10cm} \begin{flushleft}\rule{10cm}{0.3pt} \\ Comment: There is an arXiv version: https://arxiv.org/abs/2306.08511 \end{flushleft} \end{minipage}  \\ 
\midrule
157 & \begin{minipage}[t]{3cm}\begin{flushleft}\fontsize{9}{9.5}\selectfont Ties in Multiwinner Approval Voting \vspace{2mm} \newline IJCAI-2023  \vspace{2mm}\newline  \citeapp{jan-fal:c:ties-approval-voting} \end{flushleft}\end{minipage} & \begin{minipage}[t]{3cm}\begin{flushleft}\fontsize{9}{9.5}\selectfont L.~Janeczko, P.~Faliszewski\end{flushleft}\end{minipage} & \fontsize{9}{9.5}\selectfont \begin{minipage}[t]{10cm} \begin{flushleft} A1: Euclidean 1D, PabuLib, Resampling \newline 1000 samples/datapoint, election sizes: \{30\}x[20, 100] \vspace{2mm} \end{flushleft} \end{minipage} \begin{minipage}[t]{10cm} \begin{flushleft}  Euclidean: $k/2$ or $k$ approved on average ($k$ is the commitee size) \vspace{2mm} \end{flushleft} \end{minipage} \begin{minipage}[t]{10cm} \begin{flushleft}  Note (parameters): Resampling parameters: $\phi=0.75$, $k$, $k/2$, or $2k$ candidates approved on average ($k$ is the committee size) \vspace{2mm} \end{flushleft} \end{minipage} \begin{minipage}[t]{10cm} \begin{flushleft}\rule{10cm}{0.3pt} \\ Comment: There is an arXiv version: https://arxiv.org/abs/2305.01769 \end{flushleft} \end{minipage}  \\ 
\midrule
158 & \begin{minipage}[t]{3cm}\begin{flushleft}\fontsize{9}{9.5}\selectfont Learning to Design Fair and Private Voting Rules (Extended Abstract) \vspace{2mm} \newline IJCAI-2023  \vspace{2mm}\newline  \citeapp{moh-liu-che-ros-xia:c:fair-and-private} \end{flushleft}\end{minipage} & \begin{minipage}[t]{3cm}\begin{flushleft}\fontsize{9}{9.5}\selectfont F.~Mohsin, A.~Liu, P.~Chen, F.~Rossi, L.~Xia\end{flushleft}\end{minipage} & \fontsize{9}{9.5}\selectfont \begin{minipage}[t]{10cm} \begin{flushleft} O1: Impartial Culture, Plackett-Luce \newline election sizes: \{?\}x\{140\} \vspace{2mm} \end{flushleft} \end{minipage} \begin{minipage}[t]{10cm} \begin{flushleft}  Samples note: Not stated \newline \vspace{2mm} \end{flushleft} \end{minipage} \begin{minipage}[t]{10cm} \begin{flushleft}  Note (general): Setup description very limited due to extended abstract version. \vspace{2mm} \end{flushleft} \end{minipage} \begin{minipage}[t]{10cm} \begin{flushleft}\rule{10cm}{0.3pt} \\ Comment: There is a journal version: Farhad Mohsin, Ao Liu, Pin-Yu Chen, Francesca Rossi, Lirong Xia:
Learning to Design Fair and Private Voting Rules. J. Artif. Intell. Res. 75: 1139-1176 (2022) \end{flushleft} \end{minipage}  \\ 
\midrule
159 & \begin{minipage}[t]{3cm}\begin{flushleft}\fontsize{9}{9.5}\selectfont Convergence in Multi-Issue Iterative Voting under Uncertainty \vspace{2mm} \newline IJCAI-2023  \vspace{2mm}\newline  \citeapp{kav-mei-ros-xia:convergence-iterative} \end{flushleft}\end{minipage} & \begin{minipage}[t]{3cm}\begin{flushleft}\fontsize{9}{9.5}\selectfont J.~Kavner, R.~Meir, F.~Rossi, L.~Xia\end{flushleft}\end{minipage} & \fontsize{9}{9.5}\selectfont \begin{minipage}[t]{10cm} \begin{flushleft} O1: Impartial Culture \newline 10000 samples/datapoint, election sizes: \{2, 4, 8, 16, 32\}x\{7, 11, 15, 19\} \vspace{2mm} \end{flushleft} \end{minipage} \begin{minipage}[t]{10cm} \begin{flushleft}  Note (general): Iterative voting with uncertainty. There are multiple binary issues but agents have preferences over all values of all issues, so 3 binary issues means 8 candidates. \vspace{2mm} \end{flushleft} \end{minipage} \begin{minipage}[t]{10cm} \begin{flushleft}\rule{10cm}{0.3pt} \\ Comment: There is arXiv version: https://arxiv.org/abs/2301.08873 \end{flushleft} \end{minipage}  \\ 
\midrule
160 & \begin{minipage}[t]{3cm}\begin{flushleft}\fontsize{9}{9.5}\selectfont Diversity, Agreement, and Polarization in Elections \vspace{2mm} \newline IJCAI-2023  \vspace{2mm}\newline  \citeapp{fal-kac-sor-szu-was:c:div-agr-pol-map} \end{flushleft}\end{minipage} & \begin{minipage}[t]{3cm}\begin{flushleft}\fontsize{9}{9.5}\selectfont P.~Faliszewski, A.~Kaczmarczyk, K.~Sornat, S.~Szufa, T.~Was\end{flushleft}\end{minipage} & \fontsize{9}{9.5}\selectfont \begin{minipage}[t]{10cm} \begin{flushleft} O1: Euclidean 1D, Euclidean 2D, Euclidean 3D-or-more, Group-Separable, Impartial Culture, Mallows, PrefLib, Single-Crossing, Single-Peaked (Conitzer/Random Peak), Single-Peaked (Walsh/Uniform), Urn Model \newline 292 samples/datapoint, election sizes: \{8\}x\{96\} \vspace{2mm} \end{flushleft} \end{minipage} \begin{minipage}[t]{10cm} \begin{flushleft}  Samples note: This is an experiment about the map of elections. 292 is the number of elections in the map in total \newline \vspace{2mm} \end{flushleft} \end{minipage} \begin{minipage}[t]{10cm} \begin{flushleft}  Mallows: Normalized $\phi$, chosen uniformly at random \vspace{2mm} \end{flushleft} \end{minipage} \begin{minipage}[t]{10cm} \begin{flushleft}  Urn: choose the contagion parameter a according to the Gamma distribution with shape parameter $k = 0.8$ and scale parameter $\theta = 1$ (not describedf in the paper—inferred from a previous map paper, “Putting a Compass on the Map of Elections”) \vspace{2mm} \end{flushleft} \end{minipage} \begin{minipage}[t]{10cm} \begin{flushleft}  Euclidean: Uniform 1D ($[0,1]$); Uniform 2D ($[0,1]^2$); Uniform 3D ($[0,1]^3$); Uniform 5D ($[0,1]^5$); Uniform 10D ($[0,1]^10$); Uniform 2D Sphere; Uniform 3D Sphere \vspace{2mm} \end{flushleft} \end{minipage} \begin{minipage}[t]{10cm} \begin{flushleft}  Note (general): Preflib: Irish, Sushi, Grenoble \vspace{2mm} \end{flushleft} \end{minipage} \begin{minipage}[t]{10cm} \begin{flushleft} \rule{10cm}{0.3pt} \\ O2: Euclidean 1D, Euclidean 2D, Euclidean 3D-or-more, Identity, Impartial Culture, Mallows, PrefLib, Single-Peaked (Conitzer/Random Peak), Single-Peaked (Walsh/Uniform), Urn Model \newline 1 samples/datapoint, election sizes: [8]x[1000] \vspace{2mm} \end{flushleft} \end{minipage} \begin{minipage}[t]{10cm} \begin{flushleft}  Samples note: In some sense, here each vote is a sample, so one could write “1000” as the number of samples. \newline \vspace{2mm} \end{flushleft} \end{minipage} \begin{minipage}[t]{10cm} \begin{flushleft}  Mallows: norm-$\phi \in \{0.05, 0.2, 0.5\}$ \vspace{2mm} \end{flushleft} \end{minipage} \begin{minipage}[t]{10cm} \begin{flushleft}  Urn: $\alpha \in \{0.2,1\}$ \vspace{2mm} \end{flushleft} \end{minipage} \begin{minipage}[t]{10cm} \begin{flushleft}  Euclidean: Uniform 1D ($[0,1]$); Uniform 2D ($[0,1]^2$); Uniform 3D ($[0,1]^3$); Uniform 2D Sphere; Uniform 3D Sphere \vspace{2mm} \end{flushleft} \end{minipage} \begin{minipage}[t]{10cm} \begin{flushleft}  Note (general): Experiment from Figure 1. Preflib: Irish dataset \vspace{2mm} \end{flushleft} \end{minipage} \begin{minipage}[t]{10cm} \begin{flushleft}\rule{10cm}{0.3pt} \\ Comment: There is an arXiv version: https://arxiv.org/abs/2305.09780 \end{flushleft} \end{minipage}  \\ 
\midrule
161 & \begin{minipage}[t]{3cm}\begin{flushleft}\fontsize{9}{9.5}\selectfont Deliberation and Voting in Approval-Based Multi-Winner Elections \vspace{2mm} \newline IJCAI-2023  \vspace{2mm}\newline  \citeapp{meh-sre-lar:c:deliberation-approval-elections} \end{flushleft}\end{minipage} & \begin{minipage}[t]{3cm}\begin{flushleft}\fontsize{9}{9.5}\selectfont K.~Mehra, N.~Sreenivas, K.~Larson\end{flushleft}\end{minipage} & \fontsize{9}{9.5}\selectfont \begin{minipage}[t]{10cm} \begin{flushleft} A1: Mallows extension \newline 10000 samples/datapoint, election sizes: \{50\}x\{100\} \vspace{2mm} \end{flushleft} \end{minipage} \begin{minipage}[t]{10cm} \begin{flushleft}  Note (parameters): $\phi=0.2$ \vspace{2mm} \end{flushleft} \end{minipage} \begin{minipage}[t]{10cm} \begin{flushleft}\rule{10cm}{0.3pt} \\ Comment: There is an arXiv version: https://arxiv.org/abs/2305.08970 \end{flushleft} \end{minipage}  \\ 
\midrule
162 & \begin{minipage}[t]{3cm}\begin{flushleft}\fontsize{9}{9.5}\selectfont Participatory Budgeting: Data, Tools and Analysis \vspace{2mm} \newline IJCAI-2023  \vspace{2mm}\newline  \citeapp{fal-fli-pet-pie-sko-sto-szu-tal:c:participatory-budgeting-tools} \end{flushleft}\end{minipage} & \begin{minipage}[t]{3cm}\begin{flushleft}\fontsize{9}{9.5}\selectfont P.~Faliszewski, J.~Flis, D.~Peters, G.~Pierczynski, P.~Skowron, D.~Stolicki, S.~Szufa, N.~Talmon\end{flushleft}\end{minipage} & \fontsize{9}{9.5}\selectfont \begin{minipage}[t]{10cm} \begin{flushleft} A1: PabuLib \newline  \vspace{2mm} \end{flushleft} \end{minipage} \begin{minipage}[t]{10cm} \begin{flushleft}  Samples note: Experiments conducted on full data from particular cities or districts. \newline Sizes note: Sizes dependent on elections in Pabulib \newline \vspace{2mm} \end{flushleft} \end{minipage} \begin{minipage}[t]{10cm} \begin{flushleft}\rule{10cm}{0.3pt} \\ Comment: There is an arXiv version: https://arxiv.org/abs/2305.11035 \end{flushleft} \end{minipage}  \\ 
\midrule
163 & \begin{minipage}[t]{3cm}\begin{flushleft}\fontsize{9}{9.5}\selectfont An Experimental Comparison of Multiwinner Voting Rules on Approval
Elections \vspace{2mm} \newline IJCAI-2023  \vspace{2mm}\newline  \citeapp{fal-lac-sor-szu:c:comparison-multiwinner-approval} \end{flushleft}\end{minipage} & \begin{minipage}[t]{3cm}\begin{flushleft}\fontsize{9}{9.5}\selectfont P.~Faliszewski, M.~Lackner, K.~Sornat, S.~Szufa\end{flushleft}\end{minipage} & \fontsize{9}{9.5}\selectfont \begin{minipage}[t]{10cm} \begin{flushleft} A1: Disjoint, Euclidean 1D, Euclidean 2D, PabuLib, Party-list, Resampling \newline 1000 samples/datapoint, election sizes: \{100\}x\{100\} \vspace{2mm} \end{flushleft} \end{minipage} \begin{minipage}[t]{10cm} \begin{flushleft}  Euclidean: 1D: $r=0.05$; 2D: $r.0.2$ \vspace{2mm} \end{flushleft} \end{minipage} \begin{minipage}[t]{10cm} \begin{flushleft}\rule{10cm}{0.3pt} \\ Comment: There is an arXiv version: https://arxiv.org/abs/2401.11870 \end{flushleft} \end{minipage}  \\ 
\midrule

\bottomrule
\end{longtable}

\section{Skipped Papers}\label{sec:missing}
Occasionally, links associated to the papers in the DBLP website were either
missing or corrupted. It was often easy to download the papers manually after
finding them (by title) on the official webpages of the respective venues.
However, for \tocheckbox{$34$}~such troublesome papers we could not find trustworthy sources
related to the corresponding proceedings publisher to download them from.

We list these papers in the subsequent table, where we include titles, authors,
venues, tracks, and reasons for skipping the papers. The list contains:
\begin{itemize}
	\item $14$~papers from the Student Abstracts track of AAAI-2010;
	\item $15$~papers from the Doctoral Consortium of AAAI-2011;
	\item $2$~papers from the Special Track on AI and the Web of AAAI-2011;
	\item $2$~papers from the Special Track on Computational Sustainability and AI
		of AAAI-2011;
	\item $1$~paper from the Machine Learning Applications track of IJCAI-2019.
\end{itemize}

Out of the above papers, for \tocheckbox{$32$} the DBLP webpage contained links to the
Wayback Machine---a crawler that archives webpages---as the original links were
expired. However, we were unable to access the respective PDF files from the
Wayback links; instead, we only could read the abstracts. The remaining
$2$~papers had no links at all to the respective PDF files on the corresponding
official proceedings webpage.

\onecolumn

{\small
\setlength{\tabcolsep}{4pt}
\begin{longtable}{p{0.5cm}|p{3cm}|p{3cm}|p{10cm}}
    ID & Authors & Title & Note \\
\toprule
\endfirsthead

\toprule

\endhead
\endlastfoot

    \multicolumn{4}{c}{\normalsize\bf2010}\\ 
    \midrule
    1 & \begin{minipage}{3cm}\begin{flushleft} Nonparametric Bayesian Approaches for Reinforcement Learning in Partially Observable Domains \vspace{2mm} \newline AAAI-2010 \end{flushleft}\end{minipage} & \begin{minipage}{3cm}\begin{flushleft} F.~Doshi-Velez \end{flushleft}\end{minipage} & \begin{minipage}{10cm}\begin{flushleft} Track: Student Abstracts \vspace{2mm} \newline Available only via the Wayback Machine (\url{https://web.archive.org}) \end{flushleft}\end{minipage} \\
    \midrule
    2 & \begin{minipage}{3cm}\begin{flushleft} Local Optimization for Simulation of Natural Motion \vspace{2mm} \newline AAAI-2010 \end{flushleft}\end{minipage} & \begin{minipage}{3cm}\begin{flushleft} T.~Erez \end{flushleft}\end{minipage} & \begin{minipage}{10cm}\begin{flushleft} Track: Student Abstracts \vspace{2mm} \newline Available only via the Wayback Machine (\url{https://web.archive.org}) \end{flushleft}\end{minipage} \\
    \midrule
    3 & \begin{minipage}{3cm}\begin{flushleft} On Multi-Robot Area Coverage \vspace{2mm} \newline AAAI-2010 \end{flushleft}\end{minipage} & \begin{minipage}{3cm}\begin{flushleft} P.~Fazli \end{flushleft}\end{minipage} & \begin{minipage}{10cm}\begin{flushleft} Track: Student Abstracts \vspace{2mm} \newline Available only via the Wayback Machine (\url{https://web.archive.org}) \end{flushleft}\end{minipage} \\
    \midrule
    4 & \begin{minipage}{3cm}\begin{flushleft} Detecting Social Ties and Copying Events from Affiliation Data \vspace{2mm} \newline AAAI-2010 \end{flushleft}\end{minipage} & \begin{minipage}{3cm}\begin{flushleft} L.~Friedland \end{flushleft}\end{minipage} & \begin{minipage}{10cm}\begin{flushleft} Track: Student Abstracts \vspace{2mm} \newline Available only via the Wayback Machine (\url{https://web.archive.org}) \end{flushleft}\end{minipage} \\
    \midrule
    5 & \begin{minipage}{3cm}\begin{flushleft} Continual On-Line Planning \vspace{2mm} \newline AAAI-2010 \end{flushleft}\end{minipage} & \begin{minipage}{3cm}\begin{flushleft} S.~Lemons \end{flushleft}\end{minipage} & \begin{minipage}{10cm}\begin{flushleft} Track: Student Abstracts \vspace{2mm} \newline Available only via the Wayback Machine (\url{https://web.archive.org}) \end{flushleft}\end{minipage} \\
    \midrule
    6 & \begin{minipage}{3cm}\begin{flushleft} Enhancing Affective Communication in Embodied Conversational Agents \vspace{2mm} \newline AAAI-2010 \end{flushleft}\end{minipage} & \begin{minipage}{3cm}\begin{flushleft} M.~D.~Leonhardt \end{flushleft}\end{minipage} & \begin{minipage}{10cm}\begin{flushleft} Track: Student Abstracts \vspace{2mm} \newline Available only via the Wayback Machine (\url{https://web.archive.org}) \end{flushleft}\end{minipage} \\
    \midrule
    7 & \begin{minipage}{3cm}\begin{flushleft} Hierarchical Skill Learning for High-Level Planning \vspace{2mm} \newline AAAI-2010 \end{flushleft}\end{minipage} & \begin{minipage}{3cm}\begin{flushleft} J.~MacGlashan \end{flushleft}\end{minipage} & \begin{minipage}{10cm}\begin{flushleft} Track: Student Abstracts \vspace{2mm} \newline Available only via the Wayback Machine (\url{https://web.archive.org}) \end{flushleft}\end{minipage} \\
    \midrule
    8 & \begin{minipage}{3cm}\begin{flushleft} Towards a Robust Deep Language Understanding System \vspace{2mm} \newline AAAI-2010 \end{flushleft}\end{minipage} & \begin{minipage}{3cm}\begin{flushleft} M.~H.~Manshadi \end{flushleft}\end{minipage} & \begin{minipage}{10cm}\begin{flushleft} Track: Student Abstracts \vspace{2mm} \newline Available only via the Wayback Machine (\url{https://web.archive.org}) \end{flushleft}\end{minipage} \\
    \midrule
    9 & \begin{minipage}{3cm}\begin{flushleft} Framework and Schema for Semantic Web Knowledge Bases \vspace{2mm} \newline AAAI-2010 \end{flushleft}\end{minipage} & \begin{minipage}{3cm}\begin{flushleft} J.~P.~McGlothlin \end{flushleft}\end{minipage} & \begin{minipage}{10cm}\begin{flushleft} Track: Student Abstracts \vspace{2mm} \newline Available only via the Wayback Machine (\url{https://web.archive.org}) \end{flushleft}\end{minipage} \\
    \midrule
    10 & \begin{minipage}{3cm}\begin{flushleft} Multi-Agent Fault Tolerance Inspired by a Computational Analysis of Cancer \vspace{2mm} \newline AAAI-2010 \end{flushleft}\end{minipage} & \begin{minipage}{3cm}\begin{flushleft} M.~M.~Olsen \end{flushleft}\end{minipage} & \begin{minipage}{10cm}\begin{flushleft} Track: Student Abstracts \vspace{2mm} \newline Available only via the Wayback Machine (\url{https://web.archive.org}) \end{flushleft}\end{minipage} \\
    \midrule
    11 & \begin{minipage}{3cm}\begin{flushleft} Integrating Expert Knowledge and Experience \vspace{2mm} \newline AAAI-2010 \end{flushleft}\end{minipage} & \begin{minipage}{3cm}\begin{flushleft} B.~G.~Weber \end{flushleft}\end{minipage} & \begin{minipage}{10cm}\begin{flushleft} Track: Student Abstracts \vspace{2mm} \newline Available only via the Wayback Machine (\url{https://web.archive.org}) \end{flushleft}\end{minipage} \\
    \midrule
    12 & \begin{minipage}{3cm}\begin{flushleft} Integrating Reinforcement Learning into a Programming Language \vspace{2mm} \newline AAAI-2010 \end{flushleft}\end{minipage} & \begin{minipage}{3cm}\begin{flushleft} C.~L.~Simpkins \end{flushleft}\end{minipage} & \begin{minipage}{10cm}\begin{flushleft} Track: Student Abstracts \vspace{2mm} \newline Available only via the Wayback Machine (\url{https://web.archive.org}) \end{flushleft}\end{minipage} \\
    \midrule
    13 & \begin{minipage}{3cm}\begin{flushleft} Computational Social Choice: Strategic and Combinatorial Aspects \vspace{2mm} \newline AAAI-2010 \end{flushleft}\end{minipage} & \begin{minipage}{3cm}\begin{flushleft} L.~Xia \end{flushleft}\end{minipage} & \begin{minipage}{10cm}\begin{flushleft} Track: Student Abstracts \vspace{2mm} \newline Available only via the Wayback Machine (\url{https://web.archive.org}) \end{flushleft}\end{minipage} \\
    \midrule
    14 & \begin{minipage}{3cm}\begin{flushleft} Interactive Task-Plan Learning \vspace{2mm} \newline AAAI-2010 \end{flushleft}\end{minipage} & \begin{minipage}{3cm}\begin{flushleft} S.~Dong \end{flushleft}\end{minipage} & \begin{minipage}{10cm}\begin{flushleft} Track: Student Abstracts \vspace{2mm} \newline Available only via the Wayback Machine (\url{https://web.archive.org}) \end{flushleft}\end{minipage} \\
    \midrule
    
    \multicolumn{4}{c}{\normalsize\bf2011}\\ \midrule
    15 & \begin{minipage}{3cm}\begin{flushleft} The AC(C) Language: Integrating Answer Set Programming and Constraint Logic Programming \vspace{2mm} \newline AAAI-2011 \end{flushleft}\end{minipage} & \begin{minipage}{3cm}\begin{flushleft} F.~S.~Bao \end{flushleft}\end{minipage} & \begin{minipage}{10cm}\begin{flushleft} Track: Doctoral Consortium \vspace{2mm} \newline Available only via the Wayback Machine (\url{https://web.archive.org}) \end{flushleft}\end{minipage} \\
    \midrule
    16 & \begin{minipage}{3cm}\begin{flushleft} Joint Inference for Extracting Text Descriptors from Triage Images of Mass Disaster Victims \vspace{2mm} \newline AAAI-2011 \end{flushleft}\end{minipage} & \begin{minipage}{3cm}\begin{flushleft} N.~Chhaya \end{flushleft}\end{minipage} & \begin{minipage}{10cm}\begin{flushleft} Track: Doctoral Consortium \vspace{2mm} \newline Available only via the Wayback Machine (\url{https://web.archive.org}) \end{flushleft}\end{minipage} \\
    \midrule
    17 & \begin{minipage}{3cm}\begin{flushleft} Model Update for Automated Planning \vspace{2mm} \newline AAAI-2011 \end{flushleft}\end{minipage} & \begin{minipage}{3cm}\begin{flushleft} M.~V.~de~Menezes, L.~N.~de~Barros \end{flushleft}\end{minipage} & \begin{minipage}{10cm}\begin{flushleft} Track: Doctoral Consortium \vspace{2mm} \newline Available only via the Wayback Machine (\url{https://web.archive.org}) \end{flushleft}\end{minipage} \\
    \midrule
    18 & \begin{minipage}{3cm}\begin{flushleft} Long-Term Declarative Memory for Generally Intelligent Agents \vspace{2mm} \newline AAAI-2011 \end{flushleft}\end{minipage} & \begin{minipage}{3cm}\begin{flushleft} N.~Derbinsky \end{flushleft}\end{minipage} & \begin{minipage}{10cm}\begin{flushleft} Track: Doctoral Consortium \vspace{2mm} \newline Available only via the Wayback Machine (\url{https://web.archive.org}) \end{flushleft}\end{minipage} \\
    \midrule
    19 & \begin{minipage}{3cm}\begin{flushleft} Ensemble Classification for Relational Domains \vspace{2mm} \newline AAAI-2011 \end{flushleft}\end{minipage} & \begin{minipage}{3cm}\begin{flushleft} H.~Eldardiry \end{flushleft}\end{minipage} & \begin{minipage}{10cm}\begin{flushleft} Track: Doctoral Consortium \vspace{2mm} \newline Available only via the Wayback Machine (\url{https://web.archive.org}) \end{flushleft}\end{minipage} \\
    \midrule
    20 & \begin{minipage}{3cm}\begin{flushleft} Developing a Language for Spoken Programming \vspace{2mm} \newline AAAI-2011 \end{flushleft}\end{minipage} & \begin{minipage}{3cm}\begin{flushleft} B.~M.~Gordon \end{flushleft}\end{minipage} & \begin{minipage}{10cm}\begin{flushleft} Track: Doctoral Consortium \vspace{2mm} \newline Available only via the Wayback Machine (\url{https://web.archive.org}) \end{flushleft}\end{minipage} \\
    \midrule
    21 & \begin{minipage}{3cm}\begin{flushleft} A Probabilistic Trust and Reputation Model for Supply Chain Management \vspace{2mm} \newline AAAI-2011 \end{flushleft}\end{minipage} & \begin{minipage}{3cm}\begin{flushleft} Y.~Haghpanah \end{flushleft}\end{minipage} & \begin{minipage}{10cm}\begin{flushleft} Track: Doctoral Consortium \vspace{2mm} \newline Available only via the Wayback Machine (\url{https://web.archive.org}) \end{flushleft}\end{minipage} \\
    \midrule
    22 & \begin{minipage}{3cm}\begin{flushleft} Designing Water Efficient Residential Landscapes with Agent-Based Modeling \vspace{2mm} \newline AAAI-2011 \end{flushleft}\end{minipage} & \begin{minipage}{3cm}\begin{flushleft} R.~Hoenigman \end{flushleft}\end{minipage} & \begin{minipage}{10cm}\begin{flushleft} Track: Doctoral Consortium \vspace{2mm} \newline Available only via the Wayback Machine (\url{https://web.archive.org}) \end{flushleft}\end{minipage} \\
    \midrule
    23 & \begin{minipage}{3cm}\begin{flushleft} Predicting Text Quality for Scientific Articles \vspace{2mm} \newline AAAI-2011 \end{flushleft}\end{minipage} & \begin{minipage}{3cm}\begin{flushleft} A.~P.~Louis \end{flushleft}\end{minipage} & \begin{minipage}{10cm}\begin{flushleft} Track: Doctoral Consortium \vspace{2mm} \newline Available only via the Wayback Machine (\url{https://web.archive.org}) \end{flushleft}\end{minipage} \\
    \midrule
    24 & \begin{minipage}{3cm}\begin{flushleft} Scaling Up Game Theory: Achievable Set Methods for Efficiently Solving Stochastic Games of Complete and Incomplete Information \vspace{2mm} \newline AAAI-2011 \end{flushleft}\end{minipage} & \begin{minipage}{3cm}\begin{flushleft} L.~MacDermed \end{flushleft}\end{minipage} & \begin{minipage}{10cm}\begin{flushleft} Track: Doctoral Consortium \vspace{2mm} \newline Available only via the Wayback Machine (\url{https://web.archive.org}) \end{flushleft}\end{minipage} \\
    \midrule
    25 & \begin{minipage}{3cm}\begin{flushleft} Pruning Techniques in Search and Planning \vspace{2mm} \newline AAAI-2011 \end{flushleft}\end{minipage} & \begin{minipage}{3cm}\begin{flushleft} N.~Pochter \end{flushleft}\end{minipage} & \begin{minipage}{10cm}\begin{flushleft} Track: Doctoral Consortium \vspace{2mm} \newline Available only via the Wayback Machine (\url{https://web.archive.org}) \end{flushleft}\end{minipage} \\
    \midrule
    26 & \begin{minipage}{3cm}\begin{flushleft} Learning with Imprecise Classes, Rare Instances, and Complex Relationships \vspace{2mm} \newline AAAI-2011 \end{flushleft}\end{minipage} & \begin{minipage}{3cm}\begin{flushleft} S.~Ravindran \end{flushleft}\end{minipage} & \begin{minipage}{10cm}\begin{flushleft} Track: Doctoral Consortium \vspace{2mm} \newline Available only via the Wayback Machine (\url{https://web.archive.org}) \end{flushleft}\end{minipage} \\
    \midrule
    27 & \begin{minipage}{3cm}\begin{flushleft} Modeling the Effects of Emotion on Cognition \vspace{2mm} \newline AAAI-2011 \end{flushleft}\end{minipage} & \begin{minipage}{3cm}\begin{flushleft} M.~Spraragen \end{flushleft}\end{minipage} & \begin{minipage}{10cm}\begin{flushleft} Track: Doctoral Consortium \vspace{2mm} \newline Available only via the Wayback Machine (\url{https://web.archive.org}) \end{flushleft}\end{minipage} \\
     \midrule
    28 & \begin{minipage}{3cm}\begin{flushleft} Learning Sensor, Space and Object Geometry \vspace{2mm} \newline AAAI-2011 \end{flushleft}\end{minipage} & \begin{minipage}{3cm}\begin{flushleft} J.~Stober \end{flushleft}\end{minipage} & \begin{minipage}{10cm}\begin{flushleft} Track: Doctoral Consortium \vspace{2mm} \newline Available only via the Wayback Machine (\url{https://web.archive.org}) \end{flushleft}\end{minipage} \\
    \midrule
    29 & \begin{minipage}{3cm}\begin{flushleft} Incentive-Compatible Trust Mechanisms \vspace{2mm} \newline AAAI-2011 \end{flushleft}\end{minipage} & \begin{minipage}{3cm}\begin{flushleft} J.~Witkowski \end{flushleft}\end{minipage} & \begin{minipage}{10cm}\begin{flushleft} Track: Doctoral Consortium \vspace{2mm} \newline Available only via the Wayback Machine (\url{https://web.archive.org}) \end{flushleft}\end{minipage} \\
    \midrule
    30 & \begin{minipage}{3cm}\begin{flushleft} User-Controllable Learning of Location Privacy Policies With Gaussian Mixture Models \vspace{2mm} \newline AAAI-2011 \end{flushleft}\end{minipage} & \begin{minipage}{3cm}\begin{flushleft} J.~Cranshaw, J.~Mugan, N.~M.~Sadeh \end{flushleft}\end{minipage} & \begin{minipage}{10cm}\begin{flushleft} Track: Special Track on AI and the Web \vspace{2mm} \newline Available only via the Wayback Machine (\url{https://web.archive.org}) \end{flushleft}\end{minipage} \\
    \midrule
    31 & \begin{minipage}{3cm}\begin{flushleft} Dynamic Resource Allocation in Conservation Planning \vspace{2mm} \newline AAAI-2011 \end{flushleft}\end{minipage} & \begin{minipage}{3cm}\begin{flushleft} D.~Golovin, A.~Krause, B.~Gardner, S.~J.~Converse, S.~Morey   \end{flushleft}\end{minipage} & \begin{minipage}{10cm}\begin{flushleft} Track: Special Track on Computational Sustainability and AI \vspace{2mm} \newline Available only via the Wayback Machine (\url{https://web.archive.org}) \end{flushleft}\end{minipage} \\
    \midrule
    32 & \begin{minipage}{3cm}\begin{flushleft} A Whole Page Click Model to Better Interpret Search Engine Click Data  \vspace{2mm} \newline AAAI-2011 \end{flushleft}\end{minipage} & \begin{minipage}{3cm}\begin{flushleft} W.~Chen, Z.~Ji, S.~Shen, Q.~Yang \end{flushleft}\end{minipage} & \begin{minipage}{10cm}\begin{flushleft} Track: Special Track on AI and the Web \vspace{2mm} \newline No link to the pdf on the official proceedings webpage (\url{https://ojs.aaai.org/index.php/AAAI/issue/view/308}) \end{flushleft}\end{minipage}\\
    \midrule
    33 & \begin{minipage}{3cm}\begin{flushleft} Policy Gradient Planning for Environmental Decision Making with Existing Simulators \vspace{2mm} \newline AAAI-2011 \end{flushleft}\end{minipage} & \begin{minipage}{3cm}\begin{flushleft} M.~Crowley, D.~Poole \end{flushleft}\end{minipage} & \begin{minipage}{10cm}\begin{flushleft} Track: Special Track on Computational Sustainability and AI \vspace{2mm} \newline No link to the pdf on the official proceedings webpage (\url{https://ojs.aaai.org/index.php/AAAI/issue/view/308}) \end{flushleft}\end{minipage} \\
    \midrule
    \multicolumn{4}{c}{\normalsize\bf2019}\\ \midrule
    34 & \begin{minipage}{3cm}\begin{flushleft} Hierarchical Graph Convolutional Networks for Semi-Supervised Node Classification \vspace{2mm} \newline IJCAI-2019 \end{flushleft}\end{minipage} & \begin{minipage}{3cm}\begin{flushleft} F.~Hu, Y.~Zhu, S.~Wu, L.~Wang, T.~Tan \end{flushleft}\end{minipage} & \begin{minipage}{10cm}\begin{flushleft} Track: Machine Learning Applications \vspace{2mm} \newline Not present on the official proceedings webpage (\url{https://www.ijcai.org/proceedings/2019/}) \end{flushleft}\end{minipage} \\
\midrule

\bottomrule
\end{longtable}
}

\end{bibunit}

\end{document}